\documentclass[prb,
showpacs,
twocolumn,
floats,
10pt,
aps,
citeautoscript,
longbibliography,
superscriptaddress]{revtex4-2}

\usepackage{placeins}
\usepackage{lipsum}
\usepackage{xcolor}
\usepackage[normalem]{ulem}
\usepackage{comment}
\usepackage{graphicx}
\usepackage{wrapfig}
\usepackage{dcolumn}
\usepackage{bm}

\usepackage{blindtext}
\usepackage{graphics}
\usepackage{verbatim}   
\usepackage[ruled,lined]{algorithm2e}
\usepackage{amsfonts}
\usepackage{amsmath}
\usepackage{amssymb}
\usepackage{mathrsfs} 
\usepackage{multirow}
\usepackage{physics}
\usepackage{bbm}  
\usepackage{adjustbox}
\usepackage{appendix}

\usepackage{tabstackengine}
\setstackEOL{\cr}

\newcommand{\Eq}[1]{Eq.~\eqref{#1}}
\newcommand{\Eqs}[1]{Eqs.~\eqref{#1}}

\newcommand{\Fig}[1]{Fig.~\ref{#1}}
\newcommand{\Figs}[1]{Figs.~\ref{#1}}

\newcommand{\Sec}[1]{Sec.~\ref{#1}}
\newcommand{\App}[1]{App.~\ref{#1}}
\newcommand{\Alg}[1]{Alg.~\ref{#1}}

\newcommand{\mr}[1]{\mathrm{#1}}
\newcommand{\mc}[1]{\mathcal{#1}}
\newcommand{\sbra}[1]{(#1|}
\newcommand{\sket}[1]{|#1)}
\renewcommand{\vec}[1]{\mathbf{#1}}

\usepackage{enumitem}

\def\doubleI{{\mathbbm{1}}}

\def\doubleC{{\mathbbm{C}}}

\def\doubleL{{\mathbbm{L}}}
\def\doubleK{{\mathbbm{K}}}
\def\doublez{{\mathbbm{z}}}
\def\doubleW{{\mathbb{W}}}

\def\bzero{{\vec{0}}}
\def\bz{{\vec{z}}}
\def\bze{\bz} 

\def\bk{{\vec{k}}}

\def\bK{{\vec{K}}}
\def\br{{\vec{r}}}
\def\bb{{\vec{b}}}
\def\bc{{\vec{c}}}
\def\bd{{\vec{d}}}
\def\bt{{\vec{t}}}

\def\bp{{\vec{p}}}
\def\bq{{\vec{q}}}
\def\bQ{{\vec{Q}}}
\def\bx{{\vec{x}}}

\def\tbz{{\widetilde {\vec{z}}}}
\def\modified{{\mathrm{mod}}}
\def\bG{{\vec{G}}}
\def\bC{\vec{C}}
\def\bR{{\vec{R}}}
\def\bK{{\vec{K}}}
\def\bA{{\vec{A}}}
\def\bF{{\vec{F}}}
\def\bI{{\vec{I}}}
\def\bW{{\vec{W}}}
\def\bM{{\vec{M}}}
\def\bU{{\vec{U}}}
\def\bD{{\vec{D}}}
\def\bX{{\vec{X}}}
\def\tbM{{\widetilde {\vec{M}}}}
\def\tbU{{\widetilde {\vec{U}}}}
\def\tbD{{\widetilde {\vec{D}}}}
\def\tbQ{{\widetilde {\vec{Q}}}}
\def\bS{{\vec{S}}}
\def\bolde{{\boldsymbol{\epsilon}}}
\def\boldepd{{\boldsymbol{\epsilon}}} 

\def\boldone{{\boldsymbol{1}}}
\def\bSigma{{\boldsymbol{\Sigma}}}
\def\tbSigma{\widetilde {\boldsymbol{\Sigma}}}
\def\bDelta{{\boldsymbol{\Delta}}}
\def\bepsilon{{\boldsymbol{\epsilon}}}
\def\bdelta{{\boldsymbol{\delta}}}

\def\ground{\mathrm{g}}

\def\cL{\mc{L}}
\def\cP{\mc{P}}

\def\pdag{{\phantom{\dagger}}}
\def\ohat{} 

\def\discrete{\mathrm{disc}} 
\def\imp{\mathrm{imp}} 
\def\dimp{\mathrm{d}} 
\def\bbath{\mathrm{b}} 
\def\BZ{{\mathrm{BZ}}}
\def\patch{{\mathrm{p}}}
\def\Nlattice{N_\mathrm{latt}} 
\def\Ncluster{N_\mathrm{c}} 
\def\Npatchall{{N_\patch}} 
\def\Npatch{{N_\patch}} 
\def\Gpatch{{G_\patch}} 
\def\comma{,}
\def\fast{\mr{F}}
\def\slow{\mr{S}}
\def\dyn{\mr{dyn}}
\def\NKrylov{N}
\def\Nimpurity{N_\dimp} 
\def\Nbath{N_\bbath} 
\def\Hartree{\mr{HF}}
\def\Rest{\bR} 
\def\sp{\mr{c}} 
\def\spi{{\overline{\mr{c}}}} 
\def\widebpone{\widetilde{\bp}^{\scriptscriptstyle (1)}}
\def\bponeperp{\bp^{{\scriptscriptstyle (1)}\perp}}
\def\bAone{\bA^{\scriptscriptstyle (1)}}
\def\bGone{\bG^{\scriptscriptstyle (1)}}
\def\bmuone{\boldsymbol{\mu}^{\scriptscriptstyle (1)}}
\def\bpone{\bp^{\scriptscriptstyle (1)}}
\def\bponedag{\bp^{{\scriptscriptstyle (1)} \dagger}}
\def\pone{p^{\scriptscriptstyle (1)}}
\def\bponeperpdag{\bp^{{\scriptscriptstyle (1)} \perp \dagger}}
\def\bph{\bp}
\def\btd{\bt} 
\def\btpd{\bt} 
\def\fastslow{\mathrm{split}}

\def\tSigma{\widetilde \Sigma}
\def\tM{\widetilde M}
\def\tG{\widetilde G}
\def\tbG{\widetilde{\mathbf{G}}}
\def\tQ{\widetilde Q}
\def\Qbar{\overline {\bQ}}
\def\tcL{\widetilde \cL}
\def\tildet{\widetilde t}
\def\tepsilon{\widetilde \epsilon}

\newcommand{\LMUMunich}{Arnold Sommerfeld Center for Theoretical Physics, Center for NanoScience, and Munich Center for Quantum Science and Technology, Ludwig-Maximilians-Universit\"at M\"unchen, 80333 Munich, Germany}

\newcommand{\RutgersUniversity}{Department of Physics and Astronomy and Center for Condensed Matter Theory, Rutgers University, Piscataway, New Jersey 08854-8019, USA}

\definecolor{darkgreen}{rgb}{0,0.5,0}
\definecolor{darkblue}{rgb}{0,0,0.5}

 \newcommand{\rreply}[1]{{\color{black}{#1}}}
 \newcommand{\changed}[1]{{\color{black}{#1}}}

\usepackage{scalefnt}

\newenvironment{medblockmatrix}{\scalefont{0.8}\left(\begin{array}{c|c}}{\end{array}\right)}

\RequirePackage[
  hyperindex,colorlinks,bookmarksnumbered,
  plainpages=true,pdfstartview=FitH]{hyperref}
\hypersetup{linkcolor=blue,urlcolor=blue,citecolor=blue} 
\usepackage[all]{hypcap}

\begin{document}

\preprint{}

\title{Liouvillian interpolation of the self-energy of cluster dynamical mean-field theories 
}

\author{Mathias Pelz}
\email{mathias.pelz@lmu.de}
\affiliation{\LMUMunich}

\author{Jan von Delft}
\email{vondelft@lmu.de}
\affiliation{\LMUMunich}

\author{Andreas Gleis}
\email{andreas.gleis@rutgers.edu}
\affiliation{\RutgersUniversity}

\date{October 8, 2026} 

\begin{abstract}

Two widely-used non-local extensions of dynamical mean-field theory (DMFT), namely cellular DMFT (CDMFT) and the dynamical cluster approximation (DCA), both yield self-energies marred by having some unphysical properties: 
CDMFT yields real-space self-energies that are not translationally invariant, and DCA yields momentum-space self-energies with discontinuities in their momentum dependence. It is often desirable to remove these flaws by post-processing cluster DMFT results, using strategies called periodization for CDMFT and interpolation for DCA---for brevity, we refer to both cases as \textit{interpolation}. However, traditional interpolation approaches struggle to \changed{reliably} capture intricate structures such as hole pockets in the hole-doped square-lattice Hubbard model. Further, these approaches interpolate  frequency-dependent functions, which may lead to causality violations. Here, we propose \textit{Liouvillian interpolation}, a novel, intuitive, and robust scheme for interpolating   cluster DMFT results. Our key idea is to  interpolate  frequency-\textit{independent} matrix elements of the single-particle irreducible part of the Liouvillian, obtained from a continued-fraction expansion of the \changed{cluster} DMFT self-energy. We demonstrate that the ingredients of such an expansion possess a more local Fourier expansion than the functions involved in traditional interpolation schemes, and that Liouvillian interpolation conserves causality by construction. We illustrate our method for the one-dimensional Hubbard model using CDMFT, and for the two-dimensional Hubbard model using \rreply{four-site CDMFT} and four-patch DCA. For the latter, $\cL$-interpolation can (depending on doping) yield Fermi and Luttinger arcs which together \rreply{seem to} form a closed surface.

\medskip
\noindent
DOI:

\end{abstract}

\maketitle

\section{Introduction}\label{sec:intro}

Cluster dynamical mean-field theories~(cluster-DMFT)~\cite{Lichtenstein2000,Kotliar2001_CDMFT,Hettler1998_DCA1,Hettler2000_DCA2,Maier2002_DCAinterp,Maier2005,Civelli2005,Stanescu2006_CDMFTpseudogap,Stanescu2006_periodization2,Sakai2011_cumulant_periodization,Staar2013_DCAinterp,Haehner2020_DCAinterp,Senechal2002_periodization1,Biroli2002_periodization3,Biroli2004_periodization4,Verret2019,Potthoff2016_cDMFT,Klett2020_periodization5,Meixner2024} are commonly used to describe strong but short-ranged non-local correlations beyond single-site dynamical mean-field theory~(DMFT)~\cite{Georges1996}.
They effectively capture strongly correlated phenomena that may be driven by short-ranged correlations, in systems such as \changed{Mott insulators \cite{Capone2004,Sen2020,Bellomia2024},} certain types of heavy-fermion quantum criticality~\cite{DeLeo2008,DeLeo2008a,Gleis2024}, cuprate superconductors \cite{Maier2002_DCAinterp,Stanescu2006_CDMFTpseudogap,Stanescu2006_periodization2,Civelli2005,Capone2006,Kancharla2008,Sakai2011_cumulant_periodization,Sakai2013_pseudogap,Sakai2009,Sakai2010,Wu2018_pseudogap,Wu2020a,Foley2019_ED,Wu2017,Krien2022_pseudogap,Walsh2023_pseudogap,Kundu2024,Bacq-Labreuil2025,Staffieri2025_DCA,Smit2025}, 
strange metals~\cite{Wu2022,Fournier2024,Gleis2024a,Hardy2024}, Hund metals~\cite{Nomura2015_hund,Semon2017_hund,Crippa2021,FlorezAblan2025} and other multi-orbital systems~\cite{Nomura2014_mulitorbital,Harland2019_mulitorbital,Karakuzu2021,FlorezAblan2025}, \changed{and quantum spin Hall insulators \cite{Crippa2021}}.
  
\begin{figure*}
    \centering
    \includegraphics[width=0.85\linewidth]{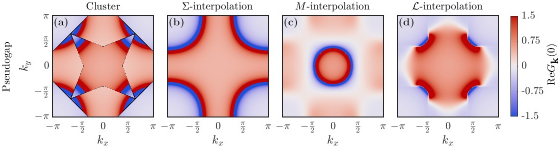}
    \caption{Real part of the Green's function at $\omega=0$, $\mr{Re}G_{\bk}(0)$, for a four-patch DCA computation of the 2D-Hubbard model in the pseudogap regime (for model details, see Sec.~\ref{sec:2dHubbard_DCA_model}). (a) DCA result, showing discontinuities in the Brillouin zone, and (b,c,d) its interpolations based on the self-energy $\Sigma$, cumulant $M$ and Liouvillian $\cL$, respectively. $\cL$-interpolation yields Fermi arcs (blue-to-red), Luttinger arcs (white), and a seemingly smooth crossover between the two. (Excerpted from Fig.~\ref{fig:DCA_logA0}(a3-d3).)
    }
    \label{fig:intro}
\end{figure*}

The most commonly used cluster-DMFT methods are cellular DMFT~(CDMFT)~\cite{Kotliar2001_CDMFT}, which couples a real-space cluster of sites to self-consistent baths, thereby breaking translation symmetry; and the dynamical cluster approximation~(DCA)~\cite{Hettler1998_DCA1,Hettler2000_DCA2}, which coarse-grains the Brillouin zone into patches and couples these to self-consistent baths, thereby introducing artificial discontinuities in momentum space. The lack of periodicity for CDMFT results (implying real-space discontinuities) and the lack of a smooth momentum dependence for DCA results are unphysical. Removing these flaws requires the cluster-DMFT results to be post-processed. CDMFT results have to be periodized  to restore real-space translational invariance; loosely speaking, this amounts to interpolating across cluster boundaries to remove the real-space discontinuities associated with non-periodic behavior. DCA results have to be interpolated  across DCA patch boundaries to restore a smooth momentum dependence ~\cite{Maier2005,Senechal2002_periodization1,Biroli2002_periodization3,Biroli2004_periodization4,Stanescu2006_periodization2,Sakai2011_cumulant_periodization,Potthoff2016_cDMFT}. For brevity, we will use ``interpolation'' as an umbrella term for both CDMFT periodization and DCA interpolation. It is to be understood that in the context of CDMFT, ``interpolation'' refers to ``periodization''. \changed{Likewise, we will use ``cluster'' as an umbrella term for the effective impurity of both CDMFT and DCA.}

\changed{Commonly-used interpolation schemes are $G$-, $\Sigma$- and $M$-interpolation, which interpolate the Green's function $G(\omega)$, self-energy $\Sigma(\omega)$, or cumulant $M(\omega)$, respectively. However, for reasons reviewed in Sec.~\ref{sec:Background},
they all have some problematic properties, sometimes including violation of causality. 

In this paper, we propose the Liouvillian, $\cL$, as a new and previously unexplored quantity for interpolation. \rreply{The Liouvillian is the 
generator of time translations in the space of all linear operators that act in Fock space (see \Eq{eq:OperatorTimeTranslation} below).}
 Its matrix elements can be obtained straightforwardly by generating a continued-fraction expansion (CFE) of the retarded self-energy. The $\cL$ matrix elements 
(i) typically have a shorter range in real space than $G(\omega)$, $ \Sigma(\omega)$ or $M(\omega)$, and (ii) unlike the latter, do not depend on frequency. These properties make them better candidates for interpolation. The key idea of $\cL$-interpolation, therefore, is to interpolate the $\cL$ matrix elements (using procedures familiar from
CDMFT or DCA interpolation). Reinserting the interpolated $\cL$ matrix elements into the self-energy CFE then yields an  interpolated self-energy that is causal by construction (with a minor caveat, addressed on p.~\pageref{p:caveat-causality}). Additionally, we find that $\cL$-interpolation mimics the properties of the cluster self-energy used as input more satisfactorily than $G$-, $\Sigma$- or $M$-interpolation. Moreover, $\cL$-interpolation is technically not more demanding than the other schemes---its only additional technical ingredient is a CFE. Consequently, we expect $\cL$-interpolation to become the method of choice for cluster-DMFT interpolation. 

The goal of this paper is mainly methodological: to give a detailed description of the concepts and computational steps involved in $\cL$-interpolation. However, to illustrate its power we also include 
one interesting physical application (\Sec{sec:2dHubbard_DCA}): a 4-patch DCA treatment of the 2-dimensional Hubbard model. The teaser \Fig{fig:intro} (excerpted from \Fig{fig:DCA_logA0}) offers a first impression of how strikingly different $\Sigma$-, $M$- and $\cL$-interpolation perform in the pseudogap regime: clearly, $\cL$-interpolation  is the scheme that  most convincingly mimics the original discontinuous cluster results. Figure~\ref{fig:intro}(d)  furthermore illustrates the main physical result of this paper: remarkably, $\cL$-interpolation can yield Fermi and Luttinger arcs that together seem to form a closed surface
 (see also Fig.~\ref{fig:FS_LS}), with a smooth crossover between Fermi and Luttinger arcs (see \Figs{fig:DCA_ReGk} and \ref{fig:DCA_Z}). The aforementioned figures vividly 
illustrate the capability of $\cL$-interpolation to 
generate intricate structures in non-trivial contexts; we invite the reader to briefly preview them before continuing.} 

This paper is structured as follows: \changed{Section~\ref{sec:BackgroundPreview} provides some background on the problematic aspects of $G$-, $\Sigma$- and $M$-interpolations and previews key aspects of $\cL$-interpolation in relation to these. In \Sec{sec:Liouvillian_basics}, we review why a CFE of the self-energy is governed by matrix elements of the Liouvillian (corresponding derivations are presented in Apps.~\ref{sec:SingleParticleIrredSubspace} to \ref{sec:QIM}).}
Section~\ref{sec:LiouvillianPeriodization} introduces $\cL$-interpolation in great detail.  In \Sec{sec:1dHubbard_benchmark}, we use the 1-dimensional Hubbard model as a testbed for $\cL$-interpolation  of CDMFT results, 
finding that $\cL$-interpolation yields faster convergence with cluster size than $\Sigma$-, $G$- or $M$-interpolation.  
\changed{As an interesting physical application, we  compare these interpolation schemes applied to cluster-DMFT results for the 2-dimensional Hubbard model, obtained using either 4-site CDMFT  and exact diagonalization (ED) (\Sec{sec:CDMFT+ED2DHubbard}), or 4-patch DCA and the numerical renormalization group (NRG) (\Sec{sec:2dHubbard_DCA}).} Section~\ref{sec:ConclusionOutlook} offers a summary and outlook. The appendices provide extensive discussions of various technical details. 

\changed{
\section{Interpolation strategies: background and preview}
\label{sec:BackgroundPreview}

This section reviews some conceptual aspects of previous interpolation schemes and previews some key elements of $\cL$-interpolation.

\subsection{Background on previous interpolation schemes}
\label{sec:Background}
}

\rreply{We start with some brief historical comments, of course without claiming completeness.
The need to reconstruct translationally invariant lattice quantities from CDMFT was already recognized in the foundational formulation of CDMFT by Kotliar \textit{et al.} in 2001 \cite{Kotliar2001_CDMFT}; a detailed analysis of self-energy periodization was given shortly thereafter by Biroli, Kotliar, and Parcollet \cite{Biroli2002_periodization3,Biroli2004_periodization4}. An example of a recent refinement is center-focused CDMFT \cite{Klett2020_periodization5,Meixner2024}. Cumulant periodization for CDMFT was pioneered by Stanescu and Kotliar in 2006 \cite{Stanescu2006_periodization2}. 
For DCA, the need for interpolation was first addressed by Jarrell \textit{et al.} in 2001 \cite{Jarrell2001}; a later refinement, called DCA${}^+$, was given by Staar, Maier and Schulthess in 2011 \cite{Staar2013_DCAinterp}, but it exhibits serious issues in strongly-correlated regimes~\cite{Haehner2020_DCAinterp}.
However, these schemes all have some problematic properties, for reasons that we will now discuss from a quite general perspective.}

\changed{In the momentum/frequency representation, the retarded single-particle Green's function, $G$, and the self-energy, $\Sigma$, and cumulant, $M$, are related by} 
\begin{subequations}
\label{subeqs:define_G_M}  
\begin{align}
    G_\bk(\omega) & = \frac{1}{\omega +\mu -\epsilon_\bk - \Sigma_\bk(\omega)} \, , \\
    \ M_\bk(\omega) & = \frac{1}{\omega +\mu - \Sigma_\bk(\omega)} \, .
\end{align}
\end{subequations}
Instead of $G$, $\Sigma$ or $M$, we will generically write $Q$ for the quantity to be interpolated. In general, $Q_\bk$ and $\tQ_\br$, 
the momentum and position representations of $Q$, are related by a Fourier transformation. However, approximate methods such as CDMFT or DCA yield $\tQ_\br$ only for a limited range of $\br$ values, say with $|\br| \le R$, where the distance cutoff $R$ is determined by the cluster size (for CDMFT) or the number of momentum patches (for DCA). Thus, interpolation schemes amount to approximating  $Q_\bk$ by a \textit{truncated} Fourier Ansatz, 
\begin{align}
\label{eq:Q_periodize}
    Q_\bk (\omega) \simeq \sum_{\br, |\mathbf{r}| \leq R} 
    \overline Q_\br (\omega) \,  \mr{e}^{-\mr{i} \bk \cdot \br} \, .
\end{align}
Here, $\overline Q_\br$ has to be estimated from quantities computed on the cluster (for CDMFT) or the Brillouin zone patches (for DCA). In the case of CDMFT, commonly used schemes for this purpose are superlattice averaging~\cite{Stanescu2006_periodization2} or center-focused extrapolation~\cite{Klett2020_periodization5}.  Other related quantities are then derived from $Q$; for example, when $\Sigma$ is interpolated   via \Eq{eq:Q_periodize}, that also determines $G$ and $M$. 

It is common knowledge among practitioners of cluster-DMFT that the quality and reliability of the interpolation scheme is quite sensitive to the choice of $Q$. 
A particularly striking example was recently pointed out by  Verret \textit{et al.}\ \cite{Verret2022_CompactTiling}. They analyzed  the limitations of existing choices for $Q$ in the context of a long-standing question on the pseudogap phase of underdoped cuprates: 
Does the Fermi surface exhibit hole or electron pockets, as deduced from various transport experiments \cite{Das2012_TransportHighTcSC1,Yangmu2019_TransportHighTcSC2}, or disconnected Fermi arcs, as indicated by angle-resolved photoemission spectroscopy (ARPES)~\cite{Damascelli2003_ARPEShighTcSC,Reber2012_ARPES_cuprates}?
\rreply{This and related questions have been studied theoretically in numerous works, using both cluster-DMFT \cite{Civelli2005,Stanescu2006_CDMFTpseudogap,Stanescu2006_periodization2,Sakai2009,Sakai2010,Sakai2011_cumulant_periodization,Chen2012a,Sakai2013_pseudogap,Wu2018_pseudogap,Wu2020a} and other approaches, such as Quantum Monte Carlo studies working in the thermodynamic limit \cite{Wu2017,Wu2018_pseudogap,Simkovic2024}, to mention but a few.}
Verret \textit{et al.}\ \cite{Verret2022_CompactTiling} have argued that ``contrary to previous claims \cite{Stanescu2006_periodization2,Stanescu2006_CDMFTpseudogap,Sakai2009,Sakai2010}, cluster-DMFT schemes cannot discriminate between Fermi arcs and hole pockets'' (with the implication that previous cluster-DMFT work attempting to answer the aforementioned question is inconclusive).
They illustrated this for a Yang-Rice-Zhang~(YRZ)~\cite{Yang2006} toy-model self-energy with built-in hole pockets: when they approximated its self-energy in CDMFT fashion and then periodized  it, their  $M$- and $G$-periodization  schemes both yielded Fermi arcs~\cite{Verret2022_CompactTiling} instead of reproducing the hole pockets. 
Verret \textit{et al.}\ argued that this failure is due to ``neglecting the self-energy between clusters''. 
They then presented a so-called \textit{compact tiling} scheme that includes estimates of these neglected parts, see Ref.~\onlinecite{Verret2022_CompactTiling} for details. For the YRZ toy model, compact tiling successfully recovers the expected hole pockets. However, their scheme was formulated in a manner not applicable to general cluster-DMFT calculations.

The fact that the quality of interpolated  cluster-DMFT results depends strongly on the choice of $Q$ can be expected on general grounds. In \Eq{eq:Q_periodize}, the Fourier truncation at $|\br|\le R$ introduces errors. To limit these, it is preferable for $\Qbar_\br(\omega)$ to have (i) a short-ranged dependence on $\br$, and (ii) a weak (or no) dependence on $\omega$. Different choices for $Q$ satisfy these criteria to different degrees, leading to errors of differing severity. Regarding (ii), we note that if $\Qbar_\br(\omega)$ does depend significantly on $\omega$, then the truncation of the Fourier series can cause unintended (even unacceptable) modifications in the analytic structure of $Q_\bk(\omega)$. An example is $\Sigma$-interpolation: if $\tSigma_\br(\omega)$ depends strongly on $\omega$, the truncation to $|\br| \le R$ can cause $\mathrm{Im} \Sigma_\bk(\omega)$ to become positive for some values of $\omega$ and $\bk$, implying a violation of causality for $G_\bk(\omega)$ \cite{Biroli2004_periodization4,Stanescu2006_CDMFTpseudogap}.

\bigskip 

\changed{
\subsection{Preview of Liouvillian interpolation}
\label{sec:Preview}}

In this paper, we propose the Liouvillian, $\cL$, as a new and previously unexplored quantity for interpolation (i.e.\ we choose $Q = \cL$). Our $\cL$-interpolation scheme is based on a general, exact relation between the self-energy and the Liouvillian: if the self-energy is expressed through a continued-fraction expansion (CFE),  its CFE coefficients are matrix elements of the single-particle irreducible part of the Liouvillian in a Krylov space  obtainable via a Lanczos scheme. \changed{Previewing a key fact  from \Sec{sec:Liouvillian_basics}, this CFE has the following form [cf.\ \Eqs{eq:Sigma=HF+dyn} and \eqref{eq:SigmadynCFE-main}]:
\begin{align}
\label{eq:previewSigmaCFE}
\bSigma(z) = 
\bSigma^\Hartree + 
\bt_1 
\cfrac{\boldone}{\bze_1 - 
{\btpd_{2}
 \cfrac{\boldone}{\bze_2 -  
 \btpd_3 
 \cfrac{\boldone}{\bze_3 - \!\! \raisebox{-0.9em}{ \ensuremath{\ddots}}
  } \btd_3} 
 \btd_{2} }} 
\bt_1 \, .
\end{align}
Here, $\bz_n = z \boldone - \bolde_n$, $z$ is a complex frequency argument, 
$\bSigma^\Hartree$ is the Hartree-Fock contribution, and  
$\bolde_n$, $\bt_n$ are matrix elements of $\cL$  in said Krylov space
[cf.~\Eq{eq:Lmatrixelements-en-tn-main}].
}  

For a lattice model, the matrix elements $\bolde_n$, $\bt_n$ are generically translation-invariant and smooth functions of $\bk$. However, when Liouvillian matrix elements are extracted from the CFE expansion of a self-energy computed using CDMFT or DCA, they either do not satisfy translation invariance or exhibit discontinuities in $\bk$-space, respectively.
Our $\cL$-interpolation scheme restores translation invariance or smooth $\bk$-dependence by interpolating  the single-particle irreducible Liouvillian matrix elements obtained from CDMFT or DCA self-energies, respectively: using superlattice averaging~\cite{Stanescu2006_periodization2} for CDMFT or a self-consistency condition for DCA, respectively, we obtain estimates 
$\overline{\cL}_\br$ for the matrix elements of $\tcL_\br$. 
\changed{Inserting these into the truncated Fourier Ansatz \eqref{eq:Q_periodize} for $\mc{L}_\bk$, we obtain interpolated
Liouvillian matrix elements, i.e.\ interpolated CFE coefficients. 
Inserting these, in turn,} into the CFE \eqref{eq:previewSigmaCFE} of the self-energy, we obtain an interpolated   self-energy and Green's function. 

\changed{$\cL$-interpolation has two properties that are advantageous
relative to conventional interpolation schemes: (i) $\tcL_\br$ appears to be shorter-ranged than $\tSigma_\br$ or $\tM_\br$; this is suggested by a case study and supported by general arguments, see \Sec{sec:1dHubbard_MPS}.
Thus the consequences of truncating the Fourier expansion
\eqref{eq:Q_periodize} are less severe.} (ii) Since the Liouvillian is, by definition, independent of frequency, its interpolated   version remains hermitian, thus ensuring causality for the resulting self-energy \changed{(with a minor caveat, see p.~\pageref{p:caveat-causality})}.
From a more general perspective: interpolating $\mc{L}$ amounts to directly interpolating the time-evolution generator for operators. This ensures unitary time evolution and therefore causality. 

For the phenomenological YRZ self-energy Ansatz of Ref.~\onlinecite{Verret2022_CompactTiling}, $\cL$-interpolation (along with a suitable choice of $\tcL_{\mathbf{r}}$) is equivalent to compact tiling; indeed,  \changed{footnote~4 of Ref.~\onlinecite{Verret2022_CompactTiling} served as inspiration for our strategy to use Liouvillian matrix elements as the objects to be interpolated.} 

\changed{This concludes the introductory sections of this paper. The next section  describes the conceptual basis for $\cL$-interpolation by reviewing the relation between the self-energy and Liouvillian matrix elements.} 

\bigskip 

\section{Liouvillian dynamics} \label{sec:Liouvillian_basics}

The use of Liouvillian methods for obtaining CFEs of 2-point functions and their self-energies has a long history \cite{Mori1965,Haydock1980a,Engel1991,Viswanath1994,Dargel2011,Dargel2012,Tiegel2014,julien2008_EROS,Auerbach2018,Auerbach2019,Foley2024_Liouvillian}. In this section we recapitulate some central ideas and results. Our presentation, building on those of Refs.~\onlinecite{julien2008_EROS,Auerbach2018,Auerbach2019,Foley2024_Liouvillian}, is self-contained and emphasizes aspects relevant for the $\cL$-interpolation strategy in later sections. \changed{We recall standard definitions for spectral representations of Green's functions and self-energies (\Sec{sec:SpectralRepresentationGSigma}); express these through the Liouvillian (\Sec{sec:ExpressingGSigmathroughL}); and show how this leads to a self-energy CFE whose coefficients are Liouvillian matrix elements (\Sec{sec:CFE-self-energy-main}). Corresponding technical details are given in Apps.~\ref{sec:SingleParticleIrredSubspace} to \ref{sec:QIM}.}

\changed{
\subsection{Spectral representation of Green's function and self-energy}
\label{sec:SpectralRepresentationGSigma}

To start, we recall standard definitions for spectral representations of Green's functions and self-energies.}

We consider a fermionic system described by a Hamiltonian, $H = H_0 + H_\textrm{int}$, with quadratic and quartic parts, 
\begin{align}
    \label{eq:Hamiltonian}
    H_0 = \sum_{\alpha \beta} \epsilon_{\alpha \beta}c^\dagger_\alpha c^\pdag_\beta , \quad  
    H_\mr{int} = \tfrac{1}{4} \sum_{\alpha \beta \gamma \delta}
    U_{\alpha \gamma , \delta \beta } \, c^\dagger_\alpha c^\dagger_\gamma c^\pdag_\delta c^\pdag_\beta . 
\end{align}
Here, $\{c^\pdag_\alpha, c^\dagger_\beta\} = \delta_{\alpha \beta}$, and $\alpha = 1, \dots, M$  is a composite single-particle index encompassing, e.g., spin, orbital, and position or momentum indices. \changed{We will use boldface notation for $M \! \times \! M$ matrices with elements labeled by these indices, e.g.\ $\boldone$ and $\bolde = \bolde^\dagger$ have elements  $\delta_{\alpha \beta}$ and $\epsilon_{\alpha \beta}
= \epsilon_{\beta \alpha}^\ast$, respectively.} Let $\bG^{\mr{R/A}}(t)$ 
and $\bG^{\mr{R/A}}(\omega)$ denote retarded/advanced single-particle Green's functions \changed{in the time domain} and their Fourier transforms \changed{to the frequency domain}, respectively,  
with elements%
\begin{subequations}%
\label{eq:defineGRA}%
\begin{align}
    \label{eq:defineGRAstandard}
    G_{\alpha \beta}^\mr{R/A}(t)
    & = \mp \mr{i} \theta(\pm t) 
    \bigl\langle \{ \ohat{c}_\alpha^\pdag (0),\ohat{c}_{\beta}^{\dagger}(-t)\} \bigr\rangle , \\ 
    G_{\alpha \beta}^\mr{R/A}(\omega) 
    & = \int_{-\infty}^\infty \mr{d}t \, \mr{e}^{\mr{i} \omega^\pm t} \, \tG_{\alpha \beta}^\mr{R/A}(t) , 
\end{align}%
\end{subequations}%
where $\omega^{\pm} = \omega \pm \mr{i} 0^+\!$, and $\bG^\mr{A}(\omega) = \bG^{\mr{R}\dagger}(\omega)$. The Fourier-domain Green's functions can be viewed as analytic continuations,  $\bG^\mr{R/A}(\omega) = \bG(z \to \omega^\pm)$, of a function $\bG(z)$ defined through the spectral representation%
\begin{subequations}%
\label{subeq:SpectralRepresentation1}%
\begin{align}
    \label{eq:defineA}
    \bG(z) & = \int_{- \infty}^\infty \mr{d} \omega \frac{\bA(\omega)}{z - \omega} , \quad z \in \mathbbm{C}, \\
    \label{eq:defineSpectralFunction}
    \bA(\omega) & = \frac{\mr{i}}{ 2 \pi } \bigl[\bG^\mr{R}(\omega) - \bG^{\mr{A}} (\omega)\bigr]  
\end{align}%
\end{subequations}%
\changed{with a normalized, hermitian spectral function, 
\begin{align}
\label{eq:Anormalization}
\int \mathrm{d} \omega \, \bA(\omega) = \boldone \, , 
 \qquad
\bA(\omega) = \bA^\dagger(\omega) \, . 
\end{align}}%
\begin{subequations}%
\label{subeq:Dyson}%
$\bG(z)$ and its non-interacting version, $\bG_0(z) = (\bz - \bolde)^{-1}$, \rreply{with $\bz = z \boldone$}, are related by the Dyson equation,
\begin{align}
\label{eq:Dyson1}
    \bG(z) & = [\bG_0^{-1}(z) - \bSigma (z)]^{-1} = \frac{\boldone}{ \bz - \bolde - \bSigma(z)} ,
    \\
    \label{eq:Sigma=HF+dyn}
    \changed{\bSigma(z)}  &
    \changed{= \bSigma^\dyn (z) + \bSigma^{\Hartree} \, , }
\end{align}%
\end{subequations}%
\changed{which defines the self-energy $\bSigma(z)$. In \Eq{eq:Sigma=HF+dyn} we split it into a dynamical part that vanishes for $|z| \to \infty$,
and a hermitian, $z$-independent part, the Hartree-Fock term. 

\changed{In later sections, we will denote the retarded functions $\bG(\omega^+)$ and
$\bSigma(\omega^+)$ by $\bG(\omega)$ and $\bSigma(\omega)$, for short.}

In this work, we will need a spectral representation for 
$\bSigma^\dyn(z)$ analogous to \Eqs{subeq:SpectralRepresentation1} 
for $\bG(z)$ above.
To this end, we introduce the following definitions.
First, we express the retarded/advanced self-energies as%
\begin{subequations}%
\label{subeq:RetardedSelfEnergy}%
\begin{align}
\bSigma^{\mr{R/A}}(\omega) & = \bSigma (z\to \omega^\pm) = \bSigma'(\omega)\mp \mr{i} \pi \bSigma''(\omega) \, ,
\end{align}
with spectral and real parts given, respectively, by 
\begin{align}
\label{eq:ImSigmaDyn}
    \bSigma'' (\omega) & = - \tfrac{1}{\pi} \mathrm{Im} \bSigma^\mr{R}(\omega) =
     \tfrac{\mr{i}}{ 2 \pi} \bigl[\bSigma^\mr{R}(\omega) - \bSigma^\mr{A} (\omega)\bigr] 
     \, ,\\
    \label{eq:ReSigmaDyn}
    \bSigma'(\omega) & = \bSigma^\Hartree + P \hspace{-1mm} \int_{- \infty}^\infty \mr{d} \omega' \frac{\bSigma''(\omega')}{\omega - \omega'} \, .
\end{align} 
\end{subequations}
We next express the spectral part  as $\bSigma'' (\omega) = \btpd_1 \bW(\omega) \btd_1 $, where the \textit{weight matrix} $\bt_1$ and
\textit{normalized self-energy spectrum} $\bW(\omega)$
satisfy%
\begin{subequations}%
\label{sub:definition_t1_W}%
    \begin{align}%
    \label{eq:definition_t1}
        \int \mr{d} \omega \,\bSigma'' (\omega)  & =\btpd_1 \btd_1 \, ,  
         \quad \bt_1 = \bt_1^\dagger\geq0 \, , 
\\ 
\label{eq:normalization_W1}
 \int \mr{d} \omega \bW (\omega) & =  \boldone \,  , 
 \qquad 
 \bW(\omega) = \bW^\dagger(\omega) \, . 
\end{align}%
\end{subequations}%
Here,  $\bt_1 = \bt_1^\dagger\geq0$ is shorthand for stating that  $\bt_1$ is a hermitian, positive semidefinite matrix. Thus, Eq.~\eqref{eq:definition_t1}  defines $\bt_1$ as the principal square root 
(defined in \App{sec:PrincipalSquareRoot}) of 
the integrated spectral weight, 
$\int \mr{d} \omega \,\bSigma'' (\omega)$,
which itself is a hermitian, positive semidefinite matrix. 
Then, $\bSigma^\dyn(z)$ has the spectral representation%
\begin{subequations}%
\label{subeq:defineS(z)first}%
    \begin{align}%
    \label{eq:defineS(z)}
    \bSigma^\dyn(z) &= \btpd_1 \bS(z) \btd_1 \, , 
    \\  
    \label{eq:define-spectral-rep-for-Sfirst}
    \bS(z) & = \int_{- \infty}^\infty \mr{d} \omega \frac{\bW(\omega)}{z - \omega} \, .
    \end{align}%
\end{subequations}%
Indeed, $-\tfrac{1}{\pi} \mathrm{Im} \bSigma^\dyn(\omega^+) = \bSigma'' (\omega)$. Thus, $\bW(\omega)$ is the spectral function for $\bS(z)$, 
the normalized dynamical part of the self-energy, just as $\bA(\omega)$ is for $\bG(z)$.

Equations~\eqref{subeq:Dyson} and \eqref{subeq:defineS(z)first} 
together imply 
\begin{align}
    \label{eq:FirstStepOfCFEforG}
    \bG(z) & = \frac{1}{\bz - \bolde - \bSigma^\Hartree - \btpd_{1} \bS(z) \btd_1} \, . 
\end{align}
This constitutes the first step of a CFE  of the Green's function.
Further steps of the CFE can be obtained by expressing $\bG(z)$ and $\bSigma(z)$ through the Liouvillian. The next section is devoted to this task.}

\subsection{Expressing $\bG(z)$ and 
$\bSigma(z)$ through $\cL$
\label{sec:ExpressingGSigmathroughL}}

We begin by recalling some basic definitions and properties involving the Liouvillian. Consider a quantum system with Hamiltonian $\ohat{H}$ acting in a Hilbert space $\mathbb{V}$. Linear operators acting on $\mathbb{V}$ form a vector space, too, to be denoted $\doubleW = \mathbb{V} \otimes \mathbb{V}^{\ast}$. We  use a round-bracket ket notation, $\sket{\;}$, for its elements,   representing a linear operator $\ohat{O}$ acting in $\mathbb{V}$ by $\sket{\ohat{O}} \in  \doubleW $ (the so-called Choi-Jamiołkowski transformation \cite{Choi1972,Jamiołkowski1972}).  

The Heisenberg equation of motion for operators, $-\mathrm{i} \partial_t \ohat{O}(t) = [ \ohat H, \ohat{O}(t)]$, is linear in $\ohat{O}(t)$. Therefore, it can be expressed in the form $-\mathrm{i} \partial_t \sket{\ohat{O}(t)} = \cL \sket{\ohat{O}(t)}$, 
\changed{where the Liouvillian $\cL \in \doubleW \otimes \doubleW^\ast$ 
is a linear operator on $\doubleW$ whose action is defined through $\mc{L} O = [H,O]$, or} 
\begin{align}
    \cL \sket{\ohat O} = \sket{[\ohat{H}, \ohat{O}]} .
\end{align}
The solution of the operator Heisenberg equation reads 
\begin{align} 
\label{eq:OperatorTimeTranslation}
    \sket{\ohat{O}(t)} = e^{i \cL t} \sket{\ohat{O}(0)} \, ,
\end{align} 
\changed{thus $\cL$ generates time translations for operators in $\doubleW$.} 

For systems in thermal equilibrium, it is useful to define an inner product on $\doubleW$ through \cite{Foley2024_Liouvillian}
\begin{align}
    \label{eq:operatornorm}
    \sbra{\ohat{A}}\ohat{B}) = \langle \{\ohat{A}^{\dagger},\ohat{B} \} \rangle \, ,
\end{align}
where $\{ \, , \, \}$ denotes an anti-commutator, $\langle \ohat O \rangle = \mathrm{Tr}[\ohat \rho \, \ohat O]$ a thermal average and $\ohat \rho = e^{- \ohat H/T}/Z$ the thermal density matrix at temperature $T \neq 0$. 
This inner product $\sbra{\, } \, )$ of \Eq{eq:operatornorm} is well-defined if $\rho$ has full rank. That is true for $T\neq 0$, but not for $T=0$. We discuss the latter case in App.~\ref{app:OperatorProductT=0}. Since $[\rho,H]=0$, the Liouvillian is hermitian w.r.t.\ this inner product, $\sbra{A} \mc{L}B) = \sbra{\mc{L}A} B)$.

We now express the Green's function $\bG(z)$ through the Liouvillian. First, note that \Eq{eq:defineGRAstandard} can be written as
\begin{align}
    \label{eq:reexpressGRAthroughLiouvillian}
    \tG_{\alpha \beta}^\mr{R/A}(t) & = \mp \mr{i} \theta(\pm t)  \sbra{c^\dagger_{\alpha}} e^{-\mr{i}\cL t} |c^\dagger_{\beta}) .
\end{align}
For notational brevity, we gather all $\sket{c^\dagger_\alpha}$  into a row  or \textit{block} of $M$ operators, denoted 
\begin{align}
    \label{eq:define-bold-c}
    \sket{\bc^\dagger} = \bigl(\sket{c_1^\dagger}, \ldots , \sket{c^\dagger_M}\bigr) \;  \in \doubleW^{\oplus M}\, .
\end{align}
(This choice is appropriate for Hamiltonians with U(1) charge symmetry, as in \Eq{eq:Hamiltonian}.
Footnote~\footnote{To keep the notation simple in the main text, we assumed U(1) charge symmetry, i.e.\ the  absence of superconductivity. 
As a result, we focused on representing $\mc{L}$ in $\doubleW_1$, the dynamically relevant subspace of operators with charge $+1$. The representation of $\mc{L}$ in $\doubleW_{-1}$, the dynamically relevant space of operators with charge $-1$, follows from $[H,c_{\alpha}^{\dagger}]^{\dagger} = -[H,c^{\pdag}_{\alpha}]$. In the presence of mean-field superconductivity, $H_\mr{int}$ also contains $c^\dagger \! c^\dagger$ and $c\, c$ terms. 
Then, the block of single-particle operators has to be extended  to include both creation and annihilation operators, $\sket{\bc^\dagger,\bc} = \bigl(\sket{c_1^\dagger}, \ldots , \sket{c^\dagger_{M}}, \sket{c_1^\pdag},  \ldots \sket{c_M^\pdag}\bigr)$, since both the Green's function and self-energy will exhibit non-zero anomalous components. With this generalization, the strategy described in our paper can also be applied in the presence of mean-field superconductivity.}
discusses the modifications required if $H_\mr{int}$ also contains $c^\dagger \! c^\dagger$ and $c\, c$, as needed to describe mean-field superconductivity.) 
The  relations $\{c^\pdag_\alpha, c_\beta^\dagger\} = \delta_{\alpha \beta}$ imply the $M \! \times \! M$ matrix equation $\sbra{\bc^\dagger} \bc^\dagger) = \boldone$. 
Then, \Eqs{eq:defineGRA} can be expressed as%
\begin{subequations}%
    \label{subeq:GRAL-retarded-defined}%
\begin{align}
    \label{eq:GRAL-retarded-defined}
    \tbG^\mr{R/A}(t) & = \mp \mr{i} \theta(\pm t) \sbra{\bc^\dagger} \mr{e}^{-\mr{i} \cL t} \sket{\bc^\dagger} \, , \\
    \bG^\mr{R/A}(\omega) & =  \sbra{\bc^\dagger} \frac{\boldone}{\omega^\pm-\cL} \sket{\bc^\dagger} \, . 
\end{align}%
\end{subequations}%
The latter \changed{are related by analytic continuation to} 
\begin{subequations}
\label{subeq:SpectralRepresentation2}
\begin{align}
    \label{eq:GL}
    \bG(z) = \sbra{\bc^\dagger} \frac{\boldone}{z-\cL} \sket{\bc^\dagger} \, ,
\end{align}
having a spectral representation of the form \eqref{subeq:SpectralRepresentation1}, with 
\begin{align}
    \label{eq:GRAL}
    \bA(\omega) = \sbra{\bc^\dagger} \delta(\omega-\cL) \sket{\bc^\dagger} \, .
\end{align}
\end{subequations}
\changed{The fact that  $\bA(\omega)$ has unit weight and is hermitian
[cf.~\Eqs{eq:Anormalization}] is self-evident from this representation.

Equations~\eqref{subeq:SpectralRepresentation2} for $\bG(z)$ and $\bA(\omega)$
are well known. Analogous equations, perhaps not as well known, 
exist for $\bS(z)$ and $\bW(\omega)$; we derive them in \App{sec:SingleParticleIrredSubspace}. This involves 
defining two operator subspaces of $\doubleW$: 
the first, $\doubleW_{\sp} = \mathrm{span}\{\sket{c_\beta^\dagger}\}$, is a subspace of single-particle creation operators; the second,
$\doubleW_{\spi}$, is the space of all single-particle irreducible
operators in $\mathrm{e}^{-\mathrm{i}\cL t} \doubleW_\sp$,
i.e.\ those components of time-evolved single-particle operators that 
are \textit{not} in $\doubleW_\sp$, i.e.\ not single-particle operators themselves. When acting 
on $\doubleW_\sp \oplus \doubleW_\spi$,  the Liouvillian 
can be decomposed into four blocks, $\cL_{\sp \sp} + \cL_{\sp \spi} + \cL_{\spi \sp} + \cL_{\spi \spi}$, representing its projections into
these subspaces. Then, the 
ingredients of the self-energy 
$\changed{\bSigma(z)}  = \bSigma^\dyn (z) + \bSigma^{\Hartree}$
can be expressed as follows (see \App{sec:SingleParticleIrredSubspace}): The Hartree-Fock contribution is given by%
\begin{align}
            \label{eq:HartreeFock1-main}
    \bSigma^{\Hartree} &= \bolde_0 - \bolde , \quad \bolde_0 = \sbra{\bc^\dagger} \cL_{\sp \sp} \sket{\bc^\dagger} \, , 
\end{align}
and the dynamical part $\bSigma^\dyn(z) = \btpd_1 \bS(z) \btd_1$
by%
\begin{subequations}%
    \label{subeq:defineS(z)}%
    \begin{align}
    \label{eq:Wz-main}
    \bS(z)  & = \sbra{\bq_1}  \cfrac{\boldone}{z - \mc{L}_{\spi \spi}}  \sket{\bq_1} \, , 
    \end{align}%
with $\bt_1 = \bt_1^\dag \ge 0$ defined through 
$\btpd_1 \btd_1 =  \sbra{\bc^\dagger} \cL_{\sp \spi} \cL_{\spi \sp} \sket{\bc^\dagger}$,  and $\sket{\bq_1} = \cL_{\spi \sp} \sket{\bc^\dagger} \btd^{-1}_1$. Thus, $\bS(z)$ is a propagator in the single-particle irreducible subspace $\doubleW_{\spi}$, and the spectral function $\bW(\omega)$
in its  spectral representation \eqref{eq:define-spectral-rep-for-Sfirst} is 
\begin{align}
 \label{eq:define-W-through-Lsbarsbar-main}
    \bW(\omega) & = \sbra{\bq_1} \delta(\omega-\cL_{\spi \spi}) \sket{\bq_1} \, .
\end{align}%
\end{subequations}
The fact that  $\bW(\omega)$ has unit weight and is hermitian
[cf.~\Eqs{eq:normalization_W1}] is self-evident from this representation.
Moreover, \Eqs{subeq:defineS(z)} immediately imply
that $\bSigma^\dyn(z)$ has the same analytic properties as $\bG(z)$ of \Eq{subeq:SpectralRepresentation2}.}

The above relations are general and exact. When dealing with fermionic many-body systems, $\doubleW_{\sp}$ can usually be handled with reasonable effort, whereas $\doubleW_{\spi}$  has to be dealt with in an approximate manner. The Hartree-Fock approximation treats the self-energy solely within $\doubleW_\sp$, approximating $\bSigma$ by $\bSigma^\Hartree$. It thus sets $\bSigma^\dyn =0$ and completely ignores $\doubleW_{\spi}$. For long-ranged interactions, $\bSigma^\Hartree$ is non-local, for on-site interactions it is local. This is also true within DMFT. Additionally, DMFT incorporates single-particle irreducible contributions from $\doubleW_\spi$ by including $\bSigma^\dyn$, but approximates it as local. Thus, $\mc{L}_{\spi \spi}$ and $\bt_1$ are approximated as local in DMFT.

Cluster extensions of DMFT additionally incorporate non-local but short-ranged matrix elements in $\mc{L}_{\spi \spi}$ and $\bt_1$. In CDMFT, matrix elements are fully considered within each real-space cluster, while inter-cluster matrix elements are neglected. As a result, translation symmetry is broken. DCA, on the other hand, tiles the Brillouin zone (BZ) into $\Npatch$ patches and approximates $\mc{L}_{\spi \spi}$ and $\bt_1$ as constant within each patch. As a result, DCA is translation invariant, but the self-energy is not a smooth function of $\bk$. Our goal in this work is to reinstate the translation invariance in CDMFT or a smooth $\bk$-dependence in DCA by interpolating  (periodizing  or interpolating, respectively) $\bt_1$ and the matrix elements of $\mc{L}_{\spi \spi}$.

\changed{This concludes our presentation of the conceptual foundations of this paper. Its central result, expressed through \Eqs{subeq:defineS(z)} and derived in \App{sec:SingleParticleIrredSubspace}, is that the dynamical part of the self-energy is governed by the single-particle irreducible part of the Liouvillian. To our knowledge, this relation has not been formulated as explicitly before, though its seeds are implicit in previous work expressing Green's functions through Liouvillians 
\cite{Viswanath1994,Uhrig1996}.}

\changed{\subsection{CFE of self-energy yields Liouvillian matrix elements}
\label{sec:CFE-self-energy-main} 

If the self-energy $\bSigma(z)$ is known analytically or numerically (obtained in whatever way---in this paper from cluster-DMFT), \Eqs{subeq:defineS(z)} can be leveraged to explicitly compute the dominant Liouvillian matrix elements of
$\cL_{\spi \spi}$. This can be done via a CFE expansion of the dynamical part of the self-energy, $\bSigma^\dyn(z)$; its CFE coefficients yield the desired matrix elements.  This task is technical rather than conceptual, hence we put  its detailed description
into appendices:  
\App{app:LKrylov} for general models,  and \App{sec:QIM}  for peculiarities arising for quantum impurity models.  
In this section we just collect five main  results. 
}

\changed{
(1) We can obtain a CFE for  $\bSigma^\dyn(z)=\btpd_1 \bS(z) \btd_1$ by representing $\bS(z)$ in a Krylov space constructed via an iterative Lanczos procedure \cite{Koch2011,Foley2024_Liouvillian}. It constructs a Krylov space spanned by a set of orthogonal operators, $\{\sket{\bq_1}, \dots, \sket{\bq}_N\}$, through repeated applications of $\cL_{\spi \spi}$, with initial vector $\sket{\bq_1}$. We describe this 
procedure in detail in \App{sec:CFE-self-energy}. This yields a CFE of the form 
\begin{align}
    \label{eq:SigmadynCFE-main}
    \bSigma^\dyn(z) & \simeq
    \btpd_1 \cfrac{\boldone}{\bze_1 - {\btpd_{2}\cfrac{\boldone}{\bze_2 - \raisebox{-0.9em}{ \ensuremath{\ddots}\raisebox{-0.9em}{ \hspace{-2.5mm} \ensuremath{- \btpd_\NKrylov \cfrac{\boldone}{\bze_N} \, \btd_\NKrylov}}}}\, \btd_{2} }} \, \btd_1\, . 
\end{align}
Here, $\bze_n = \bz - \bolde_n$, and the CFE coefficients $\bolde_n^\pdag$, $\bt_n^\pdag$ of $\bS(z)$ and hence of $\bSigma^\dyn(z)$ are matrix elements of the single-particle irreducible Liouvillian $\cL_{\spi\spi}$:
\begin{flalign}
    \label{eq:Lmatrixelements-en-tn-main}
    & \sbra{\bq_{n}} \cL_{\spi \spi} \sket{\bq_{n'}} = 
    \begin{cases}
        \boldsymbol{\epsilon}_{n}, & 
        n' = n , \\
        \bt_{n+1} , & 
        n' = n +1   , \\
        \btd_{n},  & 
        n' = n -1    ,   \\
        \bzero & \mr{otherwise} \, .
    \end{cases} \hspace{-1cm} &
\end{flalign}
The CFE \eqref{eq:SigmadynCFE-main} rigorously preserves causality. Moreover, it becomes exact for sufficiently large $N$ (under formal conditions discussed
in \App{app:MeasuresAndAllThat}).}

\changed{
(2) If  $\bW(\omega)$, the normalized spectral function of  $\bSigma^\dyn(z)$
[\Eq{subeq:defineS(z)first}], is known explicitly (analytically  or numerically),  we can compute the CFE coefficients $\bolde_n$, $\bt_n$ numerically by using an iterative Lanczos scheme analogous to that mentioned above. 
We construct a Krylov space spanned by a set of matrix polynomials, $\{ \bpone_1(\omega), \dots, \bpone_N (\omega)\}$, orthogonal w.r.t.\ the measure 
$\bW (\omega) \, \mathrm{d}\omega$, through 
repeated multiplications by $\omega$, starting with
$\bpone_1 (\omega) = \boldone$. We describe this procedure in detail in \Sec{sec:CFE-via-polynomials}. In this Krylov space,
the Liouvillian matrix elements of \Eq{eq:Lmatrixelements-en-tn-main} are expressed as [cf.~\Eq{eq:nLn'-pn-omega-pn'}]
\begin{align}
        \sbra{\bq_{n}} \cL_{\spi \spi} \sket{\bq_{n'}} & = \int \! \mr{d}\omega \, \bponedag_{n} (\omega) \,  \bW(\omega) \, \omega \, \bpone_{n'}(\omega) \, . 
\label{eq:nLn'-pn-omega-pn'-main}
\end{align}
This equation can be used to compute $\bolde_n$, $\bt_n$ explicitly. 
We give a pseudocode for doing so in  \Alg{alg:lanczos} in \App{app:MatrixElements}.}

\changed{
(3) If $\bW(\omega)$ is discrete with a finite number of $\delta$-peaks, as is the case for a finite-size system, $\bSigma^\dyn (z)$ has an exact finite-depth CFE representation of the form  \eqref{eq:SigmadynCFE-main}. This form, which encodes the full spectrum exactly in the Liouvillian matrix elements, rigorously preserves causality. Otherwise, i.e.\ if $\bW(\omega)$ is not discrete, the finite-depth representation \eqref{eq:SigmadynCFE-main} is an approximation. 

However, it is always possible to represent $\bSigma^\dyn(z)$
\textit{exactly} through a CFE of finite depth $N$, 
\begin{flalign}
    \label{eq:S(z)CFE-main}
 &    \bSigma^\dyn (z) = \btpd_1 
    \cfrac{\boldone}{\bze_1 - { \btpd_{2}\cfrac{\boldone}{\bze_2 - \raisebox{-0.9em}{\ensuremath{\ddots} \raisebox{-0.5em}{ \hspace{-2.5mm} \ensuremath{- \btpd_\NKrylov \cfrac{\boldone}{\bze_N - \Rest_\NKrylov(z)  } \, \btd_\NKrylov  \hspace{-4mm} \phantom{.}}}}}\, \btd_{2} \hspace{-2mm}} } \, \btd_1 , \hspace{-1cm} & 
\end{flalign}
utilizing a suitably chosen residual function $\Rest_N(z)$. 
In \Sec{app:ExactTerminationCFE}, we provide formulas
for computing the residual function numerically using the above-mentioned orthogonal polynomials, see \Eqs{eq:results:restfunction}.

We note in passing that a CFE for the self-energy of course implies, via the Dyson equation \eqref{eq:Dyson1}, a CFE for its Green's function, see \Eq{eq:exact-CFE-for-G}. 
}

\changed{
(4) For quantum impurity problems,  a complication arises: the impurity Green's function has the form 
\begin{align}%
\label{eq:DysonImpurityModel-main}
    \bG_{\dimp \dimp} (z) & = \Bigl[ \bz - \bepsilon_0 - \bDelta(z) - \bSigma^\dyn(z)\Bigr]^{-1} \, ,
\end{align}
where $\bDelta(z)$ is the impurity-bath hybridization function.
Some impurity solvers yield the spectral function of $\bG_{\dimp \dimp} (z)$, not the spectral function of $\bSigma^\dyn(z)$.  
Before constructing the CFE of the latter, 
the contribution of $\bDelta(z)$ has to be separated out.
Procedures for achieving this separation are described in \App{sec:SeparatingHybridizationSelfEnergy}.

(5) When NRG is used as impurity solver \cite{Bulla2008_NRG,Weichselbaum2012b,Kugler2022}, the computation of the self-energy involves log-Gaussian broadening; that introduces long tails that adversely affect the extraction of CFE coefficients. This problem can be mitigated by regularizing the CFE, preferably in a manner geared towards extracting low-energy information. To this end 
we use an iterative  ``fast/slow'' splitting scheme \cite{Bruognolo2017_OpenWilsonChain}: at each level $n$ we construct a residual function $\bR_n(z)$, split it into ``fast" (high-energy) and ``slow" (low-energy) contributions, $\Rest^\pdag_n(z) = \Rest_{n}^{\fast}(z)+\Rest_{n}^{\slow}(z)$, and use
only the slow part for the next iteration. We describe this scheme in detail in \App{sec:restfunction_splitting}. The result is an \textit{exact} CFE of the following form: 
\begin{flalign}
    \label{eq:SigmaCFE_FastSlowMode-main}
    & \bSigma^\dyn (z) = \\[-2mm]  
    & \; \; \btpd_1 \cfrac{\boldone}{\bz_1 \! - \! \bR^\fast_1 (z) - { \btpd_{2}\cfrac{\boldone}{\bz_2  \! - \! \bR^\fast_2 (z) - \! \raisebox{-0.9em}{\ensuremath{\ddots} \raisebox{-0.5em}{\hspace{-2.5mm} \ensuremath{- \btpd_\NKrylov \cfrac{\boldone}{\bz_N \!-\! \Rest_\NKrylov(z)  } \, \btd_\NKrylov  \hspace{-4mm} \phantom{.}}}}}\, \btd_{2} \hspace{-2mm}}} \, \btd_1  
     . \nonumber \hspace{-1cm} & 
\end{flalign}
Here, each $\bR_n^\fast(z)$ will, in general, contain an imaginary part; therefore, some amount of broadening is present at each level, yielding a self-energy with a continuous spectrum. 
Using splitting functions governed by two 
parameters, $\tau_\fastslow \, (> 2)$ and $\Lambda_\fastslow \, (> 1)$
[cf.~\Eq{eq:splitting_function}], we can perform the fast/slow splitting in such a manner that the resulting CFE coefficients decrease exponentially as $\bepsilon_n , \bt_n \sim \Lambda^{-n}_\fastslow$, and the overall magnitude
 of $\bR_n^\fast(z)$ likewise scales as $\sim \Lambda^{-n}_\fastslow$. We call the resulting CFE  a \textit{log-CFE}, because logarithmic scales are needed when plotting its coefficients. Their exponential decay turns out to be advantageous for $\cL$-interpolation, in that the interpolation becomes progressively more accurate for increasing depth $n$.
For numerical details on log-CFEs, see \App{app:dependence_splitting_parameters}.
}

\section{Liouvillian interpolation}
\label{sec:LiouvillianPeriodization}

In this section, we outline the basic ideas underlying Liouvillian interpolation. We start with a brief discussion of cluster-DMFT approximations (\Sec{sec:LiouvillianDMFT}) and traditional interpolation schemes (\Sec{sec:trad_periodization}), then introduce our $\cL$-interpolation scheme (\Sec{subsec:Lperiodization}).

\subsection{Cluster DMFT approximations}
\label{sec:LiouvillianDMFT}

We begin with some general remarks. For computations in a real-space basis of $\Nlattice$ lattice sites with $M$ orbitals per site, the self-energy matrix $\widetilde \bSigma(z)$ is a $\Nlattice M \!\times \!  \Nlattice M $ matrix  built from \ $\Nlattice \! \times \! \Nlattice$ blocks $[\widetilde \bSigma (z)]_{\br, \br'}$ labeled by site indices, with each block having dimension $M\times M$. If $\widetilde \bSigma(z)$  is expressed through a block-CFE, its CFE coefficients $\widetilde \bolde_n$ and $\widetilde  \bt_n$ (i.e.\ Liouvillian matrix elements)  are matrices too, built from $\Nlattice \! \times \!  \Nlattice$ blocks $[\widetilde  \bolde_n]_{\br,\br'}$ and $[\widetilde  \bt_n]_{\br,\br'}$, each of dimension $M\times M$. 

For translationally invariant systems $[\widetilde \bSigma (z)]_{\br, \br'}$ depends only on the difference vector $\bdelta =  \br-\br'$, as do $[\widetilde \bolde_n]_{\br,\br'}$ and $[\widetilde \bt_n]_{\br,\br'}$. Then the Fourier transform block-diagonalizes $\widetilde \bSigma$ w.r.t.\ spatial indices:
\begin{align}
    \label{eq:defineSigmak_Via_FT}
    \bSigma_\bk (z)& = \sum_{\bdelta} [\widetilde \bSigma (z)]_{\bdelta,\bzero}  \, \mr{e}^{-\mr{i} \bk \cdot \bdelta}. 
\end{align}
Here, $\bSigma_{\bk}(z)$ is shorthand for the diagonal elements of a block-diagonal matrix with elements $[\bSigma(z)]_{\bk, \bk'} = \delta_{\bk\bk'} \bSigma_{\bk}(z)$ obtained by viewing the Fourier transform as a basis transformation $\bSigma = \bU  \widetilde \bSigma \bU^\dagger$ with matrix entries $[\bU]_{\bk ,\br} = \mr{e}^{-\mr{i} \bk \cdot \br}/ \sqrt{\Nlattice}$.
The sum runs over all distinct difference vectors $\bdelta$ between two lattice sites. (Since the lattice includes site $\bzero$, this is equivalent to summing over all lattice sites.) For given $\bk$, $\bSigma_\bk (z)$ is an $M \times M$ matrix, as are its CFE coefficients, 
\begin{align}
    \bepsilon_{\bk n} = \sum_\bdelta [\widetilde \bolde_n]_{\bdelta,\bzero}\, \mr{e}^{-\mr{i} \bk \cdot \bdelta}\, ,\quad
    \bt_{\bk n} = \sum_\bdelta [\widetilde \bt_n]_{\bdelta,\bzero} \, \mr{e}^{-\mr{i} \bk \cdot \bdelta}\, . 
\end{align}

In conventional single-site DMFT, the ``impurity'' is a single lattice site self-consistently embedded in its surrounding, the real-space self-energy is local, $[\widetilde \bSigma(z)]_{\br,\br'}=\delta_{\br,\br'} \bSigma (z)$, and the momentum-space self-energy is $\bk$-independent, $\bSigma_\bk(z) = \bSigma (z)$. The same is true for its $n > 0$ CFE coefficients: 
$[\widetilde \bolde_n]_{\bdelta,\bzero} = \delta_{\bdelta,\bzero} \,  \bepsilon_n$ and $\bepsilon_{\bk n} = \bepsilon_n$,  likewise for $\widetilde \bt_n$ and $\bt_{\bk n}$. Single-site DMFT thus approximates the single-particle irreducible part of the Liouvillian of the lattice model by that of an effective impurity model.

For CDMFT, the real-space lattice is tiled into identical clusters, each containing $\Ncluster$ sites. The  ``impurity'' is one of these clusters, which we denote by $\mathcal{C}$, and the real-space self-energy $[\widetilde \bSigma(z)]_{\br,\br'}$ is taken to be localized on $\mc{C}$ (with open boundary conditions): it contains  non-zero contributions only when both position arguments are on the same cluster, while intercluster terms are set to zero:  $[\widetilde \bSigma(z)]_{\br,\br'} \neq 0$ only when $\br, \br' \in \mc{C}$
\changed{(yielding an $\Ncluster M \times \Ncluster M$ matrix)}. Therefore, the same is true for its $n>0$ CFE coefficients $[\widetilde \bolde_n]_{\br,\br'}$ and $[\widetilde \bt_n]_{\br,\br'}$. CDMFT thus explicitly breaks translation invariance.

DCA, by contrast, retains translational invariance at the cost of using coarse-grained momentum variables. The BZ is partitioned into equally-sized patches  labeled by patch momenta $\vec{K}$. The \changed{coarse-grained momentum-space self-energy is 
taken to have the form $\bSigma_\bk(z) = \bSigma_\bK(z)$
if $\bk \in \bK$, and zero otherwise, where each patch 
self-energy $\bSigma_\bK(z)$ is an $M\!\times \! M$ matrix. 
Thus the coarse-grained self-energy is} $\bk$-inde\-pen\-dent within patches and has discontinuous jumps at patch boundaries. $\bSigma_\bK(z)$ mimics the patch average of $\bSigma_{\bk}(z)$. The same is true for its CFE coefficients, 
$\bepsilon_{\bK n}$ and $\bt_{\bK n}$. 

Interpolation schemes seek to restore translational invariance for CDMFT self-energies or a smooth momentum dependence for DCA self-energies. (As mentioned earlier, for CDMFT, our use of ``interpolation'' should be understood to refer to ``periodization''.) In \Sec{subsec:Lperiodization}, we propose $\cL$-interpolation as a unifying scheme, applicable to both CDMFT and DCA, capable of achieving these goals. Before that, in \Sec{sec:trad_periodization} we briefly review the traditional schemes, $\Sigma$-, $M$-, and $G$-interpolation,  since we will later present results comparing the performance of all these schemes. 

\subsection{Traditional interpolation}
\label{sec:trad_periodization}

\textit{CDMFT:} When applying traditional interpolation to our  CDMFT data, we use a superlattice average \cite{Stanescu2006_periodization2} to guarantee causality. The superlattice is constructed by translating a cluster $\mc{C}$ to tile the original lattice, treating $\mathcal{C}$ as a super-site carrying internal degrees of freedom. To interpolate  a $z$-dependent $[\widetilde {\bQ}(z)]_{\br,\br'}$ obtained from CDMFT
as a real-space matrix, \changed{where} $\widetilde {\bQ}$ stands for the self-energy $\widetilde{\bSigma}$, cumulant $\widetilde{\bM}$ or Green's function $\widetilde{\bG}$, we make the CDMFT interpolation Ansatz
\begin{align}
    \label{eq:define-CDMFT-periodization}
    \bQ_{\bk}(z) & \simeq \sum_{\bdelta} \Qbar_{\bdelta}(z) \,  \mr{e}^{-\mr{i} \bk \cdot \bdelta}  \, .
\end{align}
We estimate $\Qbar_{\bdelta} (z)$ using a superlattice average~\cite{Stanescu2006_periodization2}
\begin{align}
    \label{eq:estimateQbar}
    \Qbar_{\bdelta}(z) & = \frac{1}{\Ncluster} \sum_{\br' \in \mathcal{C}} [\tbQ (z)]_{\br' + \bdelta,\br'}  \, ,
\end{align}
where the sum runs over a cluster with $\Ncluster$ sites. For $\tbQ =\tbSigma$ or $\tbM$ (but not for $\tbG)$, terms in \Eq{eq:estimateQbar} having $\br' + \bdelta \notin \mathcal{C}$ are zero. For  $\Qbar = \overline \bSigma$ or $\overline \bM$ (but not $\overline \bG$), we therefore have  $\Qbar_{\bdelta} = \boldsymbol{0}$ if  $|\bdelta|$ exceeds the cluster size. For $\bQ = \bSigma$ or $\bM$, \Eq{eq:define-CDMFT-periodization} thus is equivalent to a short-range-truncated Fourier Ansatz of the form \eqref{eq:Q_periodize}, while for $\bQ = \bG$, \Eq{eq:define-CDMFT-periodization} involves no short-ranged truncation.

Superlattice averaging as described above leads to causal self-energies or Green's functions~\cite{Stanescu2006_periodization2}. However, these may become \changed{non-causal} if the superlattice average is replaced, for instance, by a cluster average, i.e.\ if in \Eq{eq:estimateQbar} $1/\Ncluster$  is replaced by $1/N_\bdelta$, where $N_\bdelta$ is the number of pairs of
sites in the cluster having distance $\bdelta$.

\textit{DCA:} In DCA, the BZ is partitioned into $\Npatchall$ disjoint patches labeled by a set of representative momenta $\{\bK\}$. We write $V_\BZ = \cup_\bK V_\bK$ for the union of their domains, and $\sum_\bK |V_\bK| = |V_\BZ|$ for the sum of their volumes. The output of DCA computations are translationally invariant, momentum patch-averaged quantities. 
\begin{align}
    \bQ_\bK(z) & =
    \begin{cases}
        \bSigma_\bK(z) \, , \\
        \bM_\bK(z) = \big[z + \mu - \bSigma_\bK(z)\big]^{-1} .
    \end{cases}
\end{align}
These must  be combined to yield a function $\bQ_\bk(z)$ defined throughout the BZ. ``Bare'' (non-interpolated) DCA does this using step functions, $\bQ_\bk(z) = \sum_\bK \theta_{\bk \comma \bK} \bQ_{\bK}(z)$, where $\theta_{\bk \comma \bK} = 1$ if $\bk$ lies in patch $\bK$ and zero otherwise, resulting in artificial discontinuities in momentum space. The interpolation of $\bQ_\bK(z)$ amounts to smearing out the discontinuities through a DCA interpolation Ansatz of the form
\begin{align}
    \label{eq:Mscheme_DCA}
    \bQ_\bk(z) & =  \sum_\bK \alpha_{\bk \comma \bK}\,\bQ_\bK(z) \, ,
\end{align}
where for each patch $\bK$ the ``form factor'' $\alpha_{\bk \comma \bK}$ is an interpolation  function of $\bk$ defined on the entire BZ. Various choices for the form factors can be found in the literature, including splines \cite{Maier2002_DCAinterp} and low-order Fourier-Ans\"atze \cite{Ferrero2009_dimerDCA}. If $\bQ_\bK(z)$\ depends on $z$ (e.g.\ when it stands for $\bSigma_\bK(z)$ or $\bM_\bK(z)$), interpolation  can cause causality problems; a heuristic strategy for ameliorating these is to choose the form factors such that  $\alpha_{\bK \comma \bK} = 1$ \cite{Maier2005,Staar2013_DCAinterp,Haehner2020_DCAinterp}.

\textit{Canonical DCA interpolation:} For Liouvillian interpolation, the quantities to be interpolated  are $z$-\textit{independent} Liouvillian matrix elements, hence causality concerns do not arise (with one small caveat, discussed later). 
This affords us the freedom to focus on a formulation of a truncated Fourier expansion that is particularly well suited for our purposes. We will call it ``canonical'' DCA interpolation, since it ensures that the resulting $\bQ_\bk(z)$ possesses three canonical (clearly desirable) properties:  
(i) $\bQ_\bk(z)$ is invariant under the point-group symmetry operations (rotations, reflections, inversions) that leave the patches invariant. 
(ii) \label{p:patch-average} Averaging $\bQ_\bk(z)$ over patch $\bK'$ yields $\bQ_{\bK'} (z)$, i.e.\ $    \int_{V_{\bK'}} \frac{\mr{d}\bk}{|V_{\bK'}|} \bQ_{\bk} (z) = \bQ_{\bK'}(z)$ \footnote{\rreply{Property (ii) is also imposed in the  DCA${}^+$ scheme of Ref.~\onlinecite{Staar2013_DCAinterp}, see Eq.~(12) there.}}. (iii) If $\bQ_\bK(z)$ does not depend on $\bK$, $\bQ_{\bK}(z) = \bQ(z)$, then $\bQ_\bk(z) = \bQ(z)$. Below, we present a form factor construction that ensures these canonical properties. It expresses the form factors as linear combinations of lattice harmonics, using only ones that are invariant under the symmetry operations that leave all patches invariant. This construction is not new, but, for completeness, we explain its rationale nevertheless. (For some background on the definition of lattice harmonics, see \App{app:latticeharmonics}.)

For canonical DCA interpolation, we expand $\alpha_{\bk, \bK}$ as
\begin{align}
    \label{eq:Ansatz-compact}
    \alpha_{\bk \comma \bK} & = \sum_{j = 1}^\Npatchall \phi_{\bk \comma j} \Psi_{j \comma \bK} \, . 
\end{align}
Here, the $\phi_{\bk,j}$ are real-valued lattice harmonics chosen according to a criterion described below; $j$ enumerates the chosen ones in order of increasing complexity, with $\phi_{\bk,1} = 1$; and the $\Psi_{j \comma \bK}$ are real expansion coefficients, to be determined in a manner described below.

To ensure property (i), we include in \Eq{eq:Ansatz-compact} only lattice harmonics $\phi_{\bk,j}$ that are invariant under the point group symmetries of the \textit{patches}; that immediately implies the same for $\alpha_{\bk,\bK}$. (The choice of $\phi_{\bk, j}$ may also depend on additional details of the chosen patching scheme and the physics it attempts to capture. A concrete example is given in  \Sec{sec:2dHubbard_DCA},  where we consider 4-patch DCA results for a Hubbard model on a 2D square lattice and use $\phi_{\bk,j}$ functions constructed via \Eq{eq:latticeharmonicconstruction-patchsymmetry} in \App{app:latticeharmonics}.)
 
To ensure property (ii), we impose the condition
\begin{align}
    \label{eq:alpha-self-consistency}
    \int_{V_{\bK'}} \frac{\mr{d}\bk}{|V_{\bK'}|} \alpha_{\bk \comma \bK} & = \delta_{\bK' \comma \bK} \, .  
\end{align}
Inserting Ansatz \eqref{eq:Ansatz-compact} into condition \eqref{eq:alpha-self-consistency} we obtain 
\begin{align}
    \label{eq:PhiPsi=1}
    \sum_{j=1}^\Npatch \Phi_{\bK' \comma j} \Psi_{j \comma \bK} & = \delta_{\bK' \comma  \bK}, 
\end{align}
where the coefficients $\Phi_{\bK' \comma  j}  = \int_{V_{\bK'}} \frac{\mr{d}\bk}{|V_{\bK'}|} \, \phi_{\bk \comma j}$ encode information about patch shapes. Viewing \Eq{eq:PhiPsi=1} as an $\Npatch \times \Npatch$ matrix equation, $\Phi \cdot \Psi = \doubleI$,  we find $\Psi_{j \comma \bK} = (\Phi^{-1})_{j \comma \bK}$ (assuming $\Phi^{-1}$ exists). This completes the construction of the form factors $\alpha_{\bk \comma \bK}$. Note that they are unchanged if the $\phi_{\bk,j}$ functions are replaced by linearly transformed combinations  $\sum_{j'}\phi_{\bk,j'}T_{j'j}$ (assuming $T$ is invertible), because by \Eq{eq:PhiPsi=1} that would just result in inversely transformed coefficients $\sum_{j''} T^{-1}_{j,j''} \Psi_{j'',\bK}$.

Property (iii) holds, too, since the form factors satisfy
\begin{align}
    \label{eq:interpolation-sum=1}    
   \sum_\bK \alpha_{\bk \comma \bK} & = 1
\end{align}
by construction. This can be seen as follows. The only lattice harmonic whose BZ integral is nonzero is the constant one: $\int_\BZ \mr{d} \bk \, \phi_{\bk,j} = |V_\BZ| \, \delta_{j,1}$. Therefore,
\begin{align}
    \label{eq:sumK_Phi_Kr}
    \sum_{\bK'} |V_\bK'| \Phi_{\bK' \comma j} =  \sum_{\bK'} \int_{V_\bK'} \mathrm{d} \bk \,  \phi_{\bk \comma  j} = |V_\BZ| \, \delta_{1 \comma j} \, .
\end{align}
Moreover, defining $\psi_j = \sum_\bK \Psi_{j \comma \bK}$, we find from \Eq{eq:PhiPsi=1},
\begin{align}
    \label{eq:sumK'}
    \sum_{\bK'} |V_\bK'| \sum_{j=1}^\Npatch \Phi_{\bK' \comma j} \psi_j= \sum_{\bK' \comma \bK} |V_\bK'| \, \delta_{\bK' \comma  \bK} = |V_\BZ| .
\end{align}
Inserting \Eq{eq:sumK_Phi_Kr} into \eqref{eq:sumK'} we find $\psi_1 = 1$. Summing \Eq{eq:PhiPsi=1} over $\sum_\bK$ we obtain $\sum_j  \Phi_{\bK' \comma  j}  \psi_j  = 1$, and since $\Phi_{\bK' \comma 1} \psi_1 = 1 \cdot 1$, this implies $    \sum_{j > 1} \Phi_{\bK' \comma  j}  \psi_j = 0$. This linear combination of columns of $\Phi$, which are linearly independent since $\Phi^{-1}$ was assumed to exist, can vanish only if all coefficients $\psi_{j>1}$ vanish, implying  $\psi_j = \delta_{j \comma 1}$. Thus  $\sum_\bK \alpha_{\bk \comma \bK} = \sum_{j} \phi_{\bk \comma j} \psi_j = \phi_{\bk,1} = 1$, proving \Eq{eq:interpolation-sum=1}.

Finally, we remark that the interpolation Ansatz of \Eqs{eq:Mscheme_DCA} and \eqref{eq:Ansatz-compact} for $\bQ_\bk(z)$ and $\alpha_{\bk \comma \bK}$ can be expressed as a truncated Fourier expansion having the form of \Eq{eq:Q_periodize}, see \App{app:latticeharmonics} for details.

If canonical DCA interpolation is used to interpolate   the self-energy or the cumulant, the resulting functions \rreply{$\bSigma_\bk(z)$} and $\bM_\bk(z)$ may exhibit frequency-dependent ringing effects that lead to violations of causality \cite{Okamoto2003}. A possible remedy is to use modified versions of Eqs.~(\ref{eq:Mscheme_DCA},\ref{eq:Ansatz-compact}):%
\begin{subequations}%
\label{subeq:modified-interpolation}%
\begin{align}
    \label{eq:modified-interpolation}
\bQ_\bk(z) & =  \sum_\bK \alpha^\modified_{\bk \comma \bK}\,\bQ_\bK(z) \, ,
\\
\label{eq:Ansatz-compact-modified} 
    \alpha^\modified_{\bk \comma \bK} & = \sum_{j = 1}^\Npatchall \frac{1}{N_j} \phi_{\bk \comma j} \Psi_{j \comma \bK} \, . 
\end{align}%
\end{subequations}%
In \Eq{eq:Ansatz-compact-modified}, $\phi$ and $\Psi$ are defined as above, and $N_j$ is the number of sites in the real-space orbit $O_j$ associated with $\phi_{\bk,j}$ (see \App{app:latticeharmonics}, \Eq{eq:latticeharmonicconstruction-patchsymmetry} for details). The factor $1/N_j$ reduces the contribution of the non-local Fourier coefficients $\phi_{\bk, j > 1}$ relative to the local one, $\phi_{\bk,1}$, thereby counteracting their causality-violating tendencies. However, it must be acknowledged that this prescription  is somewhat arbitrary and spoils the above-mentioned canonical property (ii). 

An alternative fix to ensure causality, already mentioned above, is to drop \Eq{eq:alpha-self-consistency} and instead enforce $[\alpha_{\bk \comma \bK}]_{\bk = \bK} =1$ \cite{Maier2005,Staar2013_DCAinterp,Haehner2020_DCAinterp}. However, this prescription, too, involves arbitrariness, since the choice $\bk = \bK$ singles out one point in the patch and contains no information about the actual shape of the patch. Moreover, typically it does not guarantee canonical property (ii).

By contrast, when DCA interpolation is performed by $\cL$-interpolation, as presented in the next sections, such  problems do not arise. Since $\cL$ is hermitian and its matrix elements are frequency-independent, $\cL$-interpolation respects causality by construction.  It can thus be used in conjunction with canonical DCA interpolation, thus also ensuring  the canonical properties (i)-(iii).

\changed{\subsection{$\cL$-interpolation}
\label{subsec:Lperiodization}}

Conceptually, our $\cL$-interpolation scheme is very simple: we compute the dynamical part of the self-energy, construct its CFE coefficients, and interpolate  the latter using the same strategy as in \Sec{sec:trad_periodization}---but now applied to $n$-dependent coefficients instead of $z$-dependent functions, which makes all the difference for ensuring causality. Concretely, we proceed as follows.

\bigskip 

\subsubsection{$\cL$-interpolation of CFE coefficients $\bolde_n$, $\bt_n$}
\label{sec:CFE-coefficient-interpolation}

\textit{CDMFT:} Using the Lanczos scheme of \Sec{sec:CFE-via-polynomials}, we construct the CFE of the dynamical part of the CDMFT self-energy, $[\bSigma^\dyn(z)]_{\br,\br'}$ (see \App{app:MatrixElements} for numerical details), obtaining the corresponding hermitian CFE coefficients $[\widetilde \bolde_n]_{\br, \br'}$ and $[\widetilde \bt_n]_{\br, \br'}$. From these we estimate hermitian, real-space Liouvillian matrix elements of the lattice model using  superlattice averaging as in \Eq{eq:estimateQbar}:
\begin{align}
\label{eq:estimateEpsilonbarTbar}
    \overline{\bolde}_{\bdelta n} & = \frac{1}{\Ncluster} \sum_{\br' \in \mathcal{C}} [\widetilde \bolde_n]_{\br' \changed{+ \bdelta},\br'} \, , \quad 
    \overline{\bt}_{\bdelta n} = \frac{1}{\Ncluster} \sum_{\br' \in \mathcal{C}} [\widetilde \bt_n]_{\br' \changed{+ \bdelta},\br'} \, . 
\end{align}
They are cluster-range-limited, vanishing if $|\bdelta|$ exceeds the cluster size. Note that $\overline{\bepsilon}_{\bdelta n} = \overline{\bepsilon}{}^\dag_{-\bdelta n}$ since by construction, $\bepsilon_n = \bepsilon_n^\dagger$, implying $[\widetilde \bepsilon_n]_{\br,\br'} = [\widetilde \bepsilon_n^{\dagger}]_{\br',\br} $ for its matrix elements. Likewise for $\bt_n$.

Finally, we define CDMFT-interpolated   CFE coefficients in analogy to \Eq{eq:define-CDMFT-periodization}:
\begin{align} 
    \bepsilon_{\bk n} & \simeq \sum_{\bdelta} \overline{\bolde}_{\bdelta n} \,  \mr{e}^{-\mr{i} \bk \cdot \bdelta}\, , \quad
    \bt_{\bk n} \simeq \sum_{\bdelta} \overline{\bt}_{\bdelta n} \, \mr{e}^{-\mr{i} \bk \cdot \bdelta} \, .
    \label{eq:eps_periodization-CDMFT}
\end{align}
By construction, we have $\bepsilon^\pdag_{\bk n} = \bepsilon_{\bk n}^\dagger $,  $\bt^\pdag_{\bk n} = \bt_{\bk n}^\dagger $.

The reason why in this work we use superlattice averaging in \Eq{eq:estimateEpsilonbarTbar} to estimate $\overline{\bepsilon}_{\bdelta n}$ and $\overline{\bt}_{\bdelta n}$ is that this provided the most favorable comparisons to benchmark results obtained using MPS methods (see \Sec{sec:1dHubbard_benchmark}). However, instead of superlattice averaging, one could in \Eq{eq:estimateEpsilonbarTbar} also use cluster averaging, or some other scheme, as long as $\overline{\bepsilon}_{\bdelta n} = \overline{\bepsilon}^{\dag}_{-\bdelta n}$ and $\overline{\bt}_{\bdelta n} = \overline{\bt}^{\dag}_{-\bdelta n}$, ensuring that $\bepsilon_{\bk n}$ and $\bt_{\bk n}$ and therefore also the CDMFT-interpolated   Liouvillian are hermitian, leading to causal Green's functions and self-energies after interpolation.  This is in stark contrast to periodization  schemes based on frequency-dependent quantities such as the self-energy or the cumulant, as discussed above.

\textit{DCA:} Here, we use the Lanczos scheme of \Sec{sec:CFE-via-polynomials} to construct CFE coefficients $\bepsilon_{\bK n}$ and $\bt_{\bK n}$ from the dynamical part of the patch self-energies $\bSigma^\dyn_{\bK}(z)$. We then apply the canonical DCA interpolation of \Sec{sec:trad_periodization} to the CFE coefficients, with $\alpha_{\bk \comma \bK}$ constructed according to \Eqs{eq:Ansatz-compact}--\eqref{eq:PhiPsi=1}, to define DCA-interpolated  coefficients, 
\begin{align}
    \label{eq:DCA_epsk_final_DCA}
    \bepsilon_{\bk n}  = \sum_\bK \alpha_{\bk \comma \bK} \, \bepsilon_{\bK n}\, , \quad \bt_{\bk n}  = \sum_\bK \alpha_{\bk \comma \bK} \, \bt_{\bK n}\, .     
\end{align}
They are hermitian,  since the CFE coefficients $\bepsilon_{\bK n}$ and $\bt_{\bK n}$ are hermitian by construction, and the $\alpha_{\bk \comma \bK}$ are real.

Having constructed the interpolated  coefficients $\bepsilon_{\bk n}$ and $\bt_{\bk n}$ from CDMFT or DCA as described above, we are ready to assemble the corresponding interpolated   dynamical self-energy  $\bSigma^{\dyn}_{\bk}(z)$. \rreply{The reassembly details  depend on whether the cluster-DMFT self-energy spectrum $\bW(\omega)$ used as input 
is discrete or not, as we discuss next.

\subsubsection{Broadening}
\label{sec:broadening}
}

If the impurity solver used within cluster-DMFT discretizes the bath, 
\changed{as is the case for ED and NRG}, then both spectral functions and self-energies are discrete functions of frequency. Ultimately, they have to be broadened to estimate the corresponding continuous functions of the impurity model with a continuous bath. 
\rreply{In this regard, we distinguish two protocols:
(1) ``broaden-last'' or (2) ``broaden-first''. Both take discrete spectral functions for impurity correlators as input, but protocol (1) first computes a discrete self-energy and then broadens it, whereas protocol (2) first broadens the impurity spectral functions and then computes a broadened self-energy from the latter. We now discuss these in turn.

(1) \textit{Broaden-last protocol:} Very recently, Zacinskis \textit{et al.} published the following remarkable result \cite{Zacinskis2026}: given a tool for computing discrete spectral representations of Green's functions, it is always possible to extract from them a discrete spectral representation of the self-energy, involving a self-energy spectrum $\bW(\omega)$ comprising a set of discrete poles ($\delta$-peaks). 
This has two major benefits. First, the resulting self-energy is free from broadening artifacts. Second, if the self-energy spectral function $\bW(\omega)$ consists of a finite number of 
discrete poles, then the Lanczos iteration for finding CFE coefficients will terminate automatically after, say, $\NKrylov$ steps. Thus, 
$\cL_{\spi\spi}$ is fully represented by the 
coefficients $\bolde_{n \le N}$, $\bt_{n \le N}$
[\Eq{eq:Lmatrixelements-en-tn-main}],} and the CFE \eqref{eq:SigmadynCFE-main} is an exact representation of $\bSigma^\dyn(z)$. 

We construct its  interpolated  version $\bSigma^\dyn_{\bk}(z)$, and in turn the interpolated   self-energy $\bSigma_{\bk}(z)$, by inserting the interpolated   matrix elements $\bepsilon_{\bk n}$, $\bt_{\bk n}$ into \Eq{eq:SigmadynCFE-main}: 
\begin{align}
\label{eq:SigmaCFE_periodized}
    \bSigma^{\dyn}_{\bk}(z) = \bt_{\bk1} \cfrac{\boldone}{\bz_{\bk1}-  {\bt_{\bk2} \cfrac{\boldone}{\bz_{\bk 2} -  \! \raisebox{-0.9em}{\, \ensuremath{\ddots} \raisebox{-0.5em}{ \hspace{-2mm} \ensuremath{- \bt_{\bk\NKrylov}\cfrac{\boldone}{\bz_{\bk N} } \, \bt_{\bk\NKrylov}}}}} \bt_{\bk2} \,}} \bt_{\bk1} \, ,
\end{align}
with $\bz_{\bk n} \!=\!  \bz  - \bepsilon_{\bk n}$. \rreply{This interpolated representation guarantees causality exactly by construction. However, its spectral function, $\bW_\bk(\omega)$, is still discrete. As a last step, 
it can be broadened via some broadening
scheme. 

We have employed the broaden-last protocol (1) for the CDMFT+TaSK data of \Sec{sec:1dHubbard_CDMFT} and the CDMFT+ED data of \Sec{sec:CDMFT+ED2DHubbard}; the broadening schemes used for these purposes are discussed in those sections.}

\rreply{(2) \textit{Broaden-first protocol:} Before the work of Zacinskis \textit{et al.}  \cite{Zacinskis2026}, the standard strategy for computing self-energies from discrete impurity spectral functions has long been to broaden the latter first, then use the broadened ones as input for computing the self-energy (using, e.g., the Dyson equation directly, or Bulla's asymmetric improved estimator \cite{Bulla1998}, or Kugler's symmetric improved estimator \cite{Kugler2022}). 
The resulting self-energy spectrum $\bW(\omega)$ is continuous and cannot be represented by a finite number of discrete poles.
Therefore, the matrix representation of $\cL_{\spi\spi}$ is infinite and generically cannot be handeled exactly in practice. However, the self-energy can still be represented by a finite-depth CFE, albeit terminated by residual functions, 
either $\bR_N(z)$ in \Eqs{eq:S(z)CFE-main},  or $\bR^\fast_n(z)$ in \Eq{eq:SigmaCFE_FastSlowMode-main}. These residual functions are Liouvillian resolvents, not matrix elements. Strictly speaking, interpolating them therefore constitutes a departure from pure $\cL$-interpolation.

We construct the interpolated   self-energy $\bSigma_{\bk}(z)$ by interpolating not only $\bolde_n$, $\bt_n$ but also  $\bR_N(z)$
or $\bR^\fast_n(z)$. For $\bR_N(z)$, we employ the traditional interpolation schemes for CDMFT or DCA discussed in \Sec{sec:trad_periodization}, with}
\begin{align}
    \label{eq:periodizing-rest-functions} 
    [\widetilde{\mathbf{Q}}(z)]_{\br,\br'} = [\widetilde \bR_N(z)]_{\br,\br'} \, , \quad
    \mathbf{Q}_{\bK}(z) = \bR_{\bK N}(z) \, ,
\end{align}
respectively, as input to compute $\bR_{\bk n}(z)$, and set $\bz_{\bk \NKrylov} = \bz  - \bepsilon_{\bk \NKrylov} - \bR_{\bk \NKrylov}(z)$ in \eqref{eq:SigmaCFE_periodized}. 

\changed{For a self-energy obtained via NRG spectral functions
with log-Gaussian broadening, we have to employ 
a log-CFE constructed through F/S-splitting [cf.~\Eq{eq:SigmaCFE_FastSlowMode-main}]. Then,} we similarly interpolate  the fast-mode terms $[\widetilde \bR^\fast_n(z)]_{\br,\br'}$ or $\bR^\fast_{\bK n}(z)$ for CDMFT or DCA, and set $\bz_{\bk n} = \bz  - \bepsilon_{\bk n} -  \bR_{\bk n}^{\fast}(z)$ for $n < \NKrylov$ in \eqref{eq:SigmaCFE_periodized}.

\changed{The use of a log-CFE with CFE coefficients and fast-mode residual functions that
decay exponentially $\sim \Lambda^{-n}_\fastslow$  has advantageous consequences:
the fast-mode residual functions quickly become local or equivalently $\bK$-independent. (This is  illustrated in \Fig{fig:CFE_data} in \App{app:dependence_splitting_parameters} for DCA+NRG data). Consequently, the non-trivial momentum-dependence is determined mainly by the CFE coefficients (mostly by $\epsilon_{\bK n}$, see \Fig{fig:CFE_data}).
Further, strongly correlated electron systems often exhibit several exponentially separated energy scales. 
In such cases a usual CFE would require exponential depth; by contrast, the log-CFE requires only depth $n$ to reach energy $\Lambda^{-n}$.}

\rreply{To conclude this section, we remark that our claim that $\cL$-interpolation ensures causality has a caveat:\label{p:caveat-causality} it is strictly true only 
for strict $\cL$-interpolation, which can only be performed in practice if the CFE of the input self-energy terminates exactly. If that is not the case, exact CFE termination requires $z$-dependent residual functions $\Rest_N(z)$ or $\Rest_n^\fast (z)$, as discussed above. Then, our suggestion to interpolate them} using traditional interpolation schemes, can, in principle, lead to similar causality violations for $\bSigma_{\bk}^\dyn(z)$ as those discussed in \Sec{sec:trad_periodization} for traditional periodizations of $\bSigma_\bk(z)$ or $\bM_\bk(z)$. However, for the DCA+NRG computations performed for this work, we find that the functions $\Rest_{\bK n}^{\fast}(z)$ are quite local, \changed{in that their $\bK$-dependence becomes progressively weaker with $n$, as mentioned above. Therefore,} essentially all significantly non-local features of $\bSigma^\dyn(z)$ are encoded in $\bepsilon_{\bK n}$ and $\bt_{\bK n}$, see \App{app:dependence_splitting_parameters}. As a result, we did not encounter any violations of causality. (In case DCA computations do yield causality violations, we recommend using $\alpha_{\bk\comma\bK}^{\mr{mod}}$ of \Eqs{subeq:modified-interpolation} when interpolating  $\Rest_{\bK n}^{\fast}(z)$, thereby reducing the contribution of non-local Fourier coefficients.)

\bigskip

\changed{
\subsubsection{Interpolation of $\bSigma^\Hartree$}
\label{sec:HF-interpolation}
}

DMFT defines the lattice  Green's function as 
\begin{align}
\label{eq:G-to-be-periodized}
    \bG_\bk (z) = \frac{\boldone}{\bz  - \bepsilon_\bk - \bSigma^\Hartree_\bk - \bSigma^\dyn_\bk(z)} ,
\end{align}
\changed{where $\bSigma^\dyn_\bk(z)$ is the interpolated version of the cluster-DMFT impurity self-energy,} and $\epsilon_\bk$, representing the \changed{dispersion of the} free single-particle Hamiltonian,  is exact and does not require interpolation.   The Hartree-Fock term $\bSigma^\Hartree_\bk$, unlike $\bepsilon_\bk$, depends on the density matrix used to evaluate inner products. For purely local interactions, as treated in this work, the HF term is local and identical to that of the impurity model, $\bSigma^\Hartree_\bk =\bSigma^\Hartree$, so it does not need to be interpolated, either. With non-local interactions, however, the HF term does acquire a $\bk$-dependence (even without interpolation) that is only partially captured by the impurity model: it arises from lattice expectation values computed via $\bk$-sums over corresponding lattice Green's functions. However, for CDMFT/DCA, these lattice expectation values are generally not fully periodic/continuous, leading to a non-periodic/non-continuous HF contribution. To resolve this, we propose to recompute the HF term after interpolating  the single-particle irreducible matrix elements, and to iterate this until self-consistency. 

\bigskip

\section{Benchmark: CDMFT interpolation for the 1D Hubbard model} 
\label{sec:1dHubbard_benchmark}

This section is devoted to benchmarking the $\cL$-interpolation of CDMFT data \changed{for the 1-dimensional Hubbard model. 
This model is challenging for DMFT-type approaches, which are controlled by the inverse dimensionality of the lattice model. However, provides a setting in which it is possible to obtain essentially exact benchmark data using matrix product state (MPS) techniques.} \rreply{Moreover, the dependence of CDMFT results on cluster size can be investigated systematically}. By comparing $G$-, $\Sigma$-, $M$- and $\cL$-interpolation results to \changed{our benchmark data}, we establish which interpolation schemes perform best. (Reminder: in this section, ``interpolation''\ stands for ``periodization'', i.e.\ real-space interpolation.)

In \Sec{sec:1dHubbard_MPS}, we use MPS techniques to compute and investigate the full frequency \textit{and} momentum dependence of its single-particle spectral function and self-energy. These results are numerically (almost) exact, and we will view them as the ``ground truth" for this model. We use our MPS results (i) to show that $\cL$ is indeed more short-ranged than $\Sigma$ or $M$; and (ii) as benchmark data against which we subsequently, in \Sec{sec:1dHubbard_CDMFT}, compare interpolated  CDMFT results obtained using several different schemes: $G$-, $\Sigma$-, $M$- and $\cL$-interpolation.

Concretely, we consider the Hamiltonian 
\begin{align}
    \label{eq:H_1dHubbard}
    H = -\sum_{\ell\sigma} \bigl(c^{\dagger}_{\ell\sigma} c^{\phantom{\dagger}}_{\ell+1\sigma} + \mr{h.c.} \bigr) - \mu \sum_{\ell\sigma} n_{\ell\sigma} + U \sum_{\ell} n_{\ell\uparrow} n_{\ell\downarrow} ,
\end{align}
for a chain of $L=50$ sites. We set $U=5$ and $\mu = U/2$ (half-filling), yielding a  Mott insulator.

\begin{figure}[tb!]
    \includegraphics[width=0.85\linewidth]{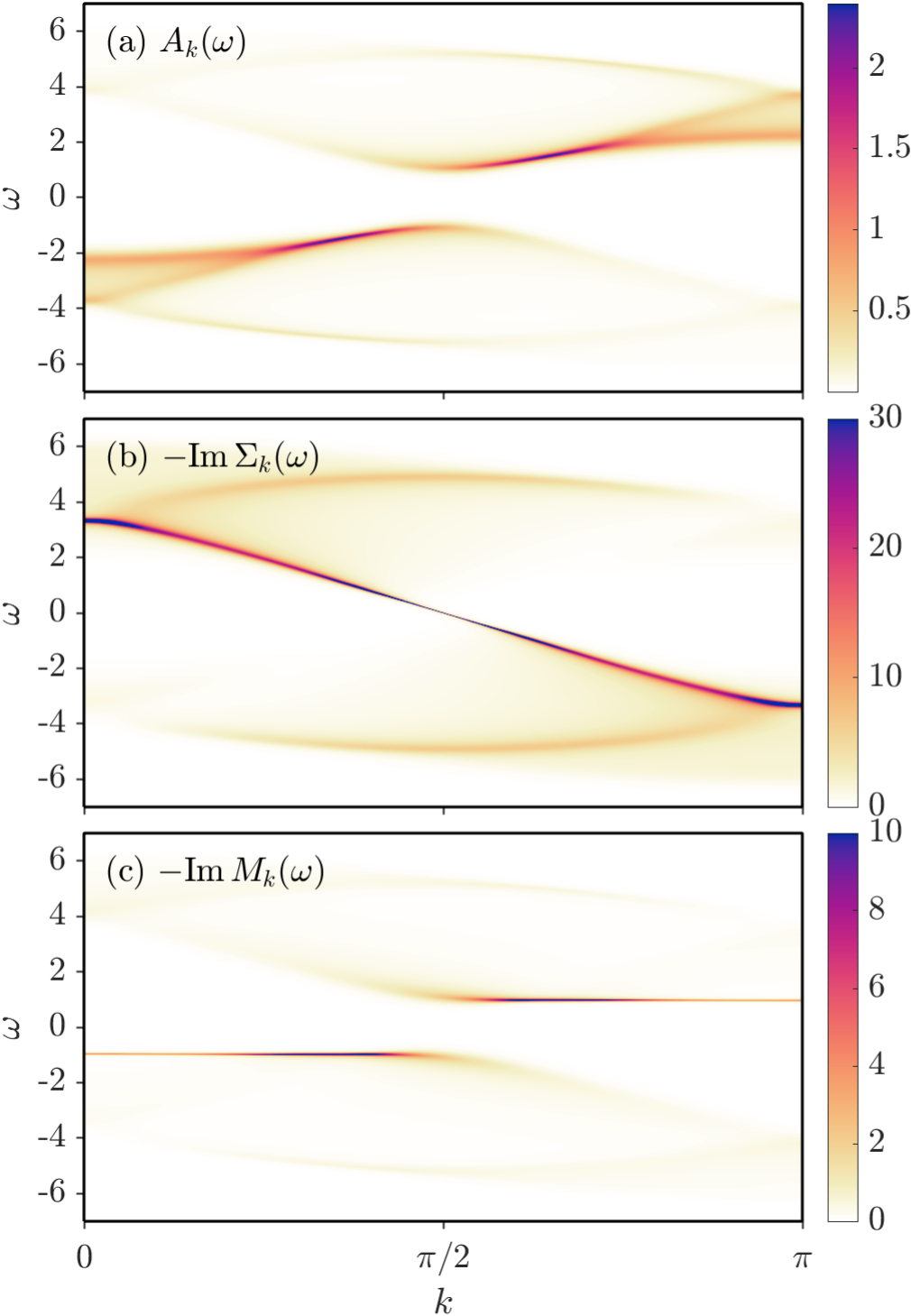} 
    \caption{\changed{TaSK results for 1D Hubbard model: (a) spectral function, (b) self-energy, and (c) cumulant.} \rreply{To obtain smooth data, we have treated $k$ as a continuous variable and for each $k$ broadened the discrete-in-$\omega$ spectrum (see \App{app:TaSK_details}). The same was done for Figs.~\ref{fig:Hubbard1d_MPS_SE_M_L_expansions} and \ref{fig:Hubbard1d_CDMFT_SE_periodized} below.}}
    \label{fig:Hubbard1d_MPS_phsym}
\end{figure}

\changed{
\subsection{TaSK for 1D Hubbard model}
\label{sec:1dHubbard_MPS}}

\changed{In this section, we obtain benchmark-quality results for the 1D Hubbard model using MPS methods.}
We compute the ground state of the Hubbard chain using the density matrix renormalization group~(DMRG). We then compute the single-particle spectral function $A_k(\omega)$ associated with the retarded correlator
\begin{align}
    \label{eq:Grk-1D-Hubbard}
    \tG_k^\mr{R}(t) = - \mr{i} \theta(t) \bigl\langle \bigl\{ c_{k\changed{\sigma}} (0),c_{k\changed{\sigma}}^{\dagger}(-t)\bigr\} \bigr \rangle ,
\end{align} 
where $c_{k\changed{\sigma}} = \tfrac{1}{\sqrt{L}}\sum_{\ell=1}^L e^{- \mr{i} k \ell}c_{\ell \sigma}$ \rreply{and $k \in (-\pi,\pi]$ is treated as a continuous variable.}
\changed{To this end, we use the tangent-space Krylov (TaSK) approach \cite{Kovalska2025}. TaSK is a  recently-developed, state-of-the-art, MPS-based method for computing real-frequency spectral functions. It} 
yields the spectral function in the form of a discrete Lehmann representation, $A^\discrete_k(\omega)$.  The eigenstates and eigenenergies involved therein are computed via a Lanczos scheme that constructs a Krylov basis within the 1-site tangent space of the ground state. $A^\discrete_k(\omega)$ is a discrete function of $\omega$ (since the chain has finite length). We broaden it to obtain $A_k(\omega)$ as a smooth function of $\omega$. For details see \App{app:TaSK_details} and Ref.~\onlinecite{Kovalska2025}.

From $A_{\bk}(\omega)$ we compute $G_{\bk}(\omega)$ [\Eq{eq:defineA}], and then $\Sigma_{\bk}(\omega)$ and $M_{\bk}(\omega)$ [\Eqs{subeqs:define_G_M}]. The resulting spectral function $A_{k}(\omega)$, self-energy $-\mr{Im} \Sigma_k(\omega)$, and cumulant $-\mr{Im} M_k(\omega)$ are shown in Figs.~\ref{fig:Hubbard1d_MPS_phsym}(a-c).
As expected, $A_{k}(\omega)$ exhibits a Mott gap while $-\mr{Im} \Sigma_k(\omega)$ has a dispersive pole. The main feature of $-\mr{Im} M_k(\omega)$, on the other hand, is a $k$-independent peak at negative frequencies for $0\leq k \leq \pi/2$ and another flat peak at positive frequencies at $\pi/2 \leq k \leq \pi$. Thus, the features of $-\mr{Im} M_k(\omega)$ are significantly less dispersive than those of both $A_k(\omega)$ and $-\mr{Im} \Sigma_k(\omega)$. As a result, $M$-interpolation is expected to yield better results than $\Sigma$-interpolation for the present parameters where the model is in a Mott phase. This aligns with findings in the literature~\cite{Stanescu2006_periodization2}.

In order to test more quantitatively which interpolation scheme may yield the best outcomes, we Fourier-transform to position space.
Since the Hubbard model is inversion symmetric, we analyze and truncate its cosine coefficients
\begin{subequations}
\label{eq:Fourier-coefficients-r}
\begin{align}
  \tSigma_{r}(\omega) &= \int_{-\pi}^{\pi} \frac{\mr{d} k}{2\pi}  (2-\delta_{r,0})\cos(k r) \, \Sigma_k(\omega) \, ,
  \\
 \tM_{r}(\omega) &= \int_{-\pi}^{\pi} \frac{\mr{d} k}{2\pi}  (2-\delta_{r,0})\cos(k r) \, M_k(\omega) \, ,
\end{align}
\end{subequations}
instead of the Fourier coefficients. Truncating  
the former is equivalent to truncating the latter 
if the truncation should respect inversion symmetry. Below, we refer to cosine expansions if we mean the current case, and to Fourier expansions for more generic statements.

\begin{figure}[t] 
    \includegraphics[width=\linewidth]{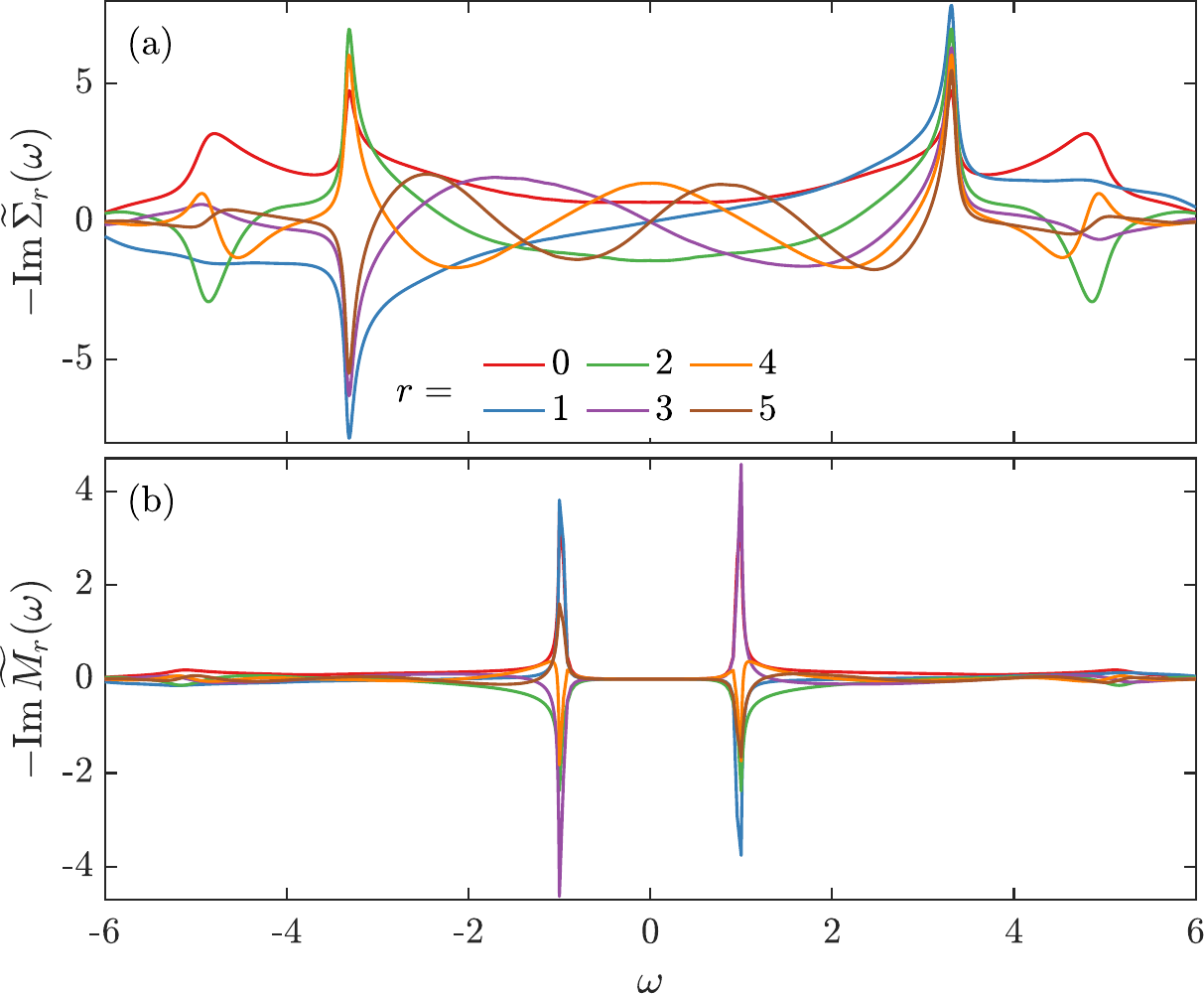} 
    \caption{First six Fourier coefficients of the imaginary part of (a) the self-energy of \Fig{fig:Hubbard1d_MPS_phsym}(b) and (b) cumulant of \Fig{fig:Hubbard1d_MPS_phsym}(c).}
    \label{fig:Hubbard1d_MPS_SE_M_Fourier_phsym}
\end{figure}

For $r\leq5$, the cosine coefficients $\tSigma_{r} (\omega)$ and $\tM_{r}(\omega)$ are shown in \Fig{fig:Hubbard1d_MPS_SE_M_Fourier_phsym}(a) and (b), respectively. The maximal absolute amplitude of $-\mr{Im} \tM_{r}(\omega)$ clearly decays faster with increasing $r$ than that of 
$-\mr{Im} \tSigma_{r}(\omega)$. This suggests that for this application, $M$-interpolation will outperform $\Sigma$-interpolation.  Indeed, we confirm this explicitly below. 

\begin{figure}[t] 
    \includegraphics[width=\linewidth]{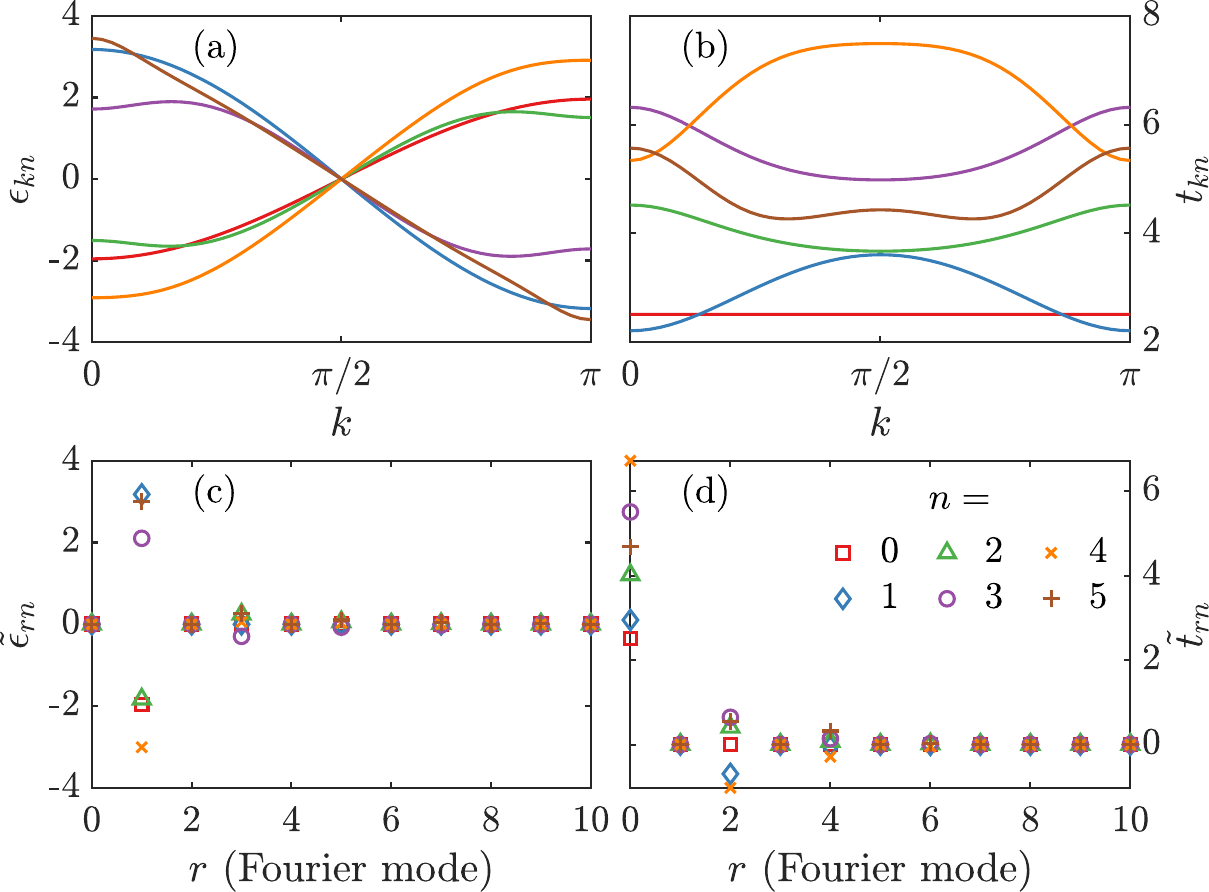}
    \caption{(a,b) Matrix elements $\epsilon_{kn}$ and $t_{kn}$ of $\cL$, extracted via a CFE from the self-energy spectrum of \Fig{fig:Hubbard1d_MPS_phsym} (b). (c,d) The corresponding cosine coefficients $\tepsilon_{rn}$ and $\tildet_{rn}$ up to order $r=10$. The color scheme of panel (d) applies to all panels.}
    \label{fig:Hubbard1d_MPS_LiouvilleMatEle}
\end{figure}

\begin{figure*}[t] 
    \includegraphics[width=0.85\linewidth]{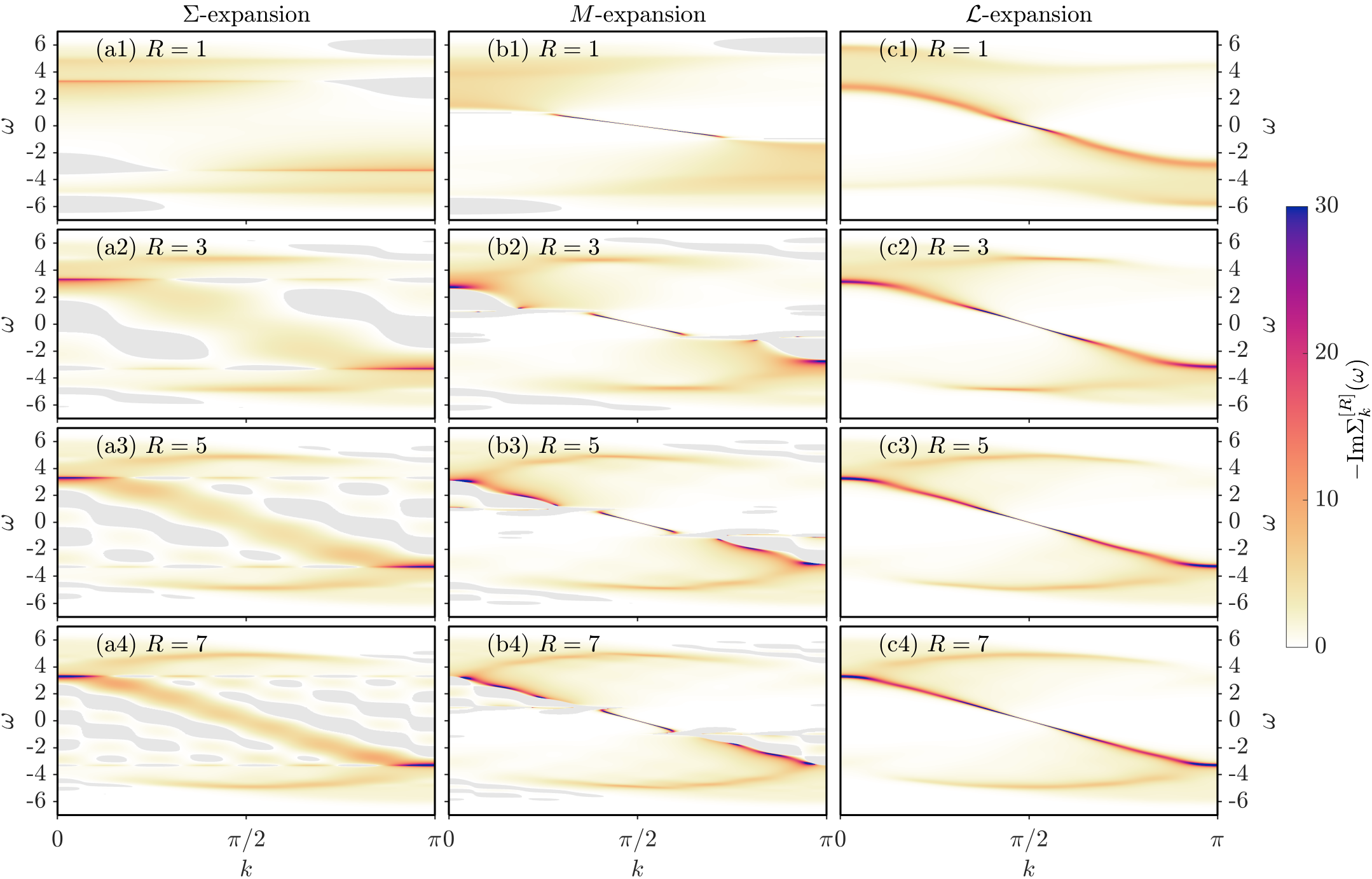} 
    \caption{\changed{TaSK results for 1D Hubbard model:} self-energy $- \mr{Im} \Sigma^{[R]}_k(\omega)$ obtained from truncating the Fourier series of $\Sigma$-, $M$- or $\cL$-expansions at $r \le R$, as described in the text. Regions where $-\mr{Im} \, \Sigma^{[R]}_{k}(\omega) < 0$ (non-causal behavior) are depicted in grey. \vspace{2mm}}    \label{fig:Hubbard1d_MPS_SE_M_L_expansions}
\end{figure*}

Now consider the truncated cosine series, 
\begin{subequations}
\label{eq:truncatedFourier}
\begin{align}
    \label{eq:truncatedFourier-a}
    \Sigma^{[R]}_k(\omega) &= \sum_{r = 0}^{R} \tSigma_{r} (\omega) \cos(k r) \, , \\
    \label{eq:truncatedFourier-b}
    M^{[R]}_k(\omega) &= \sum_{r = 0}^{R} \tM_{r} (\omega) \cos(k r)\, . 
\end{align} 
\end{subequations}
Here, $r$ plays the role of $\bdelta$ from Sec.~\ref{sec:LiouvillianPeriodization}. (Compared to Eq.~\eqref{eq:defineSigmak_Via_FT}, we combined terms with $\mr{e}^{\pm \mr{i} k r}$ in the sum above, so that the magnitude of the non-local coefficients $\tSigma_{r} (\omega)$ with $r > 0$ is directly comparable to the local coefficient $\tSigma_{0} (\omega)$, and similarly for $M$.) Such truncated Fourier series will generically lead to non-causal features (positive imaginary parts for $\Sigma_k(\omega)$ or $M_k(\omega)$)  due to ringing~\cite{Okamoto2003} because the truncation discards $\omega$-dependent terms. Consider, for instance, approximating $\Sigma_k(\omega)$ by the first two terms of its cosine expansion,
\begin{align}
    \label{eq:SigmaFourier_R=1}
    \Sigma^{[1]}_k(\omega) =\tSigma_0(\omega) + \cos(k) \,\tSigma_1(\omega) \, .
\end{align}
As shown in Fig.~\ref{fig:Hubbard1d_MPS_SE_M_Fourier_phsym}, there exist frequencies  where $-\mr{Im} \tSigma_0(\omega) < -\mr{Im} \tSigma_1(\omega)$. At these frequencies, the approximation $\Sigma^{[1]}_k(\omega)$  will be non-causal at $k=\pi$, because $-\mr{Im} \Sigma^{[1]}_{k=\pi}(\omega) = -\mr{Im} [\tSigma_0(\omega) - \tSigma_1(\omega)] < 0$. 
In general, $\Sigma^{[1]}_k(\omega)$ will have non-causal features whenever $|\mr{Im} \tSigma_0(\omega)| < |\mr{Im} \tSigma_1(\omega)|$, and likewise for $M^{[1]}_k(\omega)$. Less severe truncation by keeping more cosine coefficients (thus considering longer-ranged terms) gradually reduces non-causal features. We discuss this in more detail below, together with \Fig{fig:Hubbard1d_MPS_SE_M_L_expansions}.

As just discussed, non-causal behavior can arise when truncating Fourier series with $\omega$-dependent coefficients. This problem can be avoided by adopting a strategy in which all Fourier series to be truncated involve only $\omega$-\textit{independent} coefficients.  To this end, we consider the CFE representation of $G_k(\omega^+)$, which amounts to a change of variables from $\omega$ to $n$: it encodes the $\omega$ dependence explicitly in a continued fraction involving  $\omega$-\textit{independent} CFE coefficients $\epsilon_{kn}$ and $t_{kn}$ . Fourier representations of the $k$ dependence of $\epsilon_{kn}$ and $t_{kn}$ can then be safely truncated without yielding non-causal behavior.  

To illustrate this, we have computed a CFE for $G_k(\omega^+)$ starting from our TaSK result for the discrete spectral function $A^\discrete_k(\omega)$, using the strategy discussed after \Eq{eq:exact-CFE-for-G}. Figures~\ref{fig:Hubbard1d_MPS_LiouvilleMatEle}(a) and \ref{fig:Hubbard1d_MPS_LiouvilleMatEle}(b) show the resulting Liouvillian matrix elements $\epsilon_{kn}$ and $t_{kn}$, respectively, for $n \leq 5$. For fixed $n$ they are smooth functions of $k$. Their cosine coefficients, 
\begin{subequations}
\label{eq:fourier_coeff}
\begin{align}
        \tepsilon_{rn} &= \int_{-\pi}^{\pi} \frac{\mr{d} k}{2\pi}  (2-\delta_{r,0})\cos(k r) \, \epsilon_{kn}\, ,\quad \\
        \tildet_{rn} &= \int_{-\pi}^{\pi} \frac{\mr{d} k}{2\pi}  (2-\delta_{r,0})\cos(k r) \,  t_{kn} 
\end{align}
\end{subequations}
turn out to decay quickly with $r$. This can be seen from \Figs{fig:Hubbard1d_MPS_LiouvilleMatEle}(c) and \ref{fig:Hubbard1d_MPS_LiouvilleMatEle}(d), showing the cosine coefficients $\tepsilon_{rn}$ and $\tildet_{rn}$, respectively. We find that $\tepsilon_{rn} = 0$ for even $r$ while $\tildet_{rn} = 0$ for odd $r$ (because particle-hole symmetry implies $\epsilon_{k,n} = -\epsilon_{k+\pi,n}$ and $t_{k,n} = t_{k+\pi,n}$). We find that the magnitude of the cosine coefficients decays monotonically with $r$, $|\tepsilon_{1,n}| > |\tepsilon_{3,n}| > \dots$ and $|\tildet_{0,n}| > |\tildet_{2,n}| > \dots$\, . This suggests that the cosine series, defined as 
\begin{align}
    \label{eq:Fourier-Series-Liouvillian}
    \epsilon^{[R]}_{kn} &= \! \sum_{r = 0}^{R} \tepsilon_{rn}  \cos(kr) , \quad t^{[R]}_{kn} = \sum_{r = 0}^{R} \tildet_{rn}  \cos(kr) , 
\end{align}
converge smoothly, and that truncating them will not cause significant ringing artifacts.   Note that even if ringing does occur, the resulting spectral functions and self-energies are still guaranteed to be causal, in stark contrast to the truncated cosine series of $\Sigma_{k}(\omega)$ or $M_{k}(\omega)$. The reason is that \textit{any} approximation of $\epsilon_{kn}$ or $t_{kn}$ that leaves $\epsilon_{kn}$ real leads to a hermitian $\cL$ and therefore to causal spectra.

Our discussion so far suggests that a truncated Fourier expansion of $\cL$ has much more favorable convergence properties than a truncated expansion of $\Sigma$ or $M$. We verify this explicitly in \Fig{fig:Hubbard1d_MPS_SE_M_L_expansions}, where we investigate how $\Sigma^{[R]}_k(\omega)$, obtained via $\Sigma$, $M$ or $\cL$ expansions involving truncated ($r \le R$) summations, evolves when the expansion order $R$ is increased.

The first column, \Figs{fig:Hubbard1d_MPS_SE_M_L_expansions}(a1-a4), shows the evolution with $R$ of the self-energy $\Sigma^{[R]}_k(\omega)$,  represented by the truncated cosine series \Eq{eq:truncatedFourier-a}. In \Fig{fig:Hubbard1d_MPS_SE_M_L_expansions}(a1) with $R=1$, only the first two coefficients are retained, i.e.\ \Eq{eq:SigmaFourier_R=1}.
This corresponds to neglecting any contributions to the self-energy beyond nearest neighbors, similar to 2-site CDMFT/DCA. Clearly, the approximated self-energy in \Fig{fig:Hubbard1d_MPS_SE_M_L_expansions}(a1) does not capture the main features of the actual self-energy shown in \Fig{fig:Hubbard1d_MPS_phsym}(b).
The dispersive Mott pole, the most distinctive feature of the self-energy in \Fig{fig:Hubbard1d_MPS_phsym}(b), is not captured at all by the truncated Fourier expansion of $\Sigma_k(\omega)$.
Further, as discussed earlier, the truncated cosine expansion yields non-causal artifacts (positive imaginary part) due to ringing, shown as grey areas.
With increasing expansion order [\Fig{fig:Hubbard1d_MPS_SE_M_L_expansions}(a2-a4)], the correct features are better captured, but convergence is very slow in the expansion order.
Even for $R = 7$, which corresponds to considering up to 7 nearest-neighbors (as one would do for 8-site CDMFT/DCA), the self-energy is not fully converged and still shows non-causal features due to ringing. 

The second column, \Fig{fig:Hubbard1d_MPS_SE_M_L_expansions}(b1-b4), shows the evolution with $R$ of the self-energy $\Sigma^{[R]}_k(\omega) = \omega + \mu - [M^{[R]}_k(\omega)]^{-1}$ when $M^{[R]}_k(\omega)$ is represented by the cosine expansion \eqref{eq:truncatedFourier-b} truncated at $r \le R$. In contrast to the truncated $\Sigma$ expansion, the truncated $M$ expansion captures the dispersive pole of the self-energy already at the lowest order shown, $R=1$.
However, a low-order truncation also leads to significant non-causal ringing artifacts, appearing prominently in the vicinity of the dispersive pole. Further, the shape of the pole dispersion is not very well captured at low orders. Even at the highest order shown, $R =7$ in \Fig{fig:Hubbard1d_MPS_SE_M_L_expansions}(b4), the self-energy is not captured very well, and there are still non-causal ringing artifacts left (though to a lesser extent than for the $\Sigma$ expansion). Overall, truncating the Fourier expansion of $M$ leads to somewhat better results than truncating that of $\Sigma$.

The third column, \Fig{fig:Hubbard1d_MPS_SE_M_L_expansions}(c1-c4), shows a self-energy $\Sigma^{[R]}_k(\omega)$ obtained as follows: using $A^\discrete_k(\omega)$ as input, we performed a finite-depth CFE expansion of $G_k(\omega)$ to obtain the Liouvillian matrix elements $\epsilon_{kn}$ and $t_{kn}$; these were cosine transformed via \Eq{eq:fourier_coeff} to obtain $\tepsilon_{rn}$ and $\tildet_{rn}$, from which we constructed $R$-truncated Fourier expansions $\epsilon^{[R]}_{kn}$ and $t^{[R]}_{kn}$ [\Eq{eq:Fourier-Series-Liouvillian}]. We used these to compute $\Sigma^{[R]}_k(\omega)$ via the finite-depth CFE \eqref{eq:SigmadynCFEexplicitfiniteN} while using the broadening scheme of Ref.~\onlinecite{Kovalska2025}, Sec.~S-2.2, henceforth to be called ``CFE+Gaussian broadening''. \changed{It uses the last part of 
the  CFE to define a discrete residual function and applies Gaussian broadening to the latter.}

As can be seen from \Fig{fig:Hubbard1d_MPS_SE_M_L_expansions}(c1-c4), non-causal features are absent at any truncation order, in stark contrast to the truncated $\Sigma$ or $M$ expansion. Further, the correct dispersion of the self-energy pole is already captured at the lowest order, $R = 1$, with higher orders revealing more and more details. Qualitative convergence is reached at $R = 3$ (as one would use for 4-site CDMFT/DCA). At order $R = 5$ and beyond, the self-energy appears to be quite well converged, changes beyond this order are only minor.

To summarize, our analysis of numerically (almost) exact MPS data reveals that truncating the Fourier expansion of $\cL$ matrix elements leads to faster and more controlled convergence than truncating that of $\Sigma$ or $M$. Additionally, truncating the $\cL$ expansion leads to causal spectra at all expansion orders, in stark contrast to truncating the $\Sigma$ or $M$ expansion where non-causal features appear at low orders. Therefore, we can expect $\cL$-interpolation to outperform $\Sigma$ or $M$-interpolation.  We explicitly explore this in \Sec{sec:1dHubbard_CDMFT}.

\rreply{While the above discussion focused on the specific case study of the 1D Hubbard model, we expect the real-space Liouvillian to be generically shorter-ranged than the real-space self-energy,
based on the following arguments: 
acting with the Liouvillian $n$ times on a local operator $\sket{c_{\br \sigma}}$ involves real-space Liouvillian matrix 
elements $\widetilde{\bolde}_{\bdelta m}$ and $\widetilde{\bt}_{\bdelta m}$
(shorthand for $[\widetilde \bolde_m]_{\br'+ \bdelta,\br'}$ and $[\widetilde \bt_m]_{\br'+ \bdelta,\br'}$ with $\br' \in \mathcal{C}$), 
with $m \leq n$. This amounts to $n$ applications of the Hamiltonian. If the latter has range $R$, $\mc{L}^{n} \sket{c_{\br \sigma}}$ has non-trivial support on at most $R n$ sites, i.e.\ the matrix elements $\widetilde{\bolde}_{\bdelta m}$ and $\widetilde{\bt}_{\bdelta m}$ smear the local operator over at most $R n$ sites.
We note that due to orthogonalization and normalization, the range of the matrix elements $\widetilde{\bolde}_{\bdelta n}$ and $\widetilde{\bt}_{\bdelta n}$ can nevertheless exceed $R n$, but we expect them to generically decay quickly with distance. By contrast, the real-space self-energy can have an arbitrarily long range if the real-space Liouvillian has a nearest-neighbor contribution. For instance, consider a momentum-space self-energy of the form $\Sigma_{\bk}(z) \propto (z - \epsilon_{\bk 1})^{-1}$, where the Liouvillian matrix element $\epsilon_{\bk 1}$ stems from a real-space Liouvillian containing 
contains nearest-neighbor contributions. 
Its high-frequency expansion is $\Sigma_{\bk}(z) \propto z^{-1} \sum_{n \geq 0} (\epsilon_{\bk 1}/z)^n$.
Since $(\epsilon_{\bk 1})^n$ has a range of up to $n$ sites in real space, $\widetilde{\Sigma}_{\bdelta}(z)$ has an arbitrarily long range.}

\begin{figure*}[p!]
    \includegraphics[width=0.85\linewidth]{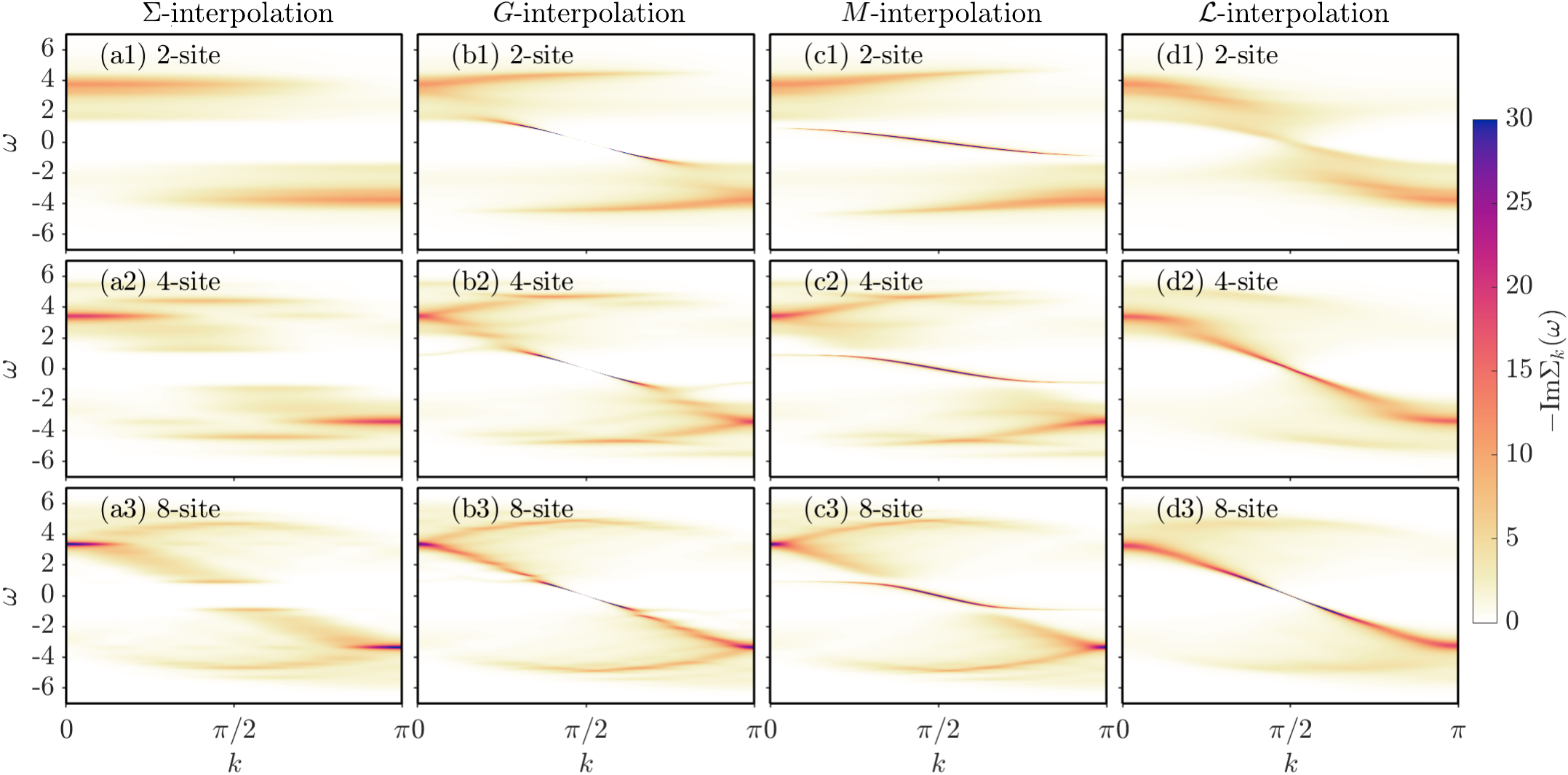} 
    \includegraphics[width=0.85\linewidth]{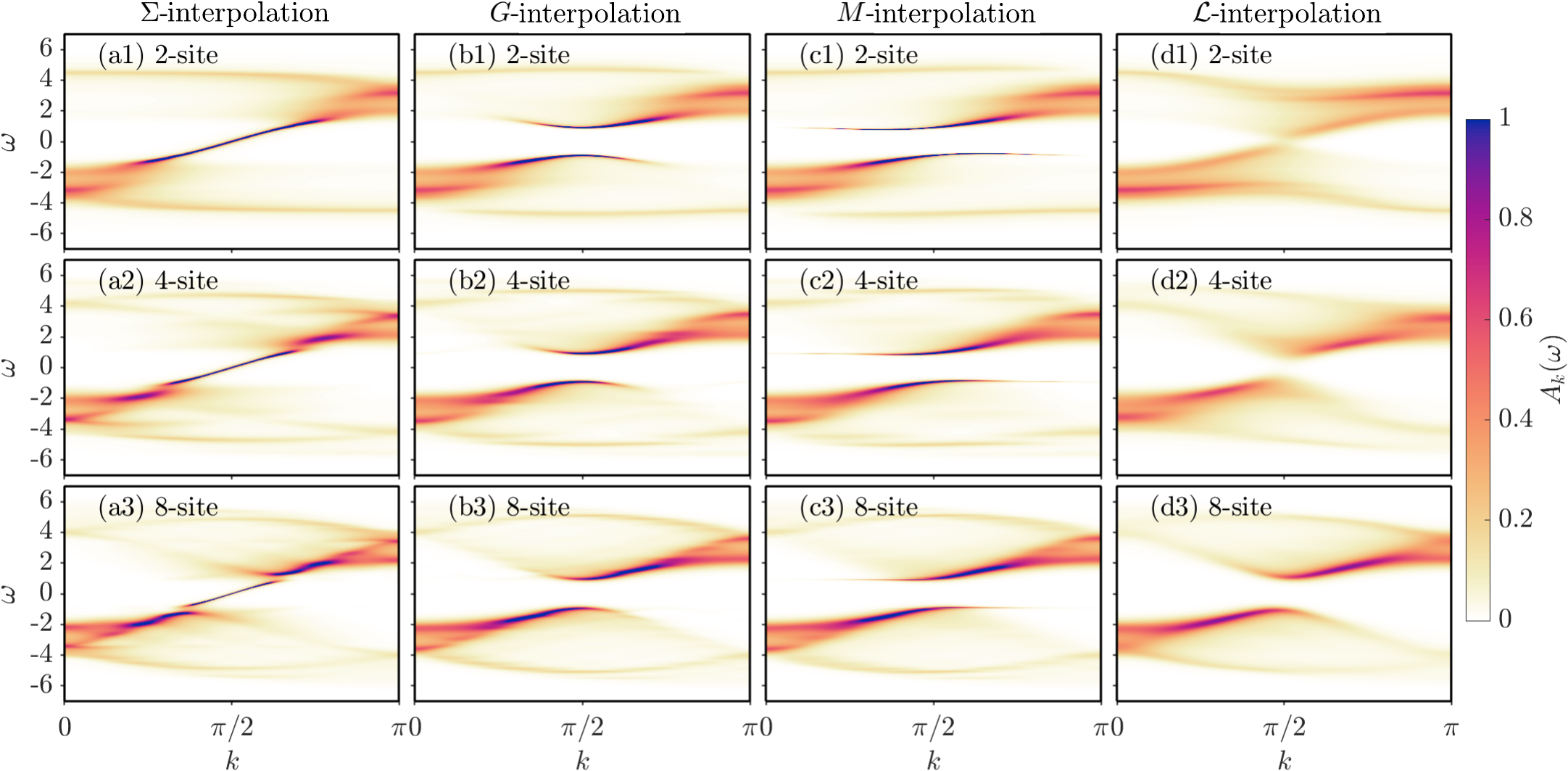} 
    \includegraphics[width=0.85\linewidth]{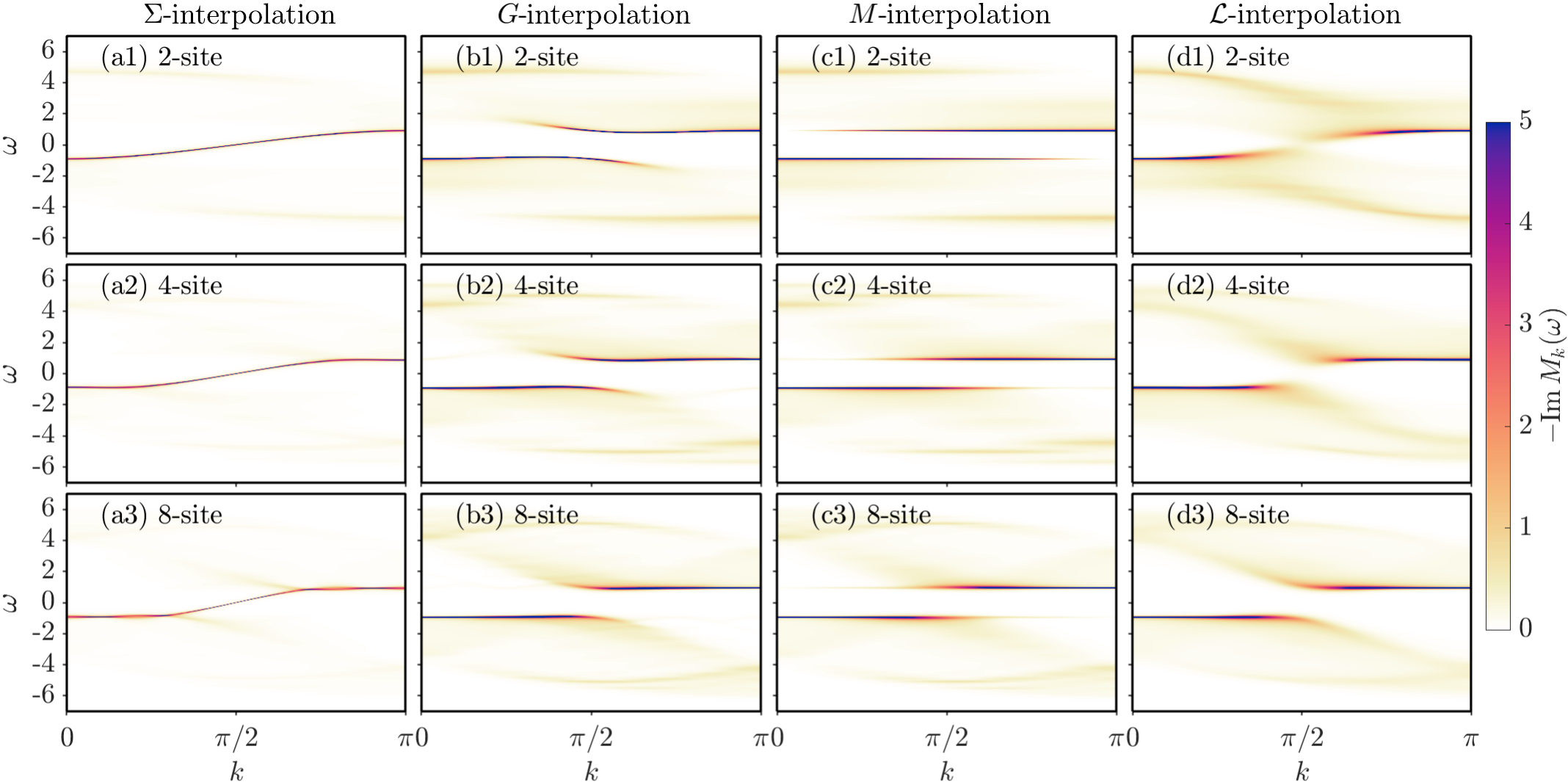} 
    \vspace{-2mm}
        \caption{CDMFT+TaSK results for 1D Hubbard model: \changed{self-energy $-\mr{Im} \Sigma_k(\omega)$ (top part), spectral function $A_k(\omega)$ (center part), and cumulant  $-\mr{Im} M_k(\omega)$ (bottom part)} for different cluster sizes $\Ncluster$, obtained using four different interpolation schemes.  
        }    \label{fig:Hubbard1d_CDMFT_SE_periodized}
\end{figure*} 

\bigskip 

{\changed{\subsection{CDMFT+TaSK for 1D Hubbard model}
\label{sec:1dHubbard_CDMFT}}

To illustrate how $\mc{L}$-interpolation performs compared to $M$- or $\Sigma$- interpolation when applied to CDMFT data, we perform CDMFT+TaSK calculations for the 1D Hubbard model analyzed in the previous section. \rreply{(For a pioneering study on CDMFT for 1D models, see Ref.~\cite{Capone2004}.)} 

\changed{The cluster impurity model that we solve with CDMFT+TaSK} contains as impurity an open-boundary Hubbard chain of $\Ncluster$ sites (thus $\Ncluster$ is the cell size); its two outermost sites on the left and right hybridize with two self-consistent baths with continuous spectra \cite{Kotliar2001_CDMFT}. \changed{Having converged the CDMFT loop, we compute the discrete impurity self-energy using TaSK (see \App{app:TaSK-impuritySolver} for details). The result is a $\Ncluster \times \Ncluster$ matrix-valued function $\widetilde \bSigma{}^\discrete_\imp (\omega)$ with a discrete frequency spectrum.} 

\bigskip 

\changed{
\subsubsection{CDMFT+TaSK interpolation details}
\label{sec:CDMFT+TaSK-interpolation-details}}

We use $\widetilde \bSigma{}^\discrete_\imp (\omega)$ as input for two purposes:

(1) \changed{For $\cL$-interpolation, we construct a CFE following the broaden-last protocol (cf.~\Sec{sec:broadening}):} First, we compute the CFE of the discrete $\widetilde \bSigma{}^\discrete_\imp (\omega)$ via the Lanczos scheme of \App{sec:CFE-via-polynomials} to obtain the single-particle irreducible Liouvillian matrix elements $\widetilde \bepsilon_n$ and $\widetilde \bt_n$. Next, we $\mc{L}$-interpolate  them, using the superlattice average \eqref{eq:estimateEpsilonbarTbar} to obtain $\overline \bepsilon _{rn}$, $\overline \bt_{rn}$, and \Eq{eq:eps_periodization-CDMFT} to obtain $\bepsilon_{kn}$ and $\bt_{kn}$. We use these in \Eq{eq:SigmaCFE_periodized} to construct an $\cL$-interpolated   self-energy $\bSigma_k(\omega)$. \changed{Lastly, we broaden $\bSigma_k(\omega)$ for individual $k$-values using the CFE+Gaussian broadening scheme of Ref.~\cite{Kovalska2025}, Sec.~S-2.2,
using $\sigma = 0.3$ and $\mc{N}_{\mr{CFE}} = 1$.

(2) \changed{For $\Sigma$-, $G$-, or $M$-interpolation, we follow the broaden-first protocol:}
First, we use CFE+Gaussian broadening \cite{Kovalska2025} to obtain a smooth version of the impurity self-energy, $\widetilde \bSigma_\imp (\omega)$. The latter serves as input for traditional $\Sigma$-, $G$-, $M$-interpolation. For this, we employ superlattice averaging [\Eqs{eq:define-CDMFT-periodization}, \eqref{eq:estimateQbar}], using for $\widetilde \bQ(\omega)$ either the broadened $\widetilde \bSigma (\omega)$, or corresponding functions $\widetilde \bG (\omega)$ or $\widetilde \bM (\omega)$, obtained from $\widetilde \bSigma (\omega)$ in standard manner (Eqs.~(7) and (6) of Ref.~\onlinecite{Verret2022_CompactTiling}). 

The self-energies thus obtained from $\cL$-, $\Sigma$-, $G$-, and $M$-interpolation can be compared to the self-energy computed directly with TaSK in Sec.~\ref{sec:1dHubbard_MPS}, as described next.} 

\bigskip 

\changed{
\subsubsection{CDMFT+TaSK interpolation results}
\label{sec:CDMFT+TaSK-interpolation-results}}

\changed{Figure~\ref{fig:Hubbard1d_CDMFT_SE_periodized} compares the lattice self-energy (top part), spectral function (center part), 
and cumulant (bottom part)} for three cluster sizes, $\Ncluster \in \{2,4,8\}$, in each case obtained in four different ways, via $\Sigma$-, $G$-, $M$- and $\mc{L}$-interpolation. \rreply{None of the 36 subpanels in these figures contain any causality-violating regions, because we used superlattice averaging for $\Sigma$-, $G$- and $M$-interpolations, as detailed in the preceding paragraph.}

\changed{Consider first the self-energies $-\mathrm{Im} \Sigma_k(\omega)$ [\Fig{fig:Hubbard1d_CDMFT_SE_periodized}, top part]. The first column [\Fig{fig:Hubbard1d_CDMFT_SE_periodized}, top, (a1-a3)]} shows the results of $\Sigma$-interpolation. Even at the largest cluster size, this scheme fails to reproduce the dispersive pole of the self-energy, i.e.\ it fails to correctly capture the Mott physics in this model. This behavior is in line with the analysis of the truncated cosine series in Fig.~\ref{fig:Hubbard1d_MPS_SE_M_L_expansions}, where we have seen that a large number of cosine coefficients are required to capture the dispersive pole. $\Sigma$-interpolation is based on the assumption that the self-energy is weakly $k$-dependent, and for the 1D Hubbard model, this is simply not the case.

\changed{The next two columns  [\Fig{fig:Hubbard1d_CDMFT_SE_periodized}, top, (b1-b3) and (c1-c3)]} show the self-energy resulting \changed{from} $G$- and $M$-interpolation, respectively. Both schemes work very well in producing the dispersive Mott pole, even for the small 2-site cluster. 
However, for the larger clusters (4-site and 8-site), stripy structures show up which are artifacts of the interpolation.  These stripy structures are more pronounced for $G$-interpolation than for $M$-interpolation. For the 2-site cluster, $G$-interpolation seems to perform better than $M$-interpolation in reproducing the features of the self-energy, in particular the slope of the dispersive pole.
The reason for that exceptional performance of the $G$ and $M$ schemes at small $\Ncluster$ is the very weak $k$-dependence of the cumulant in this model. However, if the cumulant were strongly $k$-dependent (for instance, if it had a dispersive pole), then $G$ and $M$-interpolation are expected to encounter issues similar to $\Sigma$-interpolation in the present case. The 1D Mott insulator is therefore a case where the $G$ and $M$-interpolation perform very well, but there is no guarantee that this is generically the case.

\changed{The last column [\Fig{fig:Hubbard1d_CDMFT_SE_periodized}, top, (d1-d3)]} shows the self-energy resulting from $\mc{L}$-interpolation. The dispersive pole of the self-energy, which is one of its main features, is captured for all cluster sizes. However, the pole strength is not correctly captured for the small clusters, which results in an incorrect size of the Mott gap. The information on the correct self-energy pole weight therefore seems to reside in longer-ranged matrix elements of $\mc{L}$, a conclusion which can also be drawn from Fig.~\ref{fig:Hubbard1d_MPS_SE_M_L_expansions}.

\changed{The middle and bottom parts of \Fig{fig:Hubbard1d_CDMFT_SE_periodized}} show, respectively,  the spectral functions $A_k(\omega)$ and cumulants 
$-\mathrm{Im} M_k(\omega)$ corresponding to the self-energies 
\changed{in the top part.} The $\Sigma$-interpolated   spectral functions and cumulants display a dispersive pole even for the largest cluster size, a result that is evidently incorrect. The $G$- and $M$- schemes both produce the correct Mott gap in  \changed{$A_k(\omega)$ and $-\mathrm{Im} M_k(\omega)$}, even for the smallest cluster size. 
Further, when using these interpolation schemes, the qualitative features of \changed{$A_k(\omega)$ and $-\mathrm{Im} M_k(\omega)$} compare quite well with the TaSK result for the 1D Mott insulator [\Figs{fig:Hubbard1d_MPS_phsym}(a,c)]. $\mc{L}$-interpolation,  on the other hand, has trouble reproducing the correct size of the Mott gap at small cluster sizes, due to the incorrect weight of its self-energy pole. 
For the largest $\Ncluster = 8$ cluster, however, $\mc{L}$-interpolation reproduces the correct Mott gap and the qualitative features of both \changed{$A_k(\omega)$ and $-\mathrm{Im} M_k(\omega)$} compare very well with the TaSK result. For $\Ncluster = 8$, $\mc{L}$-interpolation
yields smoother spectral features, with fewer interpolation artifacts, than $G$- or $M$-interpolation. 

\changed{Let us summarize the main conclusions of 
\Sec{sec:1dHubbard_benchmark}. First, Fourier coefficients of matrix elements $\bepsilon_{kn}$ and $\bt_{kn}$ obtained via a pure (numerically almost exact) TaSK computation decay more rapidly than Fourier coefficients of $\tSigma$ and $\tM$ obtained from the same data (Figs.~\ref{fig:Hubbard1d_MPS_SE_M_Fourier_phsym}, \ref{fig:Hubbard1d_MPS_LiouvilleMatEle}). Second,
self-energies reconstructed from truncated Fourier expansions of $\Sigma$, $M$, or $\cL$ remain causal 
only for $\cL$-expansions (\Fig{fig:Hubbard1d_MPS_SE_M_L_expansions}). 
Third, when comparing interpolated CDMFT+TaSK results to pure TaSK benchmark results, $\cL$-interpolation has some problems with quantitatively capturing the Mott gap for small clusters,
but eventually converges remarkably well for larger cluster sizes.}

\bigskip

\begin{figure}[tb!]
    \includegraphics[width=0.7\linewidth]{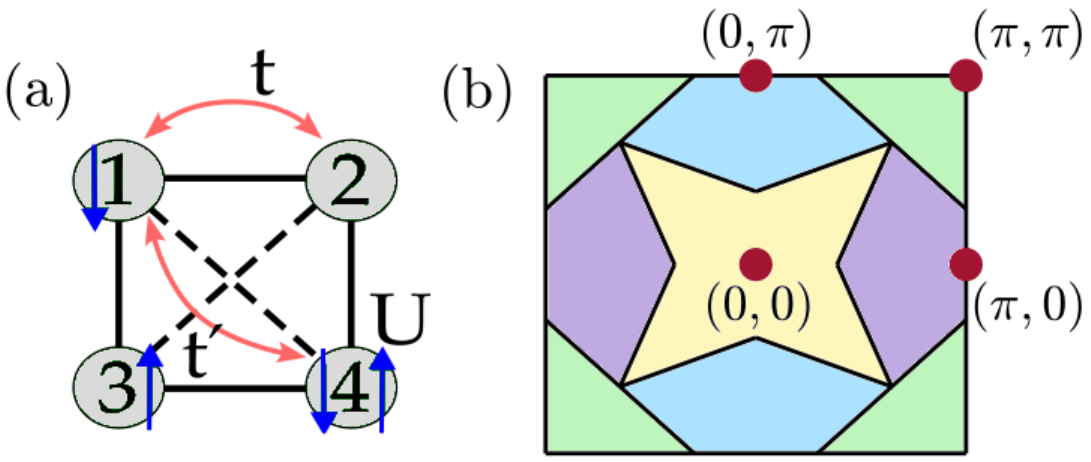} 
    \caption{(a) Illustration of hopping and interaction terms in the 2D Hubbard model. (b) Star patching geometry with patch momenta $\vec{K}\!\in \!\{(0,0),  (0,\pi), (\pi,0), (\pi,\pi)\}$.} \vspace{-5mm}
    \label{fig:star_patching}
\end{figure}

\begin{figure*}[t!]
    \centering
    \includegraphics[width=0.85\linewidth]{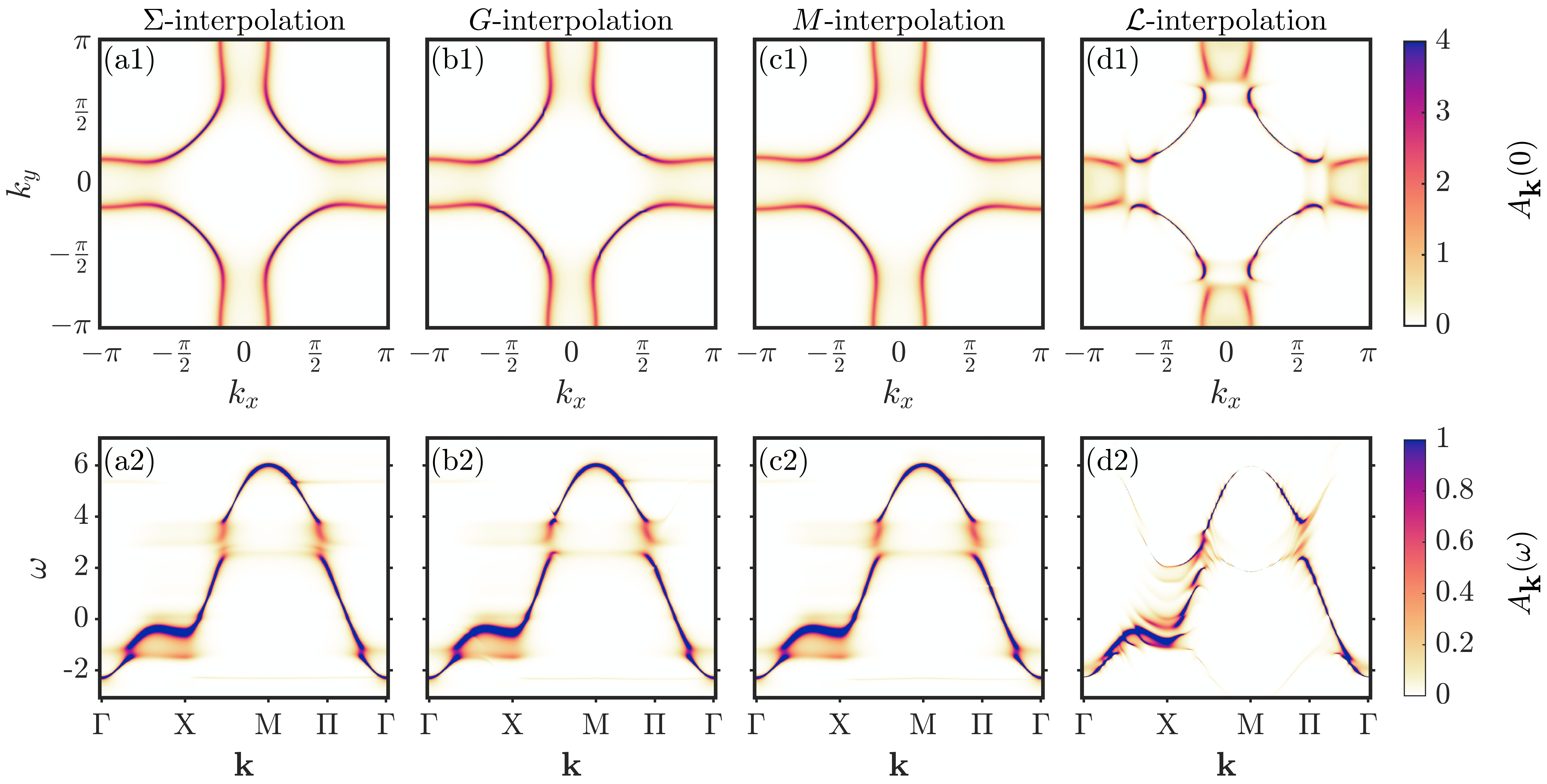} 
    \caption{Interpolated  CDMFT+ED results for a 4-site cluster impurity with $t=1$, $t^{\prime} = -0.3t$, $t^{\prime\prime} = 0.2t$, and $U=8t$, in the pseudogap phase at doping $\delta=0.065$. Top row: The spectral function at the Fermi surface, $A_\bk(\omega=0)$. Bottom row: the spectral function $A_\bk(\omega)$ along the path 
     ${\Gamma=(0,0)} \!\rightarrow\! {\mr{X}=(0,\pi)} \! \rightarrow\! {\mr{M}=(\pi,\pi)}\! \rightarrow\! {\mr{\Pi}=(\pi/2,\pi/2)} \!\rightarrow\! {\Gamma}$ in the Brillouin zone.}
    \label{fig:A_ED}
\end{figure*}

\changed{
\section{CDMFT+ED for 2D Hubbard model}
\label{sec:CDMFT+ED2DHubbard}

In Sections \ref{sec:CDMFT+ED2DHubbard} and 
\ref{sec:2dHubbard_DCA} we compare various interpolation schemes for cluster-DMFT results for the 2D Hubbard model. We consider the Hamiltonian 
\begin{align}
    \label{eq:H_2dHubbard}
    H &\!= -\!\sum_{\br\br'\sigma} t_{\br\br'}\,c^{\dagger}_{\br\sigma} c^{\pdag}_{\br'\sigma} \!-\! \mu \sum_{\br\sigma} n_{\br\sigma} \!+\! U \sum_{\br} n_{\br\uparrow} n_{\br\downarrow} \, ,
\end{align}
where $t_{\br\br'}$ contains nearest-, next-nearest-, and next-next-nearest neighbor hopping,  $t$, $t'$ and $t''$, respectively. In the present \Sec{sec:CDMFT+ED2DHubbard}, the cluster-DMFT results are obtained using CDMFT, thus ``interpolation'' means real-space periodization; in \Sec{sec:2dHubbard_DCA}, they are obtained using DCA; there, ``interpolation'' means $\bk$-space interpolation.

The CDMFT results used in this section were obtained 
employing the 4-site cluster shown in \Fig{fig:star_patching}(a),
with exact diagonalization (ED) as impurity solver. The CDMFT+ED results were kindly provided to us by the authors of Ref.~\onlinecite{Verret2022_CompactTiling}. They used $t = 1$, $t^\prime = -0.3t$, $t^{\prime\prime} = 0.2t$,  $U = 8t$ and a doping level of $\delta = 0.065$, corresponding to the pseudogap phase. Details regarding their numerical methodology are given in Refs.~\cite{Verret2022_CompactTiling,Foley2019_ED}. 

CDMFT+ED yields a discrete spectral representation of the 
cluster  Green's function $\bG_{\dimp \dimp}(z)$, a $4 \times 4$ matrix whose indices label the cluster sites. It is computed for a discrete (self-consistently determined) hybridization function $\bDelta(z)$. 
Since the number of poles is finite, the CFE of the self-energy is likewise finite.  Thus, it is not necessary to compute a residual function.

Before the CFE of the self-energy can be constructed, we first need to separate the contributions of $\bSigma^\dyn(z)$ and $\bDelta(z)$. We do this using the explicit separation scheme described
in \App{sec:SeparatingHybridizationSelfEnergy}, \Eqs{eq:PoleSubtraction}. We find that a clean separation is indeed achieved for the CDMFT+ED data, 
yielding a discrete self-energy $\widetilde \bSigma{}^\discrete_\imp (\omega)$
represented by a discrete set of pole weights and locations 
[cf.~\Eq{eq:DiscreteSigmaDyn}].}

\changed{
\subsubsection{CDMFT+ED interpolation details}
\label{sec:CDMFT+ED-interpolation-details}}

We use $\widetilde \bSigma{}^\discrete_\imp (\omega)$ as input for two purposes.

\changed{(1) For $\cL$-interpolation, we construct a CFE following the broaden-last protocol. We proceed just
as for CDMFT+TaSK in \Sec{sec:CDMFT+TaSK-interpolation-details}  above, but use a different broadening scheme for the last step: we broaden the $\cL$-interpolated   self-energy $\bSigma_\bk(\omega)$ for individual $\bk$-points 
via Gaussian broadening with broadening width $\sigma = 0.2$.}

(2) \changed{For traditional $\Sigma$-, $G$-, or $M$-interpolation, we follow the broaden-first protocol:}
First, we use a Gaussian broadening kernel with standard deviation $\sigma=0.2$ to obtain a smooth version of the impurity self-energy, $\widetilde \bSigma_\imp (\omega)$.
The latter serves as input for traditional $\Sigma$-, $G$-, $M$-interpolation, using superlattice averaging just as for CDMFT+TaSK in \Sec{sec:CDMFT+TaSK-interpolation-details} above.}

\bigskip 

\changed{
\subsubsection{CDMFT+ED interpolation results}
\label{sec:CDMFT+ED-interpolation-results}}

Figure~\ref{fig:A_ED} compares the results of $\Sigma$-, $G$-,  $M$- and $\cL$-interpolation. The top row shows the spectral function at the Fermi surface, $A_\bk(\omega=0)$.
\changed{$\Sigma$-, $G$-, and $M$-interpolation all produce} a hole-like Fermi surface, with the spectral weight decaying strongly towards the BZ boundary. By contrast, $\cL$-interpolation unveils a more refined structure: rather than arcs, it reveals an intersected arc suggestive of a hole pocket. This highlights the ability of $\cL$-interpolation to capture finer details of the Fermi surface, including potential hole pockets that may be overlooked by other interpolation methods.
\changed{Remarkably, the Fermi surface obtained through $\cL$-interpolation closely resembles the Fermi surface of the toy model used in Ref.~\cite{Verret2022_CompactTiling}.
Indeed, by comparing the cluster spectral properties of their CDMFT + ED results to that of their toy model, the authors of Ref.~\cite{Verret2022_CompactTiling} have argued that the Fermi surface should resemble that of their toy model.
    However, they could not show this directly using the interpolation methods available at that time; $\cL$-interpolation closes this gap.}

\changed{The bottom row of \Fig{fig:A_ED} shows the spectral function $A_{\bk}(\omega)$ along the path $\Gamma\mr{X}\mr{M}\Pi\Gamma$ in the BZ. $\cL$-interpolation yields a distinct gap at $\mr{X}$ and $\omega=0$; by contrast, the $\Sigma$-, $G$- and $M$-interpolation do not. At high energies, all four schemes yield several distinct branches separated by small gaps, which together represent the full spectral band. We attribute the substructure to finite-size effects from the ED impurity solver. 
For $\cL$-interpolation, the substructure is resolved more finely. 
However, due to the finite-size effects inherent in ED data, it remains uncertain which of the spectral features are physical rather than artifacts.}

\changed{Said finite-size artifacts can be avoided by using a more powerful impurity solver, such as NRG. We employ the latter in the next section, devoted to DCA+NRG.} 

\bigskip 

\section{DCA+NRG for 2D Hubbard model}
\label{sec:2dHubbard_DCA}

\changed{Next, we again consider the 2D Hubbard model, but now treated with DCA+NRG. A thorough analysis of our DCA+NRG results for this model will be published separately \cite{Pelz2026b,Pelz2026c}. Here, we focus on interpolation rather than physical implications.} 

We compare the non-interpolated discontinuous DCA results to smooth interpolated versions obtained from $M$-, $\Sigma$- and $\cL$-interpolations, focusing on data sets for which different interpolation schemes turn out to yield strikingly different results. Our analysis contains a striking physical result: $\cL$-interpolation can (depending on doping) resolve Fermi and Luttinger arcs which together \rreply{seem to} form a closed surface, with a smooth crossover between Fermi and Luttinger arcs.

\changed{The Hamiltonian is again given by \Eq{eq:H_2dHubbard}. In this section, we use $t=1$ as energy unit, set $t^{\prime} = -0.3t$, $t''=0$, $U=7t$, and use temperature $T = 10^{-10}t$. We use the chemical potential $\mu$ to tune the doping $\delta$ and consider three choices of doping, corresponding to 
a Fermi-liquid regime ($\delta=0.155$, $\mu=1.10$),
a pseudogap regime ($\delta=0.125$, $\mu=1.28$), and 
a Mott insulator regime ($\delta=0.00$, $\mu=3.50$). We use DCA, using the ``star patching'' geometry of \Fig{fig:star_patching}(b) \cite{Gull2010_patching}. It has four patches
($\Npatch = 4$), labeled by the  patch momenta $\vec{K}_1=(0,0)$, $\vec{K}_2=(\pi,0)$, $\vec{K}_3=(0,\pi)$, $\vec{K}_4=(\pi,\pi)$.

We solve the 4-patch impurity model using NRG (see \App{sec:nrg_details_app} for NRG details). NRG uses log-Gaussian broadening and yields patch self-energies $\bSigma_{\bK}(z)$ having continuous spectral functions $\bW_\bK(\omega)$.}

\changed{
\subsection{DCA+NRG: interpolation details}
\label{sec:2dHubbard_DCA_model}}

\changed{We use $\bSigma{}_\bK (\omega)$ as input for all three interpolation schemes used in this section, all performed using the broaden-first protocol (\Sec{sec:broadening}). DCA interpolation requires the use of form factors. For $\cL$-interpolation, we use the form factors $\alpha_{\bk \comma \bK}$
of \Eq{eq:Ansatz-compact}. For $\Sigma$- and $M$-interpolation,
we use the modified form factors $\alpha^\modified_{\bk \comma \bK}$ of \Eqs{subeq:modified-interpolation}, to sidestep causality issues, as discussed below. We now discuss these form factors in detail
for the star patching geometry of \Fig{fig:star_patching}(b).} 

The 4 patches are invariant under $x$- and $y$-reflections, but not under $\pi/2$ rotations about the origin. The lowest few lattice harmonics that are invariant under these symmetry operations are constructed explicitly in \App{app:latticeharmonics}. They have the form
\begin{gather}%
\label{eq:DCA-Fourier-modes}
\begin{array}{lll}
    \phi_{\bk,1} & = 1\, ,  \quad \; &  \phi_{\bk,3}  = \cos k_x - \cos k_y\, , \\ 
    \phi_{\bk,2} &  = \cos k_x + \cos k_y \, , & \phi_{\bk,4} =  \cos k_x  \cos k_y \, . 
\end{array}
\end{gather}
leading to 
\begin{align}
    \Psi_{j,\bK} \simeq
    \begin{pmatrix}
        0.2500 &   0.2500 &   0.2500 &   0.2500 \\
        0.4010 &   0.0092 &   0.0092 &  -0.4194 \\
        0.0000 &   -0.4050 &  0.4050 &   0.0000 \\ 
        0.7591 &  -0.6645 &  -0.6645 &   0.5698
    \end{pmatrix} 
\end{align}
for the coefficients appearing in Eq.~\eqref{eq:Ansatz-compact} for the canonical DCA interpolation functions $\alpha_{\bk \comma \bK}$. The latter are then used to interpolate $\epsilon_{\bK n}$, $t_{\bK n}$, and $R^{\fast}_{\bK n}(z)$ via \Eq{eq:Mscheme_DCA}.

\begin{figure}
    \centering
    \includegraphics[width=\linewidth]{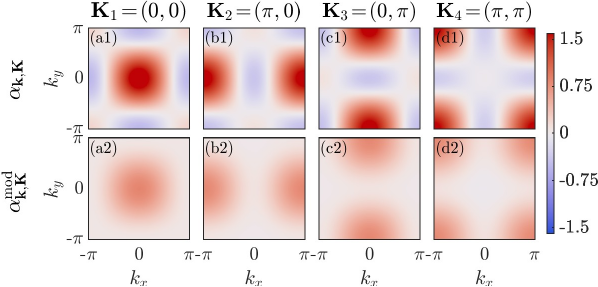} 
    \caption{(a1-d1) The canonical DCA interpolation functions  $\alpha_{\bk,\bK}$ of \Eq{eq:Ansatz-compact}, and (a2-d2) the modified interpolation functions $\alpha^{\mr{mod}}_{\bk,\bK}$ of \Eq{eq:Ansatz-compact-modified}, for the star patching geometry of \Fig{fig:star_patching}(b).  All plots employ the same diverging color scale. The $\alpha^{\mr{mod}}_{\bk,\bK}$ functions are \changed{predominantly} positive due to the reduced non-local contributions.}
    \label{fig:alpha_kK}
\end{figure}

The interpolation functions $\alpha_{\bk,\bK}$ are depicted in 
Figs.~\ref{fig:alpha_kK}(a1-d1). They exhibit large positive weights in the centers of their respective patches and fall off towards the other patches. Importantly, to fulfill the patch average condition \eqref{eq:alpha-self-consistency}, they become negative in the centers of some other patches. As an aside, we note that this can cause the DCA interpolation Ansatz 
$\bQ_\bk(\omega)  =  \sum_\bK \alpha_{\bk \comma \bK}\,\bQ_\bK(\omega)$ [\Eq{eq:Mscheme_DCA}] to yield negative values  for some combinations of $\bk$ and $\omega$ even if all $\bQ_\bK(\omega)$ are strictly positive functions of $\omega$. This may lead to causality violations in some situations, e.g.\ when used for $\Sigma$- or $M$-interpolation. 

\changed{To sidestep these issues, we perform 
 $\Sigma$- and $M$-interpolation using the modified interpolation Ansatz \eqref{subeq:modified-interpolation} 
 involving modified form factors $\alpha_{\bk,\bK}^\mr{mod}$.} There, the non-local Fourier modes $\phi_{\bk,j>1}$ are rescaled by a factor $1/N_j$ (defined in \App{app:latticeharmonics}), which counteracts their causality-violating tendencies. The resulting functions $\alpha_{\bk,\bK}^\mr{mod}$ are shown in Figs.~\ref{fig:alpha_kK}(a2-d2). They are 
 \changed{predominantly} positive,  as desired; however, they also exhibit a much weaker momentum dependence compared to $\alpha_{\bk,\bK}$, and they do not preserve the patch average since Eq.~\eqref{eq:alpha-self-consistency} is not fulfilled. (We have explored using other interpolation functions for $\Sigma$- and $M$-interpolation, e.g.\ ones satisfying $\alpha_{\bK,\bK}=1$, so that setting $\bk=\bK$ in the interpolated $Q_{\bk}(z)$ reproduces the input patch function $Q_\bK(z)$. These yielded minor quantitative but no qualitative differences from the results shown below, obtained using $\alpha_{\bk,\bK}^\mr{mod}$.)

\changed{Let us now turn to $\cL$-inter\-po\-lation. We generate a log-CFE of the form \eqref{eq:SigmaCFE_FastSlowMode-main} for each patch self-energy 
$\bSigma_\bK(z)$ as described in \App{sec:restfunction_splitting}, 
with fast/slow splitting parameters $\Lambda_\fastslow = 1.2$ and $\tau_\fastslow=5$. (For numerical details, a discussion of how the log-CFE coefficients depend on CFE depth $n$ and the effects of varying $\Lambda_\fastslow$ and $\tau_\fastslow$, see \App{app:dependence_splitting_parameters}.) 
This yields  CFE coefficients $\epsilon_{\vec{K}n}$, $t_{\vec{K}n}$ and residual functions $R^{\fast}_{\bK n}(z)$ for each patch. 
We interpolate them to obtain  $\epsilon_{\bk n}$ and $t_{\bk n}$
via \Eq{eq:DCA_epsk_final_DCA} and  $R^{\fast}_{\bk n}(z)$
via \Eq{eq:Mscheme_DCA}, all using the same form factors $ \alpha_{\bk \comma \bK}$.}
\rreply{This did not lead to any causality problems, since the log-CFE scheme yields $R^{\fast}_{\bK n}(z)$ functions whose $\bK$-dependence is very weak [cf.~\Fig{fig:CFE_data}(c) in \App{app:dependence_splitting_parameters}].}  

\bigskip 

\subsection{DCA+NRG interpolation results: spectra}
\label{sec:2dHubbard_DCA_periodresults}

In this section, we compare results
for spectral functions $A_{\bk}(\omega)$ and self-energies $-\mr{Im} \Sigma_{\bk}(\omega)$ obtained from different interpolation schemes. We consider three values of doping, corresponding to a Fermi liquid, a pseudogap regime, and a Mott insulator.
We plot both $\Sigma_{\bk}(\omega)$ and $A_{\bk}(\omega)$ along the path ${\Gamma=(0,0)} \!\rightarrow\! {\mr{X}=(0,\pi)} \!\rightarrow\! {\mr{M}=(\pi,\pi)}\! \rightarrow\! {\Pi=(\pi/2,\pi/2)}\! \rightarrow\! \Gamma$ in the Brillouin zone ($\Gamma\mr{X}\mr{M}\Pi\Gamma$), with momentum $\bk$ on the abscissa and frequency $\omega$ on the ordinate. The corresponding Fermi surfaces~(FS) are discussed in the next section.

\begin{figure*}[tb!]
    \centering
    \includegraphics[width=0.85\linewidth]{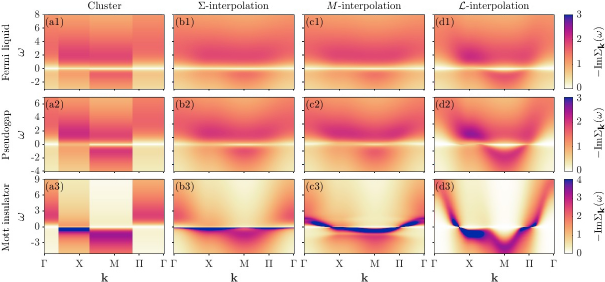} \\[2mm]
    \includegraphics[width=0.85\linewidth]{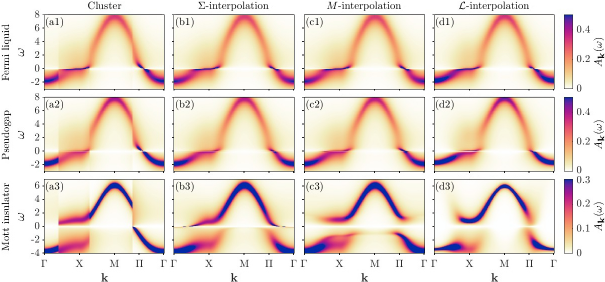} 
    \caption{\changed{DCA+NRG results for the 2D Hubbard model
    in Fermi liquid, pseudogap and Mott insulator regimes. 
    The self-energy $-\mr{Im}\Sigma_{\bk}(\omega)$ (top part) and the spectral function $A_{\bk}(\omega)$  (bottom part) are plotted} as functions of frequency $\omega$ and momentum $\bk$ along the path ${\Gamma=(0,0)} \!\rightarrow\! {\mr{X}=(0,\pi)} \! \rightarrow\! {\mr{M}=(\pi,\pi)}\! \rightarrow\! {\mr{\Pi}=(\pi/2,\pi/2)} \!\rightarrow\! {\Gamma}$ in the Brillouin zone. For column (a) no interpolation  was used; for (b-d) $\Sigma$-, $M$- and $\cL$-interpolation  was used, respectively.}
    \label{fig:DCA_Ak}
\end{figure*}

\begin{figure}[t] 
    \centering
    \includegraphics[width=0.9\linewidth]{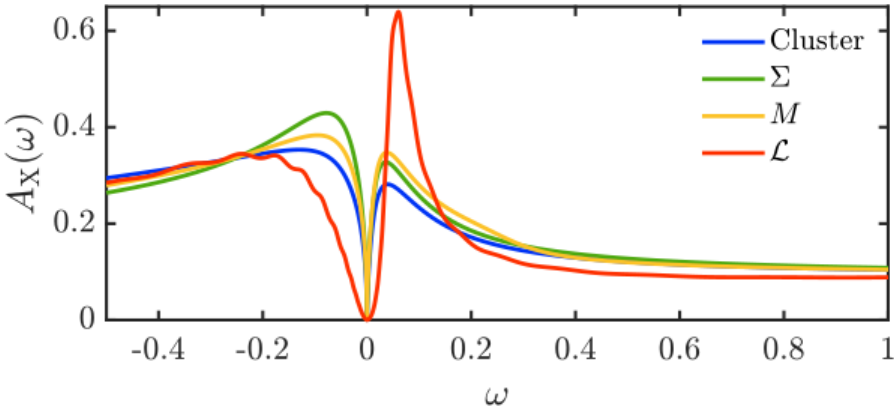} 
    \caption{\changed{The spectral function $A_{\bk = \mr{X}}(\omega)$ in the pseudogap regime of the 2D Hubbard model: different interpolation schemes, distinguished by colors, lead to pseudogaps differing in width and sharpness.}     \rreply{For $\cL$-interpolation, we used
    the fast/slow splitting parameters $\Lambda_\fastslow=1.2$, $\tau_\fastslow=5$; the effect of varying them is elucidated in \App{app:dependence_splitting_parameters}, \Fig{fig:AX_tau}.}}
    \label{fig:DCA_AkX}
\end{figure}

\changed{Figure~\ref{fig:DCA_Ak} shows self-energies $-\mr{Im}\Sigma_{\bk}(\omega)$  (top part) and spectral functions $A_\bk(\omega)$ (bottom part), 
comparing} the non-interpolated  cluster (= patch) DCA data in column (a) with that obtained after $\Sigma$-, $M$- and $\mc{L}$-interpolation  in columns (b), (c) and (d), respectively. In each part, the top, middle and bottom rows
correspond to Fermi liquid,  pseudogap, and Mott insulator regimes, respectively. (We checked that the filling is preserved by all three interpolation  schemes.) \changed{Clearly, all three interpolation schemes yield self-energies that recognizably mimic the cluster self-energy in the left-most column, though not all in the same manner, with the $\cL$-scheme exhibiting the most structure.}

Consider first the self-energy in the Fermi liquid regime
[\Fig{fig:DCA_Ak}, top part, first row].
The cluster self-energy is only weakly $\bK$-dependent [\Fig{fig:DCA_Ak}, top, (a1)]. The interpolation results 
[\Fig{fig:DCA_Ak}, top, (b1-d1)] therefore do not strongly depend on the interpolation method and closely match the non-interpolated data. \changed{The same is true for the spectral function 
$A_\bk(\omega)$ in the Fermi liquid regime [\Fig{fig:DCA_Ak}, bottom part, first row].
That said, the doping value $\delta=0.155$ used here is quite close to 
Fermi liquid-to-pseudogap transition  at $\delta^\ast = 0.14$ (see Ref.~\onlinecite{Pelz2026b}). Hence, non-local (i.e.\ $\bk$-dependent) effects are somewhat enhanced compared to the deep Fermi liquid regime (e.g.\  $\delta\gg 0.25$). In particular, the cluster self-energy is somewhat enhanced at momentum $\mr{X}$ (where the pseudogap opens when $\delta$ drops below $\delta^\ast$). This effect is mimicked better by  $\cL$- than $\Sigma$- and $M$-interpolation
[compare \Fig{fig:DCA_Ak}, top, (a1-d1)].} 

\changed{Next, we consider the self-energy in the pseudogap regime
[\Fig{fig:DCA_Ak}, top part, second row].} Here, 
the self-energy exhibits a somewhat stronger $\bK$-dependence;
correspondingly, the $\Sigma$-, $M$- and $\cL$-interpolation results differ more strongly. 
\changed{In particular, in the antinodal patches, the cluster self-energy $-\mr{Im}\Sigma_{\bK=\mathrm{X}}(\omega)$ 
[\Fig{fig:DCA_Ak}, top, (a2)] exhibits a strong but very narrow-in-$\omega$ peak near $\omega\sim0$. 
In fact, this peak is so narrow that it becomes clearly recognizable only upon zooming in very strongly (or when depicted using line cuts, as done in
\Fig{fig:DCA_CFE} in \App{app:dependence_splitting_parameters}).} All interpolation schemes manage to resolve this peak, but they yield
different amounts of $\bk$-structure. $\Sigma$- and $M$-interpolation 
turn it into \changed{broad-and-flat-in-$\bk$, narrow-in-$\omega$} structures around $\mr{X}$ and along $\mr{M}\Gamma$ \changed{(again, clearly recognizable only upon zooming in very strongly)}. By contrast, the $\cL$-scheme connects these broad and flat structures to \changed{quite prominent structures
at higher $|\omega|$-values}, yielding a more prominently structured profile for $\Sigma_{\bk}(\omega)$, i.e.\ more clearly defined bands. The appearance of such 
clearly-defined bands \changed{in the self-energy} suggests that $\cL$-interpolation  yields a more plausible representation of the pseudogap phase.

\changed{The narrow self-energy peak near $\omega = 0$ causes a small 
gap near $\omega = 0$ to open up in the spectral function $A_\bk(\omega)$ in the vicinity of momentum $\mr{X}$ [\Fig{fig:DCA_Ak}, bottom part, second row]. This gap is more pronounced for $\cL$- than $\Sigma$- and $M$-interpolation. To make this more apparent, \Fig{fig:DCA_AkX} shows line cuts of $A_{\bk = \mr{X}}(\omega)$: there, all curves dip to zero at $\omega = 0$. Note that the $\Sigma$- and $M$-curves (green, yellow) lie closer to the cluster curve (blue) than the $\cL$-curve (red) does. The reason is that for $\Sigma$- and $M$-interpolation, the frequency dependence is inherited \textit{linearly} from the cluster quantities via the interpolation formula \eqref{eq:modified-interpolation}. By contrast, for $\cL$-interpolation the $\omega$-dependence is inherited \textit{non-linearly} through the different levels of the CFE, leading to more pronounced spectral modifications.}

\rreply{In \Fig{fig:DCA_AkX}, the slight wiggles in the red 
$\mc{L}$-interpolation curve can be 
reduced by adjusting the splitting parameters
(as discussed in detail in \App{app:dependence_splitting_parameters}, 
see \Fig{fig:AX_tau}). 
We intentionally did not do so here, since the wiggles provide insight into the underlying pole structure. Importantly, the wiggles do not qualitatively affect the overall spectral features or the conclusions drawn from the spectrum.}

\changed{
Finally, consider the Mott insulator regime
[\Fig{fig:DCA_Ak}, third rows of top and bottom parts]. There, the spectral function is known to be fully gapped. 
$\Sigma$-interpolation [\Fig{fig:DCA_Ak}, top, (b3)] yields a self-energy pole that fails to cross $\omega=0$. This causes  $A_{\bk}(\omega)$ [\Fig{fig:DCA_Ak}, bottom, (b3)] to erroneously acquire some spectral weight at $\omega=0$. By contrast, $M$-interpolation works well in the Mott regime, since $M$ is inversely related to $\Sigma$, facilitating a reliable reconstruction of self-energy poles. Indeed, $M$-interpolation yields a narrow self-energy band that lies close to $\omega=0$ and crosses $\omega=0$ twice [\Fig{fig:DCA_Ak} top, (c3)]. This causes large gaps in the corresponding spectral function [\Fig{fig:DCA_Ak}, bottom, (c3)].  $\cL$-interpolation yields a similar self-energy band, however, it is much more dispersive. [\Fig{fig:DCA_Ak}, top, (d3)], due to the non-linear $\omega$-dependence of the CFE. This yields a correspondingly strong variation for the gap size of the spectral function [\Fig{fig:DCA_Ak}, bottom, (d3)]. This further demonstrates the advantage of its flexible weight distribution among multiple Liouvillian matrix elements.}

We conclude this discussion with some remarks regarding the
bandwidth of the self-energy pole.
In a 2-dimensional Mott insulator, a dispersive self-energy pole, with a dispersion roughly proportional to $-\epsilon_{\vec{k}}$, is expected on general grounds~\cite{Wagner2023}.
Qualitatively, this hallmark feature is captured by both $M$ and $\mc{L}$-interpolation, albeit with very different bandwidths.
At intermediate coupling, as studied here ($U = 7t$), the bandwidth of the self-energy pole is usually somewhat larger than the electronic bandwidth, see for instance Fig.~5 of Ref.~\cite{Pudleiner2016} ($U = 4t$) or Fig.~3 of Ref.~\cite{Kovalska2025} ($U = 7.5t$). 
Our electronic bandwidth is $8t$, and we find a self-energy pole bandwidth of $\simeq 11t$ with $\mc{L}$-interpolation, but only
 $\simeq 2t$ with $M$-interpolation. Thus $\mc{L}$-interpolation yields a self-energy pole bandwidth consistent with expectations, whereas $M$-interpolation severely underestimates it.

\subsection{DCA+NRG interpolation results: $\omega = 0$}
\label{sec:2dHubbard_DCA_periodFS}

\begin{figure*}[tb!]
    \centering
    \includegraphics[width=0.85\linewidth]{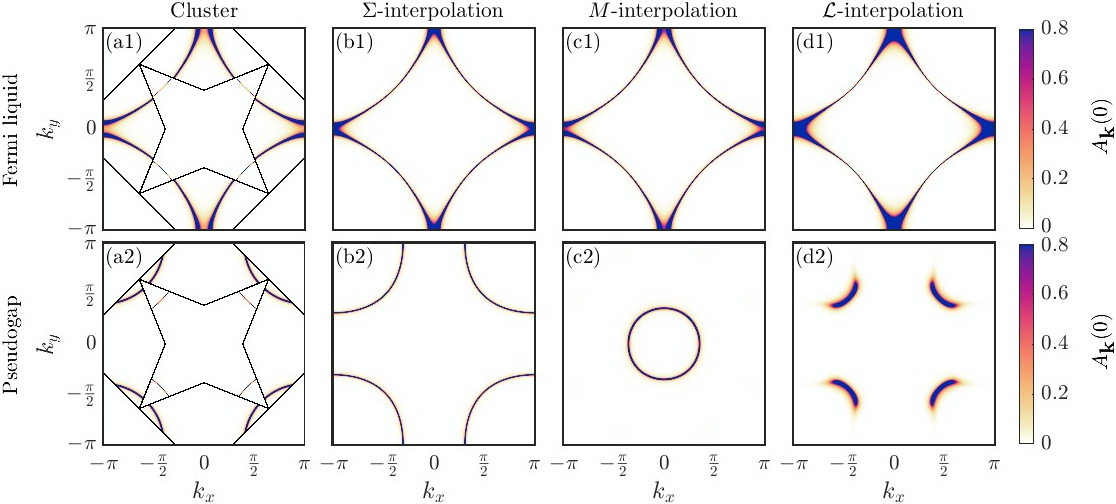} 
    \caption{DCA+NRG results for the 2D Hubbard model, 
    \changed{showing Fermi surfaces}
    in the Fermi liquid and pseudogap regimes. Spectral function at zero frequency, $A_{\bk}(0)$, for the same settings as in \Fig{fig:DCA_Ak}. Column (a):  (non-interpolated) cluster results, featuring discontinuities at the patch borders (straight black lines). Columns (b-d):  results from $\Sigma$-, $M$-, and $\cL$-interpolation.}
    \label{fig:A0}
\end{figure*}

\begin{figure*}[tb!]
    \centering
    \includegraphics[width=0.85\linewidth]{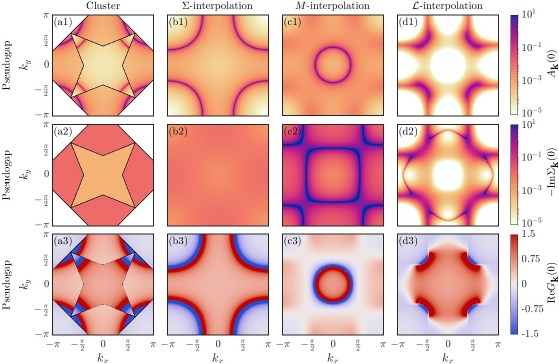} 
    \caption{DCA+NRG results for the 2D Hubbard model, showing Fermi surface properties 
    in the pseudogap regime. Top row:  $A_{\bk}(0)$ (same data as bottom row of \Fig{fig:A0}, now on a logarithmic color scale). Middle row: $-\mr{Im}\Sigma_{\bk}(0)$. Bottom row: $\mr{Re}G_{\bk}(0)$.}
    \label{fig:DCA_logA0}
\end{figure*}

\begin{figure}[t] 
    \centering
    \includegraphics[width=1\linewidth]{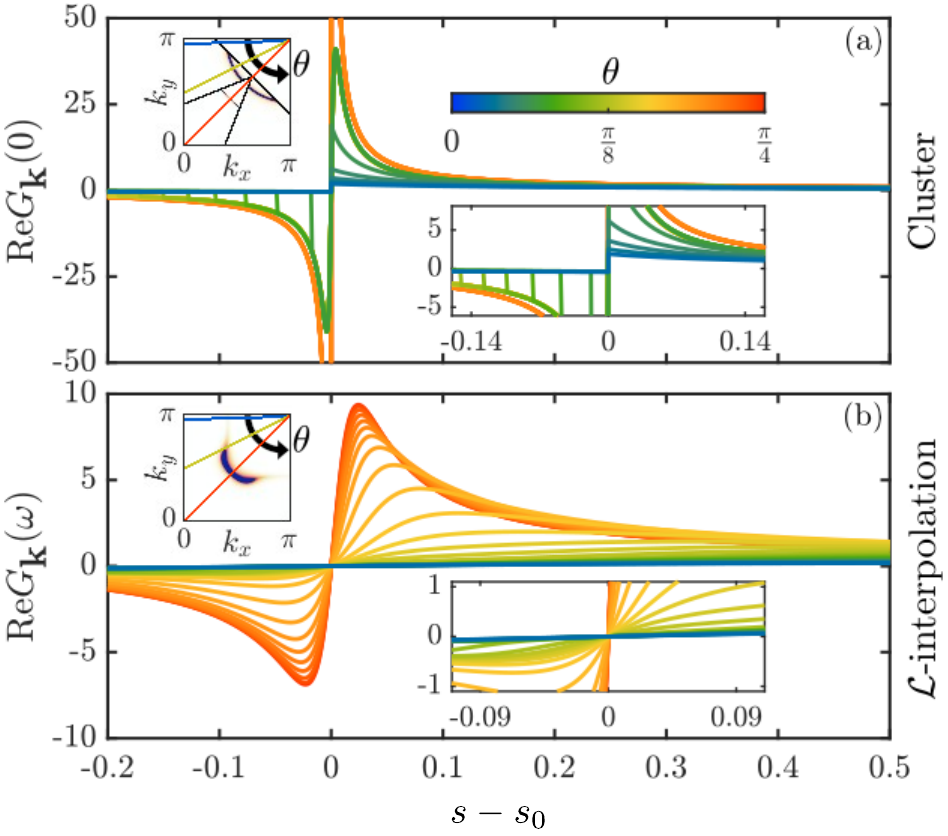} 
    \caption{Real part of the Green's function at zero frequency, $\mr{Re}G_\bk(0)$, computed  (a) without interpolation  and (b) with $\cL$-interpolation. They are plotted along several $\bk$-rays emanating from $(\pi,\pi)$ at angles $0\leq\theta\leq\frac{\pi}{4}$ (indicated by the color bar and defined in the top-left insets). 
    \changed{The position along each $\bk$-ray is parametrized by $s$ [c.f.\ Eq.~\eqref{eq:kray}]; the abscissa indicates
    $s - s_0$, where $\bk(s_0)$} is the momentum where $\mr{Re}G_\bk(0)$ changes sign.  Bottom-right insets zoom in to small values of $s - s_0$. 
    }
    \label{fig:DCA_ReGk}
\end{figure}

An examination of the Fermi surface (FS) provides additional insight into the performance of the $\cL$-scheme compared to other interpolation  methods. \Fig{fig:A0} presents a comparison of the Fermi surfaces via plots of the spectral function at zero frequency, $A_{\bk}(0)$ for both the Fermi liquid and pseudogap phases. (The Mott insulator is not shown, as $A_{\bk}(0)\!<\!10^{-5}$, \changed{except} for $\Sigma$-interpolation,  which erroneously does not show a true Mott insulator.) Panels (a1–d1) display the FS of the Fermi liquid, while panels (a2–d2) show the FS in the pseudogap phase. In column (a), the straight black lines indicate the DCA patch boundaries.

The FS for the Fermi liquid is essentially identical across all schemes and mimics the non-interpolated   result well. However, the $\cL$-scheme has enhanced intensity at the antinodal points. Apparently, any interpolation  works well for a simple Fermi liquid. 

In the pseudogap phase, the cluster result [\Fig{fig:A0}(a2)] features hole-like arcs cut off by the patch boundaries, without extending to the Brillouin zone (BZ) edge, leading to gaps around the four ``antinodal''  points  $(\pm \pi, 0)$ and $(0,\pm \pi)$. This is because the $\bK=(\pi,\pi)$ cluster patch carries no spectral weight. The traditional $\Sigma$- and $M$-schemes are unable to reproduce key aspects of the cluster result: the 
$\Sigma$-interpolated FS is hole-like but extends all the way to  the opposite BZ edges without showing gaps near the edges, while the $M$-interpolated FS is electron-like. 

\begin{figure}[!t] 
    \centering
    \includegraphics[width=\linewidth]{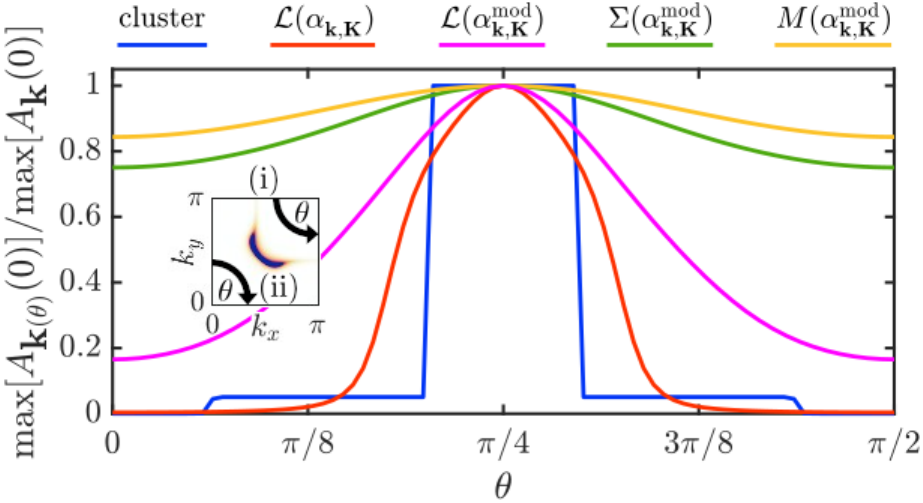}
    \caption{
    Maximum of the zero-frequency spectral function along a $\bk$-ray at angle $\theta$, $\mr{max}[A_{\bk(\theta)}(0)]$, plotted as function of $\theta$ and normalized, for ease of comparison, by its peak value at $\theta=\pi/4$. The angle $\theta$, defined in the inset, characterizes $\bk$-rays emanating (i) from $(\pi,\pi)$ for cluster results (blue), 
    $\cL$- (red, \rreply{pink}) and $\Sigma$- (green) interpolation, and (ii) from $(0,0)$ for $M$-interpolation (yellow). 
    \rreply{The legends $(\alpha_{\bk,\bK})$ or $(\alpha_{\bk,\bK}^\modified)$ at   the top indicate the form factors used for each interpolation.} }
    \label{fig:DCA_Z}
\end{figure}

The $\cL$-scheme is the only interpolation  that qualitatively captures all aspects of the cluster computation in the pseudogap phase. First, it preserves the hole-like characteristics; second, it also accumulates spectral weight around the four ``nodal'' points at $(\pm \pi/2, \pm \pi/2)$ and $(\pm \pi/2,\mp \pi/2)$, similar to the cluster result but distributed more smoothly in the shape of Fermi arcs. These are similar to those observed in photoemission spectra measured by ARPES experiments \cite{Reber2012_ARPES_cuprates}. Third,  $\mc{L}$-interpolation captures the spectral weight distribution along the Fermi surface far better than the other schemes, as discussed in more detail below (see Fig.~\ref{fig:DCA_Z}).

At first glance, the Fermi-arc structures seem to imply a non-closed FS. However, additional details become apparent when using a logarithmic color scale, as done in the first row of \Fig{fig:DCA_logA0}: The $M$-scheme yields a slightly enhanced intensity near the BZ edge, in the regions where the $\Sigma$-scheme FS  is located. The hole-like FS of the $\cL$-scheme actually does close across the BZ edges, albeit with very small spectral weight there, while an electron-like ring with even less weight is also present. On a logarithmic color scale the $\cL$-scheme thus displays structural features from both  the $\Sigma$- and $M$-schemes.

The second row of \Fig{fig:DCA_logA0} shows the imaginary part of the self-energy at zero frequency, $-\mr{Im}\Sigma_{\bk}(0)$, corresponding to the FS data in the first row. The $\Sigma$-interpolation  closely resembles the nearly constant self-energy of the non-interpolated   DCA calculation, showing no enhanced structures. The $M$-scheme produces $\Sigma$-pole rings around the nodal points; these are strongly modified in the $\cL$-interpolation. The $\cL$-scheme displays sharp pocket structures around $\mr{X}$ that suppress the FS near the BZ boundary. These pockets are connected by a less intense arc close to the FS intensity maximum. Note that for $\cL$-interpolation,  the peaks of the self-energy and the Green’s function occur very close to each other.

The last row in \Fig{fig:DCA_logA0} presents the real part of the Green's function at zero frequency, $\mr{Re}G_{\bk}(0)$. Here, a sudden blue-to-red change indicates a sign change of $\mr{Re}G_{\bk}(0)$ via a divergence, i.e.\ a Fermi surface. By contrast, a gradual blue-to-white-to-red evolution, as seen in parts of panels (c3) and (d3), indicates a sign change of $\mr{Re}G_{\bk}(0)$ via a smooth evolution through zero, i.e.\ a Luttinger surface (LS). 

The cluster result shows no true LS; instead, it shows small  discontinuities at some patch edges. The $\Sigma$-scheme interpolates these to yield a hole-like FS. The $M$-scheme yields a closed, electron-like FS and a closed LS around the antinodal points. These are well \changed{separated} and do not form arcs. The $\cL$-scheme, which most faithfully captures the characteristic structures of the cluster result, yields Fermi \textit{and} Luttinger arcs. Remarkably, these \rreply{appear to}  evolve \textit{smoothly} into each other. We describe this evolution in more detail via the next three figures by showing how three key quantities, $\mr{Re}G_{\bk}(0)$, $A_\bk(0)$ and $Z_\bk$, evolve along the Fermi and Luttinger surfaces.

Figure~\ref{fig:DCA_ReGk} shows a series of slices of $\mr{Re}G_{\bk}(0)$, taken for panel (a) from non-interpolated cluster results, and for panel (b) from  the corresponding $\cL$-interpolations. Each slice follows a ``$\bk$-ray'' (straight-line $\bk$-trajectory through the BZ) emanating from $(\pi,\pi)$ at a fixed angle, $0\leq\theta\leq\frac{\pi}{4}$, as indicated in the top-left insets of \Figs{fig:DCA_ReGk}(a,b) \changed{and described by}
\begin{align}
\label{eq:kray}
\changed{\bk(s,\theta) = (\pi,\pi)^{T} - s (\cos \theta , \sin\theta)^{T} \, .}
\end{align} 
The curves are plotted vs.\ \changed{$s -s_0$, where $\bk(s_0,\theta)$ is the momentum where, \changed{for given $\theta$,} $\mr{Re}G_{\bk}(0)$} changes sign.   
For small $\theta$, the non-interpolated  curves in (a) change sign at the patch boundaries via small discontinuities. 
\changed{As $\theta$ increases towards $\pi/4$, the location of the sign change shifts to within one of the patches (see \Fig{fig:DCA_logA0}(a3)) and the jump at the discontinuity becomes larger (see \Fig{fig:DCA_ReGk}(a)), evolving} into a divergence. The pole weight of the divergence is largest at $\theta = \pi/4$, when the $\bk$-ray cuts the nodal point. The $\cL$-scheme interpolates the small patch-boundary discontinuity to a \changed{zero crossing, i.e.\ a LS, 
in the form of a continuous curve that crosses zero with almost zero slope}. With increasing $\theta$, the curve crosses zero with increasing slope, \textit{smoothly} evolving into a shape resembling a regularized divergence which is most prominent for $\theta = \pi/4$. The regularization arises through \changed{the broadening effect of a small (non-divergent) imaginary part of the cluster self-energy, $- \mathrm{Im} \Sigma_{(0,0)}(0) \simeq 6 \times 10^{-4}$ (see \Fig{fig:DCA_logA0}(a2)).} 

\rreply{We remark that our NRG results for the self-energy may involve small inaccuracies at very low frequencies, since 4-patch DCA involves a 4-band impurity model, which is very challenging for NRG. We therefore cannot rule out the possibility that $-\mathrm{Im} \Sigma_{(0,0)}(0)$ at $T=0$
is much smaller than reported above, or even vanishes. In that case, the evolution-with-$\theta$ of $\mr{Re}G_{\bk}(0)$ from zero crossing to divergence shown in \Fig{fig:DCA_ReGk}(b) would look different, becoming more abrupt. However, for any finite temperature, $-\mathrm{Im} \Sigma_{(0,0)}(0)$ will certainly be nonzero. We therefore expect the smooth evolution-with-$\theta$ seen in \Fig{fig:DCA_ReGk}(b) to be characteristic 
for small but nonzero temperatures (e.g.\ $T \gtrsim 10^{-3}$).}

Figure~\ref{fig:DCA_Z} shows $\mr{max}[A_{\bk(\theta)}(0)]$, the maximum value assumed by the zero-frequency spectral function along a $\bk$-ray at fixed angle $\theta$, plotted as a function of $\theta$. This quantity corresponds to that measured in ARPES studies aiming to analyze the quasi-particle weight (see, e.g., Refs.~\cite{Reber2012_ARPES_cuprates,Kanigel2006,Smit2025}). The cluster result (blue) shows a step function profile. $\cL$-interpolation smooths out this step into a broader but still well-defined peak (red) whose flanks smoothly approach zero for $|\theta-\pi/4| \gtrsim \pi/8$. By contrast, $\Sigma$- or $M$-interpolation both yield only a very broad, weak maximum (green or yellow, respectively). 

\rreply{Recall from \Sec{sec:LiouvillianPeriodization} that we use the form factor $\alpha_{\bk,\bK}$ to interpolate $\cL$ (red), but a different one, $\alpha_{\bk,\bK}^{\mr{mod}}$, for  $\Sigma$/$M$ (green/yellow). For comparison, Fig.~\ref{fig:DCA_Z} includes a curve (pink) obtained by interpolating $\mc{L}$ using $\alpha_{\bk,\bK}^{\mr{mod}}$. Note that the pink $\cL(\alpha_{\bk,\bK}^{\mr{mod}})$ line lies much closer to the red $\cL (\alpha_{\bk,\bK})$ curve than the green/yellow 
$\Sigma/M(\alpha_{\bk,\bK}^{\mr{mod}})$ lines. This illustrates that the main difference between $\cL$- and $\Sigma$/$M$-interpolation lies in the interpolated object itself ($\cL$ vs.\ $\Sigma$/$M$) rather than in 
the use of $\alpha_{\bk,\bK}$ vs.\ $\alpha_{\bk,\bK}^{\mr{mod}}$.}

\begin{figure}[t!] 
    \centering
    \includegraphics[width=0.75\linewidth]{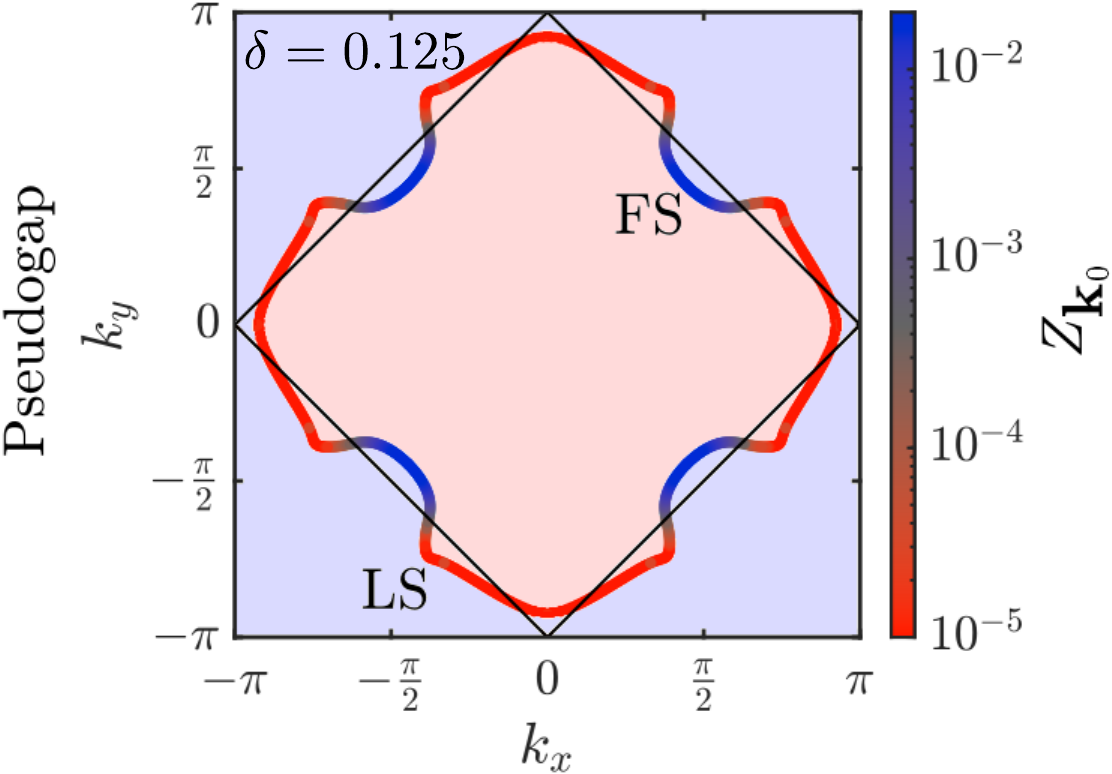} \vspace{-2mm}
    \caption{Quasi-particle weight, $Z_{\bk_0} = (1-\partial_{\omega} \mr{Re}\Sigma_{\bk_0}(0))^{-1}$, for the pseudogap phase at $\delta=0.125$, plotted along the ``sign-change trajectory'' in the BZ, i.e.\ along the locus of points $\bk_0 $ where $\mr{Re}G_{\bk}(0)$ changes sign. The color-scale shows the evolution from Fermi arcs (blue) to Luttinger arcs (red) and vice versa when crossing the antiferromagnetic zone boundary (black lines). The light blue or red background shading indicates BZ regions with 
    $\mr{Re} \, G_{\bk}(0) < 0$ or $> 0$,
    respectively. }
    \label{fig:FS_LS}
\end{figure}

Finally, \Fig{fig:FS_LS} shows the quasi-particle weight $Z_{\bk_{0}} = (1-\partial_{\omega} \mr{Re}\Sigma_{\vec{k}_0}(0))^{-1}$ along the ``sign-change trajectory'' through the BZ (the $\bk_0$-points where $\mr{Re}G_{\bk}(0)$ changes sign), i.e.\ along the  trajectory that defines the Fermi and Luttinger surfaces. We compute the slope $\partial_{\omega}\mr{Re}\Sigma_{\vec{k}_0}(0)$ via a linear fit, adjusting the fitting range along the sign-change trajectory to match appropriate fitting ranges found for our cluster data ($\omega_{\mr{fit}}\sim \pm10^{-5}$ for $\bk_0\sim(0,\pi)$ and $\omega_{\mr{fit}}\sim \pm10^{-3}$ for  $\bk_0\sim(\pi/2,\pi/2)\,$). \rreply{(For details on the fitting procedure, see \App{app:fits}.)} The resulting $Z_{\bk_0}$ is depicted along the sign-change trajectory using a logarithmic red-to-blue color-scale. The blue portions of the trajectory correspond to a sign change via a divergence, i.e.\ a FS; the red portions correspond to a sign change via a smooth zero crossing, i.e.\ a LS. The evolution from Fermi arcs to Luttinger arcs and back 
\rreply{appears to occur} \textit{smoothly}, in regions (shown in grey), that intersect the antiferromagnetic zone boundaries (straight black lines).

\changed{The apparent smooth evolution of a LS into a FS signifies that zeros and poles of the Green's function move very close together; that, combined with some broadening due to the small, constant contribution to $-\mathrm{Im} \Sigma_{(0,0)}(0)$ discussed above, yields a seemingly smooth evolution. As argued above, for $T=0$ it remains to be seen whether this broadening is physical or a numerical artifact. However, some broadening definitely is present at small but nonzero temperatures. In that case, a smooth evolution of Luttinger and Fermi arcs into each other can reliably be expected to occur---with the caveat that these surfaces are not sharply defined at finite $T$.}

(Parenthetical remark: The reason why the quasi-particle weight is very small, $Z_{\bk_{0}} \lesssim 0.02$, also along the FS parts of the sign-change trajectory, is that $\delta=0.125$ lies very close to a quantum phase transition, which will be the subject of a separate publication \cite{Pelz2026b}.)

To summarize:  In the pseudogap regime, $\cL$-interpola\-tion, in contrast to $\Sigma$- and $M$-interpolation, yields a \rreply{seemingly} smooth evolution between hole-like Fermi arcs and Luttinger arcs, realizing these two concepts on one surface. The Fermi arcs lie within the antiferromagnetic zone boundaries, and the change to Luttinger arcs occurs while crossing these boundaries. \changed{In the literature, such a smooth FS-to-LS evolution has previously been found \cite{Worm2024_FS+LS} at much higher temperatures using the  dynamical vertex approximation (D$\Gamma$A), which requires the computation of two-particle quantities  and hence is numerically very demanding.} That this behavior can be obtained via $\cL$-interpolation of computationally much cheaper DCA+NRG data  constitutes a major result of our work. It illustrates that $\cL$-interpolation is able to generate more intricate features than $\Sigma$- or $M$-interpolation, while preserving key qualitative properties  of the cluster result, such as the hole-like character of the Fermi surface.

\section{Summary and Outlook} 
\label{sec:ConclusionOutlook}

\changed{
\subsection{Summary}
\label{sec:summary}}

In this paper, we introduced a scheme for
interpolating CDMFT and DCA results through a suitable interpolation of single-particle irreducible Liouvillian matrix elements $\bolde_n$ and $\bt_n$. We pointed out (\Sec{sec:Liouvillian_basics}) that these Liouvillian matrix elements are in fact coefficients of a continued fraction expansion of the dynamical part of the self-energy, $\bSigma^\dyn(z)$ [\Eq{eq:Lmatrixelements-en-tn-main}]. They can be obtained from the spectrum $\bW(\omega)$  of the self-energy via a Lanczos scheme. Because the self-energy defines the recursion measure, this procedure yields uniquely determined matrix elements. The resulting matrix elements $\bolde_n$ and $\bt_n$ are interpolated (\Sec{sec:LiouvillianPeriodization}) using a truncated Fourier Ansatz to obtain $\bolde_{\bk n}$ and $\bt_{\bk n}$, which are reinserted into the CFE to obtain an interpolated self-energy, $\bSigma_{\bk}(z)$. 

Since the Liouvillian is independent of frequency, the same is true for its matrix elements $\bepsilon_{n}$ and $\bt_{n}$. Consequently, the interpolated   version of the Liouvillian remains hermitian, which ensures causality. This poses an inherent advantage of our $\cL$-scheme compared to traditional schemes, like $\Sigma$-, $M$-, and $G$-periodization. \changed{Moreover, $\cL$-interpolation yields results with better $\bk$-space resolution than traditional periodization schemes, because it distributes $\bk$-space information  among a large number of Liouvillian matrix elements.}  

To benchmark our $\cL$-interpolation scheme, we considered the one-dimensional Hubbard model (\Sec{sec:1dHubbard_benchmark}) and computed its spectral function using TaSK \cite{Kovalska2025}, a numerically (almost) exact MPS method. We Fourier-transformed the TaSK results for $G$, $\Sigma$, $M$, and $\cL$ to position space and found that $\cL$ shows the most local behavior (its matrix elements decay most quickly with distance). Consequently, truncating the Fourier coefficients of $\cL$ by range yields a much more faithful representation of the true spectral function and self-energy than a corresponding truncation of $G$, $\Sigma$ or $M$. This provides a fundamental advantage for momentum resolution, since larger clusters are often numerically inaccessible, particularly at low temperatures and in real frequencies.

Next, we considered the same one-dimensional Hubbard model, but now treated with CDMFT for various cluster sizes, using TaSK \cite{Kovalska2025} as an impurity solver (\Sec{sec:1dHubbard_CDMFT}). We compared the results obtained by interpolating its self-energy using the $G$-, $\Sigma$-, $M$-, and $\cL$-schemes.  We studied the convergence with cluster size of the interpolated CDMFT results to the benchmark TaSK results of \Sec{sec:1dHubbard_MPS}. For the 2-site cluster, $\cL$-interpolation arguably performed somewhat worse than $G$-\changed{interpolation}; but for  4- and 8-site clusters, $\cL$-interpolation yielded smoother spectra that resembled the benchmark TaSK result much better.

Finally, we presented  examples of cluster-DMFT interpolation for the doped, two-dimensional $t$-$t'$ Hubbard model, using \changed{four-site CDMFT with an ED impurity solver (\Sec{sec:CDMFT+ED2DHubbard})} and 
four-patch DCA with an NRG impurity solver (\Sec{sec:2dHubbard_DCA}). For the latter, we considered three choices of doping, yielding a Fermi liquid, a pseudogap phase, and a Mott insulator. For each, we compared non-interpolated   cluster results for the spectral function $A_\bk(\omega)$ and self-energy $-\mr{Im} \Sigma_\bk(\omega)$ to interpolated versions obtained via $\Sigma$-, $M$- and $\cL$-interpolation. 

We performed the CFE expansion of the DCA+NRG self-energy $\bSigma_\bK(z)$ using a fast/slow splitting scheme (\App{sec:restfunction_splitting}), implemented in such a manner that the resulting coefficients $\bolde_{\bK n}$, $\bt_{\bK n}$ and the fast-mode contributions $\Rest^{\fast}_{\bK n}(z)$ decay exponentially with CFE depth $n$. The dependence of $\bSigma_{\bK}(z)$ on the patch index $\bK$  is then mainly encoded in the matrix elements $\bolde_{\bK n}$, $\bt_{\bK n}$, while the functions $\Rest^{\fast}_{\bK n}(z)$ depend only weakly on $\bK$ and hence are easy to interpolate (\App{app:dependence_splitting_parameters}).  Consequently, $\cL$-interpolation yields  better results than $\Sigma$- or $M$-interpolation: it produces more detailed structures with higher $\bk$-space resolution, that better mimic key intrinsic features of the non-interpolated cluster results. 

This is especially pronounced for Fermi surface properties (\Sec{sec:2dHubbard_DCA_periodFS}). Comparing the spectral functions $A_\bk(\omega)$ obtained from the various interpolation schemes, we find that $\cL$-interpolation resolves its structures most sharply. While all \changed{interpolations} perform well for the Fermi liquid, the differences become strikingly apparent in the pseudogap phase. The $\cL$-scheme interpolates the FS of the pseudogap phase to Fermi arcs, qualitatively similar to those  observed in ARPES experiments. The $\cL$-scheme also is the one that best mimics the non-interpolated   cluster results, not only for the Fermi surface but also for the quasi-particle weight, as illustrated by tracking $\mr{max}[A_{\bk}(0)]$ along the Fermi surface (\Fig{fig:DCA_Z}). Perhaps most remarkably, $\cL$-interpolation yields a \rreply{seemingly} smooth evolution between Fermi and Luttinger surfaces. \changed{At much higher temperatures, orders of magnitude above the Fermi liquid scale,}
similar \changed{results have been} found previously using 
D$\Gamma$A computations \cite{Worm2024_FS+LS}. 
\changed{Our results show that these features persist to much lower temperatures (inaccessible by D$\Gamma$A), see \cite{Pelz2026b,Pelz2026c} for a more in-depth analysis.}

\changed{
\subsection{Outlook}
\label{sec:Outlook}}

\changed{As an outlook, we mention possible future uses of $\cL$-interpolation in various contexts. Points (1-6) involve applications, (7-10) further methodological progress. 

Regarding concrete applications, $\cL$-interpolation will be useful whenever a good momentum resolution of spectral features is desired. We list some examples:}

\rreply{(1) A welcome feature of $\cL$-interpolation is that it can be used to post-process both CDMFT and DCA data. However, we have deliberately refrained from adjudicating between CDMFT and DCA, because there may be problem-specific reasons for preferring one approach over the other. Nevertheless, for situations where both are feasible with comparable computational costs
(e.g.\ 4-site cluster CDMFT vs.\ 4-patch DCA),
it would be interesting to perform a systematic comparison of the results obtained when using $\cL$-interpolation for both cases. 
In particular: to what extent do the resulting Fermi surface properties agree or disagree?}

\rreply{(2) Beyond the pseudogap regime of the 2D Hubbard model studied in Sec.~\ref{sec:2dHubbard_DCA}, other aspects of this model are of high interest, too, such the competition of d-wave super\-con\-duc\-tivity and antiferromagnetism. Studying it via cluster-DMFT \cite{Capone2006,Kancharla2008} involves symmetry-breaking solutions of the cluster-DMFT self-consistency equations \cite{Foley2019_ED,Kundu2024}. To apply $\cL$-interpolation to the resulting self-energies, our formulation will have to be generalized to symmetry-broken cases---a promising goal for future work.} 

\rreply{(3)
In the context of d-wave superconductivity, multi-layered
cuprates are attracting increasing interest. One example is an 
ARPES study of a trilayer cuprate superconductor \cite{Smit2025}, supported by 
theoretical computations using a cluster-rotationally invariant slave-boson  (CRISB) method in the formulation of Ref.~\cite{Lechermann2007}, which employs $\Sigma$-periodization. Another example is a study of $n$-layer cuprate systems using
an approach that combines cluster dynamical mean-field theory and density functional theory in a charge-self-consistent manner \cite{Bacq-Labreuil2025}, employing $G$-interpolation for the self-energy. It would be fruitful to explore the benefits of $\cL$-interpolation in such challenging contexts.}

\rreply{(4)
It has recently been argued \cite{Sen2020} that the Mott transition in the standard one-band Hubbard model can be understood as a topological phase transition. The argument involved computing the lattice self-energy using single-site DMFT, mapping it to an auxiliary tight-binding chain, and analyzing the band structure of the latter. We note that the chain parameters can be obtained by a CFE expansion of the self-energy. It would be interesting to further explore this idea in the context of cluster-DMFT studies, using $\cL$-interpolation on the corresponding CFE.}

\rreply{(5) 
An interesting direction where reliable momentum resolution is required are strongly correlated insulators with topological band structures. In such systems, the self-energy can exhibit an intricate pole structure with topological properties~\cite{Wagner2023}. Since these poles must correspond to topological structure in the single-particle irreducible Liouvillian [c.f.\ Eq.~\eqref{subeq:defineS(z)}], we expect that $\cL$-interpolation will be useful to reliably capture these structures through cluster-DMFT approximations. Further, we conjecture that an analysis of the matrix elements and corresponding single-particle irreducible operators that give rise to this topological structure may lead to new insights for these systems.}

\changed{Opportunities for methodological developments:} 

\changed{(6)} In this paper, we discussed interpolation
as a post-processing tool, to be used after DMFT self-consistency has been achieved. This is to be distinguished from interpolation \textit{within} cluster-DMFT loops. The latter can yield non-causal hybridization functions if the standard self-consistency equations are used, but this causality issue has recently been solved with the generalized cavity construction of Ref.~\cite{Backes2022_genCavity}. 
In future work, it would be interesting to set up 
within-cluster-DMFT-interpolations that combine the generalized
cavity construction with $\cL$-interpolation. 

\changed{(7) The computation of two-particle properties within cluster-DMFT involves \textit{cluster} vertex functions for the impurity cluster. Their interpolation is an open problem. It would be rewarding to generalize $\cL$-interpolation for this purpose, starting, e.g., by expressing spectral representations of vertex functions \cite{Kugler2021,Lee2021,Lihm2023} through the Liouvillian.}

\changed{(8) Beyond cluster-DMFT,} $\cL$-interpolation can be used in any context requiring the interpolation of frequency-depen\-dent dynamical response functions, making it relevant also for some of the other quantum many-body embedding methods. 
\rreply{For example, it has recently been shown \cite{Giuli2026} that close relations exist between ghost rotationally invariant slave-boson theory (g-RISB) \cite{Lanata2017,Lanata2022}, the ghost-Gutzwiller approximation (g-GA) \cite{LeeTH2019}, density matrix embedding theory (DMET) \cite{Knizia2012,Sekaran2023} 
  involving ghost fermions (g-DMET) \cite{Lanata2023a}, and DMFT --- they are all complementary formulations of the same quantum embedding structure. Cluster versions of these schemes will all lead to self-energies that are defined on embedded clusters and hence may need interpolation  \cite{Lechermann2007}; $\cL$-interpolation is ideally suited for this purpose.}

(9) Finally, looking beyond cluster embedding methods, we note that the CFE representation of the Green's function or the self-energy through \textit{Liouvillian} matrix elements, as described in \Sec{sec:Liouvillian_basics}, constitutes a versatile tool that promises to be useful also in other contexts involving spectral representations. \rreply{For example, the TaSK approach for computing spectral functions using tensor network methods naturally leads to CFE representations \cite{Kovalska2025}. In the zero-temperature formulation employed in Ref.~\cite{Kovalska2025}, these CFE coefficients are identified with Hamiltonian matrix elements. However, TaSK can be reformulated in terms of operator spaces and Liouvillian matrix elements, resulting in a natural generalization to finite temperatures~\cite{Qu2026}.
Further, the insights developed in this work also enable the determination of spectral functions with high momentum resolution from explicit TaSK computations for a relatively small number of momentum points~\cite{GleisLiouvillianSampling2026}.}

\bigskip 

\subsection*{Acknowledgements}

We deeply thank Simon Verret, David S\'en\'echal and Andr\'e-Marie Tremblay for providing their CDMFT+ED data for the 2D Hubbard model, enabling us to test $\cL$-\changed{interpolation} during early stages of this work. We also thank Benjamin Bacq-Labreuil, Pierre-Olivier Downey, Antoine Georges, Maurits Haverkort, Seung-Sup Lee, Gabriel Kotliar, Andr\'e-Marie Tremblay and Aleksandrs Zacinskis for inspiring discussions. 
\changed{We are grateful to Fabian Kugler and Seung-Sup Lee for their careful reading of the manuscript and insightful comments.}
We acknowledge the Gauss Centre for Supercomputing e.V. (www.gauss-centre.eu) for funding this project by providing computing time on the GCS Supercomputer SUPERMUC-NG at Leibniz Supercomputing Centre (www.lrz.de). We also acknowledge additional computational resources provided by the Arnold Sommerfeld Center for Theoretical Physics (www.theorie.physik.uni-muenchen.de). This work was supported in part by the Deutsche Forschungsgemeinschaft under grants INST 86/1885-1 FUGG, LE 3883/2-2 and Germany’s Excellence Strategy EXC-2111 (Project No. 390814868). It is part of the Munich Quantum Valley, supported by the Bavarian state government with funds from the Hightech Agenda Bayern Plus. The National Science Foundation supported JvD in part under PHY-1748958. AG acknowledges support from the Abrahams Postdoctoral Fellowship of the Center for Materials Theory at Rutgers University.

\appendix

\begin{appendices}

\changed{
\section{Principal square root of a matrix}
\label{sec:PrincipalSquareRoot}     

\changed{In the main text, we defined  the weight matrix $\bt_1$ as the \textit{principal square root} of a hermitian, positive semidefinite matrix [cf.~\Eqs{eq:definition_t1}, \eqref{eq:t1}].} Here,  we briefly recall the definition of the principal square root and its properties.}

For any hermitian, positive semidefinite matrix $\bM$, the principal square root $\bM^\frac{1}{2}$ is the unique hermitian, positive semidefinite matrix satisfying $\bM^\frac{1}{2} \bM^{\frac{1}{2}} = \bM$. Given the eigendecomposition $\bM = \bU \bD \bU^\dagger$, the principal square root is obtained as $\bM^\frac{1}{2} = \bU \bD^\frac{1}{2} \bU^\dagger$. 

If $\bM$ is full rank (all eigenvalues nonzero), the inverse of $\bM^\frac{1}{2}$ is $\bM^{-\frac{1}{2}} = \bU \bD^{-\frac{1}{2}} \bU^\dagger$. If  $\bM$ is rank-deficient (some eigenvalues zero), the principal square root can be expressed as $\bM^\frac{1}{2} = \tbU \tbD^\frac{1}{2} \tbU^\dagger$, where $\tbD$, $\tbU$ and $\tbU^\dagger$ are truncated versions of $\bD$, $\bU$ and $\bU^\dagger$, obtained by dropping the columns and/or rows corresponding to zero eigenvalues. Then, the inverse of the principal square root can be defined only in the image space of $\tbU \tbU^\dagger$, namely as the pseudoinverse $\bM^{-\frac{1}{2}} = \tbU \tbD^{-\frac{1}{2}} \tbU^\dagger$. Moreover, then $\bM^{-\frac{1}{2}} \bM^{\frac{1}{2}} = \tbM{}^{\frac{1}{2}} \tbM^{-\frac{1}{2}} = \tbU \tbU^\dagger$ is not a unit matrix, but a projector onto the subspace spanned by the eigenvectors with nonzero eigenvalues.

\section{Operator inner product for $T=0$}
\label{app:OperatorProductT=0}

This appendix discusses subtleties arising in the definition of operator inner products at zero temperature.

The operator inner product $\sbra{\, } \, )$ of \Eq{eq:operatornorm} is well-defined only if $\rho$ has full rank, i.e.\ if $\bra{\Psi} \rho \ket{\Psi} \neq 0$ for every state $\ket{\Psi} \in \mathbb{V}$; otherwise $\ket{\Psi}\! \bra{\Psi} $ would have zero norm w.r.t.\ $\sbra{\, } \, )$ despite being a nonzero operator. The thermal density matrix $\rho$ does have full rank for $T\!\neq \! 0$. However, the zero-temperature density matrix, $\rho_\ground = \rho(T\!=\!0)$, does not have full rank, since  $\bra{\Psi} \rho_\ground \ket{\Psi} \neq 0$ only if $\ket{\Psi} \in \mathbb{V}_\ground$, the ground state subspace of $\mathbb{V}$. In this case, we restrict attention to 
$\doubleW_\ground = \{O_\ground \}$, the space of all operators with the property that either  $O_\ground$ or $O_\ground^\dagger$ does not annihilate all ground states. For any $\sket{O}\in \doubleW$, its projection into $\doubleW_\ground$ is defined as $\sket{O_\ground} = \cP_\ground \sket{O} = \sket{O - \overline{P}_\ground O \overline{P}_\ground}$. Here $\overline P_\ground = \doubleI - P_\ground$, where $P_\ground$ and $\overline{P}_\ground$ respectively are projectors onto the ground state subspace $\mathbb{V}_\ground$ and its complement, $\overline{\mathbb{V}}_{\ground}$, the space of excited states, defined such that $\mathbb{V} = \mathbb{V}_{\ground} \oplus \overline{\mathbb{V}}_{\ground}$. The projector property $\cP_\ground^2 = \cP_\ground$ is readily verified. For $T=0$, the inner product is well-defined on $\doubleW_\ground$: indeed, it yields 
\begin{align}
    \label{eq:groundstatecondition}
    \sbra{O_\ground}O_\ground)  = \mathrm{Tr} \bigl[\rho_\ground \{O_\ground^\dagger, O_\ground \}\bigr] = \mathrm{Tr} \bigl[\rho_\ground \{O^{\dagger},O \}\bigr] \, ,
\end{align}
since $\rho_\ground \overline P_\ground \!=\! \overline P_\ground \rho_\ground \!=\! 0$. If $\sket{O_\ground} \neq 0$, then $O \neq \overline{P}_\ground O \overline{P}_\ground$, hence $O$ or $O^{\dagger}$ does not annihilate $\mathbb{V}_\ground$; thus the right side of \Eq{eq:groundstatecondition}, and hence also $\sbra{O_\ground}O_\ground)$, is nonzero.

\section{Matrix elements of inverse matrices}

This appendix collects two elementary identities used to derive CFEs. They express the 11-blocks of inverses of block matrices through  Schur complements. All (block) matrix inverses appearing below are assumed to exist.

Let $A = {A_{11} \; A_{12} \choose A_{21} \; A_{22}}$ be an invertible $2\times 2$ block matrix, then the $11$-block of its inverse is given by%
\begin{subequations}%
\label{subeq:Schur2x2}%
\begin{align}
    \label{eq:(1/A)_11:2x2}
    (A^{-1})_{11} & = [A/A_{22}]^{-1} \, , \\
    \label{eq:(1/A)_11:2x2-Schur}
    [A/A_{22}] & = A_{11} - A_{12} (A_{22})^{-1} \! A_{21} \, . 
\end{align}
\end{subequations}
$[A/A_{22}]$ is known as the Schur complement of $A$ w.r.t.\  block $A_{22}$. 
\changed{It is used in \App{sec:SingleParticleIrredSubspace} to obtain \Eq{eq:GL-projected}.} 

Equations \eqref{subeq:Schur2x2} can be used to obtain similar expressions for the $11$-blocks of larger block matrices. Let $A$ be a $3 \times 3$ block matrix and $B$ a $2\times 2$ submatrix, 
\begin{align}
    A = \begin{pmatrix}
    A_{11} &  A_{12} & A_{13} \\
    A_{21} &  A_{22} & A_{23} \\
    A_{31} &  A_{32} & A_{33} \\
    \end{pmatrix}  , 
    \quad B  = \begin{pmatrix}
    A_{22} & A_{23} \\
    A_{32} & A_{33} 
    \end{pmatrix} , 
\end{align}
then the $11$-block of $A^{-1}$, found using \Eq{subeq:Schur2x2} with $A_{22}$ replaced by $B$ and $A_{12}$ by $(A_{12}, A_{13})$, etc., is given by%
\begin{subequations}%
\label{subeq:Schur3x3}%
\begin{align}
    \label{eq:(1/A)_11:3x3}
    (A^{-1})_{11} & = [A/B]^{-1}  , 
    \\
    \label{eq:(1/A)_11:3x3-Schur}
    [A/B] & = A_{11} - \begin{pmatrix} 
    A_{12}, A_{13}
    \end{pmatrix} 
    (B)^{-1} \! 
    \raisebox{-2mm}{$\begin{pmatrix} 
    A_{21} \\ A_{31} 
\end{pmatrix}.$} 
\end{align}
\end{subequations}
If $A_{23} = A_{32} = 0$, the Schur complement $[A/B]$ becomes
\begin{align}
    \label{eq:Schur-for-impurity}
    [A/B] & = A_{11} - A_{12} (A_{22})^{-1} \! A_{21} - A_{13}  (A_{33})^{-1} \! A_{31} \, .  
\end{align}
If instead $A_{13} = A_{31} = 0$, we have 
\begin{align}
\label{eq:Schur-3x3-recursive}
[A/B] & = A_{11} - 
 A_{12} (B^{-1})_{22}  A_{21}  .  
\end{align}

Finally, let $A$ be an invertible $N \times N$ tridiagonal block matrix,  defined recursively, with $n = 1, \dots, N-1$:
\begin{align}
    A = \bigl(\, \overline{A}_{11} \bigr) , \quad \; \overline{A}_{nn} = 
    \begin{medblockmatrix}
    A_{nn} &  A_{n, n+1} \;\; 0 \, \cdots \, 0 \\ \hline A_{n+1,n} &  \raisebox{-6mm}{$\overline{A}_{n+1,n+1}$} \phantom{.} \\[-5mm] 
    0 &  \\[-2mm] 
    \vdots & \\[-1mm] 
    0 & 
\end{medblockmatrix}. 
\end{align}
Then, recursive use of \Eq{eq:Schur-3x3-recursive} yields (with $n'= n+1$):%
\begin{subequations}%
\label{subeq:Schurnxn}%
\begin{align}
    \label{eq:(1/A)_11:nxn}
    (A^{-1})_{11} & = [\overline{A}_{11}/\overline{A}_{22}]^{-1} \, , \\
    \label{eq:(1/A)_11:nxn-Schur}
    [\overline{A}_{nn}/\overline{A}_{n'n'}] & = A_{nn} - A_{nn'} (\overline{A}_{n'n'})^{-1}_{11} \! A_{n'n} \, , 
\end{align}
\end{subequations}
This leads to a CFE for the $11$-block of $A^{-1}$:
\begin{align}
    \label{eq:11-block-of-inverse-tridiagonal}  \Bigl(\frac{1}{A}\Bigr)_{11} & \! = \cfrac{1}{A_{11} - \! {A_{12} \cfrac{1}{A_{22}  - \!\!\raisebox{-0.9em}{ \ensuremath{\ddots} \raisebox{-0.9em}{ \hspace{-3.5mm} \ensuremath{- A_{N-1,N} \cfrac{1}{A_{NN}} \, A_{N,N-1}  \hspace{-3mm} \phantom{.}}}}}\, A_{21} }} \, . & 
\end{align}
\changed{It is invoked in \Sec{sec:CFE-self-energy} to obtain the CFE \eqref{eq:SigmadynCFE} for $\bS(z)$.}

\changed{\section{Expressing $\bSigma(z)$ through $\cL$}
\label{sec:SingleParticleIrredSubspace}

In this appendix, we establish the conceptual foundation of this paper: we show how the self-energy $\bSigma(z)$ can be expressed through matrix elements of the Liouvillian, proving \Eqs{eq:HartreeFock1-main} and \eqref{subeq:defineS(z)} from \Sec{sec:ExpressingGSigmathroughL}.}

In \Eq{eq:GRAL-retarded-defined} for $\tbG^{R/A}(\omega)$,  
\changed{$\sket{\bc^\dagger} = \bigl( \sket{c_1^\dagger}, \dots, \sket{c_M^\dagger} \bigr)$ is a block of fermionic single-particle operators [\Eq{eq:define-bold-c}], and $\mr{e}^{-\mr{i} \cL t}$ acts on the subspace $\doubleW_{\sp} = \mathrm{span}\{\sket{c_\beta^\dagger}\}$ spanned by them.}
We call these ``charge-1'' operators, since when acting on 
particle number eigenstates, they increase the charge by $+1$. Since $\mr{e}^{-\mr{i} \cL t}$ can be expressed as an infinite power series, the Green's function is governed by the ``dynamically relevant charge-1'' subspace that is spanned by all operators in $\doubleW_{\sp}$ and all operators obtainable from them through repeated action with the Liouvillian. We denote this subspace by $\doubleW_1 (\cL) \subset \doubleW$.  Importantly (and in contrast to $\doubleW$) it depends on the Liouvillian; we will henceforth refrain from indicating this explicitly.
(For brevity, we will refer to $\doubleW_1$ simply as the ``charge-1'' subspace, even though operators with charge $+1$ may exist that are not in $\doubleW_1$ \footnote{For a simple example of charge-$1$ operators that are not in $\doubleW_{1}$, consider a free-fermion system. Since
its Hamiltonian is quadratic, acting with $\mc{L}$ on single-particle operators again yields single-particle operators. Thus $\doubleW_{\sp} = \doubleW_{1}$ and 
the single-particle irreducible subspace 
$\doubleW_\spi$ is empty. Therefore, any operator with charge $+1$ that is not single-particle, e.g.\ $\sket{c^\dagger_\alpha c_\beta^\pdag c^\dagger_\gamma}$, is not in $\doubleW_{1}$.}.)
$\doubleW_{1}$ may have additional subspace structure, depending on the symmetries of the system at hand. For example, in a system with  U(1) spin symmetry or translation symmetry, we could, in principle, define separate subspaces for spin $\uparrow$ and $\downarrow$ operators, or for operators with different momenta $\vec{k}$, respectively.

To express the self-energy through the Liouvillian, we need to exploit additional structure. We decompose the charge-1 operator space as $\doubleW_1 = \doubleW_{\sp} \oplus \doubleW_{\spi}$, where $\doubleW_{\spi}$ is the orthogonal complement of $\doubleW_\sp$ in $\doubleW_1$. It contains all charge-1 operators that are \textit{single-particle irreducible}, i.e.\ orthogonal to $\doubleW_\sp$. For example, if the Hamiltonian contains a quartic term such that $\sket{c^\dagger_\alpha c_\beta^\pdag c^\dagger_\gamma} \in \doubleW_1$, then the operator $\sket{O_{\alpha \beta \gamma}} = \sket{c^\dagger_\alpha c_\beta^\pdag c^\dagger_\gamma} - \sket{c^\dagger_\alpha} \langle c_\beta^\pdag c^\dagger_\gamma \rangle - \sket{c^\dagger_\gamma} \langle c^\dagger_\alpha c_\beta^\pdag \rangle$ is in $\doubleW_\spi$, since  $\sbra{c^\dagger_\delta} O_{\alpha \beta \gamma}) = 0$.
We denote the projectors to the operator spaces $\doubleW_{\sp}$ and $ \doubleW_{\spi}$ by 
\begin{subequations}
\begin{align}
    \label{eq:s-sbar-projectors}
    \cP_{\sp} &= \sket{\bc^\dagger} \sbra{\bc^\dagger} = {\textstyle \sum_\alpha}  \sket{c^\dagger_\alpha} \sbra{c_\alpha^\dagger}, \\
    \cP_{\spi} &= 1- \cP_{\sp} = \sket{\bar{\bc}^\dagger} \sbra{\bar{\bc}^\dagger} = {\textstyle \sum_{\bar{\alpha}}}  \sket{\bar{c}^\dagger_{\bar{\alpha}}} \sbra{\bar{c}_{\bar{\alpha}}^\dagger} \, ,
\end{align}
\end{subequations}
where $\{\sket{\bar{c}_{\bar{\alpha}}^\dagger}\}$ is an arbitrary orthonormal basis of $\doubleW_{\spi} = \mr{span} \{\sket{\bar{c}_{\bar{\alpha}}^\dagger}\}$. \changed{When acting on $\doubleW_1$}, the Liouvillian can thus be expressed as 
\begin{align}
    \label{eq:L_block_decomposition}
    \cL = \sum_{x, \tilde x \in \{\sp, \spi\}} \hspace{-2mm} \cL_{x \tilde x} \hspace{2mm} = \sum_{\bx, \tilde \bx \in \{\bc, \bar \bc\}} \hspace{-2mm}
    \sket{\mathbf{x}^{\dag}} [\cL]_{\mathbf{x}\tilde{\mathbf{x}}} \sbra{\tilde{\mathbf{x}}^{\dag}} \, .
\end{align}
Here,  $\mc{L}_{x \tilde x} = \cP_{x} \mc{L} \cP_{\tilde x}$ are projections of $\cL$ to the subspaces $\doubleW_x$, $\doubleW_{\tilde x}$, and the blocks $[\cL]_{\mathbf{x}\tilde{\mathbf{x}}} = \sbra{\mathbf{x}^{\dag}} \cL \sket{\tilde{\mathbf{x}}^{\dag}}$ are their representation in the bases spanning these subspaces. W.r.t.\ these, the Liouvillian has a $2 \times 2$ block structure, $\genfrac{(}{)}{0pt}{1}{\sp \sp \;\; \sp \spi}{\spi \sp\;\; \spi \spi}$. 

Since $\cP_{\spi} \sket{\bc^\dagger} = 0$ and $\sbra{\bc^\dagger} \cP_{\spi} = 0$, the Green's function $\bG(z)$ of \Eq{eq:GL} equals $[(z - \mc{L})^{-1}]_{\bc \bc}$, the cc block of $(z - \mc{L})^{-1}$. It  can be found using the matrix identity \eqref{subeq:Schur2x2}:
\begin{align}
    \label{eq:GL-projected}
    \bG(z) &=  \sbra{\bc^\dagger} \Bigl[ z -  \mc{L}_{\sp \sp} - \mc{L}_{\sp \spi}  \cfrac{1}{z - \mc{L}_{\spi \spi} } \mc{L}_{\spi \sp} \Bigr]^{-1} \sket{\bc^\dagger} \\ \nonumber
    &=  \Bigl[ \bz -  \sbra{\bc^\dagger} \mc{L}_{\sp \sp} \sket{\bc^\dagger}  - \sbra{\bc^\dagger} \mc{L}_{\sp \spi} \cfrac{1}{z - \mc{L}_{\spi \spi} } \mc{L}_{\spi \sp} \sket{\bc^\dagger}  \Bigr]^{-1} .
\end{align}
\rreply{with $\bz = z \boldone$.} Comparison  to the Dyson equation \eqref{subeq:Dyson} gives expressions for $\bSigma^\Hartree$ and $\bSigma^\dyn$  involving the Liouvillian:%
\begin{subequations}%
\label{subeq:Selfenergy-Liouvillian}%
\begin{align}
    \label{eq:HartreeFock1}
    \bSigma^{\Hartree} &= \bolde_0 - \bolde , \quad \bolde_0 = \sbra{\bc^\dagger} \cL_{\sp \sp} \sket{\bc^\dagger} \, ,\\ 
    \label{eq:Sigma_dyn} \bSigma^\dyn(z) & = \sbra{\bc^\dagger} \mc{L}_{\sp \spi}   \cfrac{1}{z - \mc{L}_{\spi \spi} } \mc{L}_{\spi \sp} \sket{\bc^\dagger} \, . 
\end{align}
\end{subequations}
Since  $\langle \{ c^{\pdag}_{\alpha} , [H_0, c^{\dagger}_{\beta}]\} \rangle \!=\! \epsilon_{\alpha\beta}$, the elements of $\bSigma^\Hartree$ are 
\begin{align}
    \label{eq:HartreeExplicit}
    [\bSigma^{\Hartree}]_{\alpha \beta} =  \langle \{ c^{\pdag}_{\alpha} , [H_\mr{int}, c^{\dagger}_{\beta}]\} \rangle = \sum_{\gamma \delta} \langle c^\dagger_\gamma c^\pdag_\delta \rangle U_{\alpha \gamma, \delta \beta} .
\end{align}
Evidently, they are governed by the single-particle density matrix
\footnote{Here, we exploited the fact that exchange symmetry requires $U_{\alpha \gamma , \delta \beta} = - U_{\gamma \alpha, \delta \beta}= U_{\gamma \alpha , \beta \delta }$.}. 
By contrast, the dynamical part of the self-energy, $\bSigma^\dyn(z)$, is determined by the one-particle dynamics in the single-particle irreducible subspace $\doubleW_{\spi}$.

\changed{We can simplify Eq.~\eqref{eq:Sigma_dyn} by defining 
a block (or row vector) of single-particle irreducible operators $\sket{\bq_1}$ 
that are normalized versions of $\mc{L}_{\spi \sp} \sket{\bc^\dagger}$:} 
\begin{subequations}
\label{subeq:define-q1}
\begin{align}
\label{eq:define-q1}
    \sket{\bq_1} &=  \cL_{\spi \sp} \sket{\bc^\dagger} \btd^{-1}_1 \\
    \label{eq:t1}
    \btpd_1 \btd_1 &=  \sbra{\bc^\dagger} \cL_{\sp \spi} \cL_{\spi \sp} \sket{\bc^\dagger} \, , \quad \bt_1 = \bt_1^\dagger\geq0, 
\end{align}
\end{subequations}
with $\bt_1$ the principal square root of 
$\sbra{\bc^\dagger}  \cL_{\sp \spi} \cL_{\spi \sp} \sket{\bc^\dagger}$. 
Then, \Eq{eq:Sigma_dyn} takes the form [cf.~\Eq{eq:defineS(z)}]%
\begin{subequations}%
\label{subeq:definesS(z)}%
\begin{align}
\bSigma^\dyn(z) &= \btpd_1 \bS(z) \btd_1 \, , \\
    \label{eq:Wz}
    \bS(z)  & = \sbra{\bq_1}  \cfrac{\boldone}{z - \mc{L}_{\spi \spi}}  \sket{\bq_1} \, , 
\end{align}%
\end{subequations}%
with spectral representation [cf.~\Eq{eq:define-spectral-rep-for-S}]%
\begin{subequations}%
\label{eqs:define-spectral-W}%
\begin{align}
    \label{eq:define-spectral-rep-for-S}
    \bS(z) & = \int_{- \infty}^\infty \mr{d} \omega \frac{\bW(\omega)}{z - \omega} \, ,
\\
 \label{eq:define-W-through-Lsbarsbar}
    \bW(\omega) & = \sbra{\bq_1} \delta(\omega-\cL_{\spi \spi}) \sket{\bq_1} \, .
\end{align}%
\end{subequations}
\changed{This concludes our derivation of \Eqs{eq:HartreeFock1-main} and \eqref{subeq:defineS(z)}.}

\changed{Finally, the matrix 
$\bM = \sbra{\bc^\dagger} \cL_{\sp \spi} \cL_{\spi \sp} \sket{\bc^\dagger}$ [\Eqs{subeq:define-q1}] deserves some remarks. Given the eigen\-de\-com\-position $\bM = \bU \bD \bU^\dagger$, its principal square root is  $\bt_1 = \bU \bD^\frac{1}{2} \bU^\dagger$ (\App{sec:PrincipalSquareRoot}). If $\bM$ is full rank,
then $\bt_1^{-1} = \bU \bD^{-1/2} \bU^\dagger$, 
and \Eq{eq:define-q1} yields a block (or row vector) of $M$ orthonormal operators $\sket{\bq_1} \in \doubleW_\spi^M$, with $(\bq_1 \sket{\bq_1} = \boldone$.} The right spaces of the operator blocks  $\sket{\bq_1}$ and $\sket{\bc^\dagger}$ are characterized by the same set of (physically motivated) spin and orbital quantum numbers, since  $\bU \bD^{-\frac{1}{2}}\bU^\dagger$ involves transformations out of and back into the right spaces of $\sket{\bc^\dagger}$.
This important property motivates the use of the principal square root in \Eqs{subeq:define-q1}. \

In many cases, $\bM$ is rank deficient for the simple reason that some of the single-particle orbitals are non-interacting; then, $\bM$ is zero on the entire block of non-interacting orbitals. 
Then, the rank deficiency of $\bM$ simply reflects the well-known fact that the single-particle irreducible self-energy of non-interacting orbitals is zero. 
Prominent---and for our work very relevant---examples are impurity models, where the bath orbitals are non-interacting, and $\bM$ is zero on the entire bath block. 

\changed{If $\bM$ is rank deficient in the space of interacting orbitals, we express its principal square root as $\bt_1 = \bM^\frac{1}{2} = \tbU \tbD^\frac{1}{2} \tbU^\dagger$, where $\tbU$, $\tbD$ are truncations of $\bU$, $\bD$ to that space (see \App{sec:PrincipalSquareRoot}). Then, we take $\btd_1^{-1} = \tbU \tbD^{-\frac{1}{2}} \tbU^\dagger$ to be the pseudoinverse of $\btd_1$ in that space, such that $(\bq_1 \sket{\bq_1} = \tbU \tbU^\dagger$. Note that $\sket{\bq_1}$ and $\sket{\bc^\dagger}$ again have the same quantum numbers, since $\tbU \tbD^{-\frac{1}{2}} \tbU^\dagger$ 
involves transformations out of and back into the right spaces of $\sket{\bc^\dagger}$.}

If $\bt_1$ has rank 0, then $\sket{\bq_1} = 0$. 

In the present work, we never encountered a partial rank-deficiency in the space of interacting operators: $\bt_1$ was either full rank or had rank 0 in this space. 

\changed{
\section{CFE coefficients of self-energy} 
\label{app:LKrylov}

Appendices~\ref{app:LKrylov} and \ref{sec:QIM} are complements to \Sec{sec:CFE-self-energy-main}  of the main text, providing derivations of the results collected there.
In \App{app:LKrylov} we first explain (\Sec{sec:CFE-self-energy}) why the CFE coefficients $\bolde_n$, $\bt_n$ of the dynamical part of the self-energy, $\bSigma^\dyn(z)$, correspond to matrix elements of the Liouvillian. Then, we describe  (\Sec{sec:CFE-via-polynomials}) how these matrix elements can be computed numerically if $\bW(\omega)$, the normalized spectral function of  $\bSigma^\dyn(z)$, is known explicitly. Finally, we describe (\Sec{app:ExactTerminationCFE}) how exact CFEs of finite depth can be obtained by terminating them using residual functions $\Rest_N(z)$ and derive explicit expressions for the latter. In \App{sec:QIM} we discuss some additional peculiarities arising for quantum impurity models.}

\changed{
\subsection{CFE coefficients of self-energy yield Liouvillian matrix elements} \label{sec:CFE-self-energy} 

In \Sec{sec:CFE-self-energy-main}, we stated that a CFE for the dynamical part of the self-energy, $\bSigma^\dyn(z)=\btpd_1 \bS(z) \btd_1$, 
yields  CFE coefficients that are equal to matrix elements of the single-particle irreducible part of the Liouvillian. 
Here, we show this explicitly by representing $\bS(z)$ in a Krylov space constructed via a Lanczos procedure \cite{Koch2011,Foley2024_Liouvillian}.}

According to \Eq{subeq:defineS(z)}, $\bS(z)$ involves the projection of the Liouvillian to the single-particle irreducible subspace, $\mc{L}_{\spi \spi}$.
We thus  define the  Krylov space $\doubleK_\NKrylov = \mathrm{span}\{\sket{\bq_1}, \mc{L}_{\spi \spi} \sket{\bq_1}, \dots, \mc{L}_{\spi \spi}^{\NKrylov -1} \sket{\bq_1}\}$, with  $\sket{\bq_1}$ given by \Eq{eq:define-q1}, and use an $(\NKrylov -1)$-step \textit{block}-Lanczos procedure to obtain an orthonormal basis of operator blocks such that $\doubleK_\NKrylov = \mathrm{span}\{\sket{\bq_1}$, $\sket{\bq_2}, \dots , \sket{\bq_\NKrylov} \}$ and  
\begin{align}
\label{eq:BlockOrthonormality}
    \sbra{\bq_n} \bq_{n'}) = [\doubleI_N]_{nn'} = \delta_{n n'} \boldone .
\end{align}
\begin{subequations}%
\label{subeq:operator_Krylov}%
The blocks for $n\ge 2$ are constructed iteratively by
repeating four substeps that together constitute one step:
\begin{flalign}
    \label{eq:operators_expansion}
    & (1) \hspace{-2mm} & \sket{\widetilde{\bq}_{n}} & = \cL _{\spi \spi}\sket{\bq_{n-1}} , \quad \bolde_{n-1} = \sbra{\bq_{n-1}}\widetilde{\bq}_{n} ) , & \\    
    & (2) \hspace{-2mm} & \sket{\bq^{\perp}_{n}} & = \sket{\widetilde{\bq}_{n}} - \!\! \sum_{n' = 0}^{n-1} \sket{\bq_{n'}} \sbra{\bq_{n'}} \widetilde{\bq}_{n}) , & \\
    & \label{eq:principalsquareroot-tn}
    (3) \hspace{-2mm} & \btpd_{n} \btd_{n} &  =  \sbra{\bq^{\perp}_{n}} \bq^{\perp}_{n}) , \qquad \bt_n^\pdag = \bt_n^\dagger\geq0,   & \\
    \label{eq:operators_expansion_normalize}
    & (4) \hspace{-2mm} & \sket{\bq_{n}} & = \sket{\bq^{\perp}_{n}} \btd^{-1}_n 
\end{flalign} 
\end{subequations}
(1) applies $\cL_{\spi \spi}$ to $\sket{\bq_{n-1}}$ to obtain $\sket{\widetilde{\bq}_{n}}$ and computes its overlap with $\sket{\bq_{n-1}}$; (2) orthogonalizes it w.r.t.\ all previous blocks to obtain $\sket{\bq^\perp_{n}}$; (3) defines the $M  \times  M$ weight matrix  $\bt_{n}$ as the principal square root of the matrix $\sbra{\bq^{\perp}_{n}} \bq^{\perp}_{n})$ (which is positive semidefinite, since $\cL_{\spi \spi}$ is hermitian); and (4) normalizes $\sket{\bq^\perp_{n}}$ to obtain $\sket{\bq_{n}}$, where $\btd^{-1}_n$ is the pseudoinverse of $\btd_{n}$.
The reasons for defining $\bt_n$ using the principal square root are the same as those discussed for the construction of $\sket{\bq_1}$ above. If $\bt_n$ has rank 0, the iterative procedure terminates.

Using the convenient definition $\sket{\bq_0} = 0$, the generated basis operators satisfy a three-term recursion relation,
\begin{flalign}
    \label{eq:recursion_q}
    & \cL_{\spi \spi} \sket{\bq_{n}} = \sket{\bq_{n}}\bolde_n + \sket{\bq_{n+1}}\btd_{n+1} + \sket{\bq_{n-1}}\bt_{n} \, , \; n \geq 1 . \hspace{-0.5cm} & 
\end{flalign}
If at step $N-1$ one finds $\sket{\bq_{N}} = 0$, the Lanczos iteration terminates automatically; otherwise one terminates it by hand after $N-1$ steps. Equation~\eqref{eq:recursion_q} implies that in the Krylov basis $\{\sket{\bq_1}, \dots , \sket{\bq_\NKrylov} \}$, the single-particle irreducible projection of the Liouvillian is represented by a block-tridiagonal matrix (also known as Jacobi matrix),
\begin{align}
    \label{eq:Lnn'_spi}
    [\doubleL_N]_{\spi \spi} = \begin{pmatrix}
    \boldepd_1 & \btpd_2 & \bzero & \cdots & \hspace{-2mm} \bzero & \hspace{-1mm} \bzero & \bzero \\
    \btd_2  & \boldepd_2 & \btpd_3 & \cdots & \hspace{-2mm}\bzero & \hspace{-1mm} \bzero & \bzero \\
    \bzero & \btd_3  & \boldepd_3 & \cdots & \hspace{-2mm}\bzero & \hspace{-1mm} \bzero & \bzero \\[-1mm]
    \vdots & \vdots & \vdots & \ddots & \hspace{-2mm} \vdots & \hspace{-1mm} \vdots & \vdots \\[-1mm] 
    \bzero & \bzero & \bzero & \cdots & \hspace{-2mm} \boldepd_{N-2} & \hspace{-1mm} \btpd_{\NKrylov-1} & \bzero  \\
    \bzero & \bzero & \bzero & \cdots &  \hspace{-1mm} \btd_{\NKrylov-1} & \hspace{-1mm} \boldepd_{\NKrylov-1} & \btpd_\NKrylov \\ 
    \bzero & \bzero & \bzero & \cdots & \hspace{-2mm} \bzero & \hspace{-1mm} \btd_\NKrylov & \boldepd_\NKrylov 
\end{pmatrix} \! .
\end{align}
Its $N^2$ blocks are all $M\!\times \! M$ matrices, given by  
\vspace{-0.5\baselineskip}
\begin{flalign}
    \label{eq:Lmatrixelements-en-tn}
    & \big[[\doubleL_N]_{\spi \spi}\big]_{nn'} = 
    \sbra{\bq_{n}} \cL_{\spi \spi} \sket{\bq_{n'}} = 
    \begin{cases}
        \boldsymbol{\epsilon}_{n}, & 
        n' = n , \\
        \bt_{n+1} , & 
        n' = n +1   , \\
        \btd_{n},  & 
        n' = n -1    ,   \\
        \bzero & \mr{otherwise} \, .
    \end{cases} \hspace{-1cm} &
\end{flalign}
Within $\doubleK_N$, expression \eqref{eq:Wz} for $\bS(z)$ is represented as 
\begin{align}
    \label{eq:SrepresentedInKN}
    \bS(z) \simeq \bigl[ ([\doublez_{N}]_{\spi \spi} - [\doubleL_N]_{\spi \spi})^{-1} \bigr]_{n=1,n'=1} .
\end{align}
where $\doublez_N = z \doubleI_N$ equals $z$ times a block unit matrix, with  $[\doubleI_N]_{nn'} =\delta_{nn'} \boldone$.
Using \Eq{eq:11-block-of-inverse-tridiagonal}, a formula for the $11$- (upper-left) block of the inverse of a tridiagonal block matrix, we obtain the desired CFE (with $\bze_n = \bz - \bolde_n$):
\begin{align}
    \label{eq:SigmadynCFE}
    \bS(z) & \simeq\cfrac{\boldone}{\bze_1 - {\btpd_{2}\cfrac{\boldone}{\bze_2 - \raisebox{-0.9em}{ \ensuremath{\ddots}\raisebox{-0.9em}{ \hspace{-2.5mm} \ensuremath{- \btpd_\NKrylov \cfrac{\boldone}{\bze_N} \, \btd_\NKrylov}}}}\, \btd_{2} }} \, . 
\end{align}
We have thus established that the CFE coefficients $\bolde_n^\pdag$ and $\bt_n^\pdag$ of $\bS(z)$ and hence of $\bSigma^\dyn(z)$ are matrix elements of the single-particle irreducible Liouvillian $\cL_{\spi\spi}$. The CFE  \eqref{eq:SigmadynCFE} becomes exact for sufficiently large $N$, provided that the moment problem defined by the measure $\bW (\omega) \mr{d}\omega$ is determinate. This is the case if the support of $\bW(\omega)$ is bounded, or if $\bW(\omega)$ decays fast enough at high frequencies, see \App{app:MeasuresAndAllThat} for more details.

\subsection{Computing CFE coefficients via orthogonal polynomials}
\label{sec:CFE-via-polynomials}

\changed{In \App{sec:CFE-self-energy}, the CFE coefficients of $\bS(z)$ were identified as matrix elements $\sbra{\bq_n} \cL_{\spi \spi} \sket{\bq_{n'}}$ of the single-particle-irreducible part of the Liouvillian. Now we switch perspective, assume that
the spectral function $\bW(\omega)$ of $\bS(z)$ is known (analytically or numerically), and construct a CFE for $\bS(z)$. This yields expressions for the  matrix elements of $\cL_{\spi \spi} $ as integrals involving $\bW(\omega)$.}

Since each Lanczos block $\sket{\bq_n}$ is a linear combination of powers of $\cL_{\spi \spi}$ applied to the initial block $\sket{\bq_1}$, all such matrix elements are linear combinations of diagonal matrix elements of the form $\sbra{\bq_1} \cL_{\spi \spi}^m \sket{\bq_{1}}$. Now, the latter correspond to moments of the spectral function $\bW(\omega)$ of $\bS(z)$, due to an identity following from \Eq{eq:define-W-through-Lsbarsbar}:
\begin{align}
    \label{eq:moments_of_A}
    \sbra{\bq_1} \cL_{\spi \spi}^m \sket{\bq_1} & = \int \! \mr{d}\omega \, \bW(\omega) \,\omega^m .
\end{align}
It follows that if  $\bW(\omega)$ is known explicitly (analytically or numerically), the CFE coefficients of $\bS(z)$ can be computed directly from the moments of $\bW(\omega)$. This can be done by constructing a set of orthonormal vector-valued polynomials~\cite{Viswanath1994,Auerbach2019,Foley2024_Liouvillian} using a Lanczos scheme completely analogous to that described in 
\Sec{sec:CFE-self-energy} \cite{Lanczos1950,Boley1984_Lanczos,Koch2011,pinna2025_LanczosGreensfunctions}. Below, we review this construction.

The $N$-step block Lanczos scheme of \Sec{sec:CFE-self-energy} is initialized by a length-$M$ block or row vector of operators, $\sket{\bq_1}$. Its entries, all vectors in $\doubleW_{\spi}$, are by definition mutually orthonormal, $\sbra{\bq_1} \bq_1 ) = \boldone$. Each block of Lanczos vectors $\sket{\bq_n}$ can be written as a linear combination of the form
\begin{align}
    \sket{\bq_n} = \sum_{\bar n=0}^{n-1} \cL_{\spi \spi}^{\bar n} \sket{\bq_1} \bC_{\bar n n} \, .
    \label{eq:qn_pn}
\end{align}
For each $\bar n n$, $\bC_{\bar n n}$ is an $M\!\times\! M$ matrix that right-multiplies the length-$M$ row vector of operators $\sket{\bq_1}$.
When this representation is inserted into the orthonormality relations \eqref{eq:BlockOrthonormality} and the matrix elements of \Eq{eq:Lmatrixelements-en-tn}, both $\sbra{\bq_n} \bq_n')$ and $\sbra{\bq_n} \mc{L}_{\spi \spi} \sket{\bq_n'}$ become linear combinations of matrix elements $\sbra{\bq_1} \cL_{\spi \spi}^m \sket{\bq_1}$ involving various powers $m$ of $\cL_{\spi \spi}$. Using \Eq{eq:moments_of_A} to express these through moments of $\bW(\omega)$, \Eqs{eq:BlockOrthonormality} and \eqref{eq:Lmatrixelements-en-tn} take the form
\begin{subequations}%
\label{subeq:A-integrals}%
\begin{align}
    \label{eq:orthogonal-polynomials-integral}
    [\doubleI_N]_{nn'} & = \int \! \mr{d}\omega \, \bponedag_{n} (\omega) \, \bW(\omega) \, \bpone_{n'}(\omega) ,\\ 
    \bigl[[\doubleL_N]_{\spi \spi}\bigr]_{nn'} & = \int \! \mr{d}\omega \, \bponedag_{n} (\omega) \,  \bW(\omega) \, \omega \, \bpone_{n'}(\omega) \, ,
\label{eq:nLn'-pn-omega-pn'}
\end{align}
\end{subequations}
where we have defined the matrix polynomials
\begin{align}
    \label{eq:define-block-polynomials}
    \bpone_n(\omega) = \sum_{\bar n=0}^{n-1} \omega^{\bar n} \bC_{\bar n n} . 
\end{align}
(The reason for the superscript $(1)$ on $\bpone_n(\omega)$ will become clear below.) For each $n$, $\bpone_n(\omega)$ is an $M\times M$ matrix.  Its columns, enumerated by $\alpha \!=\! 1, \dots, M$, are length-$M$ column-vector-valued polynomials of degree $n-1$, which we denote by $\pone_{n,\alpha} \! : \mathbbm{R} \to \mathbbm{C}^M$, $\omega \mapsto \pone_{n,\alpha}(\omega)$.
Note that $\pone_{n,\alpha}$ is constructed from column $\alpha$ of the matrix $\bC_{\bar n n}$, which via \Eq{eq:qn_pn} also defines entry $\alpha$ of the block $\sket{\bq_n} = \bigl(\sket{q_{n,1}}, \dots, \sket{q_{n,M}}\bigr)$. This yields  a one-to-one correspondence, $\pone_{n,\alpha} \leftrightarrow \sket{q_{n,\alpha}}$, between length-$M$ column-vector-valued polynomials and operators in $\doubleW_{\spi}$. The matrix $\bpone_n$ can thus also be viewed as a length-$M$ row vector of column-vector-valued polynomials, associated with the length-$M$  row vector of operators $\sket{\bq_n}$.

To compactify the notation, let
\begin{align}
    \label{eq:define-innerproduct-vectorvaluedfunctions}
    \langle f, g \rangle = \int \mr{d} \omega \, f^\dagger (\omega) \bW(\omega) g (\omega) \, \in \doubleC
\end{align}
denote the inner product, with matrix-valued measure $\bW(\omega) \, \mr{d}\omega$, of two length-$M$ column-vector-valued functions $f$ and $g$, with $f: \mathbbm{R} \to \mathbbm{C}^M$, $\omega \mapsto f(\omega)$, and likewise for $g$. 
Then, equations \eqref{subeq:A-integrals} can be expressed as 
\begin{subequations}%
\label{subeq:orthogonal-polynomials}%
\begin{align}
    \label{eq:orthogonal-polynomials}
    \langle \bpone_n , \bpone_{n'} \rangle & = [\doubleI_N]_{nn'} , \\
    \label{eq:polynomials-L-matrixelements}
    \langle \bpone_n , \hat \cL_{\spi \spi} \, \bpone_{n'} \rangle & = \bigl[[\doubleL_N]_{\spi \spi}\bigr]_{nn'} ,
\end{align}%
\end{subequations}%
where $\hat \cL_{\spi \spi}$ is the Liouvillian in function space, acting as $\hat \cL_{\spi \spi} f(\omega) = \omega f(\omega)$. 
For given $nn'$, \Eq{eq:orthogonal-polynomials} is an $M \times M$ matrix equation whose element $\alpha \alpha'$ reads $\langle \pone_{n,\alpha}, \pone_{n',\alpha'}\rangle = \delta_{nn'} \delta_{\alpha \alpha'}$; similarly for \Eq{eq:polynomials-L-matrixelements}.

Equation \eqref{eq:orthogonal-polynomials} states that the $N\cdot M$ vector-valued polynomials $\{ \pone_{n,\alpha} \}$ defined by the columns of all $\bpone_n(\omega)$, $n\!=\! 1, \dots, N$, are mutually orthonormal w.r.t.\ the matrix-valued measure $\bW(\omega) \mr{d} \omega$. (The superscript on $\bpone_n$ signifies that these are ``associated polynomials of the first kind'' relative to another set of orthonormal polynomials, namely those defined w.r.t.\ the measure $\bA(\omega) \mr{d} \omega$, see App. \ref{app:MeasuresAndAllThat}.)
The polynomials $\{ \pone_{n,\alpha} \}$ thus form a basis for the space of all length-$M$ column-vector-valued polynomials of degree $N-1$. 
Equation~\eqref{eq:polynomials-L-matrixelements} states that all matrix elements of $[\doubleL_N]_{\spi \spi}$, and hence all CFE coefficients of $\bS(z)$, can be expressed as matrix elements of the Liouvillian $\hat \cL_{\spi \spi}$ in this basis. 

For a given $\bW(\omega)$, the relations \eqref{subeq:orthogonal-polynomials}, together with the initial condition $\bpone_1(\omega) = \boldone$, fully define the basis $\{\pone_{n,\alpha}\}$. 
It can be constructed iteratively via a block-Lanczos procedure (in full analogy to that for $\sket{\bq_n}$) which also generates the matrix elements $[\doubleL_{\spi \spi}]_{nn'}$.
It involves (1) multiplication by $\omega$, (2) orthogonalization, (3) identification of a normalization matrix, and (4) normalization:%
\begin{subequations}%
\label{subeq:construct-pns}%
\begin{flalign}
    & (1) \hspace{-2mm} & \widebpone_{n} (\omega)  & =\omega \, \bpone_{n-1} (\omega)  , \;\;\; \bolde_{n-1} = \langle \bpone_{n-1} , \widebpone_{n}\rangle, \hspace{-1cm} & 
    \label{eq:construct-en-for-pn} \\    
    \label{eq:PolynomialOrthogonalization}
    & (2) \hspace{-2mm} &  \bponeperp_{n} (\omega)  & = \widebpone_{n} (\omega) - \!\!  \sum_{n' = 0}^{n-1} \bpone_{n'} (\omega)\langle \bpone_{n'} , \widebpone_{n} \rangle ,  \hspace{-5mm} & \\
    & (3)  \hspace{-2mm} &\btpd_{n} \btd_{n} &  =  \langle \bponeperp_{n} , \bponeperp_{n} \rangle , \qquad \bt_n^\pdag = \bt_n^\dagger\geq0, &
    \label{eq:construct-tn-for-pn}\\
    & (4)  \hspace{-2mm} & \bpone_{n} (\omega) & = \bponeperp_{n} (\omega) \btd^{-1}_n. & 
\end{flalign} 
\end{subequations}
(On the right of \Eq{eq:PolynomialOrthogonalization}, the second term involves matrix multiplication of two $M \! \times \! M$ matrices, the matrix polynomial $\bpone_{n'} (\omega)$ and the inner product matrix $\langle \bpone_{n'} , \widebpone_{n} \rangle$.) Along the way, we get the CFE coefficients $\bolde_{n-1}$ from \Eq{eq:construct-en-for-pn}, and $\bt_{n}$ as the principal square root of $\langle \bponeperp_{n} \! , \bponeperp_{n} \rangle $ in \Eq{eq:construct-tn-for-pn}, as in \Eq{eq:operators_expansion_normalize}. \changed{For a pseudocode for the Lanczos scheme \eqref{subeq:construct-pns}, see \Alg{alg:lanczos}.}

The resulting polynomials and coefficients satisfy a three-term recurrence relation analogous to \Eq{eq:recursion_q},
\begin{align}
    \label{eq:recursion_p}
    \omega \bpone_{n} = \bpone_{n} \bolde_n + \bpone_{n+1}\btd_{n+1} + \bpone_{n-1}\bt_{n} \, , \; n \geq 1 ,  
\end{align}
with $\bpone_0(\omega) = \bzero$, $\bpone_1(\omega) = \boldone$.

\label{app:MatrixElements}
\begin{algorithm}[t!]\label{alg:lanczos}
\caption{Lanczos scheme \eqref{subeq:construct-pns} for computing CFE coefficients}
    \SetKwInOut{Input}{Input}
        \Input{Self-energy $\bSigma(z)$}\vspace{2pt}
        
    \SetKwInOut{Output}{Output}
        \Output{Matrix elements $\bolde_n$, $\bt_n$}\vspace{2pt}

    \SetKwProg{Comp}{Computation:}{}{}
        \Comp{We use the same subscripts as in main text, but in code, only $\bolde_n$ and $\bt_n$ are saved as vectors.}

    \SetKwProg{Init}{Initialization}{:}{}
        \Init{}{Compute
        $\bSigma^{\prime \prime}(\omega) 
        \leftarrow \tfrac{\mr{i}}{2\pi}\big[\bSigma(\omega^+) - \bSigma^\dag(\omega^+)\big]$ and \newline
        set up $\bW(\omega) = \bt^{-1}_1\bSigma''(\omega)\bt_1^{-1}$, where \vspace{-5pt}
            \begin{align}
                \bt_1\bt_1 = \int \mathrm{d}\omega\,\bSigma''(\omega) \nonumber 
            \end{align}
            } 

    \SetKwProg{Iter}{Iteration:}{:}{}
        \Iter{Perform an $N$-step Lanczos scheme to set up an orthonormal polynomial basis $\{\bpone_n(\omega)\}_{n=1,\dots,\NKrylov }$ with starting polynomial $\bpone_1(\omega)=\boldone$}{} \vspace{-4pt}

    \SetKwProg{for}{for}{}{end}
        \for{$\ n=1,\dots,N$}{
        
            \textbf{(a)} Set $\widebpone_{n+1}(\omega) \leftarrow \omega\,\bpone_n(\omega)$ \vspace{4pt}
            
            \textbf{(b)} Compute \vspace{-5pt}
            \begin{align}
                \boldsymbol{\epsilon}_n = \int \mathrm{d}\omega\, \bponedag_{n}(\omega) \mathbf{W}(\omega) \widebpone_{n+1}(\omega) \nonumber
            \end{align}
                
            \textbf{(c)} Orthogonalize by \vspace{-5pt}
            \begin{align}
                &\bponeperp_{n+1}(\omega) \leftarrow \widebpone_{n+1}(\omega) \nonumber \\
                &\quad \ -\sum_i \bpone_i(\omega) \int \mathrm{d}\omega\, \bponedag_{i}(\omega)\bW(\omega) \widebpone_{n+1}(\omega)\nonumber
            \end{align}
                
            \textbf{(d)} Normalize $\bpone_{n+1}(\omega) = \bponeperp_{n+1}(\omega)\, \bt_{n+1}^{-1}$, where \vspace{-5pt}
            \begin{align}
                &\bt_{n+1} \bt_{n+1}= \int \mathrm{d}\omega\, [\bponeperp_{n+1}(\omega)]^\dag \bW(\omega) \bponeperp_{n+1}(\omega) \nonumber
            \end{align}
            \textit{Here, steps (c) and (d) should be repeated once for improved numerical stability.}
        }
\end{algorithm}

Equations \eqref{subeq:defineS(z)} and \eqref{eqs:define-spectral-W} together imply that the dynamical part of the self-energy can be expressed as 
\begin{align}
    \nonumber
    \bSigma^\dyn(z) & = \btpd_1 \langle \bpone_1 ,(z - \hat{\mc{L}}_{\spi \spi})^{-1} \bpone_1 \rangle \btd_1 
    \\ & \simeq  
    \btpd_1 \bigl[(\doublez -[\doubleL_N]_{\spi \spi})^{-1}\bigr]_{11} \btd_1 . 
    \label{eq:SigmaDynviaP1P1}
\end{align}
\changed{The second line is an approximate representation of the first;
conceptually, it amounts to approximating $\bW(\omega)$ 
with $N$ discrete $\delta$-weights chosen such that Eq.~\eqref{eq:define-innerproduct-vectorvaluedfunctions} remains exact if $f$ and $g$ are linear combinations of  the polynomials $\{\bpone_{n\le N}\}$.}
The same argument that led from \Eq{eq:SrepresentedInKN} to \eqref{eq:SigmadynCFE} then gives an approximate depth-$N$ CFE \eqref{eq:SigmadynCFEexplicitfiniteN} for $\bSigma^\dyn(z)$:
\begin{align}
    \label{eq:SigmadynCFEexplicitfiniteN}
    \bSigma^\dyn (z)  \simeq \btpd_1 \cfrac{\boldone}{\bze_1 - { \btpd_{2}\cfrac{\boldone}{\bze_2 - \raisebox{-0.9em}{\ensuremath{\ddots} \raisebox{-.5em}{ \hspace{-2.5mm} \ensuremath{- \btpd_\NKrylov \cfrac{\boldone}{\bze_N} \, \btd_\NKrylov }}}}\, \btd_{2}}} \btd_1 . 
\end{align}

We note in passing that a CFE for the self-energy implies a CFE for the corresponding Green's function: inserting \Eqs{eq:Sigma=HF+dyn} and \eqref{eq:SigmadynCFEexplicitfiniteN} for the self-energy into the Dyson equation \eqref{eq:Dyson1} for $\bG(z)$, we obtain
\begin{align}
    \label{eq:exact-CFE-for-G}
    \mathbf{G}(z) \simeq  \cfrac{\boldone}{\bze_0 -\btpd_{1}\cfrac{\boldone}{\bze_1 - {\btpd_{2}\cfrac{\boldone}{\bze_2 - \raisebox{-0.9em}{\ensuremath{\ddots} \raisebox{-0.5em}{ \hspace{-2.5mm} \ensuremath{- \btpd_\NKrylov \cfrac{\boldone}{\bze_N 
    } \, \btd_\NKrylov  \hspace{-4mm} \phantom{.}}}}}\, \btd_{2} 
    \hspace{-2mm}} }\, \btd_{1}}  \, .
\end{align}
This CFE can also be obtained directly by using a Lanczos scheme with measure $\bA(\omega) \mr{d}\omega$ to construct an orthogonal set of matrix polynomials $\{\bp_n(\omega), n = 0, \dots, N\}$, initiated with $\bp_0(\omega) = \boldone$. These polynomials differ from the $\{\bpone_n(\omega), n = 1, \dots, N\}$  obtained using the measure $\bW(\omega)\mr{d}\omega$, but the CFE coefficients $\bolde_{n\ge 1}$ and $\bt_{n\ge 2}$ are the same, as elaborated in \App{app:MeasuresAndAllThat}.

\changed{To conclude this section, we consider the case that the self-energy spectrum $\bW(\omega)$ is discrete with a finite number, say $N$, of $\delta$-peaks, as is the case for a finite-size system. Then, the dimension of the Liouvillian and corresponding Krylov space equals the number of discrete energies; the Lanczos scheme terminates after $\NKrylov$ steps; and the  CFE representation   \eqref{eq:SigmadynCFEexplicitfiniteN}
for $\bSigma^\dyn (z)$ is exact.}

\subsection{Exact termination of CFE}
\label{app:ExactTerminationCFE}

\changed{If the CFE for $\bSigma^\dyn(z)$ does not terminate, the finite-depth representation \eqref{eq:SigmadynCFEexplicitfiniteN} is an approximation. In that case, it may be desirable} to replace \Eq{eq:SigmadynCFEexplicitfiniteN} by an exact depth-$N$ CFE. 
To that end, we subtract an $M\!\times\!M$ matrix function, the \textit{residual  function} $\Rest_N(z)$, from the denominator at depth $N$, 
and \textit{define} it by requiring that the following expression is \textit{exact}: 
\begin{flalign}
    \label{eq:S(z)CFE}
    & \bSigma^\dyn(z) = \btpd_1 \cfrac{\boldone}{\bze_1 - { \btpd_{2}\cfrac{\boldone}{\bze_2 - \! \raisebox{-0.9em}{\ensuremath{\ddots} \raisebox{-0.5em}{ \hspace{-2.5mm} \ensuremath{ \! - \btpd_\NKrylov \cfrac{\boldone}{\bze_N \! - \! \Rest_\NKrylov(z)  } \, \btd_\NKrylov  \hspace{-4mm} \phantom{.}}}}}\, \btd_{2} \hspace{-2mm}} } \btd_1   .  \hspace{-1cm} & 
\end{flalign}
 \changed{Next, we discuss three schemes for computing residual functions, with (i) numerically less stable than (ii,iii).

Scheme (i): By} equating two exact CFEs for  $\bSigma^\dyn(z)$ of the form \eqref{eq:S(z)CFE}, having depth $N-1$ or $N$, we obtain
\begin{align}
    \label{eq:rest-invert}
    \Rest_{N} (z) &= \bze_N  - \btpd_{N}
    \Rest^{-1}_{N-1} (z) \btpd_{N} . 
\end{align}
Starting from $\bR_0(z) = \bSigma^\dyn(z)$, this recursion relation can be used to compute residual functions to any depth. 

\changed{Scheme (ii): A numerically more stable scheme is derived in \App{app:InvertingMatrices}. It takes the polynomials $\bpone_{N-1}(\omega)$ and $\bpone_N(\omega)$ as input, and computes the residual function terminating the CFE exactly at depth $N-1$ using} 
\begin{subequations}%
\label{eq:results:restfunction}%
\begin{align}
\label{eq:results:restfunction-a}
    \bR_{N-1} (z) & = \bI_N (z) - \bF^{L}_N (z) \bG_N^{-1} (z) \bF^{R}_N (z) \,  . 
\end{align}
The ingredients of this formula are defined as follows:
\begin{flalign}
    \label{eq:explicit-GFN-spectra-main-text}
    \bG_N (z) & = \! \int \!  \mr{d} \omega \, \bponedag_{N-1} (\omega) \bW(\omega) \bpone_{N-1} (\omega) /(z-\omega) \, , \hspace{-1cm}  & \\
    \bF^{L}_{N} (z)  & = \! \int \!  \mr{d} \omega \, \bponeperpdag_{N} (\omega) \bW(\omega) \bpone_{N-1} (\omega) /(z-\omega) \, , \hspace{-1cm} & \\
    \bF^{R}_N (z)  & = \! \int \!  \mr{d} \omega \, \bponedag_{N-1} (\omega) \bW(\omega) \bponeperp_{N} (\omega) /(z-\omega) \, , \hspace{-1cm} & \\
    \bI_N(z)  & = \! \int \!  \mr{d} \omega \, \bponeperpdag_{N} (\omega)  \bW(\omega) \bponeperp_{N} (\omega) /(z-\omega) \, . \hspace{-1cm}  & 
\label{eq:results:restfunction-e}
\end{flalign}
\end{subequations}

Both schemes (i) and (ii) involve a difference between two terms, one of which involves a matrix inversion. However, scheme (ii) turns out to be numerically more stable. \changed{One reason is that in \Eq{eq:results:restfunction-a} the quotient of $z$-dependent functions $\bF_N(z) \bG_N^{-1}(z)\bF_N(z)$ falls off with $z^{-1}$ at large $|z|$ by construction; by contrast, in \Eq{eq:rest-invert} the term $\btpd_{N} \Rest^{-1}_{N-1}(z) \btpd_{N}$ has to precisely cancel  $\bz_N$ to produce $z^{-1}$ behavior at large $|z|$. 
For further discussions of numerical stability, see Refs.~\onlinecite{Kugler2022} and ~\onlinecite{Zacinskis2026} (after Eq.~(13) there).
In the former,} Kugler derived an equation similar to \Eq{eq:results:restfunction-a} (his Eq.~(12)) for computing self-energies via symmetric improved estimators, and gives a detailed discussion of the numerical stability of this construction. 

\changed{Scheme (iii): A variation of scheme (ii) is to compute
$\bR_N(z)$ via} a sequence of depth-1 CFEs, each terminated by a residual function that serves as  input for the next iteration. For
initialization, we use  $\bR_0(z) = \bSigma^\dyn(z)$ as input. Then, we iteratively proceed as follows for $n\ge 1$: (a) Using $\bR_{n-1}(z)$ as input, we find its normalized spectral function $\bW_{n-1}(\omega)$ and weight matrix $\bt_n$, as follows:
\begin{subequations}%
\label{subeq:define-normalized-spectral-function}
\begin{align}
    \bR''_{n-1} (\omega) & = \frac{\mr{i}}{2\pi} \bigl[ \bR_{n-1} (\omega^+) - \bR^\dagger_{n-1} (\omega^+)\bigr] , \\
    \bt_n \bt_n & = \int \mr{d} \omega \, \bR''_{n-1} (\omega) , \quad \bt_n = \bt_n^\dagger \ge 0 , \\
    \label{eq:define-normalized-spectral-function}
    \bW_{n-1} (\omega) & = \bt_n^{-1} \bR''_{n-1} (\omega) \bt_n^{-1} .
\end{align}%
\end{subequations}
(b) We use a one-step Lanczos scheme with measure $\bW_{n-1}(\omega)\mr{d}\omega$ to  compute $\bolde_n$. (c) We express $\bR_{n-1}(z)$ via the depth-1 version of \Eq{eq:S(z)CFE}, with $\{\bSigma^\dyn,\bt_1,\bz_1, \bR_1\}$ there replaced by $\{\bR_{n-1},\bt_{n},\bz_{n},\bR_{n}$\}:
\begin{align}
    \label{eq:R1(z)CFE-nth-step}
    \bR_{n-1}(z) & = \bt_{n} \cfrac{\boldone}{\bze_{n} - \bR_{n} (z)}\bt_{n} . 
\end{align}
Concatenating $N$ depth-1 CFEs of the form \eqref{eq:R1(z)CFE-nth-step} yields a CFE of the form \eqref{eq:S(z)CFE}, with CFE coefficients identical to those obtained via an $N$-step Lanczos scheme. 

To make steps (b,c) concrete, let us put a superscript $(n)$ on polynomials and coefficients obtained from the $n$-th one-step Lanczos scheme with measure $\bW_{n-1}(\omega)\mr{d}\omega$ (\rreply{for $(n)=(1)$, we get the superscript $(1)$ encountered in \Sec{sec:CFE-via-polynomials} above,
with $\bW_0 (\omega)= \bW(\omega)$).} Step (b) yields $\bp_1^{\scriptscriptstyle (n)} (\omega) = \boldone$, $\widetilde \bp_2^{\scriptscriptstyle (n)} = \omega \boldone$, $\bolde_1^{\scriptscriptstyle (n)}$, $\bp_2^{ \scriptscriptstyle (n) \perp} = \omega \boldone - \bolde_1^{\scriptscriptstyle (n)}$.
Step (c) yields  $\bR_1^{\scriptscriptstyle (n)}$ via  \Eqs{eq:results:restfunction}, with $\{\bW,\bpone_{N-1},\bponeperp_N , \bR_{N-1}\}$ there replaced by $\{\bW_{n-1}, \boldone , \omega \boldone - \bolde_1^{\scriptscriptstyle (n)} , \bR_1^{\scriptscriptstyle (n)} \}$. Finally, we set $\bolde_{n} = \bolde_1^{\scriptscriptstyle (n)}$ and $\bR_{n} =\bR_1^{\scriptscriptstyle (n)}$.

\changed{Scheme (iii) is as stable as scheme (ii), because it likewise relies on \Eqs{eq:results:restfunction}. It forms the basis for constructing log-CFEs involving fast/slow splitting, see \App{sec:restfunction_splitting}, and the corresponding pseudocode 
of Alg.~\ref{alg:wSF} there.}

\subsection{Derivation of \Eqs{eq:results:restfunction} for residual function} 
\label{app:InvertingMatrices}

\changed{In this section, we derive formulas \eqref{eq:results:restfunction} for the numerically stable computation of residual functions. The discussion assumes some familiarity with moment problems, measures, and their relation to orthogonal polynomials, as summarized in \App{app:MeasuresAndAllThat}.}

We consider the space, $L^2(\boldsymbol{\mu})$, of  functions that are square integrable against a measure $\boldsymbol{\mu}$, that is such that $L^2(\boldsymbol{\mu})$ contains block polynomials of arbitrary degree. For definiteness, we consider the case  $\bmuone(\mr{d} \omega) = \bW(\omega)\mr{d}\omega$, where $\bW(\omega)$, the normalized spectral function of the dynamical part of the self-energy $\bSigma^\dyn(z)$, is assumed to be known analytically or numerically. (An analogous discussion applies for $\boldsymbol{\mu}( \mr{d} \omega) = \bA(\omega)\mr{d}\omega$, where $\bA(\omega)$ is the spectral function of $\bG(\omega)$.) In this space, the Liouvillian $\hat \cL$ acts on functions $\mathbf{f} \in L^2(\boldsymbol{\mu})$ according to $\hat \cL \mathbf{f}(\omega) = \omega \, \mathbf{f}(\omega)$ [cf.~\Eq{eq:GRAL}]. Since $\hat \cL$ is self-adjoint and therefore has a real spectrum, $(z-\hat \cL)$ with $\mr{Im} \, z \neq 0$ is a normal operator that is invertible.

In \Sec{sec:CFE-via-polynomials}, a Lanczos scheme is used to generate blocks of vector-valued polynomials, $\{\bpone_n(\omega), n=1, \dots, N\}$, that are orthogonal w.r.t.\ $\bmuone$ [cf.~\Eq{eq:orthogonal-polynomials}]. These polynomials span an $NM$-dimensional subspace of $L^2(\boldsymbol{\mu})$, the space of vector-valued polynomials with degree $\leq N-1$.  We call the projector onto this space $\mc{P}_N$, while $\overline{\mc{P}}_N$ projects onto its orthogonal complement (such that $\mc{P}_N + \overline{\mc{P}}_N = \doubleI$).
In the polynomial basis, the single-particle irreducible Liouvillian $\hat\cL_{\spi \spi}$ is represented \changed{exactly} by an $N \times N$ block-tridiagonal matrix [cf.~\Eq{eq:polynomials-L-matrixelements}]. This block-tridiagonal structure implies a block-CFE for $\bS(z) = \langle \bpone_{{1}} , (z - \hat \cL_{\spi \spi})^{-1} \bpone_{{1}}\rangle$, terminated by a residual function that ensures the representation is exact. \changed{We denote this residual function by
$\bR_N(z)$; it corresponds to that appearing in \Eq{eq:S(z)CFE} in \Sec{app:ExactTerminationCFE}. 

We start by constructing}
a measure $\boldsymbol{\mu}_N (\mathrm{d} \omega)$ such that 
\begin{align}
    \label{eq:mu_N_measure}
    \bR_N(z) = \int \frac{\boldsymbol{\mu}_N(\mr{d}\omega)}{z-\omega} \, .
\end{align}
Following the notation of Sec.~\ref{sec:Liouvillian_basics}, we denote projections to $\mc{P}_N$ and $\overline{\mc{P}}_N$ by indices $p$ and $\overline{p}$, and blocks of matrix representations by square brackets with indices $\bp$ and $\overline{\bp}$, respectively. By applying formula \eqref{subeq:Schur2x2}, we find
\begin{align}
    \label{eq:zLpp_inverse}
    \bigl[(z\!-\!\hat \cL_{\spi \spi})^{-1}\bigr]_{\bp\bp}  &= \\ \nonumber
    \bigl( z &\!-\! [\hat{\cL}]_{\bp\bp}  \!-\! [\hat{\cL}_{p\overline{p}} (z - \hat{\cL}_{\overline{p}\overline{p}})^{-1}  \hat{\cL}_{\overline{p}p}]_{\bp\bp}\bigr)^{-1} \, .\hspace{-0.7cm} & 
\end{align}
Here, $[\hat{\cL}]_{\bp\bp}$, represented in the polynomial basis $\bpone_n$, is the tridiagonal matrix \Eq{eq:polynomials-L-matrixelements}. From the three-term recursion for the polynomials $\bpone_n$ [\Eq{eq:recursion_p}], it follows that
\begin{align}
    \label{eq:PbarLpn}
    \overline{\mc{P}}_N \hat{\cL} \bpone_n(\omega) = \delta_{nN} \bpone_{N+1}(\omega)\bt_{N+1} \, .
\end{align}
For $\bigl[\hat{\cL}_{p\overline{p}} (z - \hat{\cL}_{\overline{p}\overline{p}})^{-1}  \hat{\cL}_{\overline{p}p}\bigr]_{nn'}$, with $n,n'\le N$, this implies%
\begin{subequations}%
\label{suqeq:DerivationRestFunction}%
\begin{flalign}
    & 
    \label{eq:NNblockofL(z-L)L}
    \langle \bpone_n , \hat{\cL}_{p\overline{p}} (z - \hat{\cL}_{\overline{p}\overline{p}})^{-1}  \hat{\cL}_{\overline{p}p}  \, \bpone_{n'} \rangle = \delta_{nN}\delta_{n'N} \bR_N(z) \, , \hspace{-1cm} & \\
    & \bR_N(z) = \btpd_{N+1} \langle \bpone_{N+1}, (z - \hat{\cL}_{\overline{p}\overline{p}})^{-1} \bpone_{N+1} \rangle \, \btd_{N+1} .  \hspace{-1cm} &
    \label{eq:DefinitionRN(z)}
\end{flalign}%
\end{subequations}%
Equation~\eqref{eq:NNblockofL(z-L)L} shows by explicit construction that the only nonzero block of $\hat{\cL}_{p\overline{p}} (z - \hat{\cL}_{\overline{p}\overline{p}})^{-1}  \hat{\cL}_{\overline{p}p}$ is its $N\!N$ block; we denote it by $\bR_N(z)$. Equation \eqref{eq:DefinitionRN(z)} expresses it through the $N\!+\!1,N\!+\!1$ block of $(z - \hat{\cL}_{\overline{p}\overline{p}})^{-1}$, exploiting \Eq{eq:PbarLpn}, whose left side provides a link (the only one available!) between the $\overline{\mc{P}}_N$ and $\mc{P}_N$ sectors. Further, \Eq{eq:mu_N_measure} immediately follows from an explicit construction of the measure,
\begin{flalign}
    & \boldsymbol{\mu}_N(\mr{d} \omega) = \btpd_{N+1} \langle \bp_{N+1}, \delta(\omega - \hat{\cL}_{\overline{p}\overline{p}}) \, \bp_{N+1} \rangle \btd_{N+1} \mr{d}\omega \, . \hspace{-1cm} & 
\end{flalign}

\changed{Next, we derive the formulas \eqref{eq:results:restfunction} for computing residual functions explicitly.} Consider the inverse of the matrix in Eq.~\eqref{eq:zLpp_inverse}, which is a block-tridiagonal matrix:
\begin{align}
    \label{eq:blockmatrix}
    \bigl[[(z-&\hat \cL_{\spi \spi})^{-1}]_{\bp\bp}\bigr]^{-1} = 
    \\ \nonumber
    &\begin{pmatrix}
    \bz_1 & -\btpd_2 & \cdots & \bzero & \bzero & \bzero \\
    -\btd_2 & \bz_2 & \cdots & \bzero & \bzero & \bzero \\
    \vdots & \vdots & \ddots & \vdots & \vdots & \vdots \\[-1mm]
    \bzero & \bzero & \cdots & \bz_{N-2}  & -\btpd_{\NKrylov-1} & \bzero \\ 
    \bzero & \bzero & \cdots & -\btd_{\NKrylov-1} & \bz_{\NKrylov-1} & -\btpd_\NKrylov \\ 
    \bzero & \bzero & \cdots & \bzero & -\btd_\NKrylov & \tbz_\NKrylov 
\end{pmatrix} \! .
\end{align}
Here, $\bz_n (z) = \bz - \bolde_n$, and $\tbz_N (z) = \bz_N - \bR_N(z)$ in block $N N$, involving the residual function $\bR_N(z)$ of \Eqs{suqeq:DerivationRestFunction}.

Using $\bigl[[(z-\hat \cL_{\spi \spi})^{-1}]_{\bp\bp}]^{-1} [(z-\hat \cL_{\spi \spi})^{-1}\bigr]_{\bp\bp} = \doubleI_{\bp\bp}$, where $\doubleI_{\bp\bp}$ is the identity in the $pp$ block, and exploiting the tridiagonal form of $\bigl[[(z-\hat \cL_{\spi \spi})^{-1}]_{\bp\bp}\bigr]^{-1}$ [cf.~\Eq{eq:blockmatrix}], we arrive at the equations
\begin{subequations}
\label{eq:txFGI_relation1}
\begin{flalign}
    \Bigl( \bigl[[(z-\hat \cL_{\spi \spi})^{-1}\bigr]_{\bp\bp}]^{-1} \bigl[(z-\hat \cL_{\spi \spi})^{-1}\bigr]_{\bp\bp} \Bigr)_{\!NN-1} &= \nonumber & \\
    -\btd_\NKrylov \bG_{\NKrylov}(z) + \tbz_\NKrylov \widetilde{\bF}^{L}_{\NKrylov}(z) 
    &= \bzero \, , \hspace{-5mm} & \\
    \Bigl( \bigl[[(z-\hat \cL_{\spi \spi})^{-1}]_{\bp\bp}\bigr]^{-1} \bigl[(z-\hat \cL_{\spi \spi})^{-1}\bigr]_{\bp\bp} \Bigr)_{\!NN} &= \nonumber & \\ 
   -\btd_\NKrylov \widetilde{\bF}^{R}_{\NKrylov}(z) + \tbz_\NKrylov \widetilde{\bI}_{\NKrylov}(z) &= \boldone .\hspace{-5mm} &
\end{flalign}
\end{subequations}%
Here, we have used the definitions%
\begin{subequations}%
\label{subeq:GFFI-definitions}%
\begin{align}
    \bG_{\NKrylov}(z) &= [(z-\hat \cL)^{-1}]_{\NKrylov-1\NKrylov-1} \\
    \widetilde{\bF}^{L}_{\NKrylov}(z) &= [(z-\hat \cL)^{-1}]_{\NKrylov\NKrylov-1} \\
    \widetilde{\bF}^{R}_{\NKrylov}(z) &= [(z-\hat \cL)^{-1}]_{\NKrylov-1\NKrylov} \\
    \widetilde{\bI}_{\NKrylov}(z) &= [(z-\hat \cL)^{-1}]_{\NKrylov\NKrylov} \, .
\end{align}%
\end{subequations}%
By rearranging Eq.~\eqref{eq:txFGI_relation1}, we arrive at the relation
\begin{align}
    \label{eq:tildex_Ninverse-explicit}
    \tbz_\NKrylov^{-1} &= \widetilde{\bI}_{\NKrylov} - \widetilde{\bF}^{L}_{\NKrylov} \bG^{-1}_{\NKrylov} \widetilde{\bF}^{R}_{\NKrylov} \, .
\end{align}

Now, the \rreply{11}-element of the inverse of \Eq{eq:blockmatrix} is  explicitly presented by a CFE that is terminated at level $N-1$ by $\bz_{N-1} - \bt_N \tbz_N^{-1} \bt_N$. Equating this to $\bz_{N-1} - \bR_{N-1}$ and using \eqref{eq:tildex_Ninverse-explicit} yields 
\begin{align}
    \label{eq:RestFunction=I-FGF}
    \bR_{N-1}   = \bt_{N}^\pdag \bigl(\widetilde{\bI}_N   - \widetilde{\bF}_N^{L}   \bG_N^{-1}  \widetilde{\bF}_N^{R}  \bigr) \bt_{N}^\pdag .
\end{align}
Defining $\bI_N = \bt_N \widetilde{I}_N \btd_N$, $\bF^L_N = \btpd_N \widetilde{F}^L_N $, $\bF^R_N = \widetilde{F}^R_N \btd_N$, and using $\bpone_N \bt_N = \bponeperp_N$, we obtain  \Eq{eq:results:restfunction-a}
of the main text. \rreply{Next, we recall \Eqs{eq:polynomials-L-matrixelements} and \eqref{eq:nLn'-pn-omega-pn'} to express the matrix elements on the right of \Eqs{subeq:GFFI-definitions} as 
\begin{align}
[(z-\hat \cL)^{-1}]_{n n'} 
     = \int \! \mr{d}\omega \, \bponedag_{n} (\omega) \,  \bW(\omega) \, (z-\omega)^{-1} \bpone_{n'}(\omega)
    \, .  
\label{eq:nLn'-pn-omega-pn'-app}
\end{align} 
so $(z-\hat \cL)^{-1}$ becomes $(z-\omega)^{-1}$ in the polynomial basis $\bpone_n$.}
\changed{Equations \eqref{subeq:GFFI-definitions} then lead to \Eqs{eq:explicit-GFN-spectra-main-text} to \eqref{eq:results:restfunction-e}.}

\section{Quantum impurity models: complications in computing self-energy CFE}
\label{sec:QIM}

\changed{For quantum impurity problems, the computation of a CFE for the self-energy  faces two complications: First, the contributions of the hybridization function and the self-energy have to be separated carefully.  Second, when NRG is used as impurity solver, the computation of the self-energy involves log-Gaussian broadening; that introduces long tails that adversely affect the extraction of CFE coefficients. In this appendix, we present strategies for dealing with these complications.}

\subsection{Separating hybridization function and self-energy}
\label{sec:SeparatingHybridizationSelfEnergy}

Quantum impurity models involve a small number $\Nimpurity$ of impurity orbitals and a large number $\Nbath$ of bath orbitals. We assemble their creation operators into length-$\Nimpurity$ and length-$\Nbath$  row vectors, denoted $\sket{\bd^{\dagger}}$ and $\sket{\mathbf{b}^{\dagger}}$, respectively, and combine them as $\sket{\bc^\dagger} = \bigl(\sket{\bd^{\dagger}}, \sket{\mathbf{b}^{\dagger}}\bigr)$, a row vector of length $M = \Nimpurity + \Nbath$.  Its entries span the single-particle subspace, $\doubleW_\sp = \doubleW_\dimp \oplus \doubleW_\bbath$, of the  impurity model. In analogy to \Eq{eq:s-sbar-projectors}, we define the projectors
\begin{align}
    \label{eq:impurity-projectors}
    \cP_{\dimp} = \sket{\bd^\dagger} \sbra{\bd^\dagger} , \quad \cP_{\bbath} = \sket{\bb^\dagger} \sbra{\bb^\dagger} , \quad \cP_{\spi} = 1- \cP_{\sp} \, , 
\end{align}
and denote the corresponding projections of the Liouvillian by $\cL_{xx'} = \cP_{x} \cL \cP_{x'}$, with $x,x' \in \{\dimp,\bbath,\spi\}$. This more finely grained decomposition $\doubleW = \doubleW_\dimp \oplus \doubleW_\bbath \oplus \doubleW_{\spi}$ induces a $3\times 3$ block structure on $\cL$, akin to Eq.~\eqref{eq:L_block_decomposition}, having blocks $[\cL]_{\mathbf{x}\tilde{\mathbf{x}}} = \sbra{\mathbf{x}^{\dag}} \cL \sket{\tilde{\mathbf{x}}^{\dag}}$ with $\bx, \tilde{\bx} \in \{\bd, \bb, \bar{\bc}\}$.
Since the bath is non-interacting, the action of $\cL$ on $\sket{\bb^\dagger}$ yields another single-particle operator,  never reaching the single-particle irreducible space $\doubleW_{\spi}$, i.e.\  $\cL \sket{\mathbf{b}^{\dagger}} \in \doubleW_{\sp}$ and $\cL_{\spi \bbath} = \cL_{\bbath \spi} = 0$.
Since $\doubleW_{\spi}$ is reached only when $\cL$ acts on impurity orbitals, the dynamical part of single-particle irreducible self-energy, $\bSigma^\dyn(z)$, is non-zero only for impurity orbitals, cf.\ Eq.~\eqref{subeq:define-q1}. For impurity models, the right space of $\sket{\bq_1}$ is thus implicitly constrained to the impurity block, where it is non-zero, and likewise for $\bSigma^\dyn(z)$.

\begin{subequations}%
\label{subeq:Gdd-impurity-model}
The full single-particle Green's function, $\bG = \sbra {\bc^\dagger}(z- \cL)^{-1} \sket{\bc^\dagger}$, is an $M\times M$ matrix containing both impurity and bath blocks. However, one typically only works with the impurity Green's function, $\bG_{\dimp \dimp} = \sbra {\bd^\dagger}(z- \cL)^{-1} \sket{\bd^\dagger}$, an $\Nimpurity \times \Nimpurity$ matrix. Its form, found via \Eq{eq:Schur-for-impurity}, is
\begin{align}%
\label{eq:DysonImpurityModel}
    \bG_{\dimp \dimp} (z) & = \Bigl[ \bz - \bepsilon_0 - \bDelta(z) - \bSigma^\dyn(z)\Bigr]^{-1} , \\ 
    \bepsilon_0 & = \sbra{\bd^{\dagger}} \cL \sket{\bd^{\dagger}} , \\ 
    \bDelta (z) & =\sbra{\bd^\dagger} \cL_{\dimp \bbath} \frac{1}{z- \cL_{\bbath \bbath }} \! \cL_{\bbath \dimp } \sket{\bd^\dagger} , \\
    \bSigma^\dyn(z) & = \sbra{\bd^\dagger} \cL_{\dimp \spi}  \frac{1}{z- \cL_{\spi\spi}} \! \cL_{\spi\dimp } \sket{\bd^\dagger} . 
\end{align}%
\end{subequations}%
The hybridization function $\bDelta(z)$ encodes all relevant information about the impurity-bath coupling. 

For $\cL$-interpolation, we are only interested in the Liouvillian matrix elements in the single-particle irreducible subspace, to be extracted from $\bSigma^\dyn(z)$. \changed{Some impurity solvers, however, yield the spectral function of $\bG_{\dimp \dimp} (z)$, not the spectral function of $\bSigma^\dyn(z)$. The single-particle irreducible Liouvillian} matrix elements cannot be extracted by a direct CFE of $\bG_{\dimp \dimp }(z)$, since that would mix single-particle reducible contributions from $\bDelta(z)$ with single-particle irreducible ones from $\bSigma^\dyn (z)$. \changed{Thus, the contribution of $\bDelta(z)$ has to be separated out. 

This separation can be performed either implicitly or explicitly. It is achieved implicitly when computing the self-energy via Kugler's symmetric improved estimator strategy \cite{Kugler2022,Zacinskis2026}. We used that strategy for the results of Secs.~\ref{sec:1dHubbard_benchmark} and \ref{sec:2dHubbard_DCA}. Alternatively, given the discrete spectra of both the hybridization function $\bDelta(z)$ and the resulting impurity Green's function $\bG_{\dimp \dimp}(z)$, the separation can be performed explicitly. Below we describe the latter approach, used for the CDMFT+ED results of \Sec{sec:CDMFT+ED2DHubbard}.} 

\changed{Suppose we are given the discrete spectral functions 
$\bDelta''(\omega)$ and $\bA_{\dimp \dimp}(\omega)$ of the  hybridization function $\bDelta(z)$ and the impurity Green's function $\bG_{\dimp \dimp} (z)$, e.g.\ } 
\begin{align}
\label{eq:DiscreteHybridization}
    \boldsymbol{\Delta}''(\omega) = \sum_b \boldsymbol{\Delta}_b \, \delta(\omega - \xi_b) \, ,
\end{align}
\changed{where $b$ enumerates bath poles and weights, 
and similarly for $\bA_{\dimp \dimp}(\omega)$.}
To extract the discrete spectrum of the self-energy, start by inverting the Dyson equation \Eq{eq:DysonImpurityModel}, 
using, e.g., the strategy of \changed{App.~B.3(i) of}  Ref.~\onlinecite{Zacinskis2026}. This yields the spectrum of $\boldsymbol{\Delta}(z) + \bSigma^\dyn(z)$ in the form 
\begin{align}
    \label{eq:spectrum-self-energy+hybridization}
    \boldsymbol{\Delta}''(\omega) + \bSigma''(\omega) = \sum_{i}  
    \changed{\overline{\bX}_i} 
    \, \delta(\omega - \changed{\overline{\alpha}_i}) \, .
\end{align}
We extract $\bSigma''(\omega)$ from  \Eq{eq:spectrum-self-energy+hybridization} by subtracting $\bDelta''(\omega)$ via the pole-subtraction strategy of Ref.~\onlinecite{Lu2014}: 
we identify matching energies $\xi_b$ and $\changed{\overline{\alpha}_i}$ 
\changed{(for every $\xi_b$, we seek the unique $\changed{\overline{\alpha}_i}$ lying closest to and within, say, $10^{-8}$ of $\xi_b$),}
and subtract $\boldsymbol{\Delta}_b$ from the corresponding 
$\changed{\overline {\bX}_i}$ to obtain 
\begin{align}
\label{eq:PoleSubtraction}
    \bSigma_i = \begin{cases}
    \changed{\overline {\bX}_i} - \boldsymbol{\Delta}_b  
    & \textrm{if } \;\; \exists \; \xi_b \; : \; |\xi_b - \changed{\overline{\alpha}_i} 
    | < 10^{-8} \, , \\
    \changed{\overline {\bX}_i} & \textrm{otherwise} \, .
\end{cases}
\end{align}
\changed{A clean subtraction of the hybridization function poles is achieved if one finds $\bSigma_i \simeq 0$ whenever the pole-matching condition in \Eq{eq:PoleSubtraction} (top line  on the right)  holds. Then, the discrete spectrum of $\bSigma^\dyn(z)$ is given by 
\begin{align}
\label{eq:DiscreteSigmaDyn}
    \btpd_1 \bW (\omega) \btd_1 \stackrel{\eqref{eq:ImSigmaDyn}}{=}
        \bSigma''(\omega) = \sum_i \bSigma_i \, 
    \delta(\omega - \overline{\alpha}_i) \, .
\end{align}} 

\subsection{CFE for NRG spectra: fast/slow splitting}
\label{sec:restfunction_splitting}

If the self-energy of a quantum impurity model is computed using NRG, as for the DCA+NRG results of \Sec{sec:2dHubbard_DCA}, a complication arises:
\changed{Standard NRG strategies for computing the self-energy with NRG \cite{Bulla1998,Weichselbaum2012b,Kugler2022} 
use broadened impurity spectral functions as input; the latter are obtained from discrete ones via log-Gaussian broadening \cite{Lee2016_NRG}. This yields a self-energy spectral function $\bW(\omega)$ possessing long tails.} 
As a result, the moment problem of the measure $\bW (\omega) \mr{d}\omega$ is indeterminate (see \App{app:MeasuresAndAllThat}), a problem artificially introduced by log-Gaussian broadening. This implies that it is not possible to uniquely determine the self-energy from its CFE coefficients without additional information. Below, we describe how such additional high-frequency information can be introduced via a regularized CFE.

Even without the problem of indeterminate moments, it is often desirable for the regularization scheme to be geared towards extracting low-energy information. Many-body systems often exhibit features that are separated by several orders of magnitude in frequency, and resolving all of those features would require a CFE with exponentially large depth $N$---unfeasible in numerical practice. Instead, we need a ``logarithmic CFE'', featuring CFE coefficients whose magnitudes decrease exponentially with depth $N$, thus naturally terminating the CFE when $N$ becomes large.

A strategy for constructing CFEs that both regularizes high-energy tails and yields exponentially decreasing CFE coefficients has been described in Ref.~\onlinecite{Bruognolo2017_OpenWilsonChain}.
There, a ``fast/slow splitting'' CFE (log-CFE) was constructed for the hybridization function of an impurity model. This was achieved by an iterative procedure analogous to using \Eq{eq:R1(z)CFE-nth-step}, with one crucial difference: after each step the spectrum of the residual function  $\bR_{n}(z)$ is split into a high-energy part and a low-energy part, called, ``fast'' (F) and ``slow'' (S) modes, respectively, and only the slow modes are used for the next iteration. Here, we adapt that scheme to construct a logarithmic CFE of our self-energy via $N$ one-step Lanczos schemes combined with fast/slow splitting.

We initialize the procedure by setting $\bR^\slow_0(z) = \bSigma^\dyn(z)$, i.e.\ we declare the input self-energy to be ``all slow modes''. For iterations $n \ge 1$, we proceed as before for steps (a) to (c), now using $\bR^{\slow}_{n-1}(z)$ as input, to compute $\bt_n$, $\bolde_n$ and $\bR_n(z)$ and   obtain the depth-1 CFE
\begin{align}
    \label{eq:R1(z)CFE-first-step-slow}
    \bR^{\slow}_{n-1} (z) & = \bt_{n} \cfrac{\boldone}{\bze_{n} - \bR_{n} (z)} \bt_{n} .
\end{align}
Thereafter comes a new step:  (d) We split $\Rest_n(z)$ into ``fast" and ``slow" contributions, $\Rest^\pdag_n(z) = \Rest_{n}^{\fast}(z)+\Rest_{n}^{\slow}(z)$, built from the high- or low-energy part of its spectrum,
\begin{align}
    \label{eq:splitR-FS}
    \Rest_{n}^{\fast/\slow} (z) = \int_{-\infty}^{\infty} \mr{d}\omega \frac{w^{\fast/\slow}_{n}(\omega) \,\Rest''_{n}(\omega)}{z - \omega} \, .
\end{align}
Here, the so-called splitting functions $w^{\fast/\slow}_{n}(\omega)$ satisfy $w^{\fast}_{n}(\omega) + w^{\slow}_{n}(\omega) = 1$ and monotonically increase/decrease with $|\omega|$ (see below). 
Then, we use only the slow-mode part, $\bR^{\slow}_n(z)$, as input for the next iteration. Iterating, we obtain an exact CFE of the following form: 
\begin{flalign}
    \label{eq:SigmaCFE_FastSlowMode}
    & \bSigma^\dyn(z) = \\[-2mm]  
    & \btpd_1 \cfrac{\boldone}{\bz_1 - \bR^\fast_1 (z) - { \btpd_{2}\cfrac{\boldone}{\bz_2  - \bR^\fast_2 (z) - \! \raisebox{-0.9em}{\ensuremath{\ddots} \raisebox{-0.5em}{\hspace{-2.5mm} \ensuremath{- \btpd_\NKrylov \cfrac{\boldone}{\bz_N \!-\! \Rest_\NKrylov(z)  } \, \btd_\NKrylov  \hspace{-4mm} \phantom{.}}}}}\, \btd_{2} \hspace{-2mm}}} \btd_1  . \nonumber \hspace{-1cm} & 
\end{flalign}
Each level contains an explicit fast-mode contribution $\bR_n^\fast(z)$, except the last level, containing the full $\bR_N(z)$. Note that $\bR_n^\fast(z)$ will, in general, contain an imaginary part; therefore, some amount of broadening is present at each level, yielding a self-energy with a continuous spectrum. The fact that $\bR_n^\fast(z)$ does not feed into the next iteration in effect regularizes the contribution from high-energy tails of the spectrum of $\bR_n(z)$. Nevertheless, the representation \eqref{eq:SigmaCFE_FastSlowMode} is exact---the fast/slow splitting does not neglect any information, it just distributes high- and low-energy information to different parts of the CFE. 

To achieve a \textit{logarithmic} CFE whose CFE coefficients decrease exponentially with $N$, we choose 
\begin{align}
    \label{eq:splitting_function}
    w^\slow_n(\omega) = \frac{1}{1+|\omega/\Omega_{n}|^{\tau_\fastslow}}, \quad \Omega_n = \Omega_{\mr{max}} \Lambda^{1-n}_\fastslow \, .
\end{align}
This box-shaped function has height 1, half-width $\Omega_n$ and box corners whose rounding is controlled by $\tau_\fastslow>2$. The scale $\Omega_n$ decreases exponentially with $n$, and we choose 
$\Lambda_\fastslow = (\Omega_{\mr{max}}/\Omega_{\mr{min}})^{1/(N-1)}$ such that $\Omega_n$ ranges between specified maximal and minimal values, $\Omega_1 = \Omega_{\mr{max}}$ and $\Omega_N = \Omega_{\mr{min}}$.
The splitting function has weight $\int \mr{d}\omega \, w^\slow_n(\omega) \sim \Lambda^{-n}_\fastslow$, resulting in exponentially decreasing CFE coefficients, $\bepsilon_n , \bt_n \sim \Lambda^{-n}_\fastslow$ and the weight of the spectrum of $\bR_n^\fast(z)$ likewise scales $\sim \Lambda^{-n}_\fastslow$ (see Fig.~\ref{fig:CFE_data}, and \App{app:SlowFastMode_numerics} for numerical details.)

\changed{In \Alg{alg:wSF} we give a pseudocode for computing the coefficients and residual functions of a log-CFE by suitably modifying scheme (iii) of \App{app:ExactTerminationCFE}
such that each iteration step includes fast/slow splitting. The starting point for fast/slow splitting can be set at an arbitrary, specified depth $n$, yielding a log-CFE whose coefficients decay exponentially when proceeding beyond that depth.
Numerical examples of how log-CFE results depend on the splitting parameters $\Lambda_\fastslow$, $\tau_\fastslow$
are given in \App{app:log-CFE-DCA-Hubbard}.}

\begin{algorithm}[h!]\label{alg:wSF}
\caption{Computation of log-CFE according to
\App{sec:restfunction_splitting}}
    \SetKwInOut{In}{Input}
        \In{Self-energy $\bSigma(z)$ and
        splitting function parameters $\Omega_{\mr{min}}$, $\Omega_{\mr{max}}$, $\Lambda_\fastslow$ and $\tau_\fastslow$.} \vspace{2pt}

    \SetKwInOut{Out}{Output}
        \Out{CFE coefficients $\bepsilon_n$, $\bt_n$, fast modes $\Rest^\fast_n(z)$.} \vspace{2pt}

    \SetKwProg{Comp}{Computation:}{}{}
        \Comp{We use the same subscripts as in main text, but in the code, only $\bolde_n$, $\bt_n$ and $\bR_n^\fast$ are saved as vectors.}

    \SetKwProg{Init}{Initialization}{:}{}
        \Init{}{CFE length $N = 1 +\ln(\Omega_{\mr{max}}/\Omega_{\mr{min}})/\ln\Lambda_\fastslow$ and the self-energy spectrum 
        $\bSigma^{\prime \prime}(\omega) 
         \leftarrow \tfrac{\mr{i}}{2\pi}\big[\bSigma(\omega^+) - \bSigma^\dag(\omega^+)\big]$.} \vspace{2pt} 

    \SetKwProg{Iter}{Iteration:}{:}{}
        \Iter{Perform $N$ successive one-step Lanczos schemes, \rreply{according to scheme (iii) of \App{app:ExactTerminationCFE}}}{} \vspace{-4pt}
    
    \SetKwProg{for}{for}{}{end}
        \for{$\ n=1,\dots,N$}{
            \textbf{(a)} 
            Normalize $\bSigma''(\omega)$ to obtain $\bt_n$ and $\bW_{n-1}$:\vspace{-5pt}
            \begin{align*}
                \bt_n\bt_n & = \int \mr{d}\omega\,\bSigma^{\prime \prime}(\omega) , \\
                \bW_{n-1}(\omega) & \leftarrow \bt_n^{-1} \bSigma^{\prime \prime}(\omega) \bt_n^{-1}\, .
            \end{align*}
                
            \textbf{(b)} Compute
            $ {\displaystyle \boldsymbol{\epsilon}_n = \int \mathrm{d}\omega\, \bW_{n-1}(\omega) \, \omega} $
                   
            \textbf{(c)} Compute $\Rest_n(\omega)$ via \Eqs{eq:results:restfunction}, \rreply{with $\bpone_{1} = \boldone$
            and $\bpone_2{}^{\perp} = (\omega \boldone - \bolde_n)$}:  \vspace{-4pt}
            \begin{align*}
                \bG_{2}(z) & \leftarrow \int \mr{d}\omega\, \frac{\bW_{n-1}(\omega)}{z-\omega} \\ 
                \widetilde{\bF}^{L}_{2}(z) & \leftarrow \int \mr{d}\omega\, (\omega \boldone - \bolde_n)\frac{\bW_{n-1}(\omega)}{z-\omega} \\
                \widetilde{\bF}^{R}_{2}(z) & \leftarrow \int \mr{d}\omega\, \frac{\bW_{n-1}(\omega) \,}{z-\omega} (\omega \boldone - \bolde_n) \\
                \widetilde{\bI}_{2}(z) & \leftarrow \int \mr{d}\omega\, (\omega \boldone - \bolde_n) \frac{\bW_{n-1}(\omega)}{z-\omega} (\omega \boldone - \bolde_n) \nonumber \\
                \Rest_n(z) & \leftarrow \widetilde{\bI}_{2}(z) - \widetilde{\bF}^{L}_{2}(z) [\bG_2(z)]^{-1} \widetilde{\bF}^{R}_{2} (z) \nonumber 
            \end{align*}
                
            \textbf{(d)} Define splitting functions: 
            $\Omega_n = \Omega_{\mr{max}}\Lambda^{1-n}_\fastslow$, \\[1mm]
            \phantom{\textbf{(d)}}$w^\slow_n(\omega) 
            \leftarrow (1+|\omega/\Omega_n|^\tau_\fastslow)^{-1}$, \\[1mm]
            \phantom{\textbf{(d)}}$w_n^\fast(\omega)
            \leftarrow1-w_n^\slow(\omega)$.\vspace{5pt}
                    
            \phantom{\textbf{(d)}} Split $\Rest_n(z)$, reset $\bSigma''(\omega)$ for next iteration:
            \begin{align*}
                \Rest_n''(\omega) & \leftarrow \tfrac{\mr{i}}{2\pi}\big[\Rest_n(\omega^+) - \Rest^\dagger_n(\omega^+)\big]
                \\
                \Rest_n^\fast(z) & = \int \mr{d}\omega \,  
                \frac{w_n^\fast(\omega)\, \Rest''_n(\omega)}{z-\omega}
                \\
                \bSigma^{\prime \prime}(\omega) & \leftarrow 
                w_n^\slow(\omega) \, \Rest''_n(\omega) 
            \end{align*}
            \textit{For the last step $N$, do not split $\bR_n(z)$}.
            }
\end{algorithm}

\section{Measures, the moment problem, and orthogonal polynomials}
\label{app:MeasuresAndAllThat}

In this Appendix, we provide some background material on moment problems, measures, and their relation to orthogonal polynomials. This material is used implicitly in \changed{\App{app:ExactTerminationCFE} on 
terminating CFEs exactly using residual functions:} it forms the basis for our statement there that the CFE coefficients for a Green's function $\bG(z)$ or its self-energy $\bSigma(z)$ are the same, although their spectral functions, $\bA(\omega)$ or $\bW(\omega)$, generate two different sets of orthogonal polynomials. All the information presented below is well-established and can be found in numerous textbooks~\cite{Ismail2005,Szegoe2012,Chihara2011,Simon2005,Achiezer2021,Stojanov2014,Dunford1988} and articles~\cite{Duran1995,Chihara1989,Simon1998,Duran2000,LopezRodriguez2001,Damanik2007}. 

Let \(\boldsymbol\mu\) be a positive semidefinite, hermitian \(M\!\times \!  M\) \emph{matrix-valued Borel measure} on \(\mathbbm{R}\); that is, $\boldsymbol\mu:\mathcal{B}(\mathbbm{R})\to\mathbb{C}^{M\times M}$ is countably additive and $\boldsymbol\mu(B)$ is positive semidefinite for all Borel sets $B$. We write, distributionally,
\begin{align}
    \boldsymbol{\mu}(\mr{d}\omega) = \mathbf{A}(\omega) \mr{d}\omega \, ,
\end{align}
with a spectral function $\mathbf{A}(\omega) = \mathbf{A}_\mathrm{c}(\omega) + \sum_{\alpha} \mathbf{A}_{\alpha} \delta(\omega - \omega_{\alpha})$ that exhibits both a continuous part $\mathbf{A}_c(\omega)$ and point masses with weights $\mathbf{A}_{\alpha}$, encoded by the $\delta$-terms. We further assume the normalization
\begin{align}
    \boldsymbol{\mu}(\mathbbm{R}) = \int_{-\infty}^{\infty} \mr{d}\omega \,  \mathbf{A}(\omega) = \boldsymbol{1} \, ,
\end{align}
where $\boldsymbol{1}$ is the $M\times M$ unit matrix.

The Stieltjes transform of $\boldsymbol{\mu}$ is defined as
\begin{align}
    \label{eq:Stieltjes_transform}
    \mathbf{G}(z) = \int_{\mathbbm{R}} \frac{\boldsymbol{\mu}(\mr{d}\omega)}{z - \omega} \, , \qquad z\in\mathbb{C}\setminus\mathbbm{R} \, .
\end{align}
$\bA(\omega)$ 
can be recovered from $\mathbf{G}(z)$ via \Eq{eq:defineSpectralFunction}. 

The (equivalence classes of) square-integrable 
vector fields $ f:\mathbbm{R}\to\mathbb{C}^{M}$ with respect to $\boldsymbol\mu$ form the Hilbert space $L^2(\boldsymbol\mu)$, with inner product
\begin{align}
    \langle g , f \rangle = \int_{\mathbbm{R}} g^{\dagger}(\omega) \boldsymbol{\mu}(\mr{d}\omega) f(\omega) \, \in \mathbb{C} \, .
\end{align}
In the following, we write $\ket{f} \in L^2(\boldsymbol{\mu})$ for the vector which is represented by the vector field $f(\omega)$. For later usage, we further define blocks of $M$ vectors by a bold symbol $\ket{\mathbf{f}} = (\ket{f_1}, ..., \ket{f_M})$. Similarly, $\mathbf{f}(\omega) = (f_1(\omega), ...,  f_M(\omega)) \in \mathbb{C}^{M\times M}$ denotes the block of representing vector fields. Furthermore, $\langle \mathbf{g}, f \rangle$ denotes a column vector with $\alpha$-th entry $\langle g_\alpha, f \rangle$, $\langle  f, \mathbf{g} \rangle = \langle \mathbf{g}, f \rangle^{\dagger}$ a corresponding row vector, and $\langle \mathbf{g}, \mathbf{f} \rangle$ a matrix with entries $\langle g_\alpha, f_\beta \rangle$. For a more formal treatment of matrix-valued measures and their $L^2$ spaces, see Ref.~\onlinecite{Dunford1988}, pages 1337 and onward.

The (matrix) moments of $\boldsymbol{\mu}$ are
\begin{align}
\label{eq:matrix_moments}
    \mathbf{m}_n = \int_{\mathbbm{R}} \boldsymbol{\mu}(\mr{d}\omega) \, \omega^n \, ,
\end{align}
with $n = 0,1,2,\dots$ and hermitian $\mathbf{m}_n$. Throughout, we assume that all moments in \eqref{eq:matrix_moments} are finite.

The (matrix-valued) \textit{moment problem} asks: given a sequence of hermitian matrices $\{\mathbf{m}_n\}_{n\ge0}$, (i) does there exist a positive semidefinite, hermitian matrix-valued Borel measure $\boldsymbol{\mu}$ on $\mathbbm{R}$ with moments \eqref{eq:matrix_moments}; and (ii) if so, is it unique? In our setting we start from $\boldsymbol{\mu}$ (equivalently, from the spectral function $\mathbf{A}(\omega)$) and compute $\{\mathbf{m}_n\}$, so existence is automatic.  Uniqueness, however, is not guaranteed: if exactly one representing measure has the moments \eqref{eq:matrix_moments}, the problem is determinate; otherwise, it is indeterminate.

A sufficient condition for a unique measure (determinate moment problem) is either a compact support of the spectral function $\mathbf{A}(\omega)$, or an exponential decay at large $|\omega|$. Spectral functions of finite-size fermionic systems, even if macroscopically large, fulfill these conditions. We therefore assume that the moment problem is determinate in our work, i.e., the spectral function $\mathbf{A}(\omega)$ having these moments is unique.

In the determinate case, \Eq{eq:matrix_moments} can be used to generate a set of orthonormal polynomials that form a complete orthonormal basis of $L^2(\boldsymbol{\mu})$ (cf.\ Eq.~\eqref{eq:block_polynomials} below): every $\ket{f} \in L^2(\boldsymbol{\mu})$ admits a unique $L^2$–expansion in these polynomials. By contrast, in the indeterminate case, there exist representing measures $\boldsymbol{\mu}$ for which the orthonormal polynomials are \textit{not} a complete basis of $L^2(\boldsymbol{\mu})$. For example, this is the case for a log-Gaussian spectral function. Nevertheless, the same moment sequence admits a family of representing measures (the Nevanlinna-extremal measures) for which the corresponding orthonormal block polynomials \textit{are} a complete orthonormal basis of $L^2(\boldsymbol{\mu})$, see, e.g., Refs.~\onlinecite{LopezRodriguez2001,Berg1995}.

We caution that some popular broadening kernels used in DMFT computations spoil determinacy or even moment existence. The log-Gaussian kernel, a standard kernel used to broaden discrete NRG spectra, has indeterminate moments, see Sec.~11.2 of Ref.~\onlinecite{Stojanov2014}. The Lorentzian kernel, often used for broadening discrete finite-size spectra, has no finite moments of order $\changed{\geq}1$. In such cases, one should either work with the underlying discrete data when possible, or use the slow–fast mode splitting described in Sec.~\ref{app:SlowFastMode}.  In these cases, the loss of determinacy is an artifact of the long high-frequency tails of the broadening kernel and is unrelated to the underlying physics.

The block Lanczos algorithm of Eqs.~\eqref{subeq:construct-pns} can be used to construct an orthonormal polynomial basis,
\begin{align}
    \label{eq:block_polynomials}
    \langle \bph_n , \bph_{n'} \rangle = \int_{\mathbbm{R}} \mr{d}\omega \, \bph_n^{\dagger}(\omega) \bA(\omega) \bph_{n'}^{\pdag}(\omega) = \delta_{nn'} \boldsymbol{1} \, .
\end{align}
Here, every column of $\bph_{n}^{\pdag}(\omega)$ is a degree $n$ polynomial $\mathbbm{C}^M$ vector field. The Lanczos algorithm also generates the hermitian matrices $\bepsilon_n$ and positive semidefinite hermitian matrices $\bt_n$. The polynomials $\bph_n(\omega)$ obey a three-term recursion relation
\begin{align}
\label{eq:three_term_recursion}
    \omega \bph_{n}(\omega) = \bph_{n}(\omega)\bolde_{n} + \bph_{n+1}(\omega)\btd_{n+1} + \bph_{n-1}(\omega) \bt_n \, ,
\end{align}
with initial conditions $\bph_{-1}(\omega) = \boldsymbol{0}$ and $\bph_{0}(\omega) = \boldsymbol{1}$.

While the Lanczos procedure takes  a given measure $\boldsymbol{\mu}$ as input to generate polynomials satisfying the recursion relation \eqref{eq:three_term_recursion}, one may ask, conversely: given such a recursion relation, are the polynomials generated from it orthonormal w.r.t.\ some measure? The answer is yes:
Favard's theorem guarantees that polynomials that obey the three-term recursion \eqref{eq:three_term_recursion}, with initial condition $\bph_{-1}(\omega) = \boldsymbol{0}$ and $\bph_{0}(\omega) = \boldsymbol{1}$, are orthogonal w.r.t.\ at least one measure $\boldsymbol{\mu}$. The measure is unique if its moment problem is determinate, a property that can also be determined from the coefficients $\bepsilon_n$ and $\bt_n$~\cite{Chihara1989,Simon1998,Duran2000}. 

The three-term recursion can be used to generate an infinite-depth CFE for $\bG(z)$, the Stieltjes transform  of $\bA(\omega)$, of the form (with $\bze_n = \bz - \bolde_n$)
\begin{align}
\label{eq:infinite-CFE-G}
     \mathbf{G}(z) = \cfrac{\boldone}{\bze_0 -
    \btpd_{1}\cfrac{\boldone}{\bze_1 -
     {\btpd_{2}\cfrac{\boldone}{\bze_2 -  \raisebox{-0.9em}{\ensuremath{\ddots}}}\, \btd_{2} \hspace{-2mm}}
    }\, \btd_{1}}  \ .
\end{align}

Another set of matrix  polynomials, called \textit{associated polynomials of the first kind} and denoted $\bpone_{n}(\omega)$, is obtained if the recursion relation Eq.~\eqref{eq:three_term_recursion} (with the same coefficients $\bepsilon_n$ and $\bt_n$) is initialized with the initial conditions $\bpone_{-1}(\omega) = \mathbf{0}$, $\bpone_{0}(\omega) = \mathbf{0}$ and $\bpone_{1}(\omega) = \boldsymbol{1}$. Then, $\bpone_{n}(\omega)$ is a polynomial of degree $n-1$. The $\bpone_{n}(\omega)$ and  $\bph_n(\omega)$ polynomials are related via the integral
\begin{align}
    \label{eq:associated_poly}
    \bpone_{n}(\omega) = \int_\mathbbm{R} \mr{d}\omega' \, \btd_1 \bA(\omega') \frac{\bph_{n}(\omega) - \bph_{n}(\omega')}{\omega - \omega'} \, ,
\end{align}
as follows readily by induction using \Eq{eq:three_term_recursion}. Now, Favard's theorem tells us that there is at least one measure w.r.t.\ which these polynomials are orthonormal, say $\bmuone$, with $ \bmuone (\mathrm{d}\omega) = \bAone (\omega) \, \mathrm{d} \omega$. Moreover, since we have assumed that the moment problem of $\boldsymbol{\mu}$ is determinate, the same is true for the measure $\bmuone$, which therefore is unique.

The three-term recursion relation also tells us that the Stieltjes transform of $\bmuone$, say $\bGone(z)$, has the infinite-depth CFE
\begin{align}
\label{eq:infinite-CFE-G1}
     \bGone(z)  =  \cfrac{\boldone}{\bze_1 - 
    \btpd_{2}\cfrac{\boldone}{\bze_2 - 
    {\btpd_{3}\cfrac{\boldone}{\bze_3 - \raisebox{-0.9em}{\ensuremath{\ddots}}}\, \btd_{3} \hspace{-2mm}}
    }\, \btd_{2}}  \, , 
\end{align}
involving the \textit{same} CFE coefficients $\bolde_{n\ge1}$,  $\bt_{n\ge2}$ as $\bG(z)$ of \Eq{eq:infinite-CFE-G}. Therefore, $\bGone (z)$ and $\bG(z)$ are related by 
\begin{subequations}
\begin{align}
    \label{eq:G_via_G1}
    \mathbf{G}(z) &= \frac{1}{\bze_0 - \bt_1 \bGone(z) \btd_{1}} \, ,\\
    \label{eq:G1_via_G}
    \bGone(z) & = \bt^{-1}_{1}  \bigl[\bze_0 - \mathbf{G}^{-1}(z) \bigr]  \, \btd_{1}^{-1} \, .
\end{align}
\end{subequations}
The  spectral density $\bAone(\omega)$ of $\bGone(z)$ can be obtained explicitly, if needed, by inserting \Eq{eq:G1_via_G} into a relation analogous to  \Eq{eq:defineSpectralFunction}. 

The above framework is directly relevant for our discussion in \Sec{sec:CFE-self-energy}. Identifying the functions $\bG(z)$ and $\bA(\omega)$ discussed above with the eponymous Green's function and spectral function there, and \Eq{eq:G_via_G1} with the Dyson equation \eqref{eq:Dyson1}, we note that $\bGone(z)$ and $\bAone(\omega)$ correspond to  the functions $\bS(z)$ and $\bW(\omega)$ characterizing the dynamical part of the self-energy, $\bSigma^\dyn(z) = \bt^\pdag_1 \bS(z) \bt_1^\dagger$. Thus, the two sets of orthogonal block polynomials generated from measures involving the spectral functions $\bA(\omega)$ or $\bW(\omega)$ of the Green's function or its self-energy, respectively, are related; the polynomials  generated from $\bW(\omega)$ are associated polynomials of the first kind w.r.t.\ the polynomials generated from $\bA(\omega)$. (That is why we put the superscript on $\bpone_n$ in \Sec{sec:CFE-self-energy}.)
Moreover, the coefficients $\bolde_{n\ge1}$ and $\bt_{n\ge2}$ generated using either $\bW(\omega)$ or $\bA(\omega)$ are identical. We also note that the computational steps involved in obtaining the CFE \eqref{eq:infinite-CFE-G} for $\bG(z)$ are entirely analogous to those presented in \Sec{sec:CFE-via-polynomials} for obtaining the CFE \eqref{eq:SigmadynCFE} for $\bS(z)$, with $\bW(\omega)$ there now replaced by $\bA(\omega)$.

\section{Lattice harmonics}
\label{app:latticeharmonics}

This appendix gives some background on lattice harmonics and their symmetry properties \cite{Altmann1965_LatticeHarmonics1,Altmann1965_LatticeHarmonics2}, invoked in \Sec{sec:trad_periodization} on canonical DCA interpolation.

Lattice harmonics are functions of $\bk$ that transform according to irreducible representations (irreps) of the symmetry point group, say $G$, of the model's real-space Bravais lattice and its BZ. (For a symmetry-broken phase, let $G$ describe the remaining symmetry.) Let $\Gamma$ be an irrep of $G$, with dimension $d_\Gamma$ and matrix representation $D^\Gamma$. Lattice harmonics forming a $\bk$-space basis for $\Gamma$ can be constructed by projecting the plane wave function $\mr{e}^{-\mr{i} \bk \cdot \br}$, for some fixed seed vector $\br$ from the real-space lattice, onto $\Gamma$ using the standard group‐theory projector:
\begin{align}
    \label{eq:latticeharmonicconstruction}
    \bigl[\phi^\Gamma_\bk \bigr]_\alpha = \frac{d_\Gamma}{|G|} \sum_{g \in G} \bigl[D^{\Gamma}(g)\bigr]_{\alpha 1} \mr{e}^{-\mr{i} \bk \cdot g(\br)} .  
\end{align}
The  $D^\Gamma$ matrices can always be chosen such that the resulting functions $\bigl[\phi^\Gamma_\bk \bigr]_\alpha$ are real, and we assume this to be the case in this work. For $\br = \bzero$, \Eq{eq:latticeharmonicconstruction} yields a $\bk$-independent function;  for seed vectors of increasing length $|\br|$, it yields lattice harmonics with an increasingly complex dependence on $\bk$. 

Consider, for example, a 2-dimensional square lattice, invariant under $G = C_{4v}$, the group of reflections about the $x$- and $y$-axes and of multiples-of-$\pi/2$ rotations about the origin. Then, the lattice harmonics obtained for seed vectors with $|\br| =1$ are linear combinations of cosine and sine functions, and those obtained for $|\br| > 1$ can include nonlinear combinations of cosine and sine functions.

For a given DCA patching scheme, let $\Gpatch$ be that subgroup of $G$ which leaves every patch individually invariant, i.e.\ for every patch $V_\bK$, if $\bk \in V_\bK$ then $g(\bk) \in V_\bK$ for all $g \in \Gpatch$. A subset of all lattice harmonics of $G$ are invariant under $\Gpatch$ (i.e.\ transform trivially under $\Gpatch$). In the main text (\Sec{sec:trad_periodization}), we denote them by $\phi_{\bk, j}$, where $j = 1, 2, 3, \dots$ enumerates them in order of increasing complexity, with $\phi_{\bk,1} = 1$. By definition, these functions obey $\phi_{g(\bk),j} = \phi_{\bk, j}$  for all $g \in \Gpatch$.

In \Sec{sec:2dHubbard_DCA} of the main text, we consider a 2-dimensional square lattice with point group symmetry $C_{4v}$, and a DCA scheme with $\Npatch = 4$ patches arranged in a star geometry, see \Fig{fig:star_patching}. These patches are invariant under $\Gpatch = C_{2v}$, the group of reflections about the $x$- and $y$-axes (but not under $\pi/2$ rotations about the origin). We therefore simply choose the $\phi_{\bk,j}$ functions to be the first $\Npatch$  irreps of $C_{4v}$
\changed{on which $C_{2v}$ acts trivially,} enumerated in order of increasing complexity:
\begin{gather}%
\label{eq:DCA-Fourier-modes-app}
\begin{array}{lll}
    \phi_{\bk,1} & = 1\, ,  \quad \; &         \phi_{\bk,3} = \cos k_x - \cos k_y\, , \\ 
    \phi_{\bk,2} & = \cos k_x + \cos k_y \, , & \phi_{\bk,4} =  \cos k_x  \cos k_y \, . 
\end{array}
\end{gather}
Remark: These functions are linearly related to the lattice harmonics of the patch-symmetry group $\Gpatch = C_{2v}$, 
\begin{gather}%
\label{eq:DCA-Fourier-modes-app-phi-prime}
\begin{array}{lll}
    \phi'_{\bk,1} & = 1\, ,  \quad &   \hspace{0.5cm} \phi'_{\bk,3}  = \cos k_y\, , \\ 
    \phi'_{\bk,2} &  = \cos k_x\, , & \hspace{0.5cm}  \phi'_{\bk,4} =  \cos k_x  \cos k_y \, . 
\end{array}
\end{gather}
These follow from a version of \Eq{eq:latticeharmonicconstruction} appropriate for  abelian groups,
\begin{align}
    \label{eq:latticeharmonicconstruction-patchsymmetry}
    \phi'_{\bk,j} = \frac{1}{|\Gpatch|} \sum_{g \in \Gpatch} \mr{e}^{-\mr{i} \bk \cdot g(\br_j)} = \frac{1}{N_j} \sum_{\br \in O_j} \mr{e}^{-\mr{i} \bk \cdot \br} , 
\end{align}
seeded by $\br_1 =  (0,0)$, $\br_{2} = (1,0)$,  $\br_3 = (0,1)$,  $\!\br_4 = (1,1)$.  On the right, the sum is over all points in the orbit $O_j$ seeded by $\br_j$, i.e.\ the set of distinct points obtained by acting with all elements of $\Gpatch$ on $\br_j$, and $N_j$ is the number of points in orbit $O_j$ (here: $N_1 = 1$, $N_2=N_3 = 2$, $N_4= 4$). We regard $N_j$ as a measure of the non-locality of the Fourier modes contributing to $\phi'_{\bk,j}$ and $\phi_{\bk,j}$. The  modified form factors $\alpha^\modified_{\bk, \bK}$ of \Eqs{subeq:modified-interpolation} aim to  ameliorate non-causality issues by reducing the weight of these non-local Fourier modes by an extra factor $1/N_j$.

To conclude this appendix, we remark that the interpolation Ansatz of \Eqs{eq:Mscheme_DCA} and \eqref{eq:Ansatz-compact} for $\bQ_\bk(z)$ and $\alpha_{\bk \comma \bK}$ can be expressed as a truncated Fourier expansion of the form of \Eq{eq:Q_periodize}: by \Eq{eq:latticeharmonicconstruction}, lattice harmonics are symmetry projections of Fourier factors $\mr{e}^{-\mr{i} \bk \cdot \br}$, and those used for $\phi_{\bk,j\le \Npatch}$ are lattice harmonics of low complexity, for which $|\br|$ is of limited range, say $|\br| \le R$.

\section{TaSK and CDMFT+TaSK for 1D Hubbard model}\label{app:TaSK_details}

\changed{The MPS results on the 1D Hubbard model presented in \Sec{sec:1dHubbard_benchmark} were all obtained 
using the tangent space Krylov ~(TaSK) scheme recently proposed in Ref.~\onlinecite{Kovalska2025}, and used as an impurity solver in 
Ref.~\onlinecite{Picoli2026}. Here, we discuss some aspects specific to our present TaSK computations that are not discussed in those two references.}

\subsection{TaSK for 1D Hubbard model}

\begin{figure}[t!]
    \centering
    \includegraphics[width=0.8\linewidth]{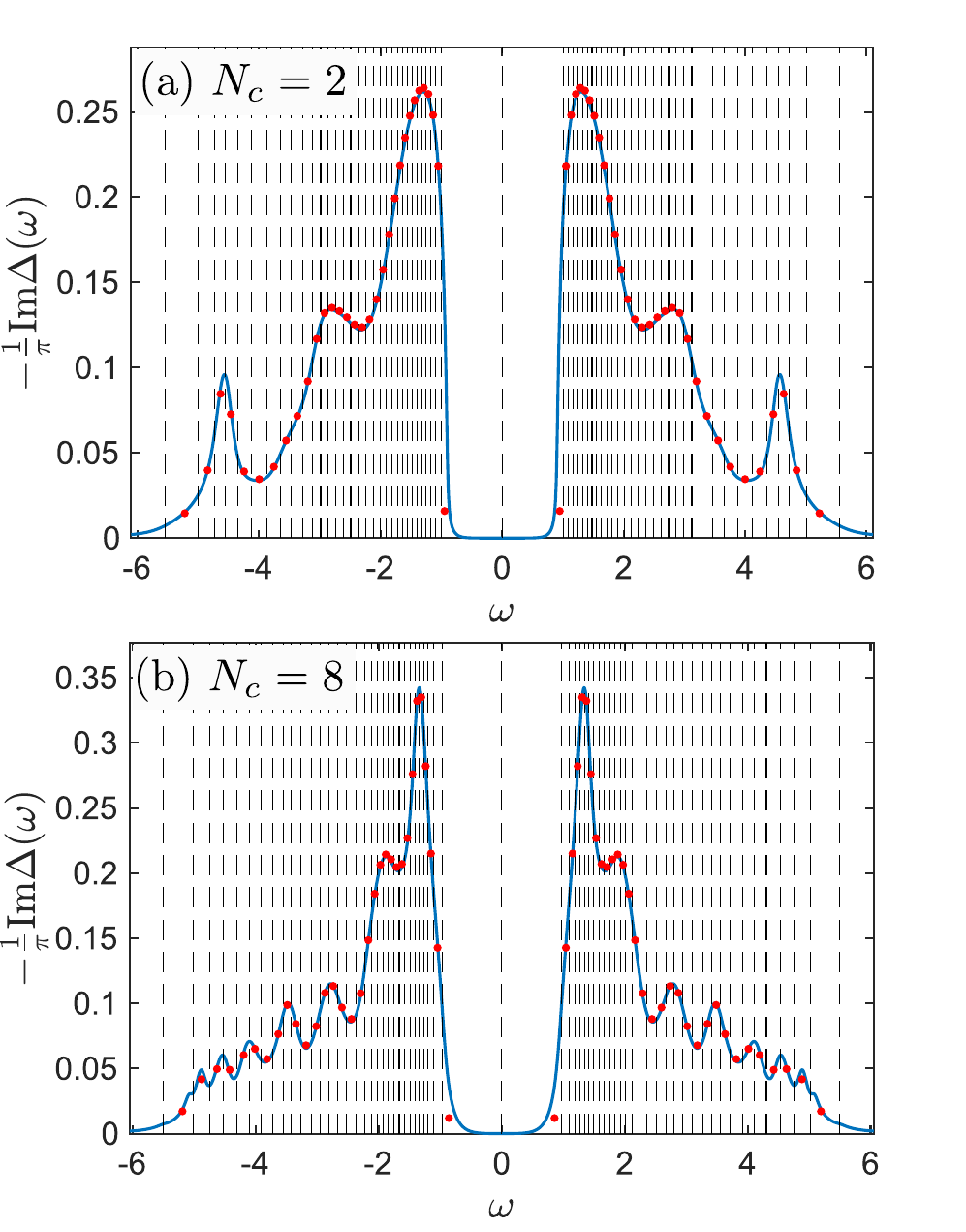}    \includegraphics[width=\linewidth]{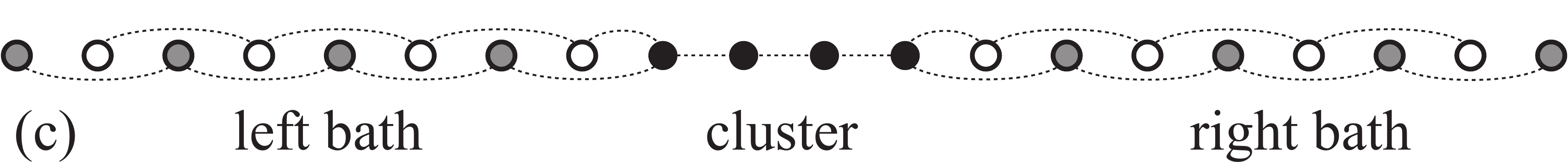} 
    \caption{\changed{CDMFT+TaSK for 1D Hubbard model:} (a,b) Hybridization functions (blue lines) at self-consistency, for a central cluster of size (a) $\Ncluster = 2$ and (b) $\Ncluster = 8$. The black dashed lines mark the boundaries $\omega_{\ell}$ between the 64 frequency intervals used for discretization. \changed{The red dots represent the discretized hybridization function $\changed{\Delta''(\omega) = 
    -\tfrac{1}{\pi} \mr{Im} \Delta (\omega) =} 
    \sum_{b} \Delta_{b} \, \delta(\omega - \xi_{b})$ of \Eq{eq:DiscreteHybridization}; the $x$-coordinate of a dot gives the location $\xi_{b} \in (\omega_{b},\omega_{b+1})$ of a delta peak, and its $y$-coordinate the weight normalized by the width of the corresponding interval, $\changed{\Delta''_{b}}/(\omega_{b+1} - \omega_{b})$.}     \changed{(c) MPS chain geometry: the impurity cluster (black sites) is coupled at each end to a discretized bath, represented by two interleaved chains describing positive-energy (white sites) and negative-energy (grey sites) bath states, respectively.}}
    \label{fig:Delta_disc}
\end{figure}

\changed{In \Sec{sec:1dHubbard_MPS}, we used TaSK to obtain benchmark results for the 1D Hubbard model. 
We consider} $L=50$ sites, with $U=5$ and $\mu = U/2$ (half-filling). To compute the spectral function $A_k(\omega)$  of the correlator $G_k(t)$ of \Eq{eq:Grk-1D-Hubbard}, we use the $k$-dependent single-particle operator
\begin{align}
    c_{k\changed{\sigma}} = \frac{1}{\sqrt{L}} \sum_{\ell = 1}^{L} \mr{e}^{-\mr{i} k \ell} c_{\ell\changed{\sigma}} \, ,
\end{align}
where we treat $k$ as a continuous variable to achieve finer $k$ resolution.
We exploit U(1)$\times$ SU(2) charge and spin symmetries. We compute the ground state using state-of-the-art single-site DMRG~\cite{White1992,Schollwock2011} with bond expansion~\cite{Gleis2022a,Gleis2022,Gleis2025}. The bond dimension of our ground state is $D^{\ast} = 500$ SU(2) multiplets.

We choose 51 equally spaced values for $k$ between $0$ and $\pi$, and for each momentum point use TaSK to compute a discrete version, $A^\discrete_k(\omega)$, of the spectral function, comprising  204 discrete $\delta$-peaks. We then compute the CFE coefficients $\epsilon_{kn}$ and $t_{kn}$ (shown in \Fig{fig:Hubbard1d_MPS_LiouvilleMatEle}) from $A^\discrete_k(\omega)$ (known at 51 $k$-values). To obtain a spectrum for arbitrary $k$, we interpolate $A^\discrete_k(\omega)$ for $k$-values between our chosen 51 ones by linear interpolation of the CFE coefficients $\epsilon_{kn}$ and $t_{kn}$. To obtain the continuous-in-frequency spectra shown in \Fig{fig:Hubbard1d_MPS_phsym}(a), 
we use the CFE+Gaussian broadening scheme of Sec.~S-2 of Ref.~\cite{Kovalska2025}, with broadening parameter $\sigma = 0.3$ and $\mathcal{N}_{\mathrm{CFE}} = 2$ there. 
For the lattice model, this is equivalent to broadening the discrete self-energy with $\sigma = 0.3$ and $\mathcal{N}_{\mathrm{CFE}} = 1$.

\begin{figure}[t!]
    \centering
    \includegraphics[width=0.8\linewidth]{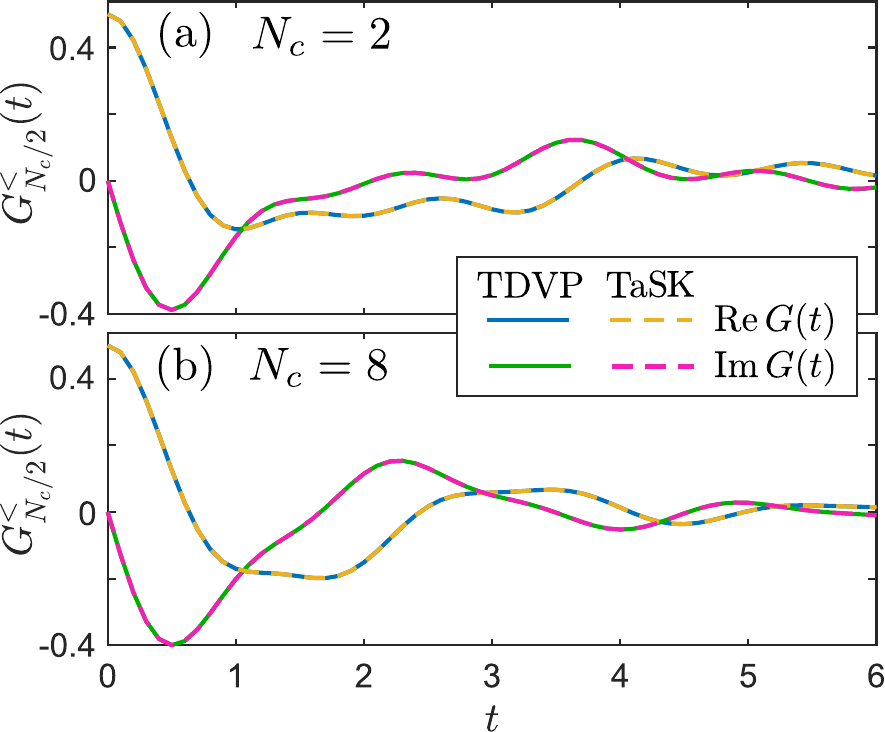} 
    \caption{Lesser Green's function $G^{<}_{\Ncluster/2}(t)$ computed via TDVP (solid lines) and TaSK (dashed lines) for (a) $\Ncluster=2$ and (b) $\Ncluster = 8$. For TDVP, we have used a time step $\delta t = 0.1$ and a third-order integrator; the bond dimension is allowed to grow to ensure a numerically exact result. With TaSK, we compute the spectrum of $G^{<}_{\Ncluster/2}(t)$ and Fourier transform to the time domain.}
    \label{fig:Gt_TDVP_vs_TaSK}
\end{figure}

\changed{
\subsection{CDMFT+TaSK for 1D Hubbard model}
\label{app:TaSK-impuritySolver}}

\changed{TaSK can also be used as an impurity solver, as demonstrated in 
Ref.~\onlinecite{Picoli2026}. We used this approach to obtain the CDMFT+TaSK results for the 1D Hubbard model presented in \Sec{sec:1dHubbard_CDMFT}.} \rreply{For background on treating the 1D Hubbard model using CDMFT, see, e.g., Ref.~\onlinecite{Bolech2003}. 

In CDMFT, only the boundary sites of the cluster couple to the baths; for a 1D model, we thus need only two baths, regardless of the cluster size \footnote{\rreply{By contrast, for DCA we would need one bath per patch, hence the computational costs increase rapidly with the number of patches. For this reason, we refrained from DCA+TaSK computations for the 1D Hubbard model.}}. 
Thus, our CDMFT} impurity model consists of an $N_c$-site Hubbard chain, whose outermost left and right sites hybridize with two spinful baths, \changed{$L$ and $R$}, described by two identical hybridization functions, $\Delta_L(\omega) = \Delta_R(\omega)$.

For the data shown in this manuscript, we use 64 spinful orbitals per bath. We discretize each bath using an adaptive grid generated via the mean method of Ref.~\onlinecite{DeVega2015}.  The resulting discretization intervals, level positions and hybridization weights at self-consistency are illustrated in Figs.~\ref{fig:Delta_disc}(a,b) for $\Ncluster = 2$ and $8$, respectively. 
\changed{For each discretized bath, we map the positive- and negative-energy states to two separate chains,
arrange these in interleaved MPS fashion,
and couple the $L$ and $R$ baths' first 
sites to the cluster's left and rights ends, respectively [\Fig{fig:Delta_disc}(c)].}

\changed{We solve this impurity model using TaSK, closely 
following the implementation presented in Ref.~\onlinecite{Picoli2026}
(but using a different discretization strategy, as described above). Each cycle of the CDMFT self-consistency loop consists 
of the following steps, taking continuous hybridization functions as input. First, we discretize them, as described above. 
Then, we compute the ground state using single-site DMRG with bond expansion \cite{Gleis2022}, truncating by singular values with a truncation threshold of $2\times 10^{-8}$.  
Next, we compute the impurity model's discrete self-energy,
a $\Ncluster \times \Ncluster$ matrix-valued function with elements $[\widetilde \bSigma{}^\discrete_\imp (\omega)]_{\ell, \ell'}$, whose $\omega$-dependence is represented by a sum of $\delta$-peaks. To this end, we use the scheme of Ref.~\cite{Zacinskis2026}, App.~B.3;
it requires several discrete spectral functions as input, which we compute  using TaSK \cite{Picoli2026}. 
Finally, to close the self-consistency loop, we broaden the discrete self-energy via the 
CFE+\-Gaussian scheme of Ref.~\cite{Kovalska2025}, using $\sigma = 0.3$ and $\mc{N}_{\mr{CFE}} = 1$, and invoke DMFT self-consistency to extract updated, continuous hybridization functions. 

Due to CDMFT self-consistency, the converged $\widetilde \bSigma{}^\discrete_\imp (\omega)$ also corresponds to $\widetilde \bSigma{}^\discrete_\mr{lattice} (\omega)$, the discrete lattice self-energy.
For $\cL$-interpolation, described in \Sec{sec:CDMFT+TaSK-interpolation-details}, we use the discrete
self-energy $\widetilde \bSigma{}^\discrete_\imp (\omega)$ as input, 
following the broaden-last protocol.}

To gauge whether TaSK provides accurate impurity spectral functions for the impurity model at hand we also conduct an independent benchmark test, formulated in the time domain. We use the single-site time-dependent variational principle~(TDVP)~\cite{Haegeman2011} with bond expansion~\cite{Li2024} to compute the lesser Green's function,
\begin{align}
    G^{<}_{\Ncluster/2}(t) = \langle \Psi_\ground | c^{\dagger}_{\Ncluster/2\sigma} e^{\mr{i}(H-E_\ground)t} c_{\Ncluster/2\sigma} | \Psi_\ground \rangle 
\end{align}
for short times. Here, $c_{\Ncluster/2\sigma}$ is the spin-$\sigma$ annihilation operator at cluster site $\Ncluster/2$, $|\Psi_\ground \rangle$ the ground state and $E_0$ its energy. We then compare the TDVP result for $G^{<}_{\Ncluster/2}(t)$ with the TaSK result, which can be easily computed from the discrete spectra. Figure~\ref{fig:Gt_TDVP_vs_TaSK} shows the result of this comparison for $\Ncluster = 2,8$. The time-dependent data obtained from both methods agree very well, demonstrating that our TaSK spectra are reliable. \vspace{5mm}

\section{log-CFE for DCA+NRG of 2D Hubbard model} 
\label{app:log-CFE-DCA-Hubbard}
\label{app:LiouvillianComputation}

\changed{This appendix gives details of our
DCA+NRG treatment of the 2D Hubbard model in \Sec{sec:2dHubbard_DCA}.
We discuss NRG settings (\Sec{sec:nrg_details_app}) and details of the log-CFE scheme underlying the $\cL$-interpolation (\Sec{app:dependence_splitting_parameters}).}

\label{app:SlowFastMode}
\label{app:SlowFastMode_numerics}

\subsection{NRG details}
\label{sec:nrg_details_app}

We study the 2D Hubbard model of \Eq{eq:H_2dHubbard} using 4-patch DCA+NRG, employing the patching scheme of \Fig{fig:star_patching}.
To solve the cluster impurity model, we use the MuNRG implementation~\cite{Lee2016_NRG,Lee2017} of the numerical renormalization group (NRG) \cite{Bulla2008_NRG,Weichselbaum2012b,Kugler2022}, which is based on the QSpace tensor library \cite{Weichselbaum2012,Weichselbaum2012b,Weichselbaum2020,Weichselbaum2024,Weichselbaum2024b}. We exploit $\mr{U}(1)$ charge and $\mr{SU}(2)$ spin rotation symmetry, use a discretization parameter of $\Lambda_\mr{NRG}=8$, no $z$-shifting, and keep up to $N_{\mr{keep}} = 10^{5}$ $\mr{U}(1) \times \mr{SU}(2)$ multiplets. Further, we use an interleaved Wilson chain geometry~\cite{Mitchell2014_iNRG,Stadler2016_iNRG}, fully interleaving the bath modes associated with the different patches (labeled by their patch momenta) in the following order: $\vec{K}_1=(0,0)$, $\vec{K}_2=(\pi,0)$, $\vec{K}_3=(0,\pi)$, $\vec{K}_4=(\pi,\pi)$. Although interleaving breaks the expected symmetry between the $(\pi,0)$ and $(0,\pi)$ patches, the resulting broadened self-energy functions $\Sigma_{(\pi,0)}(\omega)$ and $\Sigma_{(0,\pi)}(\omega)$ are essentially identical, differing by less than $10^{-3}$. Therefore, we enforced the symmetry between them in the DMFT loop, setting $\Sigma_{(0,\pi)}(\omega) \leftarrow\Sigma_{(\pi,0)}(\omega)$ before proceeding to the next DMFT iteration. 

\changed{NRG yields four discrete spectral functions, $\bA_\bK(\omega)$. 
To compute the corresponding patch self-energies $\Sigma_{\bK}(\omega)$,
we follow the broaden-first approach (cf.~\Sec{sec:broadening}): We first broaden the spectral functions  using the log-Gaussian broadening scheme of Ref.~\cite{Lee2016_NRG}, with broadening parameter $\sigma = 0.7$ there. Then, we compute self-energies using Kugler's symmetric estimator approach~\cite{Kugler2022}.} 

\begin{figure}[tb!]
    \centering
    \includegraphics[width=0.8\linewidth]{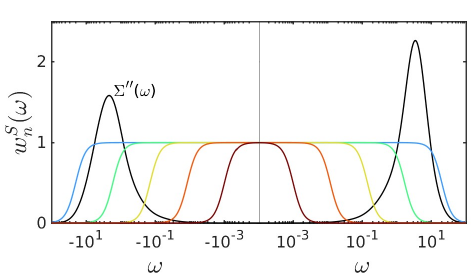} 
    \caption{Exemplary splitting functions $w_{n}^{\slow}(\omega)$ [\Eq{eq:wSF}], plotted for 
    \changed{$\Lambda_\fastslow = 12$}, $\tau_\fastslow=3$, and $n=1, \dots, 5$, resulting in logarithmically spaced box widths $\Omega_n$. The black line shows the spectral part of an arbitrary self-energy.}
    \label{fig:wSn}
\end{figure}

\subsection{Numerical details of log-CFE scheme} 
\label{app:dependence_splitting_parameters}

\changed{Using the patch self-energies $\bSigma_\bK(\omega)$ from NRG as input, we 
use the fast/slow splitting scheme of \App{sec:restfunction_splitting} to construct a log-CFE and compute its ingredients 
$\epsilon_{\vec{K}n}$, $t_{\vec{K}n}$ and $\bR^\fast_{\bK n}(z)$. Below, we illustrate their resulting exponential $n$-dependence 
and the effects of changing the shape of the splitting function.} 

\subsubsection{Exponential $n$-dependence of log-CFE coefficients}
\label{app:n-dependence_logCFE_coefficients}
\label{sec:splitting_function_again}

For convenience, we repeat here our choice for the splitting functions $w^{\fast/\slow}_n(\omega)$. They fulfill $w^{\fast}_{n}(\omega) + w^{\slow}_{n}(\omega)= 1$, and for 
$w^{\slow}_{n}(\omega)$ we use the  box function of \Eq{eq:splitting_function}, 
\begin{align}
    \label{eq:wSF}
    w^\slow_{n}(\omega) =\frac{1}{1 + |\omega/\Omega_{n}|^{\tau_\fastslow}}\, , \quad 
    \Omega_n = \Omega_{\mr{max}}\Lambda^{1-n}_\fastslow \, , 
\end{align}
with $\Lambda_\fastslow > 1$. \changed{The parameter $\tau_\fastslow$ controls the sharpness of the box. $\Omega_n$ sets its half-width, which decreases exponentially with $n$, ranging between a minimum and maximum frequency, $\Omega_{\mr{min}}$ and $\Omega_{\mr{max}}$, respectively. Figure \ref{fig:wSn} depicts $w_n^\slow(\omega)$ for five values of $n$ on a logarithmic frequency scale, so that $\Omega_n$ values appear evenly spaced. 

For the results shown in \Sec{sec:2dHubbard_DCA}, we choose
$\Lambda_\fastslow = 1.2$ and $\tau_\fastslow=5$ 
for the fast/slow splitting parameters, yielding a rather sharp split between fast and slow modes.
We set the minimal frequency $\Omega_{\mr{min}}=10^{-9}$ to 
match the numerical accuracy of the NRG impurity solver.
We choose $\Omega_{\mr{max}}$ large enough to
capture the entire support of each patch spectral function, 
\changed{$\mathbf{A}_\changed{\bK} (\omega) = \sum_i
\mathbf{A}_{\bK,i} \delta(\omega - \omega_i)$,}
such that the spectral sum rule is satisfied to within a tolerance of $10^{-5}$, \begin{align}
    \boldone-\sum_{|\omega_i|<\Omega_\mr{max}}
    \mathbf{A}_{\bK,i} (\omega_i) < 
    10^{-5} \boldone\, .
\end{align}
This cuts off over-broadened tails of  $w^\fast_1(\omega)$, thereby regularizing the spectral part of the self-energy and ensuring that the measure $\bW_\changed{\bK} (\omega)\mr{d}\omega$ is not indeterminate.}

\begin{figure}[tb!]
    \centering
    \includegraphics[width=0.85\linewidth]{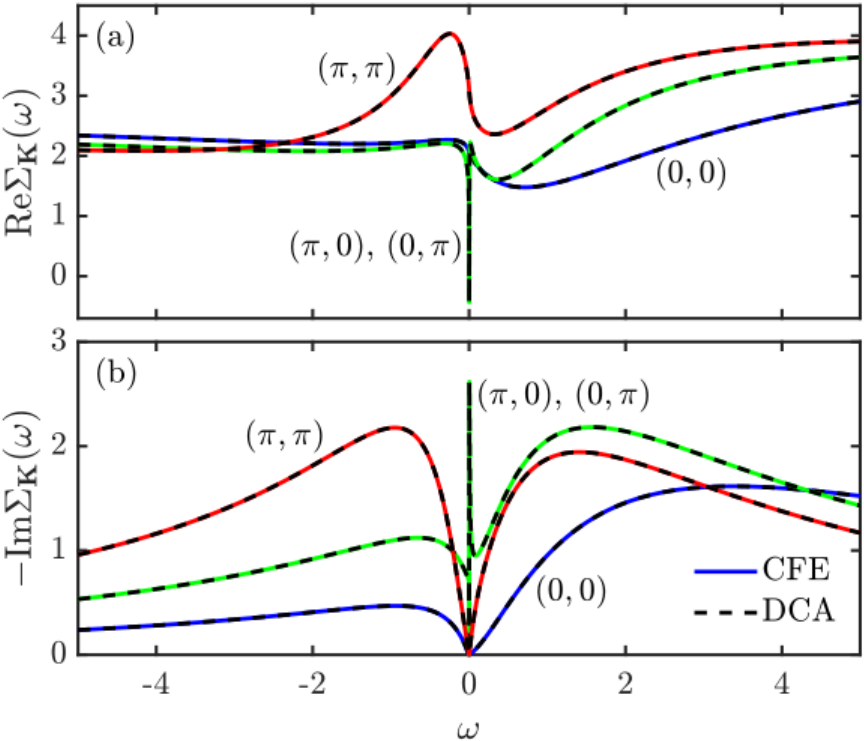} 
     \caption{\changed{DCA+NRG results for the 2D Hubbard model in the pseudogap phase.}  
    (a) Real and (b) imaginary parts of the original DCA+NRG patch self-energy $\Sigma_\bK(\omega)$ (dashed lines) and its  log-CFE representation \eqref{eq:SigmaCFE_FastSlowMode} (colored lines), for the patch momenta $\vec{K}_1\!=\!(0,0)$ (blue), $\vec{K}_2 \!=\!(\pi,0)$ and  $\bK_3\!=\!(0,\pi)$ (green), and $\bK_4\!=\!(\pi,\pi)$ (red). The excellent agreement between colored and dashed curves demonstrates that the log-CFE accurately represents the DCA+NRG results.}
    \label{fig:DCA_CFE}
\end{figure}

\begin{figure}[tb!]
    \centering
    \includegraphics[width=0.85\linewidth]{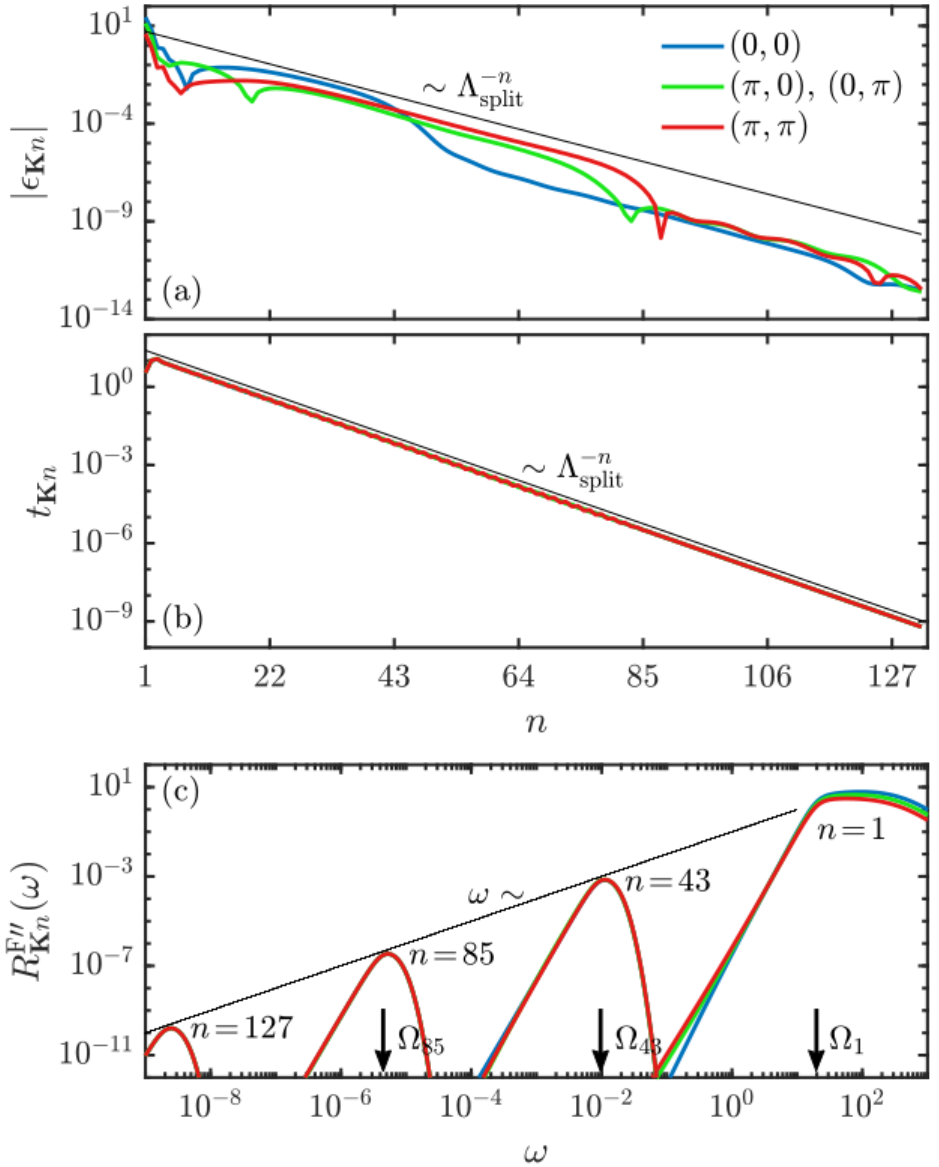} 
    \caption{(a,b) The Liouvillian matrix elements $|\epsilon_{\vec{K}n}|$ and $t_{\vec{K}n}$ obtained from a log-CFE with F/S splitting parameters $\Lambda_\fastslow = 1.2$, $\tau_\fastslow = 5$. (c) The spectral part of four fast modes $R^{\fast\prime\prime}_{\vec{K}n}(\omega)$ with $n \in \{1,\,43,\,85,\, 127\}$; arrows indicate the corresponding energy scales $\Omega_n$ [cf.~\Eq{eq:splitting_function}].
    Thin black lines indicate the scaling behavior. 
    As in \Fig{fig:DCA_CFE}, colors distinguish patch momenta. 
    }
    \label{fig:CFE_data}
\end{figure}

Figure \ref{fig:DCA_CFE} shows exemplary DCA patch self-energies (dashed lines) and their CFE representations (solid lines), confirming that our log-CFE accurately represents the cluster self-energies. Figures~\ref{fig:CFE_data}(a,b) show the corresponding Liouvillian matrix elements $\epsilon_{\vec{K}n}$ and $t_{\vec{K}n}$, demonstrating that they decrease exponentially with $n$ 
as $\sim \Lambda_\fastslow^{-n}$, as expected for a log-CFE.
Figure~\ref{fig:CFE_data}(c) shows the imaginary part of the fast modes, $R^{\fast \prime \prime}_{\vec{K}n}(\omega)= -\tfrac{1}{\pi}\mr{Im}R^{\fast}_{\vec{K}n}(\omega)$, for several choices of $n$, with individual patch momenta distinguished by line color. 
The  lines for the different patches show some slight differences for $n=1$ but overlap increasingly well with increasing $n$. This implies that the $\bK$-dependence of $R^{\fast \prime \prime}_{\vec{K}n}(\omega)$ is weak and becomes ever weaker with increasing $n$. This property, which appears to arise generically for log-CFEs, is important for reducing the likelihood of causality problems to arise when  interpolating fast-mode functions, as will be elaborated below. Moreover, the fact that $R^{\fast \prime \prime}_{\vec{K}n}(\omega)$ shows a very much weaker patch dependence than $|\epsilon_{\vec{K}n}|$ means that most of the $\bk$-dependence of the interpolation enters through the Liouvillian matrix elements. 

In the next two subsections, we illustrate how varying 
the parameters $\Lambda_\fastslow$ and $\tau_\fastslow$ that govern the shape of the splitting function $w^\slow_n(\omega)$ affects the results obtained from $\cL$-interpolation of our DCA+NRG data.

\subsubsection{Influence of sharpness of splitting function}

\begin{figure*}[tb!]
    \centering
    \includegraphics[width=0.93\linewidth]{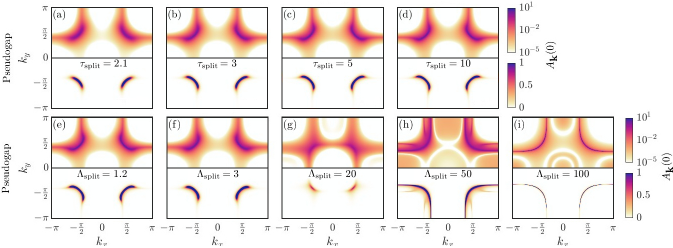} 
    \caption{$\cL$-interpolated DCA results for $A_\bk(0)$ in the pseudogap phase, computed for the same model parameters as in \Fig{fig:DCA_Ak}, shown (a-d) for four values of the parameter $\tau_\fastslow$ in \Eq{eq:wSF}, with $\Lambda_\fastslow = 1.2$ throughout, and (e-i) for five values of the parameter $\Lambda_\fastslow$ in \Eq{eq:wSF}, with $\tau_\fastslow = 5$ throughout. The top and bottom rows of each panel show the upper and lower halves of the BZ using logarithmic and linear color scales, respectively.
    } 
    \label{fig:tau_Lam}
\end{figure*}

\begin{figure}
    \centering
    \includegraphics[width=0.9\linewidth]{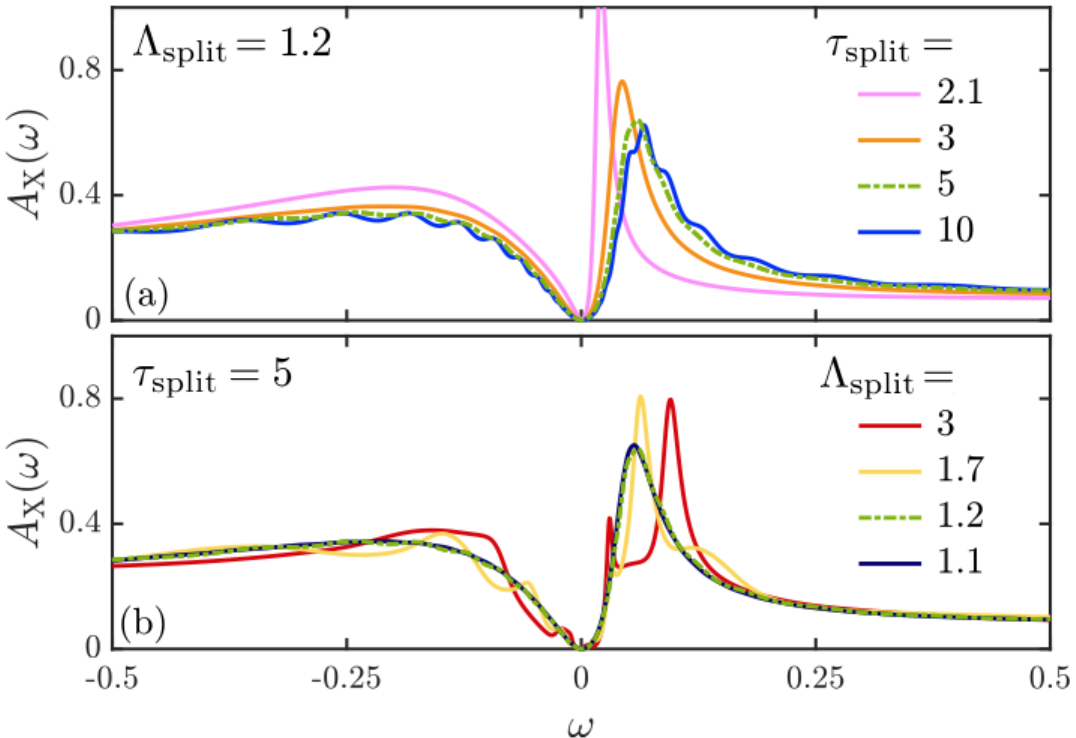} 
    \caption{$\cL$-interpolated spectral function $A_{\bk=\mr{X}}(\omega)$, (a) for the four $\tau_\fastslow$-values used for \Figs{fig:tau_Lam}(a-d), and (b) four $\Lambda_\fastslow$-values in the range used for \Figs{fig:tau_Lam}(e,f) or smaller. \rreply{The green dotted line corresponds to the 
    $\mc{L}$-line (red) in \Fig{fig:DCA_AkX}.}}
    \label{fig:AX_tau}
\end{figure}

The parameter $\tau_\fastslow>2$ controls the sharpness of $w_n^{\slow}(\omega)$. It tunes the overlap between successive fast modes and determines the smoothness of the continuous spectrum represented by the resulting CFE. The top row of \Fig{fig:tau_Lam} illustrates this effect by showing the $\cL$-interpolated $A_\bk(0)$ in the pseudogap regime for four values of $\tau_\fastslow$. Correspondingly, \Fig{fig:AX_tau}(a) displays a frequency slice at momentum $\mr{X}=(0,\pi)$ for the same values.

The impact of decreasing $\tau_\fastslow$ is clearly visible: while $\tau_\fastslow=10$ yields \changed{wiggly artifacts after interpolation due to the sharpness of the splitting function,}
reducing $\tau_\fastslow$ gradually removes the wiggles and, at the same time, sharpens the main features. However, for $\tau_\fastslow$ too small (e.g.\ $\tau_\fastslow \simeq 2$), the fast modes retain significant weight at low frequencies and the slow modes retain substantial high-frequency tails, i.e.\  $\bR^{\fast \prime\prime}_{\changed{\bK} n}(\omega)$ and $\bR^{\slow \prime \prime}_{\changed{\bK}n}(\omega)$ overlap significantly and do not achieve a clean scale separation between fast and slow modes. We find that this causes them to also retain a strong $\changed{\bK}$-dependence. This is undesirable, since Liouvillian interpolation works best if the $\changed{\bK}$-dependence is encoded entirely (or as much as possible) in the Liouvillian matrix elements $\bepsilon_{\changed{\bK}n}$ and $\bt_{\changed{\bK}n}$. For $\tau_\fastslow\leq 2$, we even observe a breakdown of tail regularization (not shown): sufficient weight remains in the tails after splitting such that $\bolde_{\changed{\bK}n}\nsim \Lambda^{-n}_\fastslow$, $\bt_{\changed{\bK}n}\nsim \Lambda^{-n}_\fastslow$, and $\Rest_{\changed{\bK}n}^\fast(z)\nsim \Lambda^{-n}_\fastslow$.

Conversely, increasing $\tau_\fastslow$ reduces the fast-mode overlap, yielding cleaner scale separation and a more stable interpolation. However, a too large  $\tau_\fastslow$ leads to underbroadening, exemplified by the wiggles in the $\tau_\fastslow=10$ curve of \Fig{fig:AX_tau}(a). The optimal choice of $\tau_\fastslow$ is therefore the largest value at which, upon decreasing $\tau_\fastslow$, these wiggles have just disappeared. We therefore choose $\tau_\fastslow = 5$.

\subsubsection{Influence of CFE length}

Due to the exponential decrease ${\Omega_n \sim \Lambda^{-n}}_\fastslow$ with $n$, the quantities $\boldsymbol{\epsilon}_{\changed{\bK}n}$, $\bt_{\changed{\bK}n}$, and $\Rest^F_{\bK n} (z)$ also decay exponentially and eventually reach the scale of numerical floating-point errors. As a result, the CFE exhibits artificial convergence. Increasing the CFE length $\NKrylov$, or equivalently decreasing the logarithmic discretization factor $\Lambda$ (while keeping $\Lambda>1$), increases the number of matrix elements in the CFE representation.

The second row of \Fig{fig:tau_Lam} shows $\cL$-interpolated results for $A_{\bk}(0)$ in the pseudogap regime for several values of $\Lambda_\fastslow$, where panels (e-i) correspond to a progressively shorter CFE. Figure~\ref{fig:AX_tau}(b) presents a frequency slice at momentum $\mr{X}=(0,\pi)$ for four values with $\Lambda_\fastslow \leq 3$. Reducing $\Lambda_\fastslow$ increases the number of Liouvillian matrix elements and thus the number of bands contributing to the interpolated  spectrum. This encodes the $\bk$-dependence through more matrix elements, thereby improving the interpolation of a continuous spectrum.

Comparing the lower row of \Fig{fig:tau_Lam} with \Fig{fig:AX_tau}(b) shows that the low-energy regime converges more rapidly ($\Lambda_\fastslow \lesssim 3$) than the high-energy regime as $\Lambda_\fastslow$ decreases, reflecting the logarithmic discretization. Accurately resolving high-energy features requires \changed{more closely-spaced $\bolde_n$ values} 
i.e., even smaller $\Lambda_\fastslow$. In our calculations, convergence at high energies is achieved for $\Lambda_\fastslow \lesssim 1.2$, as shown in \Fig{fig:AX_tau}(b).

Since we interpolate the fast modes using $\alpha_{\bk,\bK}$ (not $\alpha^\modified_{\bk, \bK}$), causality issues begin to arise for $\Lambda_\fastslow \! \gtrsim \! 3$ \changed{and large frequencies}, foreshadowing the fact that in the limit $\Lambda_\fastslow \! \gg \! 1$,  $\bR^\fast_{\changed{\bK}n}$-interpolation has similar issues as $\Sigma$-interpolation. This behavior originates from the enhanced ${\changed{\bK}}$
-dependence of the fast modes when fewer matrix elements are retained. We thus choose $\Lambda_\fastslow = 1.2$ as a good compromise between a manageable log-CFE depth and an accurate representation of the $\Lambda_\fastslow \to 1$ limit. Note that decreasing $\Lambda_\fastslow$ allows for larger values of $\tau_\fastslow$ while maintaining numerical stability.

\rreply{
\section{Evaluation of quasiparticle weight}\label{app:fits}

\begin{figure}
    \centering
    \includegraphics[width=0.9\linewidth]{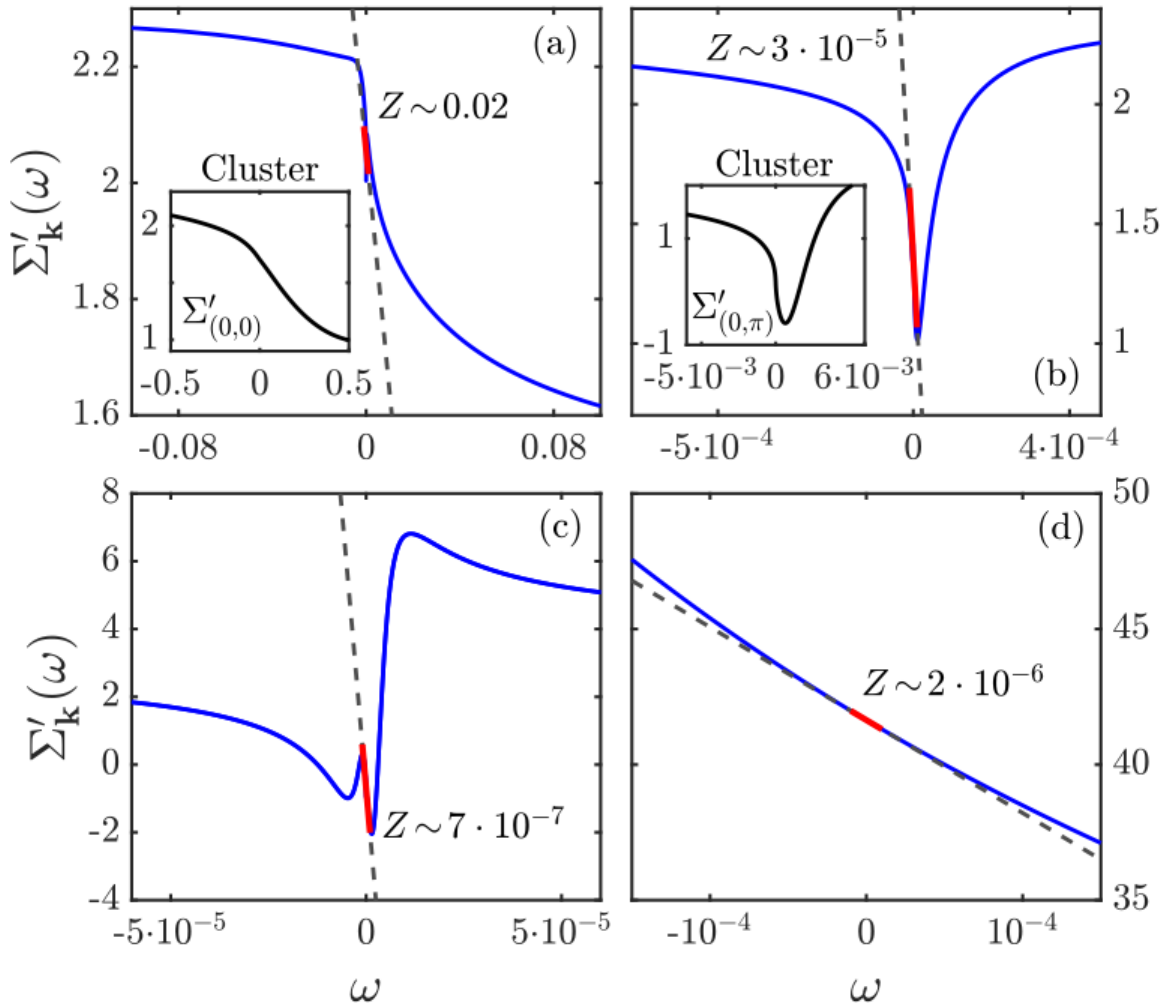}
    \caption{\rreply{The quasiparticle weight $Z_{\mathbf{k}_0}$ shown in \Fig{fig:FS_LS} was 
    extracted via linear fits to the real part of the $\mc{L}$-interpolated self-energy, Re$\Sigma_{\mathbf{k}}(\omega)$, around $\omega = 0$. Here, we illustrate this for four representative momentum points along the FS/LS trajectory, with panels (a) to (d) progressing from the nodal to the antinodal point. The grey dashed line indicates the linear fit, red color the fitting range. The two insets show two  patch  self-energies, Re$\Sigma_{\mathbf{K}}(\omega)$, for $\bK= (0,0)$
    and $(0,\pi)$.}}
    \label{fig:fits}
\end{figure}
Figure~\ref{fig:FS_LS} of the main text shows the  quasiparticle weight $Z_{\mathbf{k}_0} = (1-\partial_{\omega} \mathrm{Re}\Sigma_{\mathbf{k}_0}(0))^{-1}$ in the pseudogap phase of the 2D Hubbard model at $\delta=0.125$.
Here, we detail how we extracted $Z_{\mathbf{k}_0}$ from
our $\cL$-interpolated self-energy. To determine the slope $\partial_{\omega} \mathrm{Re}\Sigma_{\mathbf{k}_0}(0)$ without evaluating a numerical derivative, we used linear fits to the real part of the $\mathcal{L}$-interpolated self-energy, Re$\Sigma_{\mathbf{k}_0}(\omega)$, around $\omega = 0$. Nevertheless, we acknowledge that this procedure is numerically delicate, since both the interpolation and the fit are ultimately limited by the accuracy of the underlying computed cluster quantities.

\Fig{fig:fits} illustrates our strategy for four representative momentum points along the FS/LS trajectory, with panels (a) to (d) progressing from the nodal to the antinodal point. The grey dashed lines indicate the linear fits, while the red color denotes the fitting range. The insets show the corresponding cluster self-energy, $\mathrm{Re}\Sigma_{\mathbf{K}}(\omega)$. Around the nodal point, the fitting range is chosen to match the approximate shape of the cluster self-energy $\mathrm{Re}\Sigma_{(0,0)}(\omega)$ shown in the inset of panel (a). Here, we fit across numerical artifacts at very low frequencies ($\sim \pm 10^{-8}$) introduced by the NRG impurity solver in order to keep the computational cost manageable.

Toward the outer edge of the Fermi arc, the slope increases drastically and becomes more similar to the cluster self-energy $\mathrm{Re}\Sigma_{(0,\pi)}(\omega)$ shown in the inset of panel (b). Further toward the antinodal point, the proximity to the self-energy divergence leads to increasingly steep slopes and necessitates a smaller fitting range. In this regime, the extracted slope is sensitive to the chosen fitting range but remains sufficiently steep to yield a negligible quasiparticle weight.}

\end{appendices}

\FloatBarrier

\bibliography{bibliography}%

@Article{Kugler2021,
  author    = {Kugler, Fabian B. and Lee, Seung-Sup B. and von Delft, Jan},
  title     = {Multipoint Correlation Functions: Spectral Representation and Numerical Evaluation},
  doi       = {10.1103/PhysRevX.11.041006},
  issue     = {4},
  pages     = {041006},
  url       = {https://doi.org/10.1103/PhysRevX.11.041006},
  volume    = {11},
  comment   = {arXiv:2101.00707 [cond-mat.str-el]},
  journal   = {Phys. Rev. X},
  month     = {Oct},
  numpages  = {32},
  publisher = {American Physical Society},
  year      = {2021},
}

@Article{Lee2021,
  Title                    = {Computing Local Multipoint Correlators Using the Numerical Renormalization Group},
  Author                   = {Lee, Seung-Sup B. and Kugler, Fabian B. and von Delft, Jan},
  Journal                  = {Phys. Rev. X},
  Year                     = {2021},

  Month                    = {Oct},
  Pages                    = {041007},
  Volume                   = {11},

  Comment                  = {arXiv:2101.00708 [cond-mat.str-el]},
  Doi                      = {10.1103/PhysRevX.11.041007},
  Issue                    = {4},
  Numpages                 = {40},
  Url                      = {https://link.aps.org/doi/10.1103/PhysRevX.11.041007}
}

@Article{Meixner2024,
  author    = {Meixner, Michael and Menke, Henri and Klett, Marcel and Heinzelmann, Sarah and Andergassen, Sabine and Hansmann, Philipp and Schäfer, Thomas},
  title     = {Mott transition and pseudogap of the square-lattice {H}ubbard model: Results from center-focused cellular dynamical mean-field theory},
  doi       = {10.21468/SciPostPhys.16.2.059},
  pages     = {059},
  url       = {https://scipost.org/10.21468/SciPostPhys.16.2.059},
  volume    = {16},
  journal   = {SciPost Phys.},
  publisher = {SciPost},
  year      = {2024},
}

@Article{Qu2026,
  author       = {Dai-Wei Qu and Qiaoyi Li and Wei Li and Jan von Delft and Andreas Gleis},
  date         = {2026},
  journal = {to be published},
}

@Article{GleisLiouvillianSampling2026,
  author       = {Andreas Gleis},
  date         = {2026},
  journal = {to be published},
}

@Article{Kundu2024,
  author    = {Kundu, Sarbajaya and S\'en\'echal, David},
  title     = {{CDMFT+HFD}: An extension of dynamical mean field theory for nonlocal interactions applied to the single band extended {H}ubbard model},
  doi       = {10.21468/SciPostPhysCore.7.2.033},
  pages     = {033},
  url       = {https://scipost.org/10.21468/SciPostPhysCore.7.2.033},
  volume    = {7},
  journal   = {SciPost Phys. Core},
  publisher = {SciPost},
  year      = {2024},
}

@Article{Knizia2012,
  author    = {Knizia, Gerald and Chan, Garnet Kin-Lic},
  title     = {Density Matrix Embedding: A Simple Alternative to Dynamical Mean-Field Theory},
  doi       = {10.1103/PhysRevLett.109.186404},
  issue     = {18},
  pages     = {186404},
  url       = {http://link.aps.org/doi/10.1103/PhysRevLett.109.186404},
  volume    = {109},
  journal   = {Phys. Rev. Lett.},
  month     = {Nov},
  numpages  = {5},
  publisher = {American Physical Society},
  year      = {2012},
}

@Article{Sekaran2023,
  author       = {Sajanthan Sekaran and Oussama Bindech and Emmanuel Fromager},
  date         = {2023},
  journal = {J. Chem. Phys.},
  title        = {A unified density matrix functional construction of quantum baths in density matrix embedding theory beyond the mean-field approximation},
  pages        = {034107},
  url          = {https://doi.org/10.1063/5.0157746},
  volume       = {159},
}

@Article{Lanata2017,
  author    = {Lanat\`a, Nicola and Yao, Yongxin and Deng, Xiaoyu and Dobrosavljevi\ifmmode \acute{c}\else \'{c}\fi{}, Vladimir and Kotliar, Gabriel},
  title     = {Slave Boson Theory of Orbital Differentiation with Crystal Field Effects: Application to ${\mathrm{UO}}_{2}$},
  doi       = {10.1103/PhysRevLett.118.126401},
  issue     = {12},
  pages     = {126401},
  url       = {https://link.aps.org/doi/10.1103/PhysRevLett.118.126401},
  volume    = {118},
  journal   = {Phys. Rev. Lett.},
  month     = {Mar},
  numpages  = {6},
  publisher = {American Physical Society},
  year      = {2017},
}

@Article{Lanata2022,
  author    = {Lanat\`a, Nicola},
  title     = {Operatorial formulation of the ghost rotationally invariant slave-boson theory},
  doi       = {10.1103/PhysRevB.105.045111},
  issue     = {4},
  pages     = {045111},
  url       = {https://link.aps.org/doi/10.1103/PhysRevB.105.045111},
  volume    = {105},
  journal   = {Phys. Rev. B},
  month     = {Jan},
  numpages  = {9},
  publisher = {American Physical Society},
  year      = {2022},
}

@Article{Lanata2023a,
  author    = {Lanat\`a, Nicola},
  title     = {Derivation of the ghost {G}utzwiller approximation from quantum embedding principles: Ghost density matrix embedding theory},
  doi       = {10.1103/PhysRevB.108.235112},
  issue     = {23},
  pages     = {235112},
  url       = {https://link.aps.org/doi/10.1103/PhysRevB.108.235112},
  volume    = {108},
  journal   = {Phys. Rev. B},
  month     = {Dec},
  numpages  = {9},
  publisher = {American Physical Society},
  year      = {2023},
}

@Article{Jarrell2001,
  author    = {Jarrell, M. and Maier, Th. and Huscroft, C. and Moukouri, S.},
  title     = {Quantum Monte Carlo algorithm for nonlocal corrections to the dynamical mean-field approximation},
  doi       = {10.1103/PhysRevB.64.195130},
  issue     = {19},
  pages     = {195130},
  url       = {https://link.aps.org/doi/10.1103/PhysRevB.64.195130},
  volume    = {64},
  journal   = {Phys. Rev. B},
  month     = {Oct},
  numpages  = {23},
  publisher = {American Physical Society},
  year      = {2001},
}

@Article{Lichtenstein2000,
  author    = {Lichtenstein, A. I. and Katsnelson, M. I.},
  title     = {Antiferromagnetism and d-wave superconductivity in cuprates: A cluster dynamical mean-field theory},
  doi       = {10.1103/PhysRevB.62.R9283},
  issue     = {14},
  pages     = {R9283--R9286},
  url       = {https://link.aps.org/doi/10.1103/PhysRevB.62.R9283},
  volume    = {62},
  journal   = {Phys. Rev. B},
  month     = {Oct},
  numpages  = {0},
  publisher = {American Physical Society},
  year      = {2000},
}

@Article{Karakuzu2021,
  author    = {Karakuzu, Seher and Johnston, Steven and Maier, Thomas A.},
  title     = {Superconductivity in the bilayer {H}ubbard model: Two {F}ermi surfaces are better than one},
  doi       = {10.1103/PhysRevB.104.245109},
  issue     = {24},
  pages     = {245109},
  url       = {https://link.aps.org/doi/10.1103/PhysRevB.104.245109},
  volume    = {104},
  journal   = {Phys. Rev. B},
  month     = {Dec},
  numpages  = {12},
  publisher = {American Physical Society},
  year      = {2021},
}

@Article{Simkovic2024,
  author    = {Fedor Simkovic and Riccardo Rossi and Antoine Georges and Michel Ferrero},
  title     = {Origin and fate of the pseudogap in the doped {H}ubbard model},
  note      = {arXiv:2209.09237 [cond-mat.str-el]},
  pages     = {eade9194},
  url       = {https://www.science.org/doi/10.1126/science.ade9194},
  volume    = {385},
  journal   = {Science},
  year      = {2024},
}

@Article{Chen2012a,
  author    = {Chen, K.-S. and Meng, Z. Y. and Pruschke, T. and Moreno, J. and Jarrell, M.},
  title     = {Lifshitz transition in the two-dimensional {H}ubbard model},
  doi       = {10.1103/PhysRevB.86.165136},
  issue     = {16},
  pages     = {165136},
  url       = {https://link.aps.org/doi/10.1103/PhysRevB.86.165136},
  volume    = {86},
  journal   = {Phys. Rev. B},
  month     = {Oct},
  numpages  = {8},
  publisher = {American Physical Society},
  year      = {2012},
}

@Article{Gleis2022,
  author    = {Gleis, Andreas and Li, Jheng-Wei and von Delft, Jan},
  title     = {Controlled Bond Expansion for Density Matrix Renormalization Group Ground State Search at Single-Site Costs},
  doi       = {10.1103/PhysRevLett.130.246402},
  issue     = {24},
  pages     = {246402},
  url       = {https://doi.org/10.1103/PhysRevLett.130.246402},
  volume    = {130},
  journal   = {Phys. Rev. Lett.},
  month     = {Jun},
  notes     = {arXiv:2207.14712 [cond-mat.str-el]},
  numpages  = {8},
  publisher = {American Physical Society},
  year      = {2023},
}

@Article{Picoli2026,
  author = {Felipe D. Picoli and Ming Huang and Andreas Gleis and Jan von Delft},
  year   = {2026},
  journal = {to be published},
  title  = {Tangent-Space Krylov Solver for Quantum Impurity Models},
}

@Article{Zacinskis2026,
  author = {Zacinskis, Aleksandrs and Ebel, Frank T. and Pelz, Mathias and Kugler, Fabian B. and Held, Karsten and von Delft, Jan and Haverkort, Maurits W. and Gleis, Andreas},
  year = {2026},
  url = {https://arxiv.org/abs/2605.04974},
  journal = {arXiv:2605.04974 [cond-mat.str-el]},
  title = {Symmetric estimator for discrete self-energy of discrete many-body systems},
}

@Article{Kugler2022,
  Title                    = {Improved estimator for numerical renormalization group calculations of the self-energy},
  Author                   = {Kugler, Fabian B.},
  Journal                  = {Phys. Rev. B},
  Year                     = {2022},
  Month                    = {Jun},
  Pages                    = {245132},
  Volume                   = {105},
  Doi                      = {10.1103/PhysRevB.105.245132},
  Issue                    = {24},
  Numpages                 = {11},
  Publisher                = {American Physical Society},
  Url                      = {https://link.aps.org/doi/10.1103/PhysRevB.105.245132}
}

@Article{Bulla1998,
  author    = {Ralf Bulla and Alex C. Hewson and Thomas Pruschke},
  title     = {Numerical renormalization group calculations for the self-energy of the impurity {A}nderson model},
  doi       = {10.1088/0953-8984/10/37/021},
  pages     = {8365},
  url       = {http://iopscience.iop.org/0953-8984/10/37/021},
  volume    = {10},
  journal   = {J. Phys.: Condens. Matter},
  year      = {1998},
}

@Article{Lu2014,
  author    = {Lu, Y. and H\"oppner, M. and Gunnarsson, O. and Haverkort, M. W.},
  title     = {Efficient real-frequency solver for dynamical mean-field theory},
  doi       = {10.1103/PhysRevB.90.085102},
  issue     = {8},
  pages     = {085102},
  url       = {http://link.aps.org/doi/10.1103/PhysRevB.90.085102},
  volume    = {90},
  journal   = {Phys. Rev. B},
  month     = {Aug},
  numpages  = {18},
  publisher = {American Physical Society},
  year      = {2014},
}

@InCollection{Koch2011,
  author    = {Erik Koch},
  booktitle = {The {LDA}+{DMFT} approach to strongly correlated materials, Forschungszentrum Jülich},
  title     = {The {L}anczos Method},
  editor    = {Eva Pavarini and Erik Koch and Dieter Vollhardt and Alexander Lichtenstein},
  publisher = {Verlag des Forschungszentrum J\"ulich},
  comment   = {ISBN 978-3-89336-734-4},
  year      = {2011},
  URL = {https://www.cond-mat.de/events/correl11/manuscripts/koch.pdf}
}

@Article{Lihm2023,
  author    = {Lihm, Jae-Mo and Halbinger, Johannes and Shim, Jeongmin and von Delft, Jan and Kugler, Fabian B. and Lee, Seung-Sup B.},
  title     = {Symmetric improved estimators for multipoint vertex functions},
  doi       = {10.1103/PhysRevB.109.125138},
  issue     = {12},
  note      = {arXiv:2310.12098 [cond-mat.str-el]},
  pages     = {125138},
  url       = {https://doi.org/10.1103/PhysRevB.109.125138},
  volume    = {109},
  journal   = {Phys. Rev. B},
  month     = {Mar},
  numpages  = {32},
  publisher = {American Physical Society},
  year      = {2024},
}

@Article{Tiegel2014,
  author    = {Tiegel, Alexander C. and Manmana, Salvatore R. and Pruschke, Thomas and Honecker, Andreas},
  title     = {Matrix product state formulation of frequency-space dynamics at finite temperatures},
  doi       = {10.1103/PhysRevB.90.060406},
  issue     = {6},
  pages     = {060406},
  url       = {http://link.aps.org/doi/10.1103/PhysRevB.90.060406},
  volume    = {90},
  journal   = {Phys. Rev. B},
  month     = {Aug},
  numpages  = {5},
  publisher = {American Physical Society},
  year      = {2014},
}

@Article{Dargel2011,
  author   = {Dargel, P. E. and Honecker, A. and Peters, R. and Noack, R. M. and Pruschke, T.},
  title    = {Adaptive {L}anczos-vector method for dynamic properties within the density matrix renormalization group},
  doi      = {10.1103/PhysRevB.83.161104},
  number   = {16},
  pages    = {161104},
  volume   = {83},
  journal  = {Phys. Rev. B},
  month    = {Apr},
  numpages = {4},
  year     = {2011},
}

@Article{Dargel2012,
  author    = {Dargel, P. E. and W\"ollert, A. and Honecker, A. and McCulloch, I. P. and Schollw\"ock, U. and Pruschke, T.},
  title     = {{L}anczos algorithm with matrix product states for dynamical correlation functions},
  doi       = {10.1103/PhysRevB.85.205119},
  issue     = {20},
  pages     = {205119},
  url       = {https://link.aps.org/doi/10.1103/PhysRevB.85.205119},
  volume    = {85},
  journal   = {Phys. Rev. B},
  month     = {May},
  numpages  = {11},
  publisher = {American Physical Society},
  year      = {2012},
}

@article{Verret2022_CompactTiling,
  title = {{F}ermi arcs versus hole pockets: Periodization of a cellular two-band model},
  author = {Verret, S. and Foley, A. and S\'en\'echal, D. and Tremblay, A.-M. S. and Charlebois, M.},
  journal = {Phys. Rev. B},
  volume = {105},
  issue = {3},
  pages = {035117},
  numpages = {8},
  year = {2022},
  month = {Jan},
  publisher = {American Physical Society},
  doi = {10.1103/PhysRevB.105.035117},
  url = {https://link.aps.org/doi/10.1103/PhysRevB.105.035117}
}

@article{Damascelli2003_ARPEShighTcSC,
  title = {Angle-resolved photoemission studies of the cuprate superconductors},
  author = {Damascelli, Andrea and Hussain, Zahid and Shen, Zhi-Xun},
  journal = {Rev. Mod. Phys.},
  volume = {75},
  issue = {2},
  pages = {473--541},
  numpages = {0},
  year = {2003},
  month = {Apr},
  publisher = {American Physical Society},
  doi = {10.1103/RevModPhys.75.473},
  url = {https://link.aps.org/doi/10.1103/RevModPhys.75.473}
}

@article{Reber2012_ARPES_cuprates,
  author = {Reber, T. J. and Plumb, N. C. and Sun, Z. and Cao, Y. and Wang, Q. and McElroy, K. and Iwasawa, H. and Arita, M. and Wen, J. S. and Xu, Z. J. and Gu, G. and Yoshida, Y. and Eisaki, H. and Aiura, Y. and Dessau, D. S.},
  title = {The origin and non-quasiparticle nature of {F}ermi arcs in $\mr{Bi}_2\mr{Sr}_2\mr{CaCu}_2\mr{O}_{8+\delta}$},
  journal = {Nature Phys.},
  volume = {8},
  pages = {606-610},
  doi = {https://doi.org/10.1038/nphys2352},
  year = {2012}
}

@article{Yangmu2019_TransportHighTcSC2,
  author = {Yangmu Li  and W. Tabis  and Y. Tang  and G. Yu  and J. Jaroszynski  and N. Barišić  and M. Greven },
  title = {Hole pocket–driven superconductivity and its universal features in the electron-doped cuprates},
  journal = {Science Advances},
  volume = {5},
  number = {2},
  pages = {eaap7349},
  year = {2019},
  doi = {10.1126/sciadv.aap7349}
}

@article{Das2012_TransportHighTcSC1,
  title = {Visualizing electron pockets in cuprate superconductors},
  author = {Das, Tanmoy and Markiewicz, R. S. and Bansil, A. and Balatsky, A. V.},
  journal = {Phys. Rev. B},
  volume = {85},
  issue = {22},
  pages = {224535},
  numpages = {9},
  year = {2012},
  month = {Jun},
  publisher = {American Physical Society},
  doi = {10.1103/PhysRevB.85.224535},
  url = {https://link.aps.org/doi/10.1103/PhysRevB.85.224535}
}

@article{Senechal2002_periodization1,
  title = {Cluster perturbation theory for {H}ubbard models},
  author = {S\'en\'echal, David and Perez, Danny and Plouffe, Dany},
  journal = {Phys. Rev. B},
  volume = {66},
  issue = {7},
  pages = {075129},
  numpages = {11},
  year = {2002},
  month = {Aug},
  publisher = {American Physical Society},
  doi = {10.1103/PhysRevB.66.075129},
  url = {https://link.aps.org/doi/10.1103/PhysRevB.66.075129}
}

@article{Stanescu2006_periodization2,
  title = {{F}ermi arcs and hidden zeros of the {G}reen function in the pseudogap state},
  author = {Stanescu, Tudor D. and Kotliar, Gabriel},
  journal = {Phys. Rev. B},
  volume = {74},
  issue = {12},
  pages = {125110},
  numpages = {6},
  year = {2006},
  month = {Sep},
  publisher = {American Physical Society},
  doi = {10.1103/PhysRevB.74.125110},
  url = {https://link.aps.org/doi/10.1103/PhysRevB.74.125110}
}

@article{Biroli2002_periodization3,
  title = {Cluster methods for strongly correlated electron systems},
  author = {Biroli, Giulio and Kotliar, Gabriel},
  journal = {Phys. Rev. B},
  volume = {65},
  issue = {15},
  pages = {155112},
  numpages = {5},
  year = {2002},
  month = {Apr},
  publisher = {American Physical Society},
  doi = {10.1103/PhysRevB.65.155112},
  url = {https://link.aps.org/doi/10.1103/PhysRevB.65.155112}
}

@article{Biroli2004_periodization4,
  title = {Cluster dynamical mean-field theories: Causality and classical limit},
  author = {Biroli, G. and Parcollet, O. and Kotliar, G.},
  journal = {Phys. Rev. B},
  volume = {69},
  issue = {20},
  pages = {205108},
  numpages = {20},
  year = {2004},
  month = {May},
  publisher = {American Physical Society},
  doi = {10.1103/PhysRevB.69.205108},
  url = {https://link.aps.org/doi/10.1103/PhysRevB.69.205108}
}

@article{Kovalska2025,
      title={Tangent space Krylov computation of real-frequency spectral functions: Influence of density-assisted hopping on 2{D} {M}ott physics}, 
      author={Oleksandra Kovalska and Jan von Delft and Andreas Gleis},
      year={2025},
      journal={arXiv:2510.07279 [cond-mat.str-el]},
      url={https://arxiv.org/abs/2510.07279}, 
}

@article{Verret2019,
  title = {Intrinsic cluster-shaped density waves in cellular dynamical mean-field theory},
  author = {Verret, S. and Roy, J. and Foley, A. and Charlebois, M. and S\'en\'echal, D. and Tremblay, A.-M. S.},
  journal = {Phys. Rev. B},
  volume = {100},
  issue = {22},
  pages = {224520},
  numpages = {14},
  year = {2019},
  month = {Dec},
  publisher = {American Physical Society},
  doi = {10.1103/PhysRevB.100.224520},
  url = {https://link.aps.org/doi/10.1103/PhysRevB.100.224520}
}

@article{Klett2020_periodization5,
  title = {Real-space cluster dynamical mean-field theory: Center-focused extrapolation on the one- and two particle-levels},
  author = {Klett, Marcel and Wentzell, Nils and Sch\"afer, Thomas and Simkovic, Fedor and Parcollet, Olivier and Andergassen, Sabine and Hansmann, Philipp},
  journal = {Phys. Rev. Res.},
  volume = {2},
  issue = {3},
  pages = {033476},
  numpages = {11},
  year = {2020},
  month = {Sep},
  publisher = {American Physical Society},
  doi = {10.1103/PhysRevResearch.2.033476},
  url = {https://link.aps.org/doi/10.1103/PhysRevResearch.2.033476}
}

@article{Bruognolo2017_OpenWilsonChain,
  title = {Open {W}ilson chains for quantum impurity models: Keeping track of all bath modes},
  author = {Bruognolo, B. and Linden, N.-O. and Schwarz, F. and Lee, S.-S. B. and Stadler, K. and Weichselbaum, A. and Vojta, M. and Anders, F. B. and von Delft, J.},
  journal = {Phys. Rev. B},
  volume = {95},
  issue = {12},
  pages = {121115},
  numpages = {5},
  year = {2017},
  month = {Mar},
  publisher = {American Physical Society},
  doi = {10.1103/PhysRevB.95.121115},
  url = {https://link.aps.org/doi/10.1103/PhysRevB.95.121115}
}

@article{Worm2024_FS+LS,
  title = {Fermi and {L}uttinger Arcs: Two Concepts, Realized on One Surface},
  author = {Worm, Paul and Reitner, Matthias and Held, Karsten and Toschi, Alessandro},
  journal = {Phys. Rev. Lett.},
  volume = {133},
  issue = {16},
  pages = {166501},
  numpages = {7},
  year = {2024},
  month = {Oct},
  publisher = {American Physical Society},
  doi = {10.1103/PhysRevLett.133.166501},
  url = {https://link.aps.org/doi/10.1103/PhysRevLett.133.166501}
}

@article{Foley2024_Liouvillian,
  title={{L}iouvillian recursion method for the electronic {G}reen's function}, 
  author={Alexandre Foley},
  year={2024},
  journal={arXiv:2401.02527 [cond-mat.str-el]},
  url={https://arxiv.org/abs/2401.02527},
}

@article{Foley2019_ED,
  title = {Coexistence of superconductivity and antiferromagnetism in the {H}ubbard model for cuprates},
  author = {Foley, A. and Verret, S. and Tremblay, A.-M. S. and S\'en\'echal, D.},
  journal = {Phys. Rev. B},
  volume = {99},
  issue = {18},
  pages = {184510},
  numpages = {11},
  year = {2019},
  month = {May},
  publisher = {American Physical Society},
  doi = {10.1103/PhysRevB.99.184510},
  url = {https://link.aps.org/doi/10.1103/PhysRevB.99.184510}
}

@article{Stanescu2006_CDMFTpseudogap,
title = {A cellular dynamical mean field theory approach to {M}ottness},
journal = {Ann. Phys.},
volume = {321},
number = {7},
pages = {1682-1715},
year = {2006},
note = {Special Issue},
issn = {0003-4916},
doi = {https://doi.org/10.1016/j.aop.2006.03.009},
url = {https://www.sciencedirect.com/science/article/pii/S0003491606000972},
author = {Tudor D. Stanescu and Marcello Civelli and Kristjan Haule and Gabriel Kotliar}
}

@article{Krien2022_pseudogap,
  title = {Explaining the pseudogap through damping and antidamping on the {F}ermi surface by imaginary spin scattering},
  author = {Krien, F. and Worm, P. and Chalupa-Gantner, P. and others},
  journal = {Commun. Phys.},
  volume = {5},
  pages = {336},
  year = {2022},
  doi = {10.1038/s42005-022-01117-5},
  url = {https://doi.org/10.1038/s42005-022-01117-5}
}

@Article{Haydock1980a,
  author       = {Roger Haydock},
  year         = {1980},
  journal = {Solid state physics},
  title        = {The recursive solution to the {S}chr\"odinger equation},
  pages        = {215--294},
  url          = {https://www.sciencedirect.com/science/article/abs/pii/S0081194708605056},
  volume       = {35}
}

@article{Wu2018_pseudogap,
  title = {Pseudogap and {F}ermi-Surface Topology in the Two-Dimensional {H}ubbard Model},
  author = {Wu, Wei and Scheurer, Mathias S. and Chatterjee, Shubhayu and Sachdev, Subir and Georges, Antoine and Ferrero, Michel},
  journal = {Phys. Rev. X},
  volume = {8},
  issue = {2},
  pages = {021048},
  numpages = {17},
  year = {2018},
  month = {May},
  publisher = {American Physical Society},
  doi = {10.1103/PhysRevX.8.021048},
  url = {https://link.aps.org/doi/10.1103/PhysRevX.8.021048}
}

@article{Walsh2023_pseudogap,
  title = {Superconductivity in the two-dimensional {H}ubbard model with cellular dynamical mean-field theory: A quantum impurity model analysis},
  author = {Walsh, C. and Charlebois, M. and S\'emon, P. and Tremblay, A.-M. S. and Sordi, G.},
  journal = {Phys. Rev. B},
  volume = {108},
  issue = {7},
  pages = {075163},
  numpages = {17},
  year = {2023},
  month = {Aug},
  publisher = {American Physical Society},
  doi = {10.1103/PhysRevB.108.075163},
  url = {https://link.aps.org/doi/10.1103/PhysRevB.108.075163}
}

@article{Sakai2013_pseudogap,
  title = {Raman-Scattering Measurements and Theory of the Energy-Momentum Spectrum for Underdoped {B}i$_2${S}r$_2${C}a{C}u{O}$_{8+\ensuremath{\delta}}$ Superconductors: Evidence of an $s$-Wave Structure for the Pseudogap},
  author = {Sakai, S. and Blanc, S. and Civelli, M. and Gallais, Y. and Cazayous, M. and M\'easson, M.-A. and Wen, J. S. and Xu, Z. J. and Gu, G. D. and Sangiovanni, G. and Motome, Y. and Held, K. and Sacuto, A. and Georges, A. and Imada, M.},
  journal = {Phys. Rev. Lett.},
  volume = {111},
  issue = {10},
  pages = {107001},
  numpages = {5},
  year = {2013},
  month = {Sep},
  publisher = {American Physical Society},
  doi = {10.1103/PhysRevLett.111.107001},
  url = {https://link.aps.org/doi/10.1103/PhysRevLett.111.107001}
}

@article{Kotliar2001_CDMFT,
  title = {Cellular Dynamical Mean Field Approach to Strongly Correlated Systems},
  author = {Kotliar, Gabriel and Savrasov, Sergej Y. and P\'alsson, Gunnar and Biroli, Giulio},
  journal = {Phys. Rev. Lett.},
  volume = {87},
  issue = {18},
  pages = {186401},
  numpages = {4},
  year = {2001},
  month = {Oct},
  publisher = {American Physical Society},
  doi = {10.1103/PhysRevLett.87.186401},
  url = {https://link.aps.org/doi/10.1103/PhysRevLett.87.186401}
}

@Article{Hettler2000_DCA2,
  Title                    = {Dynamical cluster approximation: Nonlocal dynamics of correlated electron systems},
  Author                   = {Hettler, M. H. and Mukherjee, M. and Jarrell, M. and Krishnamurthy, H. R.},
  Journal                  = {Phys. Rev. B},
  Year                     = {2000},
  Month                    = {May},
  Pages                    = {12739--12756},
  Volume                   = {61},
  Doi                      = {10.1103/PhysRevB.61.12739},
  Issue                    = {19},
  Numpages                 = {0},
  Publisher                = {American Physical Society},
  Url                      = {https://link.aps.org/doi/10.1103/PhysRevB.61.12739}
}

@Article{Hettler1998_DCA1,
  Title                    = {Nonlocal dynamical correlations of strongly interacting electron systems},
  Author                   = {Hettler, M. H. and Tahvildar-Zadeh, A. N. and Jarrell, M. and Pruschke, T. and Krishnamurthy, H. R.},
  Journal                  = {Phys. Rev. B},
  Year                     = {1998},
  Month                    = {Sep},
  Pages                    = {R7475--R7479},
  Volume                   = {58},
  Doi                      = {10.1103/PhysRevB.58.R7475},
  Issue                    = {12},
  Numpages                 = {0},
  Publisher                = {American Physical Society},
  Url                      = {https://link.aps.org/doi/10.1103/PhysRevB.58.R7475}
}

@Article{Bulla2008_NRG,
  Title                    = {Numerical renormalization group method for quantum impurity systems},
  Author                   = {Bulla, Ralf and Costi, Theo A. and Pruschke, Thomas},
  Journal                  = {Rev. Mod. Phys.},
  Year                     = {2008},
  Month                    = {Apr},
  Pages                    = {395--450},
  Volume                   = {80},
  Doi                      = {10.1103/RevModPhys.80.395},
  Issue                    = {2},
  Numpages                 = {0},
  Publisher                = {American Physical Society}
}

@Article{Lee2016_NRG,
  Title                    = {Adaptive broadening to improve spectral resolution in the numerical renormalization group},
  Author                   = {Lee, Seung-Sup B. and Weichselbaum, Andreas},
  Journal                  = {Phys. Rev. B},
  Year                     = {2016},
  Month                    = {Dec},
  Pages                    = {235127},
  Volume                   = {94},
  Author+an                = {1=SSBL},
  Doi                      = {10.1103/PhysRevB.94.235127},
  Issue                    = {23},
  Numpages                 = {15},
  Publisher                = {American Physical Society},
  Url                      = {http://link.aps.org/doi/10.1103/PhysRevB.94.235127}
}

@article{Lee2017,
  title = {Doublon-Holon Origin of the Subpeaks at the {H}ubbard Band Edges},
  author = {Lee, Seung-Sup B. and von Delft, Jan and Weichselbaum, Andreas},
  journal = {Phys. Rev. Lett.},
  volume = {119},
  issue = {23},
  pages = {236402},
  numpages = {5},
  year = {2017},
  month = {Dec},
  publisher = {American Physical Society},
  doi = {10.1103/PhysRevLett.119.236402},
  url = {https://link.aps.org/doi/10.1103/PhysRevLett.119.236402}
}

@article{Gull2010_patching,
  title = {Momentum-space anisotropy and pseudogaps: A comparative cluster dynamical mean-field analysis of the doping-driven metal-insulator transition in the two-dimensional {H}ubbard model},
  author = {Gull, E. and Ferrero, M. and Parcollet, O. and Georges, A. and Millis, A. J.},
  journal = {Phys. Rev. B},
  volume = {82},
  issue = {15},
  pages = {155101},
  numpages = {14},
  year = {2010},
  month = {Oct},
  publisher = {American Physical Society},
  doi = {10.1103/PhysRevB.82.155101},
  url = {https://link.aps.org/doi/10.1103/PhysRevB.82.155101}
}

@Article{Lanczos1950,
  Title                    = {An iteration method for the solution of the eigenvalue problem of linear differential and integral operators},
  Author                   = {Cornelius Lanczos},
  Number                   = {4},
  Pages                    = {255--282},
  Volume                   = {45},
  year                     = {1950},
  Doi                      = {10.6028/jres.045.026},
  Journal             = {Journal of Research of the National Bureau of Standards},
  Publisher                = {United States Government}
}

@article{Sakai2011_cumulant_periodization,
  title = {Cluster-size dependence in cellular dynamical mean-field theory},
  author = {Sakai, Shiro and Sangiovanni, Giorgio and Civelli, Marcello and Motome, Yukitoshi and Held, Karsten and Imada, Masatoshi},
  journal = {Phys. Rev. B},
  volume = {85},
  issue = {3},
  pages = {035102},
  numpages = {11},
  year = {2012},
  month = {Jan},
  publisher = {American Physical Society},
  doi = {10.1103/PhysRevB.85.035102},
  url = {https://link.aps.org/doi/10.1103/PhysRevB.85.035102}
}

@Article{Sakai2009,
  author    = {Sakai, Shiro and Motome, Yukitoshi and Imada, Masatoshi},
  title     = {Evolution of Electronic Structure of Doped {M}ott Insulators: Reconstruction of Poles and Zeros of {G}reen's Function},
  doi       = {10.1103/PhysRevLett.102.056404},
  issue     = {5},
  pages     = {056404},
  url       = {https://link.aps.org/doi/10.1103/PhysRevLett.102.056404},
  volume    = {102},
  journal   = {Phys. Rev. Lett.},
  month     = {Feb},
  numpages  = {4},
  publisher = {American Physical Society},
  year      = {2009},
}

@Article{Sakai2010,
  author    = {Sakai, Shiro and Motome, Yukitoshi and Imada, Masatoshi},
  title     = {Doped high-${T}_{c}$ cuprate superconductors elucidated in the light of zeros and poles of the electronic {G}reen's function},
  doi       = {10.1103/PhysRevB.82.134505},
  issue     = {13},
  pages     = {134505},
  url       = {https://link.aps.org/doi/10.1103/PhysRevB.82.134505},
  volume    = {82},
  journal   = {Phys. Rev. B},
  month     = {Oct},
  numpages  = {16},
  publisher = {American Physical Society},
  year      = {2010},
}

@article{Ferrero2009_dimerDCA,
  title = {Pseudogap opening and formation of {F}ermi arcs as an orbital-selective {M}ott transition in momentum space},
  author = {Ferrero, Michel and Cornaglia, Pablo S. and De Leo, Lorenzo and Parcollet, Olivier and Kotliar, Gabriel and Georges, Antoine},
  journal = {Phys. Rev. B},
  volume = {80},
  issue = {6},
  pages = {064501},
  numpages = {21},
  year = {2009},
  month = {Aug},
  publisher = {American Physical Society},
  doi = {10.1103/PhysRevB.80.064501},
  url = {https://link.aps.org/doi/10.1103/PhysRevB.80.064501}
}

@article{Okamoto2003,
  title = {Fictive impurity models: An alternative formulation of the cluster dynamical mean-field method},
  author = {Okamoto, S. and Millis, A. J. and Monien, H. and Fuhrmann, A.},
  journal = {Phys. Rev. B},
  volume = {68},
  issue = {19},
  pages = {195121},
  numpages = {8},
  year = {2003},
  month = {Nov},
  publisher = {American Physical Society},
  doi = {10.1103/PhysRevB.68.195121},
  url = {https://link.aps.org/doi/10.1103/PhysRevB.68.195121}
}

@article{Choi1972,
  title = {Positive Linear Maps on {C}*-Algebras},
  volume={24},
  DOI={10.4153/CJM-1972-044-5},
  number={3},
  journal={Canadian Journal of Mathematics},
  author={Choi, Man-Duen},
  year={1972},
  pages={520–529}
}

@article{Jamiołkowski1972,
  title = {Linear transformations which preserve trace and positive semidefiniteness of operators},
  journal = {Reports on Mathematical Physics},
  volume = {3},
  number = {4},
  pages = {275-278},
  year = {1972},
  issn = {0034-4877},
  doi = {https://doi.org/10.1016/0034-4877(72)90011-0},
  url = {https://www.sciencedirect.com/science/article/pii/0034487772900110},
  author = {A. Jamiołkowski}
}

@book{Viswanath1994,
    author = {Viswanath, V. S. and Müller, Gerhard},
    title = {The Recursion Method},
    publisher = {Springer Berlin, Heidelberg},
    year = {1994},
    doi = {https://doi.org/10.1007/978-3-540-48651-0}
}

@article{Mori1965,
    author = {Mori, Hazime},
    title = {A Continued-Fraction Representation of the Time-Correlation Functions},
    journal = {Progress of Theoretical Physics},
    volume = {34},
    number = {3},
    pages = {399-416},
    year = {1965},
    month = {09},
    issn = {0033-068X},
    doi = {10.1143/PTP.34.399},
    url = {https://doi.org/10.1143/PTP.34.399},
}

@article{Maier2005,
  title = {Quantum cluster theories},
  author = {Maier, Thomas and Jarrell, Mark and Pruschke, Thomas and Hettler, Matthias H.},
  journal = {Rev. Mod. Phys.},
  volume = {77},
  issue = {3},
  pages = {1027--1080},
  numpages = {0},
  year = {2005},
  month = {Oct},
  publisher = {American Physical Society},
  doi = {10.1103/RevModPhys.77.1027},
  url = {https://link.aps.org/doi/10.1103/RevModPhys.77.1027}
}

@article{Maier2002_DCAinterp,
  title = {Angle-resolved photoemission spectra of the {H}ubbard model},
  author = {Maier, Th. A. and Pruschke, Th. and Jarrell, M.},
  journal = {Phys. Rev. B},
  volume = {66},
  issue = {7},
  pages = {075102},
  numpages = {8},
  year = {2002},
  month = {Aug},
  publisher = {American Physical Society},
  doi = {10.1103/PhysRevB.66.075102},
  url = {https://link.aps.org/doi/10.1103/PhysRevB.66.075102}
}

@article{Georges1996,
  title = {Dynamical mean-field theory of strongly correlated fermion systems and the limit of infinite dimensions},
  author = {Georges, Antoine and Kotliar, Gabriel and Krauth, Werner and Rozenberg, Marcelo J.},
  journal = {Rev. Mod. Phys.},
  volume = {68},
  issue = {1},
  pages = {13--125},
  numpages = {0},
  year = {1996},
  month = {Jan},
  publisher = {American Physical Society},
  doi = {10.1103/RevModPhys.68.13},
  url = {https://link.aps.org/doi/10.1103/RevModPhys.68.13}
}

@article{DeLeo2008a,
  title = {${T}\!=\!0$ Heavy-Fermion Quantum Critical Point as an Orbital-Selective {M}ott Transition},
  author = {De Leo, Lorenzo and Civelli, Marcello and Kotliar, Gabriel},
  journal = {Phys. Rev. Lett.},
  volume = {101},
  issue = {25},
  pages = {256404},
  numpages = {4},
  year = {2008},
  month = {Dec},
  publisher = {American Physical Society},
  doi = {10.1103/PhysRevLett.101.256404},
  url = {https://link.aps.org/doi/10.1103/PhysRevLett.101.256404}
}

@article{DeLeo2008,
  title = {Cellular dynamical mean-field theory of the periodic {A}nderson model},
  author = {De Leo, Lorenzo and Civelli, Marcello and Kotliar, Gabriel},
  journal = {Phys. Rev. B},
  volume = {77},
  issue = {7},
  pages = {075107},
  numpages = {7},
  year = {2008},
  month = {Feb},
  publisher = {American Physical Society},
  doi = {10.1103/PhysRevB.77.075107},
  url = {https://link.aps.org/doi/10.1103/PhysRevB.77.075107}
}

@article{Gleis2024,
  title = {Emergent Properties of the Periodic {A}nderson Model: A High-Resolution, Real-Frequency Study of Heavy-Fermion Quantum Criticality},
  author = {Gleis, Andreas and Lee, Seung-Sup B. and Kotliar, Gabriel and von Delft, Jan},
  journal = {Phys. Rev. X},
  volume = {14},
  issue = {4},
  pages = {041036},
  numpages = {48},
  year = {2024},
  month = {Nov},
  publisher = {American Physical Society},
  doi = {10.1103/PhysRevX.14.041036},
  url = {https://link.aps.org/doi/10.1103/PhysRevX.14.041036}
}

@article{Gleis2024a,
  title = {Dynamical Scaling and {P}lanckian Dissipation Due to Heavy-Fermion Quantum Criticality},
  author = {Gleis, Andreas and Lee, Seung-Sup B. and Kotliar, Gabriel and von Delft, Jan},
  journal = {Phys. Rev. Lett.},
  volume = {134},
  issue = {10},
  pages = {106501},
  numpages = {9},
  year = {2025},
  month = {Mar},
  publisher = {American Physical Society},
  doi = {10.1103/PhysRevLett.134.106501},
  url = {https://link.aps.org/doi/10.1103/PhysRevLett.134.106501}
}

@article{Hardy2024,
  title = {Enhanced Strange Metallicity due to {H}ubbard-${U}$ {C}oulomb Repulsion},
  author = {Hardy, Andrew and Parcollet, Olivier and Georges, Antoine and Patel, Aavishkar A.},
  journal = {Phys. Rev. Lett.},
  volume = {134},
  issue = {3},
  pages = {036502},
  numpages = {6},
  year = {2025},
  month = {Jan},
  publisher = {American Physical Society},
  doi = {10.1103/PhysRevLett.134.036502},
  url = {https://link.aps.org/doi/10.1103/PhysRevLett.134.036502}
}

@article{Nomura2014_mulitorbital,
  title = {Multiorbital cluster dynamical mean-field theory with an improved continuous-time quantum {M}onte {C}arlo algorithm},
  author = {Nomura, Yusuke and Sakai, Shiro and Arita, Ryotaro},
  journal = {Phys. Rev. B},
  volume = {89},
  issue = {19},
  pages = {195146},
  numpages = {12},
  year = {2014},
  month = {May},
  publisher = {American Physical Society},
  doi = {10.1103/PhysRevB.89.195146},
  url = {https://link.aps.org/doi/10.1103/PhysRevB.89.195146}
}

@article{Harland2019_mulitorbital,
  title = {Electronic correlations and competing orders in multiorbital dimers: A cluster {DMFT} study},
  author = {Harland, Malte and Poteryaev, Alexander I. and Streltsov, Sergey V. and Lichtenstein, Alexander I.},
  journal = {Phys. Rev. B},
  volume = {99},
  issue = {4},
  pages = {045115},
  numpages = {11},
  year = {2019},
  month = {Jan},
  publisher = {American Physical Society},
  doi = {10.1103/PhysRevB.99.045115},
  url = {https://link.aps.org/doi/10.1103/PhysRevB.99.045115}
}

@article{Semon2017_hund,
  title = {Validity of the local approximation in iron pnictides and chalcogenides},
  author = {S\'emon, Patrick and Haule, Kristjan and Kotliar, Gabriel},
  journal = {Phys. Rev. B},
  volume = {95},
  issue = {19},
  pages = {195115},
  numpages = {6},
  year = {2017},
  month = {May},
  publisher = {American Physical Society},
  doi = {10.1103/PhysRevB.95.195115},
  url = {https://link.aps.org/doi/10.1103/PhysRevB.95.195115}
}

@article{Nomura2015_hund,
  title={Nonlocal correlations induced by {H}und’s coupling: A cluster {DMFT} study},
  volume={91},
  pages={235107},
  ISSN={1550-235X},
  url={http://dx.doi.org/10.1103/PhysRevB.91.235107},
  number={23},
  journal={Phys. Rev. B},
  publisher={American Physical Society (APS)},
  author={Nomura, Yusuke and Sakai, Shiro and Arita, Ryotaro},
  year={2015}
}

@article{Fournier2024,
  title={Two ${T}$-linear scattering-rate regimes in the triangular lattice {H}ubbard model},
  volume={17},
  pages={072},
  ISSN={2542-4653},
  url={http://dx.doi.org/10.21468/SciPostPhys.17.3.072},
  number={3},
  journal={SciPost Physics},
  author={Fournier, Jérôme and Downey, Pierre-Olivier and Hébert, Charles-David and Charlebois, Maxime and Tremblay, André-Marie},
  year={2024}
}

@Article{Wu2020a,
  author    = {Wu, Wei and Scheurer, Mathias S. and Ferrero, Michel and Georges, Antoine},
  title     = {Effect of Van Hove singularities in the onset of pseudogap states in {M}ott insulators},
  doi       = {10.1103/PhysRevResearch.2.033067},
  issue     = {3},
  pages     = {033067},
  url       = {https://link.aps.org/doi/10.1103/PhysRevResearch.2.033067},
  volume    = {2},
  journal   = {Phys. Rev. Res.},
  month     = {Jul},
  numpages  = {6},
  publisher = {American Physical Society},
  year      = {2020},
}

@article{Wu2022,
  title={Non-{F}ermi liquid phase and linear-in-temperature scattering rate in overdoped two-dimensional {H}ubbard model},
  volume={119},
  pages={e2115819119},
  ISSN={1091-6490},
  url={http://dx.doi.org/10.1073/pnas.2115819119},
  number={13},
  journal={Proceedings of the National Academy of Sciences},
  author={Wú, Wéi and Wang, Xiang and Tremblay, André-Marie},
  year={2022}
}

@article{Yang2006,
  title = {Phenomenological theory of the pseudogap state},
  author = {Yang, Kai-Yu and Rice, T. M. and Zhang, Fu-Chun},
  journal = {Phys. Rev. B},
  volume = {73},
  issue = {17},
  pages = {174501},
  numpages = {10},
  year = {2006},
  month = {May},
  publisher = {American Physical Society},
  doi = {10.1103/PhysRevB.73.174501},
  url = {https://link.aps.org/doi/10.1103/PhysRevB.73.174501}
}

@article{Weichselbaum2012,
  title = {Non-{A}belian symmetries in tensor networks: {A} quantum symmetry space approach},
  journal = {Annals of Physics},
  volume = {327},
  number = {12},
  pages = {2972-3047},
  year = {2012},
  issn = {0003-4916},
  doi = {https://doi.org/10.1016/j.aop.2012.07.009},
  url = {https://www.sciencedirect.com/science/article/pii/S0003491612001121},
  author = {Andreas Weichselbaum},
}

@article{Weichselbaum2012b,
  title = {Tensor networks and the numerical renormalization group},
  author = {Weichselbaum, Andreas},
  journal = {Phys. Rev. B},
  volume = {86},
  issue = {24},
  pages = {245124},
  numpages = {17},
  year = {2012},
  month = {Dec},
  publisher = {American Physical Society},
  doi = {10.1103/PhysRevB.86.245124},
  url = {https://link.aps.org/doi/10.1103/PhysRevB.86.245124}
}

@article{Weichselbaum2020,
  title = {X-symbols for non-{A}belian symmetries in tensor networks},
  author = {Weichselbaum, Andreas},
  journal = {Phys. Rev. Res.},
  volume = {2},
  issue = {2},
  pages = {023385},
  numpages = {16},
  year = {2020},
  month = {Jun},
  publisher = {American Physical Society},
  doi = {10.1103/PhysRevResearch.2.023385},
  url = {https://link.aps.org/doi/10.1103/PhysRevResearch.2.023385}
}

@Article{Weichselbaum2024,
  title={{QSpace - An open-source tensor library for Abelian and non-Abelian symmetries}},
  author={Andreas Weichselbaum},
  journal={SciPost Phys. Codebases},
  pages={40},
  year={2024},
  publisher={SciPost},
  doi={10.21468/SciPostPhysCodeb.40},
  url={https://scipost.org/10.21468/SciPostPhysCodeb.40},
}

@Article{Weichselbaum2024b,
  title={{Codebase release 4.0 for QSpace}},
  author={Andreas Weichselbaum},
  journal={SciPost Phys. Codebases},
  pages={40-r4.0},
  year={2024},
  publisher={SciPost},
  doi={10.21468/SciPostPhysCodeb.40-r4.0},
  url={https://scipost.org/10.21468/SciPostPhysCodeb.40-r4.0},
}

@article{Auerbach2018,
  title = {Hall Number of Strongly Correlated Metals},
  author = {Auerbach, Assa},
  journal = {Phys. Rev. Lett.},
  volume = {121},
  issue = {6},
  pages = {066601},
  numpages = {6},
  year = {2018},
  month = {Aug},
  publisher = {American Physical Society},
  doi = {10.1103/PhysRevLett.121.066601},
  url = {https://link.aps.org/doi/10.1103/PhysRevLett.121.066601}
}

@article{Auerbach2019,
  title = {Equilibrium formulae for transverse magnetotransport of strongly correlated metals},
  author = {Auerbach, Assa},
  journal = {Phys. Rev. B},
  volume = {99},
  issue = {11},
  pages = {115115},
  numpages = {16},
  year = {2019},
  month = {Mar},
  publisher = {American Physical Society},
  doi = {10.1103/PhysRevB.99.115115},
  url = {https://link.aps.org/doi/10.1103/PhysRevB.99.115115}
}

@article{Stadler2016_iNRG,
  title = {Interleaved numerical renormalization group as an efficient multiband impurity solver},
  author = {Stadler, K. M. and Mitchell, A. K. and von Delft, J. and Weichselbaum, A.},
  journal = {Phys. Rev. B},
  volume = {93},
  issue = {23},
  pages = {235101},
  numpages = {16},
  year = {2016},
  month = {Jun},
  publisher = {American Physical Society},
  doi = {10.1103/PhysRevB.93.235101},
  url = {https://link.aps.org/doi/10.1103/PhysRevB.93.235101}
}

@article{Mitchell2014_iNRG,
  title = {Generalized {W}ilson chain for solving multichannel quantum impurity problems},
  author = {Mitchell, A. K. and Galpin, M. R. and Wilson-Fletcher, S. and Logan, D. E. and Bulla, R.},
  journal = {Phys. Rev. B},
  volume = {89},
  issue = {12},
  pages = {121105},
  numpages = {5},
  year = {2014},
  month = {Mar},
  publisher = {American Physical Society},
  doi = {10.1103/PhysRevB.89.121105},
  url = {https://link.aps.org/doi/10.1103/PhysRevB.89.121105}
}

@InCollection{Potthoff2016_cDMFT,
  author    = {Potthoff, M.},
  booktitle = {{DMFT}: From Infinite Dimensions to Real Materials Modeling and Simulation, Forschungszentrum J\"ulich},
  title     = {Cluster Extensions of Dynamical Mean-Field Theory},
  editor    = {Eva Pavarini and Erik Koch and Dieter Vollhardt and Alexander Lichtenstein},
  year      = {2018},
  URL = {https://www.cond-mat.de/events/correl18/manuscripts/potthoff.pdf},
  publisher = {Verlag des Forschungszentrum J\"ulich},
}

@article{julien2008_EROS,
  title={Extended Recursion in Operator Space ({EROS}), a new impurity solver for the single impurity {A}nderson model}, 
  author={Jean-Pierre Julien and R. C. Albers},
  year={2008},
  journal={arXiv:0810.3302 [cond-mat.str-el]},
  url={https://arxiv.org/abs/0810.3302}
}

@article{pinna2025_LanczosGreensfunctions,
  title={Approximation theory for {G}reen's functions via the {L}anczos algorithm}, 
  author={Gabriele Pinna and Oliver Lunt and Curt von Keyserlingk},
  year={2025},
  journal={Phys. Rev. B},
  volume={112},
  pages={054435},
  doi={10.1103/lsl4-4lb4},
  url={https://doi.org/10.1103/lsl4-4lb4}
}

@article{Boley1984_Lanczos,
  title = {The {L}anczos-{A}rnoldi algorithm and controllability},
  journal = {Systems \& Control Letters},
  volume = {4},
  number = {6},
  pages = {317-324},
  year = {1984},
  issn = {0167-6911},
  doi = {https://doi.org/10.1016/S0167-6911(84)80072-9},
  url = {https://www.sciencedirect.com/science/article/pii/S0167691184800729},
  author = {D.L. Boley and G.H. Golub}
}

@Book{Dunford1988,
  author    = {Dunford, Nelson and Schwartz, Jacob T.},
  publisher = {Wiley Interscience Publ.},
  title     = {Linear Operators, Part 2: Spectral Theory, Self Adjoint Operators in Hilbert Space},
  year      = {1988},
  address   = {New York},
  isbn      = {0471608475},
  volume    = {2},
  pagetotal = {859192317},
  ppn_gvk   = {1088657877},
  subtitle  = {Nelson},
  url = {https://www.wiley.com/en-us/shop/general-introductory-mathematics/linear-operators-part-2-spectral-theory-self-adjoint-operators-in-hilbert-space-p-9780471608479},
}

@Book{Stojanov2014,
  author    = {Stojanov, Jordan M.},
  publisher = {Dover Publ.},
  title     = {Counterexamples in probability},
  year      = {2014},
  address   = {Mineola, NY},
  edition   = {3. ed.},
  isbn      = {0486499987},
  note      = {Includes bibliographical references (pages 317-364) and index},
  series    = {Dover books on mathematics},
  pagetotal = {368},
  ppn_gvk   = {756955777},
}

@article{Damanik2007,
      title={The Analytic Theory of Matrix Orthogonal Polynomials}, 
      author={David Damanik and Alexander Pushnitski and Barry Simon},
      year={2008},
      journal={Surveys in Approximation Theory},
      volume={4},
      pages={1--85},
      url={https://arxiv.org/abs/0711.2703}
}

@Book{Simon2005,
  author    = {Simon, Barry},
  publisher = {American Mathematical Society},
  title     = {Orthogonal polynomials on the unit circle},
  year      = {2005},
  address   = {Providence, Rhode Island},
  isbn      = {9781470431990},
  note      = {Includes bibliographical references and indexes. Description based on online resource; title from PDF title page (ebrary, viewed April 11, 2017).},
  number    = {Volume 54, Part 1},
  series    = {Colloquium Publications},
  pagetotal = {1496},
  ppn_gvk   = {1749550644},
}

@Book{Achiezer2021,
  author    = {Achiezer, Naum I.},
  editor    = {N. Kemmer},
  publisher = {Society for Industrial and Applied Mathematics},
  title     = {The classical moment problem and some related questions in analysis},
  year      = {2021},
  address   = {Philadelphia, Pennsylvania},
  isbn      = {9781611976397},
  number    = {82},
  series    = {Classics in applied mathematics},
  pagetotal = {1252},
  ppn_gvk   = {1759478040},
}

@Article{LopezRodriguez2001,
  author    = {Lopez-Rodriguez, Pedro},
  journal   = {Mathematica Scandinavica},
  title     = {The {N}evanlinna parametrization for a matrix moment problem},
  year      = {2001},
  issn      = {0025-5521},
  month     = dec,
  number    = {2},
  pages     = {245},
  volume    = {89},
  doi       = {10.7146/math.scand.a-14340},
  publisher = {Det Kgl. Bibliotek/Royal Danish Library},
}

@Book{Szegoe2012,
  author    = {Szeg\"o, Gábor},
  publisher = {American mathematical society},
  title     = {Orthogonal polynomials},
  year      = {2012},
  address   = {New York city},
  edition   = {Online-Ausg.},
  isbn      = {9781470431716},
  note      = {"List of references": p. 378-393. - Electronic reproduction; Providence, Rhode Island; American Mathematical Society; 2012. - Description based on print version record},
  series    = {AMS ebook collection},
  pagetotal = {1401},
  ppn_gvk   = {1654541184},
}

@Book{Chihara2011,
  author    = {Chihara, Theodore S.},
  publisher = {Dover Publications, Inc.},
  title     = {An introduction to orthogonal polynomials},
  year      = {2011},
  address   = {Mineola},
  edition   = {First published},
  isbn      = {0486479293},
  note      = {This Dover edition, first published in 2011, is an unabridged republication of the work orginally published in 17978 by Gordon and Breach, Science Publisher, Inc., New York},
  pagetotal = {249},
  ppn_gvk   = {1620990997},
}

@Book{Ismail2005,
  author    = {Ismail, Mourad E. H.},
  editor    = {Walter Van Assche},
  publisher = {Cambridge University Press},
  title     = {Classical and quantum orthogonal polynomials in one variable},
  year      = {2005},
  address   = {Cambridge ;},
  isbn      = {9781107101333},
  note      = {Includes bibliographical references (p. [661]-696) and indexes.},
  number    = {98},
  series    = {Encyclopedia of mathematics and its applications},
  pagetotal = {706},
  ppn_gvk   = {1748751069},
}

@Article{Duran1995,
  author    = {Duran, Antonio J.},
  journal   = {Canadian Journal of Mathematics},
  title     = {On Orthogonal Polynomials With Respect to a Positive Definite Matrix of Measures},
  year      = {1995},
  issn      = {1496-4279},
  month     = feb,
  number    = {1},
  pages     = {88--112},
  volume    = {47},
  doi       = {10.4153/cjm-1995-005-8},
  publisher = {Canadian Mathematical Society},
}

@Article{Duran2000,
  author    = {Duran, Antonio J. and Lopez-Rodriguez, Pedro},
  journal   = {Journal of Functional Analysis},
  title     = {N-Extremal Matrices of Measures for an Indeterminate Matrix Moment Problem},
  year      = {2000},
  issn      = {0022-1236},
  month     = jul,
  number    = {2},
  pages     = {301--321},
  volume    = {174},
  doi       = {10.1006/jfan.2000.3585},
  publisher = {Elsevier BV},
}

@Article{Chihara1989,
  author    = {Chihara, T. S.},
  journal   = {Transactions of the American Mathematical Society},
  title     = {Hamburger moment problems and orthogonal polynomials},
  year      = {1989},
  issn      = {1088-6850},
  number    = {1},
  pages     = {189--203},
  volume    = {315},
  doi       = {10.1090/s0002-9947-1989-0986686-1},
  publisher = {American Mathematical Society (AMS)},
}

@Article{Simon1998,
  author    = {Simon, Barry},
  journal   = {Advances in Mathematics},
  title     = {The Classical Moment Problem as a Self-Adjoint Finite Difference Operator},
  year      = {1998},
  issn      = {0001-8708},
  month     = jul,
  number    = {1},
  pages     = {82--203},
  volume    = {137},
  doi       = {10.1006/aima.1998.1728},
  publisher = {Elsevier BV},
}

@article{DeVega2015,
  title = {How to discretize a quantum bath for real-time evolution},
  author = {de Vega, In\'es and Schollw\"ock, Ulrich and Wolf, F. Alexander},
  journal = {Phys. Rev. B},
  volume = {92},
  issue = {15},
  pages = {155126},
  numpages = {14},
  year = {2015},
  month = {Oct},
  publisher = {American Physical Society},
  doi = {10.1103/PhysRevB.92.155126},
  url = {https://link.aps.org/doi/10.1103/PhysRevB.92.155126}
}

@Article{Berg1995,
  author    = {Berg, Christian},
  journal   = {Journal of Computational and Applied Mathematics},
  title     = {Indeterminate moment problems and the theory of entire functions},
  year      = {1995},
  issn      = {0377-0427},
  month     = dec,
  number    = {1–3},
  pages     = {27--55},
  volume    = {65},
  doi       = {10.1016/0377-0427(95)00099-2},
  publisher = {Elsevier BV},
}

@article{Haehner2020_DCAinterp,
  title = {Continuous momentum dependence in the dynamical cluster approximation},
  author = {H\"ahner, Urs R. and Maier, Thomas A. and Schulthess, Thomas C.},
  journal = {Phys. Rev. B},
  volume = {101},
  issue = {19},
  pages = {195114},
  numpages = {10},
  year = {2020},
  month = {May},
  publisher = {American Physical Society},
  doi = {10.1103/PhysRevB.101.195114},
  url = {https://link.aps.org/doi/10.1103/PhysRevB.101.195114}
}

@article{Staar2013_DCAinterp,
  title = {Dynamical cluster approximation with continuous lattice self-energy},
  author = {Staar, Peter and Maier, Thomas and Schulthess, Thomas C.},
  journal = {Phys. Rev. B},
  volume = {88},
  issue = {11},
  pages = {115101},
  numpages = {16},
  year = {2013},
  month = {Sep},
  publisher = {American Physical Society},
  doi = {10.1103/PhysRevB.88.115101},
  url = {https://link.aps.org/doi/10.1103/PhysRevB.88.115101}
}

@article{Backes2022_genCavity,
  title = {Nonlocal correlation effects in fermionic many-body systems: Overcoming the noncausality problem},
  author = {Backes, Steffen and Sim, Jae-Hoon and Biermann, Silke},
  journal = {Phys. Rev. B},
  volume = {105},
  issue = {24},
  pages = {245115},
  numpages = {13},
  year = {2022},
  month = {Jun},
  publisher = {American Physical Society},
  doi = {10.1103/PhysRevB.105.245115},
  url = {https://link.aps.org/doi/10.1103/PhysRevB.105.245115}
}

@article{Bolech2003,
  title = {Cellular dynamical mean-field theory for the one-dimensional extended {H}ubbard model},
  author = {Bolech, C. J. and Kancharla, S. S. and Kotliar, G.},
  journal = {Phys. Rev. B},
  volume = {67},
  issue = {7},
  pages = {075110},
  numpages = {9},
  year = {2003},
  month = {Feb},
  publisher = {American Physical Society},
  doi = {10.1103/PhysRevB.67.075110},
  url = {https://link.aps.org/doi/10.1103/PhysRevB.67.075110}
}

@article{Gleis2022a,
  title = {Projector formalism for kept and discarded spaces of matrix product states},
  author = {Gleis, Andreas and Li, Jheng-Wei and von Delft, Jan},
  journal = {Phys. Rev. B},
  volume = {106},
  issue = {19},
  pages = {195138},
  numpages = {14},
  year = {2022},
  month = {Nov},
  publisher = {American Physical Society},
  doi = {10.1103/PhysRevB.106.195138},
  url = {https://link.aps.org/doi/10.1103/PhysRevB.106.195138}
}

@article{Gleis2025,
      title={Reply to comment on "Controlled bond expansion for Density Matrix Renormalization Group ground state search at single-site costs"}, 
      author={Andreas Gleis and Jheng-Wei Li and Jan von Delft},
      year={2025},
      journal={arXiv:2501.12291 [cond-mat.str-el]},
      url={https://arxiv.org/abs/2501.12291}, 
}

@article{Li2024,
  title = {Time-Dependent Variational Principle with Controlled Bond Expansion for Matrix Product States},
  author = {Li, Jheng-Wei and Gleis, Andreas and von Delft, Jan},
  journal = {Phys. Rev. Lett.},
  volume = {133},
  issue = {2},
  pages = {026401},
  numpages = {9},
  year = {2024},
  month = {Jul},
  publisher = {American Physical Society},
  doi = {10.1103/PhysRevLett.133.026401},
  url = {https://link.aps.org/doi/10.1103/PhysRevLett.133.026401}
}

@article{Schollwock2011,
title = {The density-matrix renormalization group in the age of matrix product states},
journal = {Annals of Physics},
volume = {326},
number = {1},
pages = {96-192},
year = {2011},
issn = {0003-4916},
doi = {https://doi.org/10.1016/j.aop.2010.09.012},
url = {https://www.sciencedirect.com/science/article/pii/S0003491610001752},
author = {Ulrich Schollwöck}
}

@article{White1992,
  title = {Density matrix formulation for quantum renormalization groups},
  author = {White, Steven R.},
  journal = {Phys. Rev. Lett.},
  volume = {69},
  issue = {19},
  pages = {2863--2866},
  numpages = {0},
  year = {1992},
  month = {Nov},
  publisher = {American Physical Society},
  doi = {10.1103/PhysRevLett.69.2863},
  url = {https://link.aps.org/doi/10.1103/PhysRevLett.69.2863}
}

@article{Haegeman2011,
  title = {Time-Dependent Variational Principle for Quantum Lattices},
  author = {Haegeman, Jutho and Cirac, J. Ignacio and Osborne, Tobias J. and Pi\ifmmode \check{z}\else \v{z}\fi{}orn, Iztok and Verschelde, Henri and Verstraete, Frank},
  journal = {Phys. Rev. Lett.},
  volume = {107},
  issue = {7},
  pages = {070601},
  numpages = {5},
  year = {2011},
  month = {Aug},
  publisher = {American Physical Society},
  doi = {10.1103/PhysRevLett.107.070601},
  url = {https://link.aps.org/doi/10.1103/PhysRevLett.107.070601}
}

@Article{Pelz2026b,
   author        = {Mathias Pelz and Gabriel Kotliar and Jan von Delft and Andreas Gleis},
   title         = {Quantum criticality in the two-dimensional {H}ubbard model},
   url           = {https://arxiv.org/abs/2605.15059},
   archiveprefix = {arXiv},
   journal       = {arXiv:2605.15059 [cond-mat.str-el]},
   primaryclass  = {cond-mat.str-el},
   year          = {2026},
}

@Article{Pelz2026c,
   author        = {Mathias Pelz and Gabriel Kotliar and Jan von Delft and Andreas Gleis},
   title         = {Dynamical scaling near the pseudogap quantum critical point of the two-dimensional {H}ubbard model},
   url           = {https://arxiv.org/abs/2605.15060},
   archiveprefix = {arXiv},
   journal       = {arXiv:2605.15060 [cond-mat.str-el]},
   primaryclass  = {cond-mat.str-el},
   year          = {2026},
}

@article{Altmann1965_LatticeHarmonics1,
  title = {Lattice Harmonics {I.} {C}ubic Groups},
  author = {Altmann, S. L. and Cracknell, A. P.},
  journal = {Rev. Mod. Phys.},
  volume = {37},
  issue = {1},
  pages = {19--32},
  numpages = {0},
  year = {1965},
  month = {Jan},
  publisher = {American Physical Society},
  doi = {10.1103/RevModPhys.37.19},
  url = {https://link.aps.org/doi/10.1103/RevModPhys.37.19}
}

@article{Altmann1965_LatticeHarmonics2,
  title = {Lattice Harmonics {II}. Hexagonal Close-Packed Lattice},
  author = {Altmann, S. L. and Bradley, C. J.},
  journal = {Rev. Mod. Phys.},
  volume = {37},
  issue = {1},
  pages = {33--45},
  numpages = {0},
  year = {1965},
  month = {Jan},
  publisher = {American Physical Society},
  doi = {10.1103/RevModPhys.37.33},
  url = {https://link.aps.org/doi/10.1103/RevModPhys.37.33}
}

@Article{Wagner2023,
author={Wagner, N.
and Crippa, L.
and Amaricci, A.
and Hansmann, P.
and Klett, M.
and K{\"o}nig, E. J.
and Sch{\"a}fer, T.
and Sante, D. Di
and Cano, J.
and Millis, A. J.
and Georges, A.
and Sangiovanni, G.},
title={{M}ott insulators with boundary zeros},
journal={Nature Communications},
year={2023},
month={Nov},
day={20},
volume={14},
number={1},
pages={7531},
issn={2041-1723},
doi={10.1038/s41467-023-42773-7},
url={https://doi.org/10.1038/s41467-023-42773-7}
}

@article{Pudleiner2016,
  title = {Momentum structure of the self-energy and its parametrization for the two-dimensional {H}ubbard model},
  author = {Pudleiner, P. and Sch\"afer, T. and Rost, D. and Li, G. and Held, K. and Bl\"umer, N.},
  journal = {Phys. Rev. B},
  volume = {93},
  issue = {19},
  pages = {195134},
  numpages = {12},
  year = {2016},
  month = {May},
  publisher = {American Physical Society},
  doi = {10.1103/PhysRevB.93.195134},
  url = {https://link.aps.org/doi/10.1103/PhysRevB.93.195134}
}

@article{Staffieri2025_DCA,
  title = {Signatures of the {F}ermi surface reconstruction of a doped {M}ott insulator in a slab geometry},
  author = {Staffieri, Gregorio and Fabrizio, Michele},
  journal = {Phys. Rev. B},
  volume = {112},
  issue = {15},
  pages = {155155},
  numpages = {12},
  year = {2025},
  month = {Oct},
  publisher = {American Physical Society},
  doi = {10.1103/hyth-626t},
  url = {https://link.aps.org/doi/10.1103/hyth-626t}
}

@article{Engel1991,
  title = {Calculation of the {GW} self-energy in semiconducting crystals},
  author = {Engel, G. E. and Farid, Behnam and Nex, C. M. M. and March, N. H.},
  journal = {Phys. Rev. B},
  volume = {44},
  issue = {24},
  pages = {13356--13373},
  numpages = {0},
  year = {1991},
  month = {Dec},
  publisher = {American Physical Society},
  doi = {10.1103/PhysRevB.44.13356},
  url = {https://link.aps.org/doi/10.1103/PhysRevB.44.13356}
}

@article{Crippa2021,
  title = {Local versus nonlocal correlation effects in interacting quantum spin {H}all insulators},
  author = {Crippa, L. and Amaricci, A. and Adler, S. and Sangiovanni, G. and Capone, M.},
  journal = {Phys. Rev. B},
  volume = {104},
  issue = {23},
  pages = {235117},
  numpages = {10},
  year = {2021},
  month = {Dec},
  publisher = {American Physical Society},
  doi = {10.1103/PhysRevB.104.235117},
  url = {https://link.aps.org/doi/10.1103/PhysRevB.104.235117}
}

@article{FlorezAblan2025,
  title = {Bath parameterization in multiband cluster dynamical mean-field theory},
  author = {Florez-Ablan, Diego and Mejuto-Zaera, Carlos and Capone, Massimo},
  journal = {Phys. Rev. B},
  volume = {112},
  issue = {8},
  pages = {085153},
  numpages = {16},
  year = {2025},
  month = {Aug},
  publisher = {American Physical Society},
  doi = {10.1103/lnhv-kvy4},
  url = {https://link.aps.org/doi/10.1103/lnhv-kvy4}
}

@article{Capone2004,
  title = {Cluster-dynamical mean-field theory of the density-driven {M}ott transition in the one-dimensional {H}ubbard model},
  author = {Capone, M. and Civelli, M. and Kancharla, S. S. and Castellani, C. and Kotliar, G.},
  journal = {Phys. Rev. B},
  volume = {69},
  issue = {19},
  pages = {195105},
  numpages = {5},
  year = {2004},
  month = {May},
  publisher = {American Physical Society},
  doi = {10.1103/PhysRevB.69.195105},
  url = {https://link.aps.org/doi/10.1103/PhysRevB.69.195105}
}

@article{LeeTH2019,
  title = {Rotationally invariant slave-boson and density matrix embedding theory: {U}nified framework and comparative study on the one-dimensional and two-dimensional {H}ubbard model},
  author = {Lee, Tsung-Han and Ayral, Thomas and Yao, Yong-Xin and Lanata, Nicola and Kotliar, Gabriel},
  journal = {Phys. Rev. B},
  volume = {99},
  issue = {11},
  pages = {115129},
  numpages = {9},
  year = {2019},
  month = {Mar},
  publisher = {American Physical Society},
  doi = {10.1103/PhysRevB.99.115129},
  url = {https://link.aps.org/doi/10.1103/PhysRevB.99.115129}
}

@article{Lechermann2007,
  title = {Rotationally invariant slave-boson formalism and momentum dependence of the quasiparticle weight},
  author = {Lechermann, Frank and Georges, Antoine and Kotliar, Gabriel and Parcollet, Olivier},
  journal = {Phys. Rev. B},
  volume = {76},
  issue = {15},
  pages = {155102},
  numpages = {20},
  year = {2007},
  month = {Oct},
  publisher = {American Physical Society},
  doi = {10.1103/PhysRevB.76.155102},
  url = {https://link.aps.org/doi/10.1103/PhysRevB.76.155102}
}

@article{Giuli2026,
      title={Unifying Variational and Dynamical Quantum Embedding: {F}rom Ghost {G}utzwiller Approximation to Dynamical Mean-Field Theory}, 
      author={Samuele Giuli and Tsung-Han Lee and Yong-Xin Yao and Gabriel Kotliar and Andrei E. Ruckenstein and Olivier Gingras and Nicola Lanat\`a},
      year={2026},
      journal={arXiv:2603.20559 [cond-mat.str-el]},
      url={https://arxiv.org/abs/2603.20559}, 
      journal={}
}

@article{Civelli2005,
  title = {Dynamical Breakup of the {F}ermi Surface in a Doped {M}ott Insulator},
  author = {Civelli, M. and Capone, M. and Kancharla, S. S. and Parcollet, O. and Kotliar, G.},
  journal = {Phys. Rev. Lett.},
  volume = {95},
  issue = {10},
  pages = {106402},
  numpages = {4},
  year = {2005},
  month = {Sep},
  publisher = {American Physical Society},
  doi = {10.1103/PhysRevLett.95.106402},
  url = {https://link.aps.org/doi/10.1103/PhysRevLett.95.106402}
}

@article{Wu2017,
  title = {Controlling {F}eynman diagrammatic expansions: {P}hysical nature of the pseudogap in the two-dimensional {H}ubbard model},
  author = {Wu, Wei and Ferrero, Michel and Georges, Antoine and Kozik, Evgeny},
  journal = {Phys. Rev. B},
  volume = {96},
  issue = {4},
  pages = {041105(R)},
  numpages = {5},
  year = {2017},
  month = {Jul},
  publisher = {American Physical Society},
  doi = {10.1103/PhysRevB.96.041105},
  url = {https://link.aps.org/doi/10.1103/PhysRevB.96.041105}
}

@article{Capone2006,
  title = {Competition between $d$-wave superconductivity and antiferromagnetism in the two-dimensional {H}ubbard model},
  author = {Capone, M. and Kotliar, G.},
  journal = {Phys. Rev. B},
  volume = {74},
  issue = {5},
  pages = {054513},
  numpages = {5},
  year = {2006},
  month = {Aug},
  publisher = {American Physical Society},
  doi = {10.1103/PhysRevB.74.054513},
  url = {https://link.aps.org/doi/10.1103/PhysRevB.74.054513}
}

@article{Kancharla2008,
  title = {Anomalous superconductivity and its competition with antiferromagnetism in doped {M}ott insulators},
  author = {Kancharla, S. S. and Kyung, B. and S\'en\'echal, D. and Civelli, M. and Capone, M. and Kotliar, G. and Tremblay, A.-M. S.},
  journal = {Phys. Rev. B},
  volume = {77},
  issue = {18},
  pages = {184516},
  numpages = {12},
  year = {2008},
  month = {May},
  publisher = {American Physical Society},
  doi = {10.1103/PhysRevB.77.184516},
  url = {https://link.aps.org/doi/10.1103/PhysRevB.77.184516}
}

@article{Sen2020,
  title = {The {M}ott transition as a topological phase transition},
  author = {Sen, Sudeshna and Wong, Patrick J. and Mitchell, Andrew K.},
  journal = {Phys. Rev. B},
  volume = {102},
  issue = {8},
  pages = {081110(R)},
  numpages = {6},
  year = {2020},
  month = {Aug},
  publisher = {American Physical Society},
  doi = {10.1103/PhysRevB.102.081110},
  url = {https://link.aps.org/doi/10.1103/PhysRevB.102.081110}
}

@article{Bellomia2024,
  title = {Quasilocal entanglement across the {M}ott-{H}ubbard transition},
  author = {Bellomia, Gabriele and Mejuto-Zaera, Carlos and Capone, Massimo and Amaricci, Adriano},
  journal = {Phys. Rev. B},
  volume = {109},
  issue = {11},
  pages = {115104},
  numpages = {13},
  year = {2024},
  month = {Mar},
  publisher = {American Physical Society},
  doi = {10.1103/PhysRevB.109.115104},
  url = {https://link.aps.org/doi/10.1103/PhysRevB.109.115104}
}

@article{Bacq-Labreuil2025,
  title = {Toward an \textit{Ab Initio} Theory of High-Temperature Superconductors: A Study of Multilayer Cuprates},
  author = {Bacq-Labreuil, Benjamin and Lacasse, Benjamin and Tremblay, A.-M. S. and S\'en\'echal, David and Haule, Kristjan},
  journal = {Phys. Rev. X},
  volume = {15},
  issue = {2},
  pages = {021071},
  numpages = {44},
  year = {2025},
  month = {May},
  publisher = {American Physical Society},
  doi = {10.1103/PhysRevX.15.021071},
  url = {https://link.aps.org/doi/10.1103/PhysRevX.15.021071}
}

@article{Smit2025,
      title={Enhanced coherence and layer-selective charge order in a trilayer cuprate superconductor}, 
      author={S. Smit and M. Bluschke and P. Moen and N. Heinsdorf and E. Zavatti and G. Bellomia and S. Giuli and S. K. Y. Dufresne and C. T. Suen and V. Zimmermann and C. Au-Yeung and S. Zhdanovich and J. I. Dadap and M. Zonno and S. Gorovikov and H. Lee and C-T. Kuo and J-S. Lee and D. Song and S. Ishida and H. Eisaki and B. Keimer and M. Michiardi and I. S. Elfimov and G. Levy and D. J. Jones and M. Capone and A. Damascelli},
      year={2025},
      journal={arXiv:2506.01448 [cond-mat.str-el]},
      url={https://arxiv.org/abs/2506.01448}, 
      journal={}
}

@Article{Kanigel2006,
author={Kanigel, A.
and Norman, M. R.
and Randeria, M.
and Chatterjee, U.
and Souma, S.
and Kaminski, A.
and Fretwell, H. M.
and Rosenkranz, S.
and Shi, M.
and Sato, T.
and Takahashi, T.
and Li, Z. Z.
and Raffy, H.
and Kadowaki, K.
and Hinks, D.
and Ozyuzer, L.
and Campuzano, J. C.},
title={Evolution of the pseudogap from {F}ermi arcs to the nodal liquid},
journal={Nature Physics},
year={2006},
month={Jul},
day={01},
volume={2},
number={7},
pages={447-451},
issn={1745-2481},
doi={10.1038/nphys334},
url={https://doi.org/10.1038/nphys334}
}

@article{Uhrig1996,
  title = {Exact Single Spin Flip for the Hubbard Model in $\mathit{d}\phantom{\rule{0ex}{0ex}}=\phantom{\rule{0ex}{0ex}}\ensuremath{\infty}$},
  author = {Uhrig, G\"otz S.},
  journal = {Phys. Rev. Lett.},
  volume = {77},
  issue = {17},
  pages = {3629--3632},
  numpages = {0},
  year = {1996},
  month = {Oct},
  publisher = {American Physical Society},
  doi = {10.1103/PhysRevLett.77.3629},
  url = {https://link.aps.org/doi/10.1103/PhysRevLett.77.3629}
}

\end{document}